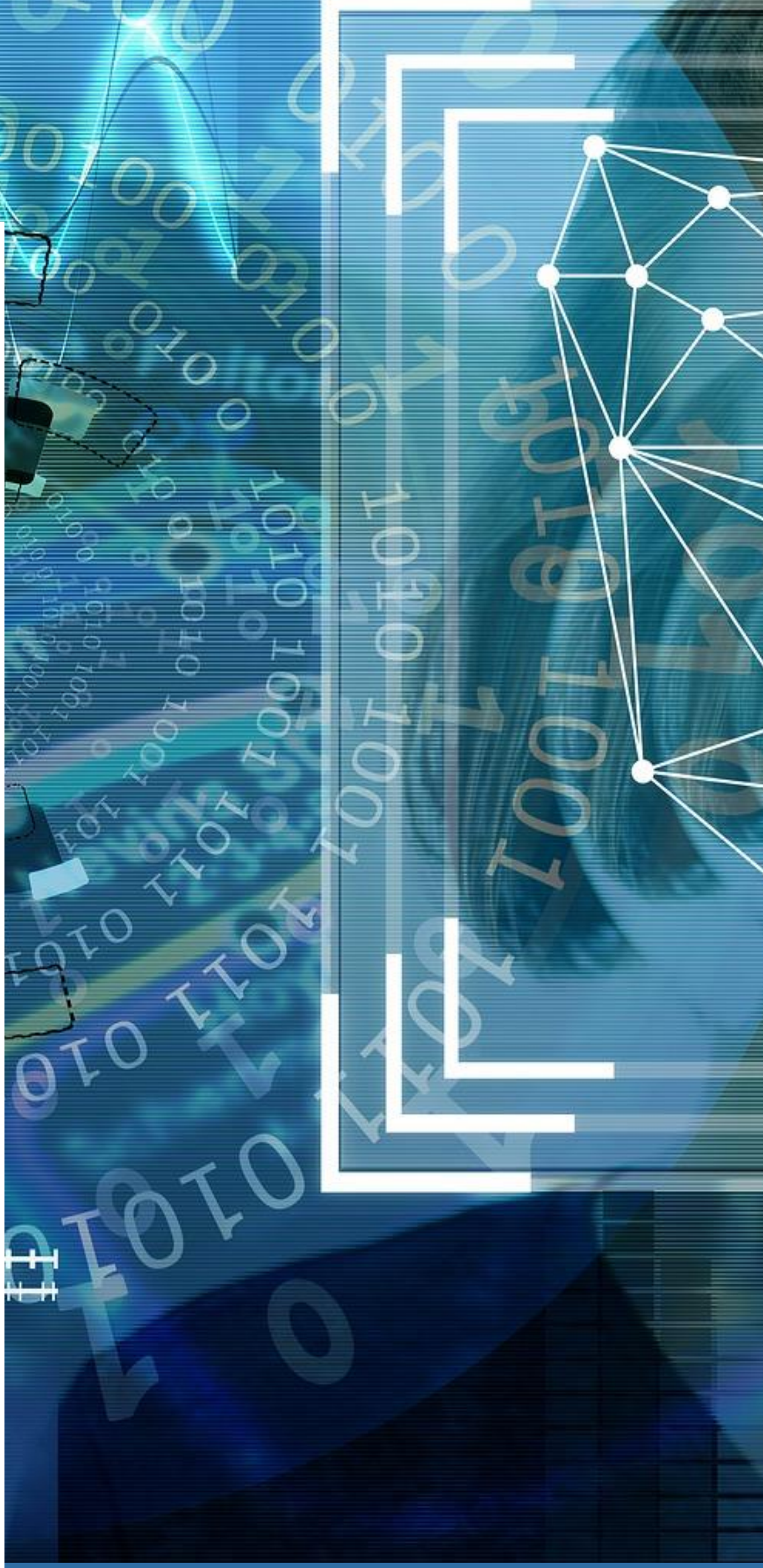

# State of the Art of Audio- and Video-Based Solutions for AAL

**Working Group 3.**

Audio- and Video-based AAL Applications

March 2022

COST EUROPEAN COOPERATION IN SCIENCE & TECHNOLOGY

goodbrother

This publication is based upon work from COST Action GoodBrother – Network on Privacy-Aware Audio- and Video-Based Applications for Active and Assisted Living, supported by COST (European Cooperation in Science and Technology).

COST (European Cooperation in Science and Technology) is a funding agency for research and innovation networks. Our Actions help connect research initiatives across Europe and enable scientists to grow their ideas by sharing them with their peers. This boosts their research, career and innovation.

www.cost.eu

Funded by the European Union

Other documents published by the Goodbrother COST Action are available at https://goodbrother.eu/deliverables.

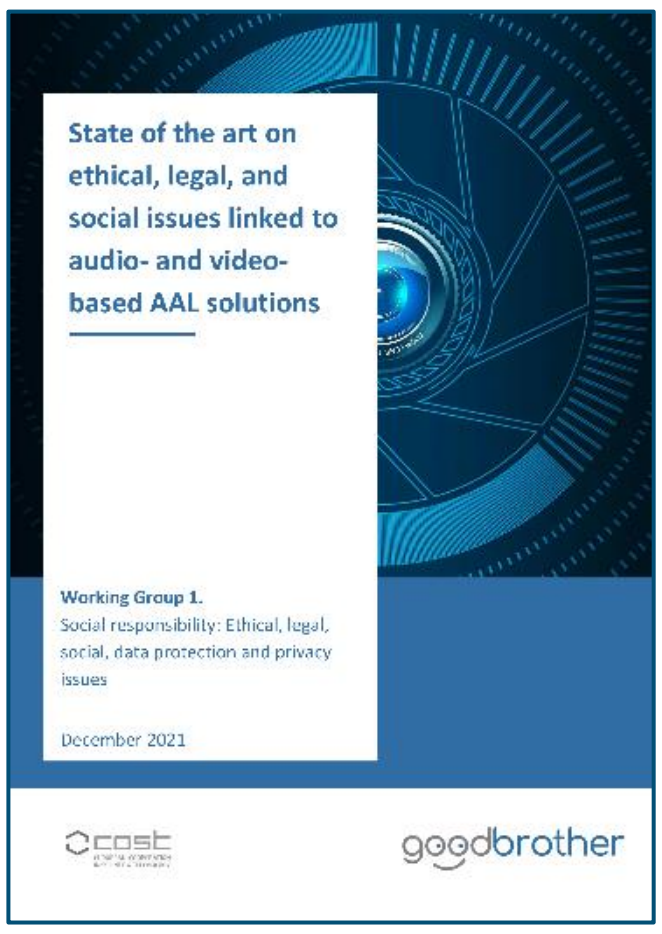

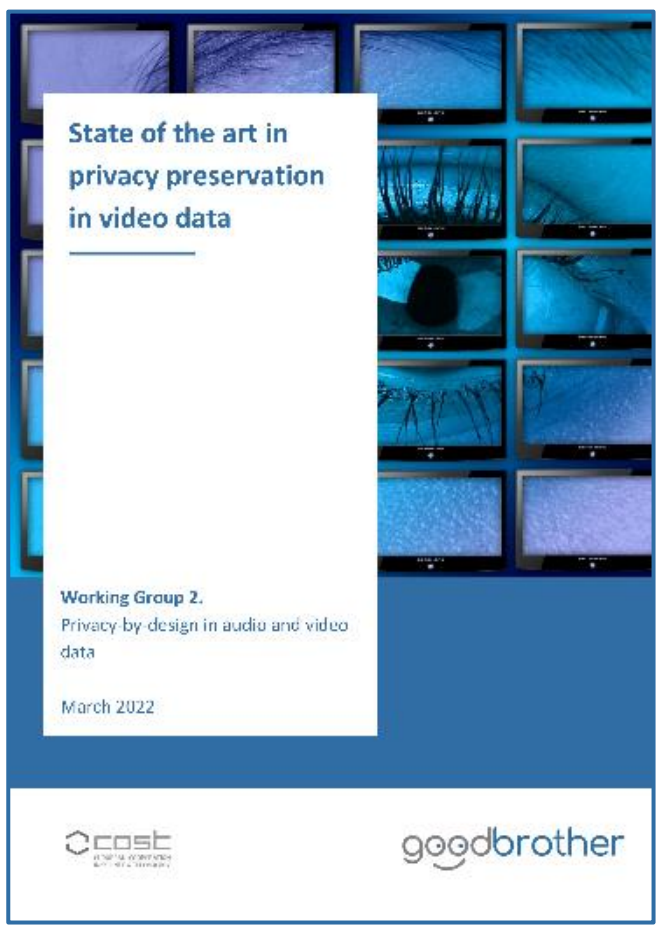

# Details about this white paper

## Executive summary

It is a matter of fact that Europe is facing more and more crucial challenges regarding health and social care due to the demographic change and the current economic context. The recent COVID-19 pandemic has stressed this situation even further, thus highlighting the need for taking action. Active and Assisted Living (AAL) technologies come as a viable approach to help facing these challenges, thanks to the high potential they have in enabling remote care and support. Broadly speaking, AAL can be referred to as the use of innovative and advanced Information and Communication Technologies to create supportive, inclusive and empowering applications and environments that enable older, impaired or frail people to live independently and stay active longer in society. AAL capitalizes on the growing pervasiveness and effectiveness of sensing and computing facilities to supply the persons in need with smart assistance, by responding to their necessities of autonomy, independence, comfort, security and safety. The application scenarios addressed by AAL are complex, due to the inherent heterogeneity of the end-user population, their living arrangements, and their physical conditions or impairment. Despite aiming at diverse goals, AAL systems should share some common characteristics. They are designed to provide support in daily life in an invisible, unobtrusive and user-friendly manner. Moreover, they are conceived to be intelligent, to be able to learn and adapt to the requirements and requests of the assisted people, and to synchronise with their specific needs. Nevertheless, to ensure the uptake of AAL in society, potential users must be willing to use AAL applications and to integrate them in their daily environments and lives. In this respect, video- and audio-based AAL applications have several advantages, in terms of unobtrusiveness and information richness. Indeed, cameras and microphones are far less obtrusive with respect to the hindrance other wearable sensors may cause to one's activities. In addition, a single camera placed in a room can record most of the activities performed in the room, thus replacing many other non-visual sensors. Currently, video-based applications are effective in recognising and monitoring the activities, the movements, and the overall conditions of the assisted individuals as well as to assess their vital parameters (e.g., heart rate, respiratory rate). Similarly, audio sensors have the potential to become one of the most important modalities for interaction with AAL systems, as they can have a large range of sensing, do not require physical presence at a particular location and are physically intangible. Moreover, relevant information about individuals' activities and health status can derive from processing audio signals (e.g., speech recordings). Nevertheless, as the other side of the coin, cameras and microphones are often perceived as the most intrusive technologies from the viewpoint of the privacy of the monitored individuals. This is due to the richness of the information these technologies convey and the intimate setting where they may be deployed. Solutions able to ensure privacy preservation by context and by design, as well as to ensure high legal and ethical standards are in high demand. After the review of the current state of play and the discussion in GoodBrother, we may claim that the first solutions in this direction are starting to appear in the literature. A multidisciplinary debate among experts and stakeholders is paving the way towards AAL ensuring ergonomics, usability, acceptance and privacy preservation. The DIANA, PAAL, and VisuAAL projects are examples of this fresh approach.

This report provides the reader with a review of the most recent advances in audio- and video-based monitoring technologies for AAL. It has been drafted as a collective effort of WG3 to supply an introduction to AAL, its evolution over time and its main functional and technological underpinnings. In this respect, the report contributes to the field with the outline of a new generation of *ethical-aware AAL technologies* and a proposal for a novel comprehensive taxonomy of AAL systems and applications. Moreover, the report allows non-technical readers to gather an overview of the main components of an AAL system and how these function and interact with the end-users.

The report illustrates the state of the art of the most successful AAL applications and functions based on audio and video data, namely (i) lifelogging and self-monitoring, (ii) remote monitoring of vital signs, (iii) emotional state recognition, (iv) food intake monitoring, activity and behaviour recognition, (v) activity and personal assistance, (vi) gesture recognition, (vii) fall detection and prevention, (viii) mobility assessment and frailty recognition, and (ix) cognitive and motor rehabilitation. For these application scenarios, the report illustrates the state of play in terms of scientific advances, available products and research project. The open challenges are also highlighted.

The report ends with an overview of the challenges, the hindrances and the opportunities posed by the uptake in real world settings of AAL technologies. In this respect, the report illustrates the current procedural and technological approaches to cope with acceptability, usability and trust in the AAL technology, by surveying strategies and approaches to co-design, to privacy preservation in video and audio data, to transparency and explainability in data processing, and to data transmission and communication. User acceptance and ethical considerations are also debated. Finally, the potentials coming from the silver economy are overviewed.

## Disclaimer and acknowledgements

This report was created within the Working Group 3 on Audio- and Video-based AAL Applications, in cooperation with the Working Group 1 and Working Group 2 of the COST Action 19121 Good Brother funded by the COST Action.

This report's contents are the authors' sole responsibility and do not necessarily reflect the views of the European Commission. The opinions expressed here are those of the authors only and do not necessarily reflect the authors' organisations.

The authors would like to thank the COST Action for the provided support.

## Author details

The CA 19121 WG 3 Review document and the White Paper have been prepared in collaboration with members of the CA 19121 WG3, WG1 and WG2. They have diverse backgrounds, ranging from computer science, engineering, law, ethics, philosophy and psychology.

The following authors (here listed alphabetically) contributed to the White Paper on the State of the Art of Audio- and Video-based technologies for AAL:

- Slavisa Aleksic, Leipzig University of Applied Sciences, Germany
- Michael Atanasov, TU Wien, Austria
- Jean Calleja Agius, Department of Anatomy, University of Malta, Malta
- Kenneth Camilleri, Centre for Biomedical Cybernetics, Faculty of Engineering, Department of Systems and Control Engineering, Malta
- Anto Čartolovni, Digital healthcare ethics laboratory (Digit-HeaL), Catholic University of Croatia, Croatia
- Pau Climent-Pérez, Universidad de Alicante, Spain
- Sara Colantonio, Institute of Information Science and Technologies of the National Research Council of Italy, Italy
- Stefania Cristina, Department of Systems and Control Engineering, University of Malta, Malta
- Vladimir Despotovic, University of Luxembourg, Luxembourg
- Hazım Kemal Ekenel, Dept. of Computer Engineering, Istanbul Technical University, Turkey
- Ekrem Erakin, Dept. of Computer Engineering, Istanbul Technical University, Turkey
- Francisco Florez-Revuelta, Universidad de Alicante, Spain

- Danila Germanese, Institute of Information Science and Technologies of the National Research Council of Italy, Italy
- Nicole Grech, Department of Anatomy, University of Malta, Malta
- Steinunn Gróa Sigurðardóttir, Reykjavik University, School of Computer Science, Iceland
- Murat Emirzeoğlu, Sağlık Bilimleri Fakültesi, Fizyoterapi ve Rehabilitasyon, Karadeniz Technical University Research Information System, Turkey
- Ivo Iliev, TU Sofia, Bulgaria
- Mladjan Jovanovic, Singidunum University, Belgrade, Serbia
- Martin Kampel, TU Wien, Austria
- William Kearns, University of South Florida, United States
- Andrzej Klimczuk, SGH Warsaw School of Economics, Poland
- Lambros Lambrinos, Department of Communication and Internet Studies, Cyprus University of Technologies, Cyprus
- Jennifer Lumetzberger, TU Wien, Austria
- Wiktor Mucha, Technical University of Vienna, Austria
- Sophie Noiret, TU Wien, Austria
- Zada Pajalic, VID Specialized University, Oslo, Norway
- Rodrigo Rodriguez Pérez, Getafe University Hospital, Spain
- Galidiya Petrova, TU Sofia, Bulgaria
- Sintija Petrovica, Riga Technical University, Latvia
- Peter Pocta, University of Zilina, Slovakia
- Angelica Poli, Università Politecnica delle Marche, Italy
- Mara Pudane, Riga Technical University, Latvia
- Susanna Spinsante, Università Politecnica delle Marche, Italy
- Albert Ali Salah, Utrecht University, Netherlands
- Maria Jose Santofimia, Universidad de Castilla-La Mancha, Spain
- Anna Sigríður Islind, Reykjavik University, Iceland
- Lacramioara Stoicu-Tivadar, University Politehnica Timisoara, Romania
- Hilda Tellioğlu, TU Wien, Austria
- Andrej Zgank, University of Maribor, Slovenia

# List of abbreviations

| | |
|---|---|
| 2D | Two dimensional |
| 3D | Three dimensional |
| 5tSTS | 5 times Sit-to-stand |
| AAL | Active and Assisted Living |
| AALIANCE2 | European Next Generation Ambient Assisted Living Innovation Alliance |
| ADL | Activities of the Daily Living |
| AF | Atrial Fibrillation |
| AI | Artificial Intelligence |
| AmI | Ambient Intelligence |
| ANN | Artificial Neural Networks |
| ANOVA | Analysis of variance |
| AR | Autoregressive |
| ASD | Autism Spectrum Disorders |
| ASL | American Sign Language |
| BADLs | Basic Activities of Daily Living |
| BER | bit error rate |
| BoW | Bag-of-words |
| BP | Blood Pressure |
| C3D | Convolutional 3D |
| CAM | Class Activation Maps |
| CCD | Charge-coupled device |
| CHMM | Couple Hidden Markov Model |
| CNNs | Convolutional Neural Networks |
| COPD | Chronic obstructive pulmonary disease |
| CSI | Channel State Information |
| CSI | Channel State Information |
| CSL | Chinese Sign Language |
| CV | Computer Vision |
| CVNN | Complex-Valued Neural Network |
| DL | Deep Learning |
| DMM | Depth Motion Maps |
| DR | data rate |
| DTs | Decision Trees |
| DV | data volume |
| E2E-D | end-to-end delay |
| EC | European Commission |
| ECG | Electrocardiography |
| EDA | Electrodermal Activity |
| EEG | Electroencephalography |
| EMG | Electromyography |
| eSIM | embedded SIM |
| FACS | Facial Action Coding Scheme |
| FAU | Facial Action Units |
| fMRI | functional Magnetic Resonance Imaging |
| FPV | First-person video |
| FV | Fisher Vector |
| G20 | Group of Twenty |
| GDPR | General Data Protection Regulation |
| GMM | Gaussian Mixture Model |
| GSR | Galvanic Skin Response |
| HAR | Human activity recognition |
| HB | Haemoglobin |
| $HBO_2$ | Oxygenated haemoglobin |
| HCI | Human-Computer Interaction |
| HHI | Human-Human Interaction |
| HMA | hang motion analysis |
| HMM | Hidden Markov Model |
| HOG | Histogram of Oriented Gradient |
| HON4D | Histogram of Oriented 4D Surface normal |
| HOOF | Histogram of Oriented Optical Flow |
| HR | heart rate |
| HRV | Heart Rate Variability |
| HS | Heel Strike |
| IADLs | Instrumental Activities of Daily Living |
| ICA | Independent Component Analysis |
| ICE | Individual Conditional Expectation |
| ICT | Information and Communication Technologies |
| ICU | Intensive Care Unit |
| IMU | inertial measurement units |
| IoHT | Internet of Healthy Things |
| IoT | Internet of Things |
| iPPG | Image photoplethysmography |
| iPTT | Image-based Pulse Transit Time |
| IR | Infrared |
| IUV | Index (I), and the (u, v) coordinates within the body part surface |
| KLT | Kanade-Lucas-Tomasi |
| k-NN | $K^{th}$-Nearest Neighbour |
| LBP | Local Binary Pattern |
| LDA | Linear Discriminant Analysis |
| LED | light emitting diode |
| LIME | Local Interpretable Model-agnostic Explanation |
| LOSOCV | Leave-One-Subject-Out-Cross-Validation |
| LPC | Linear Predictive Coding |
| LRCN | Long-term Recurrent Convolutional Network |
| LSTM | Long-Short Term Memory |
| LTC | Long-term Temporal Convolutions |
| MAE | Mean Absolute Error |
| MAR | mean absolute error |
| MCI | Mild Cognitive Impairment |
| MDR | Minimum Bounding Rectangle |
| MET | Metabolic Equivalent Task |
| MEWS | Modified Early Warning Score |
| MFCC | Mel Frequency Cepstral Coefficients |
| mHealth | Mobile Health |
| MLN | Markov Logic Network |
| ML | Machine Learning |
| MoPAct | Mobilising the Potential of Active Ageing in Europe |
| MoSIFT | Motion Scale-Invariant Feature Transform |
| MPDMM | Multi-directional Projected Depth Motion Maps |
| mpFR | Mobile Phone Food Record |
| MRI | Magnetic Resonance Imaging |
| ms | milliseconds |
| NB | Naïve-Bayes |
| NGO | non-governmental organisation |
| OECD | Organisation for Economic Co-operation and Development |
| OSA | Obstructive Sleep Apnoea |
| PCA | Principal Component Analysis |
| PD | Parkinson’s disease |
| PDP | Partial Dependence Plots |
| PET | Positron Emission Tomography |
| PIR | Passive Infrared |
| Pixel | Picture element |
| PL | packet loss |
| PPG | Photoplethysmography |
| PTT | Pulse Transit Time |
| R&D | Research and Development |
| RF | Random Forest |
| RGB | Red, Green, Blue |
| RGBD or RGB+D | Red, Green, Blue, + Depth |
| ROI | Region of Interest |
| rPPG | Remote photoplethysmography |
| RR | Respiratory Rate |
| RVNN | Real-Valued Neural Network |
| $SaO_2$ | arterial oxygen saturation |
| SC | Skin Conductance |
| ScReK | Scenario Recognition based on knowledge |
| sEMG | surface electromyograph |
| SHAP | Shapley Additive Explanation |
| SIFT | Scale-Invariant Feature Transform |
| SKT | Skin Temperature |
| SLAM | Simultaneous Localisation and Mapping |
| SLM | Spatial Light Modulator |
| SLM | Spatial Light Modulator |
| SLR | Sign Language Recognition |
| SMOTE | synthetic minority over-sampling technique |
| SMPL | Skinned Multi-Person Linear Mode |
| SNN | Switching Neural Network |
| SNR | Signal-to-Noise Ratios |
| SPCA | Supervised Principal Component Analysis |
| $SpO_2$ | peripheral arterial oxygen saturation |
| STIP | Space–Time Interest Points |
| STS | Sit-To-Stand |
| STW | Sit-To-Walk |
| SURF | Speeded Up Robust Features |
| SVMs | Support Vector Machines |
| TDD | Trajectory-pooled Deep-convolutional Descriptor |
| TO | Toe Off |
| TSN | Temporal Segment Network |
| TUG | Timed Up and Go |
| TUGT | Timed Up and Go test |
| TV | Tidal Volume |
| UBM | Universal background model |
| UNECE | United Nations Economic Commission for Europe |
| UWB | ultra-wideband |
| VLC | visible light communication |
| WEF | World Economic Forum |
| WHO | World Health Organization |
| WoS | Web of Science |
| WPAN | wireless personal area network |
| XGBoost | eXtreme Gradient Boosting |

# List of Figures

# List of Tables

# Table of contents

# 1. Cost Action 19121 and this White paper

## 1.1 Cost Action 19121: GoodBrother

The European Cooperation in Science and Technology (COST) is a funding organisation for the creation of research networks, called COST Actions (CA). These networks offer an open space for collaboration among scientists across Europe (and beyond) and thereby give impetus to research advancements and innovation. Many institutions around Europe participate actively in the CA19121 - Network on Privacy-Aware Audio- and Video-Based Applications for Active and Assisted Living, also called GoodBrother.

GoodBrother's main focus is opening a research forum and raising awareness on the ethical, social, technological and scientific impact of technologies used to support the Active and Assisted Living (AAL) paradigm [I.1].

It is a matter of fact that Europe is facing more and more crucial challenges regarding health and social care due to the demographic change and the current economic context. The recent COVID-19 pandemic has stressed this situation even further, thus highlighting the need for taking action. AAL solutions come as a viable approach to help facing these challenges. AAL aims at improving the health, quality of life, and wellbeing of older, impaired, and frail people, by using different sensors to monitor the environment and its dwellers. Among the various technologies underpinning AAL, cameras and microphones are among the most frequently used. Indeed, they enable monitoring an environment thus gathering actionable information on it in the most straightforward and natural way. They permit detecting and describing events, persons, objects, actions, and interactions. Recent technological advances (e.g., in Artificial Intelligence) have given these devices the ability to 'see' and 'hear' what happens. Nevertheless, the increasing power and capabilities of these technologies is sometimes seen as intrusive by some end-users (e.g., the assisted persons, professional and informal caregivers.)

The General Data Protection Regulation (GDPR) establishes the obligation for technologies to meet data protection principles by design and data protection by default. Therefore, AAL solutions must consider privacy-by-design methodologies in order to protect the fundamental rights of those being monitored.

GoodBrother aims to increase the awareness of the ethical, legal, and privacy issues associated with audio- and video-based monitoring and to propose privacy-aware working solutions for assisted living by creating an interdisciplinary community of researchers and industrial partners from different fields (computing, engineering, healthcare, law, sociology) and other stakeholders (users, policymakers, public services), stimulating new research and innovation. GoodBrother aims to offset the "Big Brother" sense of continuous monitoring by increasing user acceptance, exploiting these new solutions, and improving market reach.

## 1.2 Working Group 3 on Audio- and video-based AAL applications

Working Group 3 brings together researchers and industry people working in different aspects of audio- and video-based AAL applications. The WG offers a forum for the discussion and comparison of diverse technological approaches, their integration with crosscutting disciplines, such as robotics, Artificial Intelligence and Big Data, as well as the contamination with health economics and business modelling for the silver economy.

Taking advantage of intergroup interactions, WG3 will leverage the guidelines emanated from WG1 and the methods for privacy preservation proposed in WG2 in order to adapt existing AAL systems and develop new AAL solutions compliant with users' requirements and the legal regulation.

The main objectives of WG3 can be summarised as follows:

- Review the state-of-the-art of audio- and video-based monitoring technologies and potential applications for AAL
- Study of the economic opportunities of these applications in the Silver Economy
- Exchange knowledge on different computer vision, audio processing, and artificial intelligence techniques used by the participants in order to find commonalities, which could lead to the development of more robust, accurate and reliable systems
- Integrate knowledge from WG1 and WG2 in the design and development of AAL applications
- Foster the collaboration among participants in R&I projects.

Although the clear technological focus of this WG, the team of researchers and scientists involved is highly multidisciplinary. The group consists of more than 60 members, whose background and expertise have been surveyed at the beginning of the activities. Figure 1 shows the distribution of the background among the members as well as the various fields of expertise of the ICT members. These diverse background and expertise have been instrumental to define the strategy to address the first-year activities and to plan the structure of this deliverable.

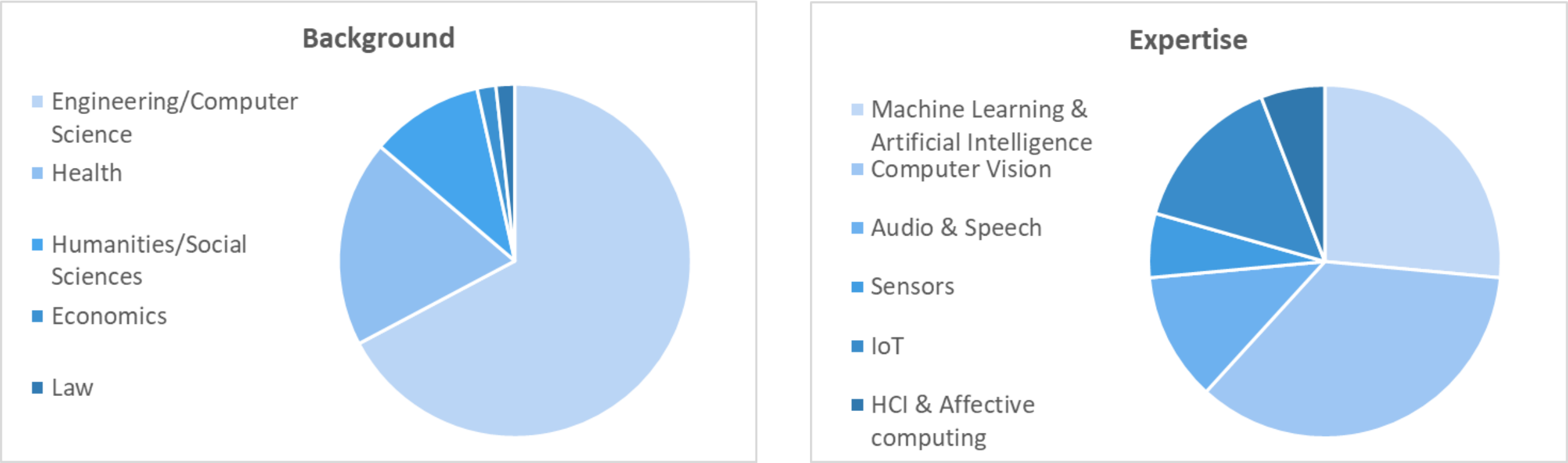

Figure 1. Background and ICT expertise of the WG3 members as resulting from the initial survey

The activities of the first year comprised periodical online meetings (i.e., one each month), which enabled initial acquaintance among WG3 members as well as the discussion and agreement on the most effective and suitable strategy to deliver the WG's first year outcome. Such an outcome comprises, besides the networking and knowledge exchange activities, the present review document and the White Paper obtained by summarising it.

In the following section, we provide the reader with an overview of how the drafting work has been organized, containing its structure and its main content.

## 1.3 Scope of this document

This document aims to review the current state of the art of audio- and video-based monitoring technologies for AAL applications. The review activity was organized by taking advantage of the expertise and the previous experiences of the various members of the Working Group. The work started by collecting inputs and entries about the relevant AAL applications to be considered and by discussing the possible strategies to structure the document and organise the drafting activities.

AAL is a large and growing domain of scientific and technological solutions, which is continuously fed by new results presented in the literature. A systematic review of the whole domain is a challenging task, which has not been addressed yet so comprehensively, as it requires a huge effort in terms of time and resources (see Section 1.4 on existing previous works). Stemming from these considerations, the strategy agreed by the WG3 members was based on a "*divide et impera*" approach, which consisted in firstly identifying the structure and the main sections of the document, and then in assigning the drafting task of each section to a group of members having the needed expertise and previous experience.

## 1.4 Previous works

A growing body of literature has examined the state of the art of AAL by surveying the work done from different viewpoints. Initial surveys belong to the very beginning of AAL and they mainly introduce the driving needs and the potentials of AAL [I.2]-[I.4]. More recent works focus on specific AAL technological or methodological underpinnings, such as specific environments, specific sensor technologies or specific AAL scopes/functions. In [I.5] and [I.6], the authors concentrate on technologies for ambient intelligence and smart homes. As for the sensing technologies, the work in [I.7] presents a review of mobile technologies (i.e., smart phones, watches, or wristbands) for promoting active and healthy lifestyles. In [I.8], the authors survey the environmental and ambient technologies for AAL, for older-adults assistance and care. IoT sensors are reviewed in [I.9] and [I.10]. Video-based sensing technologies for lifelogging purposes are surveyed and discussed in [I.11]. The requirements and implementation challenges for AAL systems are analysed in [I.12], along with the attempt to systematise system architectures.

Some more comprehensive surveys include the one presented in [I.13], which identifies the three generation of AAL systems; the very recent one in [I.14], which discusses the main issues related to the development of a complete AAL system. Finally, several books have been published to cover the entire AAL domain, by collating research and development advances on the main technological and methodological issues [I.15], [I.16].

## 1.5 Structure and main contribution of the document

This report has been structured to provide an overview of the main technological and methodological advances in video- and audio-based AAL.

Initially, an introduction to AAL, its evolution over time and its main functional underpinnings are overviewed in Section 2. The main contributions of the Section comprise the call for a new generation of ethical-aware AAL technologies and a proposal for a novel comprehensive taxonomy of AAL systems and applications.

Afterwards, in Section 3 the main technological underpinnings of video- and audio-based AAL are presented. In this Section, the video and audio sensing technologies are overviewed, by describing the types sensors most commonly used and their functionalities. A brief survey of the methods for analysing and understanding the various types of data and their multimodal fusion is also provided. The main contribution of this Section consists in evidencing the various functionalities that various sensors can offer and highlighting the methodological challenges still open for data processing and understanding.

The longest part of the document is dedicated to presenting the most recent advances in the state of the art of the most successful AAL applications that leverage video and audio data. This is done in Section 4, which covers a wide range of AAL scopes and functions, namely lifelogging and self-monitoring, remote monitoring of vital signs, emotional state recognition, food intake monitoring, activity and behaviour recognition, activity and personal assistance, gesture recognition, fall detection and prevention, mobility assessment and frailty recognition, cognitive and motor rehabilitation. Each section provides a broad overview of the corresponding assistive and supporting functions, discussing open challenges and available datasets, where appropriate. Most of the works revised deal with visual-based AAL, as they have attracted a big body of research in the last years. Audio-based solutions are revised and discussed where a significant corpus of research is present.

Finally, Section 5 contains an overview of the challenges, hindrances and opportunities posed by the uptake in real world settings of AAL technologies. More precisely, procedural approaches based on co-design, privacy preservation in video and audio data, transparency in data processing, and data transmission and communication are discussed. User acceptance and ethical considerations are also included. Finally, the potentials coming from the silver economy are overviewed.

The document ends with Section 6, which draws the conclusions of the work.

# 2. A gentle introduction to AAL and the proposal for a taxonomy

This Section reports the main concepts behind AAL. It starts with an introduction to the AAL concept (Section 2.1); then, it moves on to overview the history of AAL (Section 2.2), and to describe the functional architecture of a typical AAL system (Section 2.3). The Section ends with a proposal for a taxonomy for AAL technologies and applications (Section 2.4).

## 2.1. The AAL concept

The concept of "*Active and Assisted Living*" (AAL) emerged in the early 2000s, by evolving the idea of assistive technologies to address the societal challenges of health, demographic change and wellbeing. Since then, numerous research and development endeavours have fuelled the field, stimulated by various funding opportunities, among which the European Framework Programmes (e.g., FP7 and Horizon 2020), the former Ambient Assisted Living and the current Active and Assisted Living Joint Programmes, EIT Health, and the Innovation Partnership on Active and Healthy Ageing are some of the most notable ones.

Several definitions of AAL have appeared so far in the literature [II.1] - [II.4]. Broadly speaking, AAL can be referred to the use of innovative and advanced Information and Communication Technologies (ICT) to create supportive, inclusive and empowering applications and environments that enable older, impaired or frail people to live independently and stay active longer in society [II.5]. AAL has capitalized on the growing pervasiveness and effectiveness of ICT systems to supply the persons in need with smart assistance, by responding to their necessities of autonomy, independence, comfort, security and safety. A plethora of solutions has been developed so far to respond to these core needs by

- supporting and ensuring a sustained wellbeing, quality of life and safety of people with any kind of impairment,
- alleviating the burden of chronic diseases, also by ensuring a continuous and remote monitoring and contrasting the shortage of health personnel,
- contributing towards more sustainable health, care and social services, by reducing the pressure on formal health and care infrastructures thanks to remote monitoring and tele-assistance,
- preventing ageing and impaired community from social isolation,
- supporting and relieving the burden of formal and informal caregivers,
- promoting better and healthier lifestyles for the individuals at risk,
- enacting disease prevention strategies based on personalised risk assessment and continuous monitoring.

With respect to the last issue, in the middle 2010s, the so-called *Quantified Self movement* has emerged thanks to the upsurge of consumer-friendly wearable sensors (e.g., smart watches, smart bracelets or wrist bands). The idea behind this movement is to promote the self-tracking and self-monitoring of a person's health information to increase the self-awareness on her or his health status. Quantified Self and AAL technologies strongly overlap, especially when considering the lifelogging and vital signs monitoring application scenarios. Similarly, AAL can make extensive use of mobile-Health (mHealth) applications as well as of the so-called Internet of Healthy Things (IoHT) to monitor and track health-related information [II.6].

From our viewpoint, AAL is an umbrella term that comprises the whole range of adaptive and intelligent ICT solutions that fit into the daily living and working environments, to offer unobtrusive, effective and ubiquitous support to ageing, impaired or frail people. From an initial version, mainly related to indoor environments, AAL concept has broadened to include also outdoor and on-the-move facilities, to ensure the continuum of support and care, thus increasing autonomy and independence of the assisted individuals. See Figure 2 and Figure 3 for some sample illustrations of AAL technologies and services.

The application scenarios addressed by AAL are complex, due to the inherent heterogeneity of the end-user population, their living arrangements, and their physical conditions or impairment [II.7]. Some of the most common and effective functionalities include:

- improving safety at home by preventing accidents and incidents that might occur in an assisted environment via, for instance, fall detection systems, alarms and warnings,
- maintaining under control chronic diseases or medication compliance, with connected devices for vital data measuring as well as medication reminders,
- maintaining physical and mental abilities, with the support of intelligent mobility aids, coaching systems and brain-training activities,
- maintaining interaction with other people with dedicated apps and online community platforms,
- improving quality of life for caregivers with technology for information sharing, and better coordination,
- early detecting risks in care homes, thus reducing the number of accidents and improving communication between caregivers and their patients.

Despite aiming at diverse goals, AAL systems should share some common characteristics [II.8]. They are designed to provide support in daily life in an invisible, unobtrusive and user-friendly manner. Moreover, they are conceived to be intelligent, to be able to learn and adapt to the requirements and requests of the assisted people, and to synchronise with their specific needs.

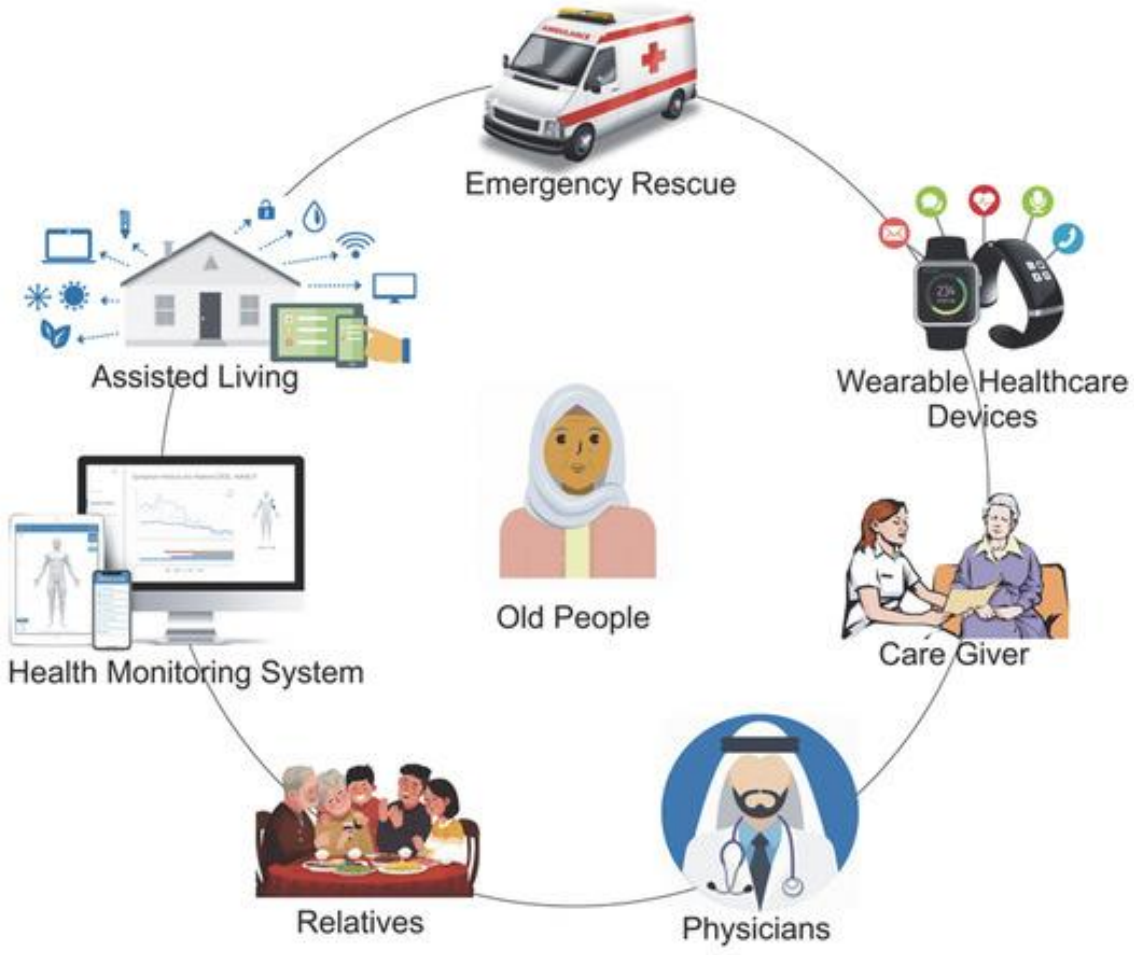

Figure 2. An example of home-based AAL system from [II.9]

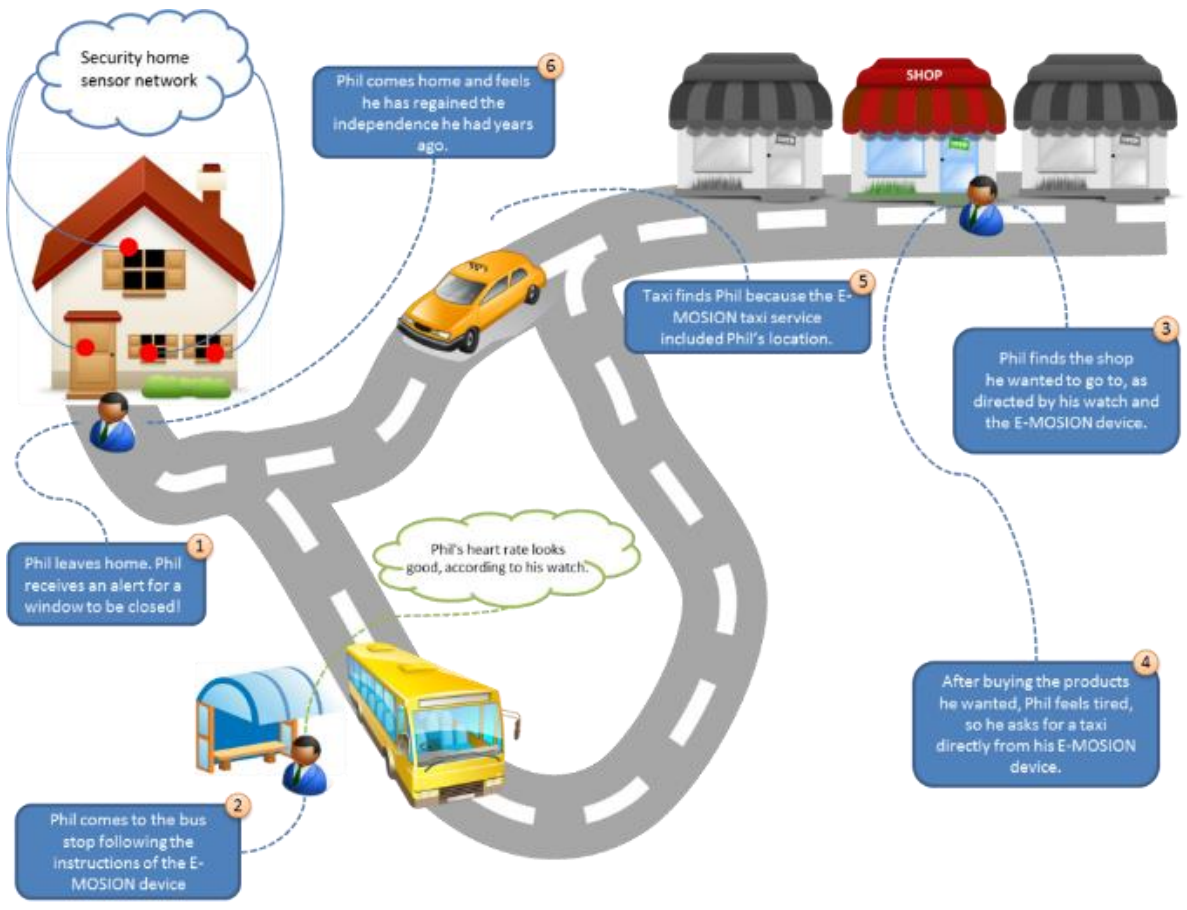

Figure 3. An example of indoor and outdoor AAL system from the AAL-funded project CLOCKWORK

In this respect, the field of AAL technologies overlaps, from the technological viewpoint, with many other fields, such as those of Assistive and Supportive Technologies, Ambient Intelligence (AmI), Pervasive Computing, Personal Informatics, e-textiles, IoHT and Robotics. Indeed, depending on the embodiment, AAL can include a wide range of robots, especially physically or socially assistive robotics that help users perform certain tasks [II.10].

Nevertheless, to ensure the uptake of AAL in society, potential user groups must be willing to use AAL applications and to integrate them in their daily environment and lives [II.11]. In this respect, participatory design is a cornerstone of technology acceptance, as it consists in involving the eventual end-users in the development process. Another big contribution to acceptance comes from a clear and comprehensive consideration of ethical and legal issues related to privacy, autonomy preservation, data protection and security [II.12].

This Section provides the reader firstly with an overview of the evolution in time of AAL technologies; then, it sketches a typical functional architecture of an AAL system, which is instrumental to introduce the main R&D advances reviewed in this report. Finally, this Section describes a proposal for a taxonomy of AAL applications.

## 2.2. History of AAL Technology

The paradigm of AAL technology emerged initially with the idea to use ICT-based solutions to advance assistive technologies into a more comprehensive endeavour. Starting from tech devices able to assist people with disabilities with just one task or to ensure their safety, the goal was to evolve towards encompassing completely the living areas and the person's needs [II.13]. The name "Ambient and Assisted Living" emerged at that time, showing overlaps with the idea of Ambient Intelligence. Lately, as already mentioned, the concept has broadened, by addressing the needs of healthy and active ageing, and disease prevention and including the support outdoor and on-the-move, thus taking the name of Active and Assisted Living.

Previous works have identified three generations of AAL technologies, which correspond to assistive solutions with increasing levels of automation and technological facilities [II.13]-[II.15]. Considering the most recent advances in the field and the debate on ethical and legal issues, we argue that a fourth generation of AAL technologies is emerging. In this fourth wave, more advanced functionalities based on Artificial Intelligence (AI) and Robotics as well as context awareness are being fully integrated in AAL. Moreover, more attention is being paid to the co-design process and to address the social, ethical and legal issues, as we are aiming at in GoodBrother. Figure 4 illustrates these four waves of AAL technology. In the following, a very brief overview of these generations is provided.

***First-Generation Technologies***: the first generation of AAL technology mainly consisted of alert and alarm systems coming in the form of a pendant or alarm device. In case of an emergency, the assisted individual could press the button or pendant in order to send an alarm to a call centre [II.15]. The first successful examples of this type of devices date back to the 1990s and include LifeAlert or SafeLife Beghelli. Experimental studies demonstrated that these first-generation alarms were able to ensure several benefits, such as safety and security of the assisted individuals and a reduction of stress levels of them, their families and caregivers [II.16]. On the other hand, they exhibited a number of limitations, as the assisted individual needed to remember to carry or wear the device and to be able to press the button.

***Second-Generation Technologies***: the second generation of AAL technology was more technologically advanced as it integrated sensors able to perceive risky conditions and react accordingly, as in case of environmental hazards [II.16]. An example of these solutions, particularly useful for older adults with mild cognitive impairment, was the device able to detect a gas leak and send an automatic alarm by contacting the appropriate authorities [II.16]. These technologies emerged by the end of 90s, early 2000s, and some of them are available on the market.

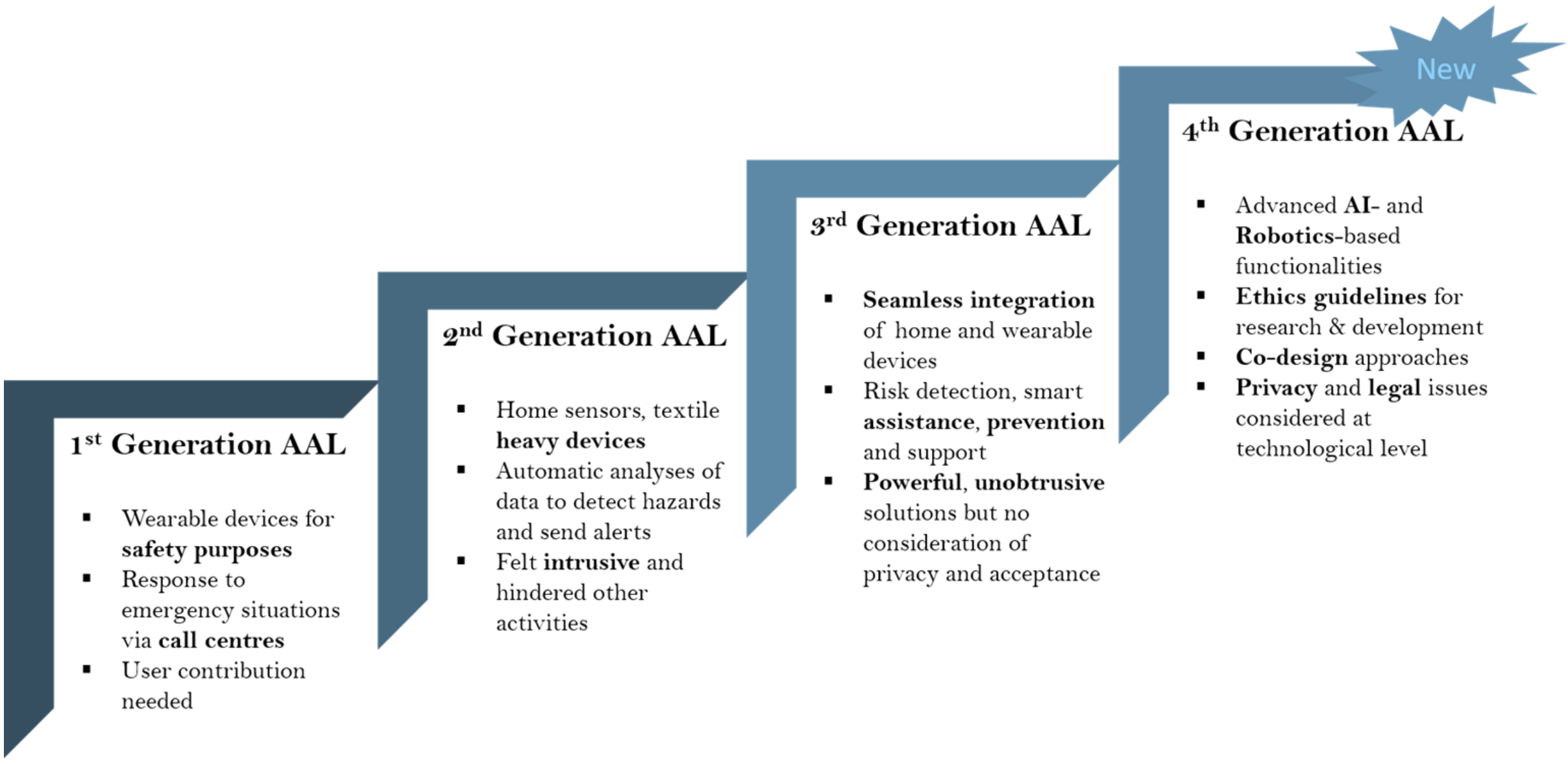

Figure 4. The four generations of AAL technology. The fourth one is a brand-new wave, whose launched is testified by the "AAL Guidelines for Ethics, Data Privacy and Security", the GoodBrother initiative and the likes

***Third-Generation Technologies:*** The third generation of AAL technology has been boosted by the advancements of ICT moving towards a more comprehensive concept of AAL. The solutions in this case are complex systems that integrate diverse sensors and computing facilities, in some cases seamlessly incorporated into home appliances, to monitor the home environment for any types of risk or accident as well as to track the activity and health status of the assisted person. In this wave, the technologies are conceived to not only detect and report problems, but also to prevent worst scenarios by using preventative and prediction strategies. Moreover, they include actuators to provide the assisted individuals with assistance, and smart interfaces to provide them with information, support, and encouragement [II.17]-[II.19]. In the last years, there has been an explosion of these third-wave AAL solutions. Some of the products currently available are included in the AAL catalogue online, categorised according to the AAL applications they support as well as the aspects of the quality of life they address. Figure 5. A group of AAL products published on AAL Catalogue online. These are categorized under the heading "Vitality & Abilities" shows a group of AAL products for supporting vitality and abilities.

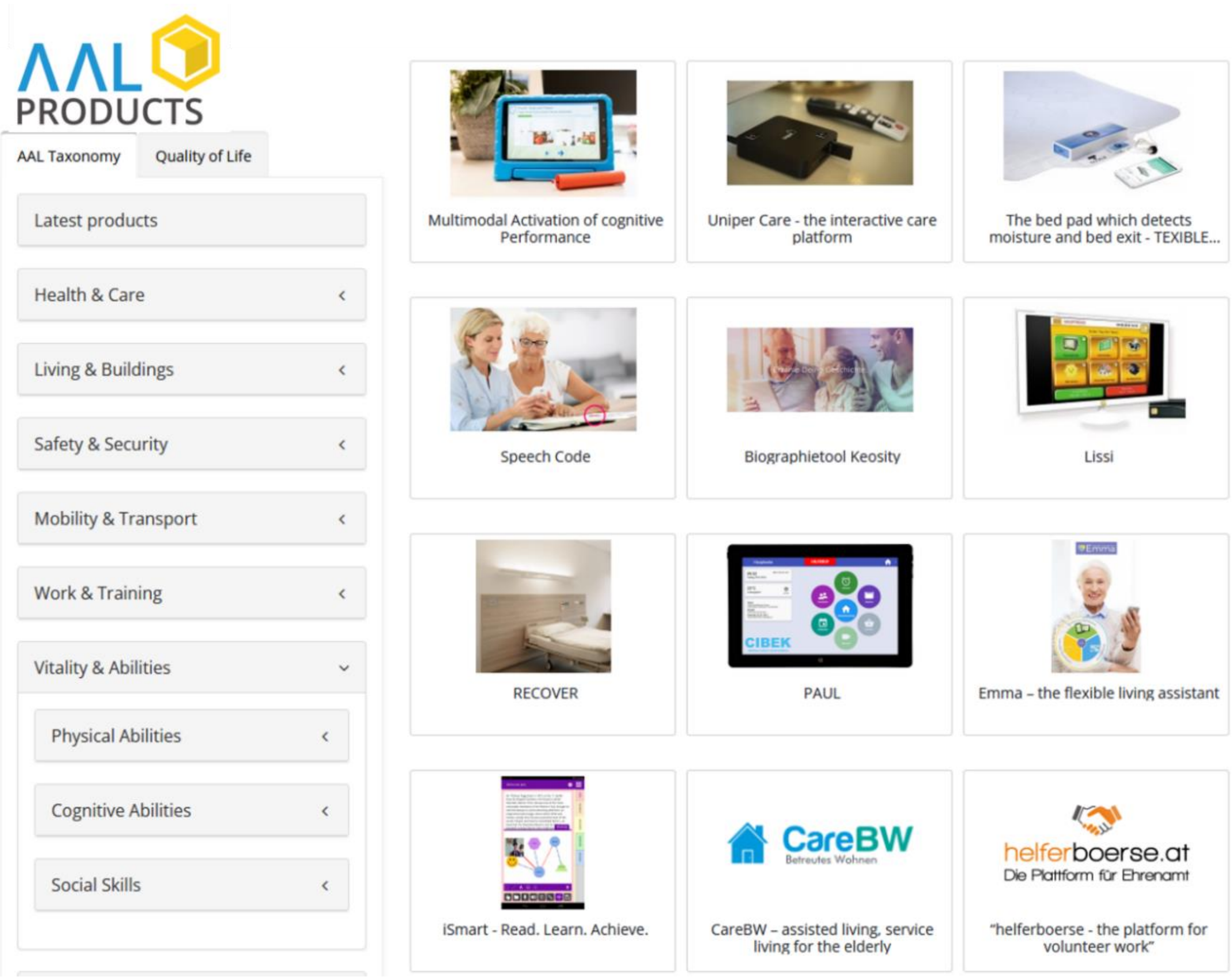

Figure 5. A group of AAL products published on AAL Catalogue online. These are categorized under the heading "Vitality & Abilities"

***Fourth-generation Technologies***: We argue that a new generation of AAL technology has started. The technological advances and the successes of the most recent Computer Vision (CV) and Signal Processing techniques (based mainly on Deep Learning) has strongly boosted the field of AAL. More powerful and effective solutions are being developed and experimented, as, for instance, those based on conversational agents and robotic assistants. Furthermore, in consideration of this increasing power and pervasiveness of technological solutions, the scientific communities, policy makers and funding bodies have started discussing strategies and approaches to ensure that such solutions respond to the highest ethical, legal and privacy standards and requirements. The AAL Programme has issued the "AAL Guidelines for Ethics, Data Privacy and Security" [II.20], with which "AAL goes one step further and demands more than just legal and ethical compliance, – it proposes Ethical Excellence, by fostering the implementation of the ethical dialogue and integrating relevant values in an iterative process of discussion. And applies this method not only during projects' lifetime but also for solutions already in the market". In this respect, the co-design of the AAL solution with the end-users is currently considered as an indispensable step that accompanies the whole development process [II.21]. Moreover, the issues pertaining to privacy preservation, especially in the case of audio and video data, have started to be handled also at the technological level, with different approaches now appearing in the literature [II.22], [II.23], as extensively surveyed in GoodBrother's WG2 deliverable.

In the following sections, we provide an overview of the main technological components of AAL solutions.

## 2.3. A functional view of an AAL system architecture

An AAL system can serve several functions or application scenarios, by integrating different ICT and sensor technologies to address the needs of diverse end-users [II.7], [II.24]. Consequently, an AAL system can come in different forms and with different architectures. Previous works [II.25] have attempted to find uniformities in existing solutions towards a standard design strategy. Nevertheless, diving in depth into the technical details of AAL software is over-encompassing and out of the scope of this deliverable. A more high-level, functional view of the typical architecture of an AAL system is much more pertinent as it allows the reader to have an overview of how an AAL system functions and interacts with the end-users. Essentially, a functional view of an AAL system derives from the compulsory components and functionalities it encompasses. Foremost, the system should incorporate sensor technologies that enable the acquisition of useful data about the ambient settings and the environment as well as the psychophysical conditions and the activities of the assisted individuals. The sensor technologies constitute *the sensing or perception layer* of the system. Dedicated algorithms process the data acquired by the sensors with the goal to understand, detect or measure the conditions of interest, for instance to recognise the activities of the assisted person to identify a risky situation (e.g., a fall) or to assist her in performing a task (e.g., dressing up). These algorithms compose the *data-processing* or *understanding layer* of an AAL system. The data can be stored locally and the algorithms can be embedded in the living environment or in mobile devices (this situation is currently referred to as computing or processing *at the edge or edge processing/computing*) or the processing layer can be located on distant servers on the cloud. Currently, the data processing layer is often a mixture of algorithms at the edge and on the cloud. In accordance with the information inferred by interpreting the sensed data, the AAL system can provided the end-users with feedback, for instance on suggestions on what to do next when performing a task, alarms or alerts when a risky situation is detected, or engaging them in a game. This happens through the *actuator/application* or *interaction layer*. This layer comprises the user interfaces and may require additional computing facilities to support the interactions, such as planning and controlling [II.26]. Figure 6 shows the three layers and their components.

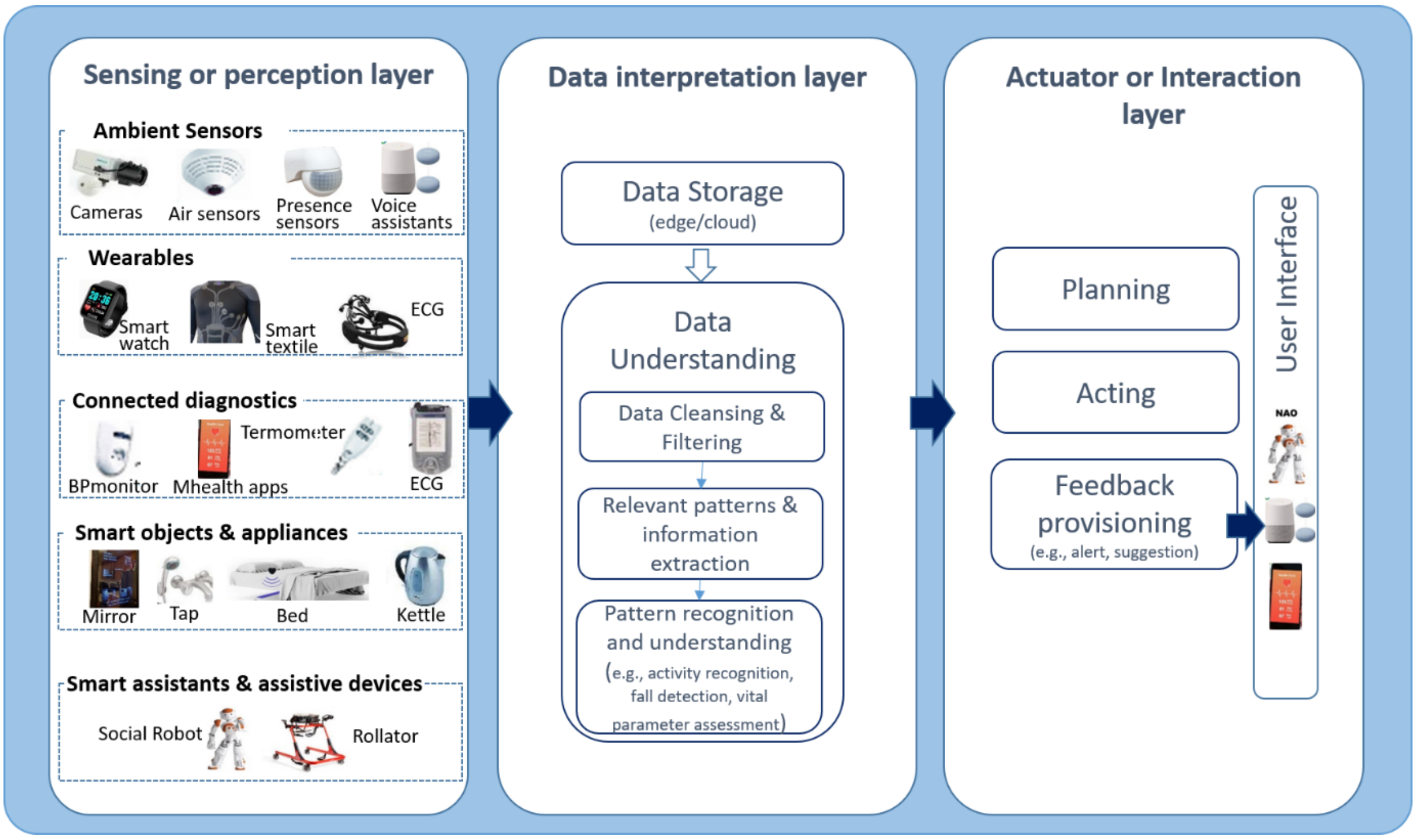

Figure 6. A functional view of the main components of an AAL system. The various sensors networked in the sensing layer need a suitable orchestration to regulate data recording and storage. The algorithms in the data-understanding layer can work on data coming from a single sensor or on multiple sensors. The user interface could be embedded in any type of smart appliance (i.e., could be tangible or intangible)

As shown in the figure, the sensor layer can be composed of simple smart and connected devices (i.e., the so-called Internet-of-Things) or it can correspond to a more complex sensor network composed by environmental sensors, intelligent devices, video cameras, or audio-based home assistants [II.24].

See Figure 7 for different types of sensors that could be deployed in a living environment.

Environmental and appliance sensors can acquire information about temperature, humidity, air quality as well as the status of the appliance. Moreover, non-invasive sensors, such as cameras or infrared sensors, have been integrated in various devices (i.e., the so-called *smart objects*) and appliances (e.g., mirrors, TVs, rings, bracelets, watches), to monitor individuals in a non-obtrusive and easily acceptable manner, without affecting their normal activities. Accelerometers, gyroscopes, infrared or radar sensors can be embedded into smartphones or wearable devices, such as smart watches, fitness bands, clothing and fabrics to continuously monitoring people in both indoor and outdoor applications. Other medical devices (e.g., pulse oximeters, blood pressure monitors) can be connected to the network, thus allowing for the automatic transfer of vital parameters. This whole set of smart and connected devices used to monitor the vital parameters of an individual currently goes under the name of IoHT.

The variety of technologies automatically implies a greater complexity of data, as they can change considerably in terms of size, heterogeneity, and sampling frequency. Consequently, several issues related to data management are also relevant in AAL technology (e.g., communication protocols, security controls, energy consumption, failure detection, interoperability among diverse vendor devices, and so on).

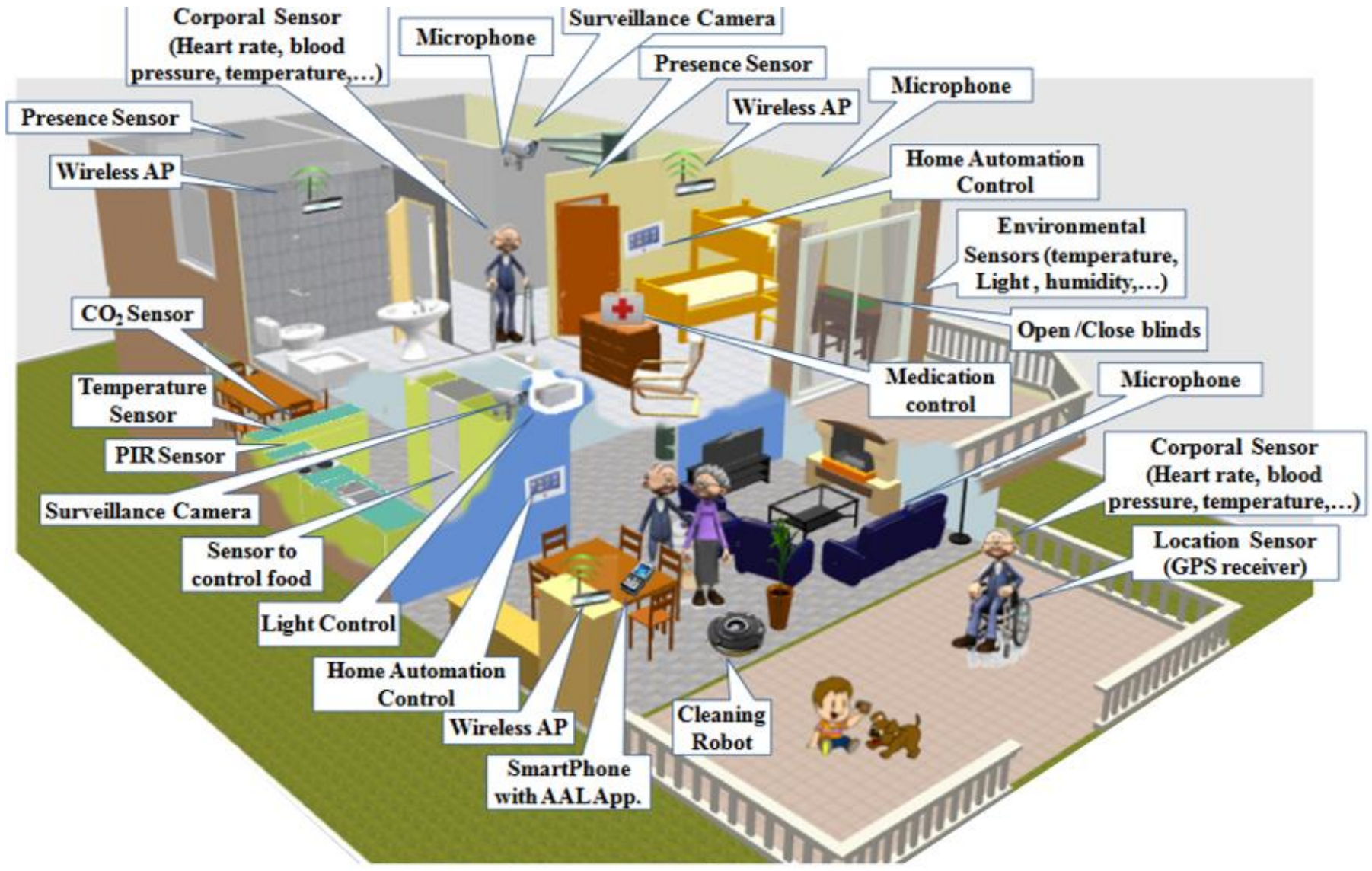

Figure 7. An example of the various types of sensors that could be deployed in a house or assisted-living environment (reprinted from [II.23])

Among the various typologies introduced above, the video and audio sensors are among the most powerful ones in terms of the information they convey. For instance, a single camera placed in a room can record most of the activities performed in the room, thus replacing many other non-visual sensors. As the costs of cameras dropped a lot, a plethora of works have used cameras and CV techniques to address most of the AAL applications scenarios, boosting notably the field. Currently, video-based applications are able to recognise and monitor the activities, the movements, and the overall conditions of the assisted individuals as well as to assess their vital parameters (e.g., heart rate, respiratory rate). Similarly, audio sensors have a potential to become one of the most important modalities for interaction with AAL systems, as they can have a large range of sensing, do not require physical presence at a particular location and are physically intangible. Moreover, using voice is a more natural way of interaction than tactile interfaces.

Overall, audio and video-based sensors appear as less obtrusive with respect to the hindrance other wearable sensors can cause to one's activities. Nevertheless, they are often perceived as the most intrusive technologies from the viewpoint of the privacy of the monitored individuals. This is due to the richness of the information these technologies convey and the intimate setting where they are inserted.

In the following section, we propose a first attempt to categorise the various axes that characterise an AAL system.

## 2.4 A taxonomy of AAL technologies and applications

Considering the plethora of diverse systems developed in the last decades, systematising the various AAL technologies and applications is a challenging endeavour, which has been undertaken so far only partially, by providing only very limited classification schemes. Indeed, while several scoping reviews or scientific surveys have analysed the existing literature in the field of AAL, only a couple of attempts produced an AAL taxonomy and both of them considered mainly the AAL application scenarios.

In 2014, the project TAALXONOMY[1], funded by the Austrian Research Promotion Agency FFG, proposed a taxonomy for the practical and effective classification of AAL products and services, by taking into account international definitions, initiatives and standards as well as feedbacks from relevant stakeholders, users and experts. Nevertheless, the TAAXONOMY classification scheme[2] mainly distinguishes the various living environments and life aspects that are addressed by the AAL system or service. The primary categorization axis is according to the application scope and groups AAL systems into "Health & Care", "Living & Buildings", "Safety & Security", "Mobility & Transport", "Work & Training", "Vitality & Abilities", "Leisure & Culture", and "Information & Communication" classes. Within each of these classes, a secondary classification is applied based on the field of application. For instance, the "Health & Care" class is further split in "Health Care and Prevention", "Body and Vital Data", "Telecare and Telehealth", "Electronic Health Record", "Nutrition & Diet", "Personal Hygiene", "Therapy", "Drugs and Pharmaceuticals", and "Care".

The TAAXONOMY classification scheme has served the creation of the AAL-products[3] online catalogue, put together by the Department of Strategic Management, Marketing and Tourism from the University of Innsbruck and EURAC Bolzano. AAL Products is conceived to give solutions provider the opportunity to place their AAL solutions cost-free, thus enhancing visibility and awareness of AAL solutions. Moreover, it provides the visitors with an overview of assistive- and smart-technology products and services available on the market.

Another AAL taxonomy appears in the work by Byrne *et al*. [II.27], which provides a classification framework based on four core categories: "smart homes", "intelligent life assistants", "wearables", and

1 https://www.taalxonomy.eu/en/ (last accessed: 23/02/2022)
2 https://www.taalxonomy.eu/wp-content/uploads/Downloads/D4.1-ANNEX-TAALXONOMY-final-oeffentlich.xlsx (last accessed: 23/02/2022)
3 https://www.aal-products.com/index.php/frontend/start?categorie=-1 (last accessed: 23/02/2022)

“robotics”. Each of them has distinctive sub-classes that categorise AAL systems in accordance with their primary function. For instance, Smart Homes are further distinguished in “General Health Monitoring” and “Platforms”; whilst “Intelligent Life Assistants” are further split into “Wandering Prevention Tools”, “Electronic Home Control Systems” and “Fall Detection Systems”.

In GoodBrother WG3, we have worked on an AAL taxonomy with the idea to distinguish the various technological facilities, ICT components, main functionalities and application scopes of an AAL system. The idea of such a taxonomy is to illustrate the wide spectrum of options that characterise assistive and supportive technologies and systems. The resulting scheme is a tree-like structure that is described in the following paragraphs, starting from the root and the primary branches moving towards the leaves.

An AAL application or system is characterised by its “*Technological underpinnings*” and “*Assistive & supportive undertakings*”, as shown in Figure 8. The “*Technological underpinnings*” include the ICT facilities for “*Data acquisition*” and “*Data processing & understanding*” as well as the “*Enabling infrastructure*” and the facilities for the “*Human-computer interaction*”. An AAL application can implement one or more “*Assistive & supportive scopes*”, which can be characterised by defining their “*End-user & beneficiaries*”, the “*Ambient settings*” where the application is deployed, the “*Care needs*” that the application addresses, the “*Assistive and/or supportive functionalities*” it provides, the “*Other involved actors*” if any, and the type of the “*Output & feedback provided*” during the interaction with the end-users.

Delving more in depth in the technological underpinnings, “*Data acquisition*” can be further specialised as shown in Figure 9. More specifically, it encompasses diverse modalities for acquiring data about the end-users and the environment where they live, work or move. Reflecting the major interest of GoodBrother in video- and audio-based AAL, among the “*Data sensing devices*”, we highlight video and audio sensors. “*Video and image sensors*” include “*RGB cameras*”, “*Depth sensors*”, “*Thermal cameras*” and “*Multispectral cameras*” (in this we include also hyperspectral ones). Among the “*Other types of sensors*”, we include the various sensors that use different modalities to acquire data, such as gyroscopes and accelerometers that fall within “*Motion-based”* sensors or Galvanic Skin Response sensors that fall within the “*Electrodermal”* sensors, and so on so forth. The sensor devices can be deployed in various modalities: they can be installed in the environment (e.g., on the ceiling); they can be worn by the user when they are integrated into wearables; they can be integrated into social robots or home assistants (e.g., Google Home or Amazon Alexa), or they can be embedded in smart appliances (e.g., in the fridge or behind a smart mirror) or integrated into the smart phone. Data can come also in the form of manual input from the users, via a mobile or tablet applications, by replying to questionnaires, or filling forms or from diagnostic examinations.

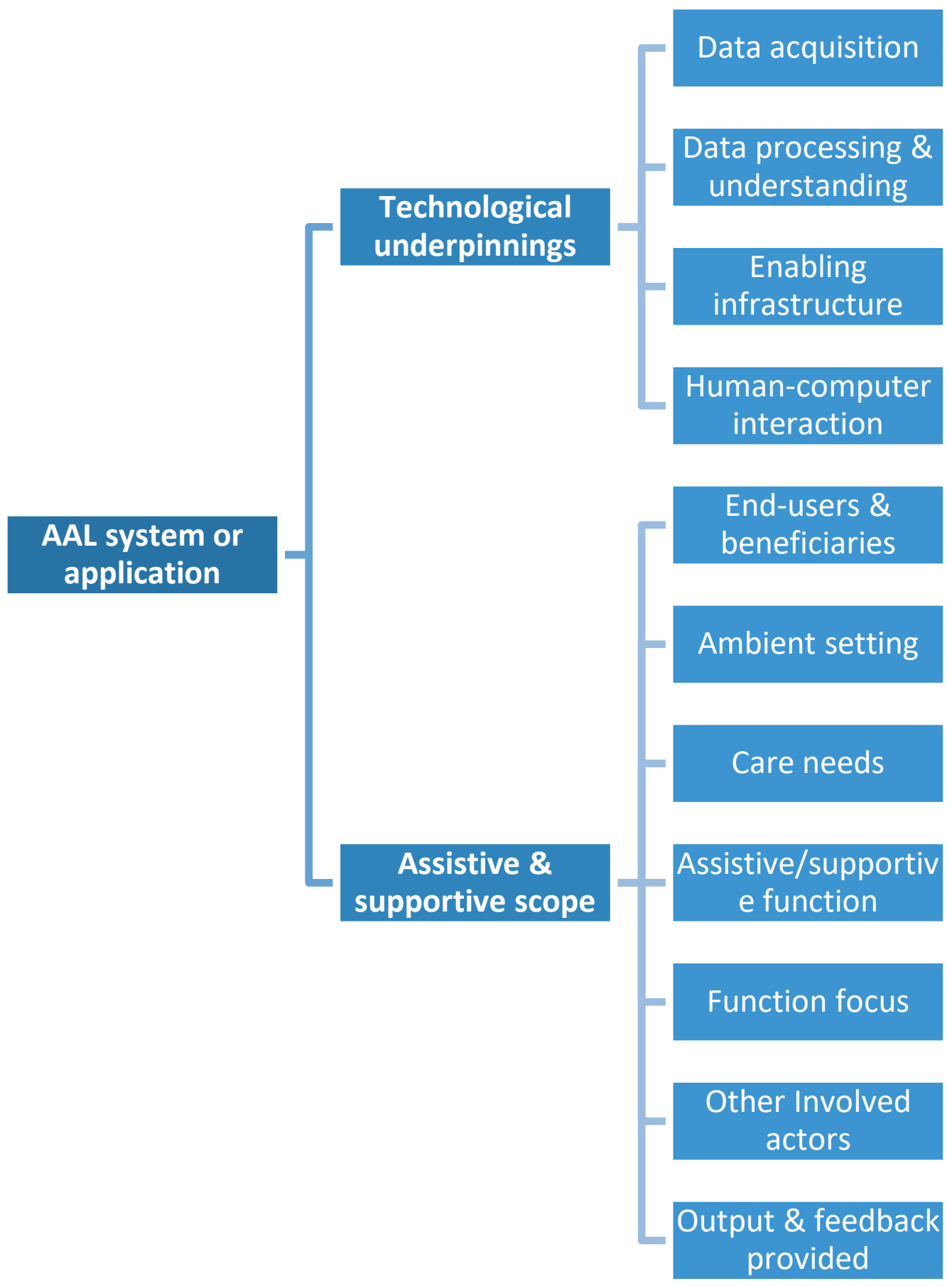

Figure 8. The primary layers of the proposed AAL taxonomy

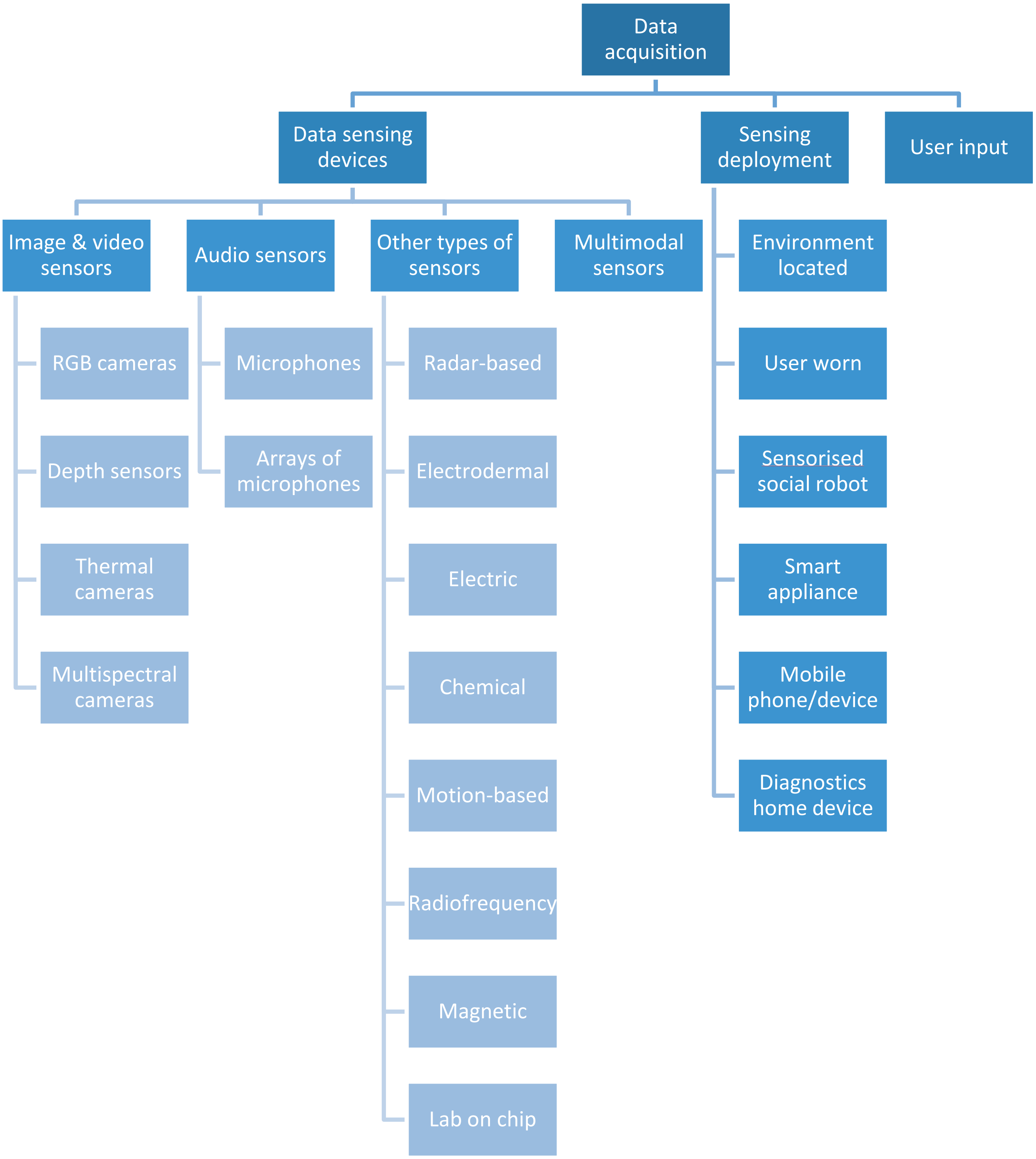

Figure 9. The specialization of data acquisition

“Data processing & understanding” encompasses methods and technologies belonging to disciplines such as CV, Signal Processing, Multimodal and Big data analytics, Artificial Intelligence and Machine Learning. These include a wide range of methods whose listing is out of the scope of this report. The “Computing infrastructure” comprises all the ICT facilities for the storage, the transmission and the communication of data, as well as for orchestrating the sensor devices and the computing resources. Computing localization may be can be “on-site” or on the cloud. Finally, the “Human-computer interaction” comprises the interfaces through which the support and feedback are provided to and the commands or inputs are received from the end-users. Figure 10 summarises further this specialization.

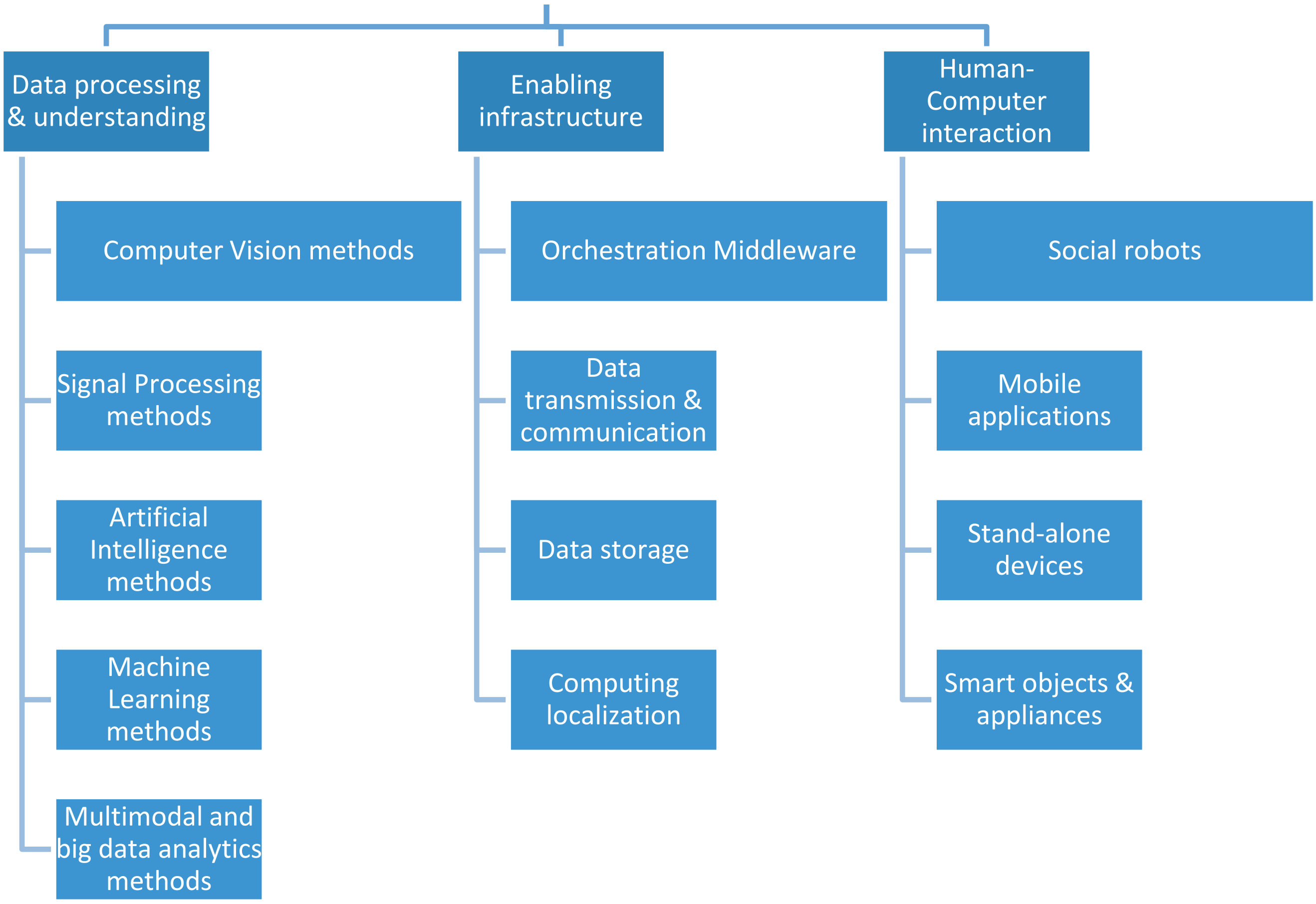

Figure 10. The underpinnings of “Data processing & understanding”, “Computing infrastructure” and the “Human-computer interaction” categories

When considering the “Assistive & supportive scope”, an AAL system can address various types of “*End-users & beneficiaries*”. These include frail or impaired individuals, chronic and multi-morbidity patients, and ageing healthy subjects. Among the beneficiaries, we can also include formal and informal caregivers, as the assistive and supportive environments can help them by alleviating their work and burden. Moreover, clinicians and General Practitioners can be addressed as end-users with the provision of information and analyses of physical, cognitive and behaviour information of chronic patients recorded remotely. The “Care needs” addressed by an AAL application include the primary prevention, for preventing diseases for at-risk subjects, the diagnosis or risk detection for the detection of disease exacerbation events, therapy or treatment for enacting and monitoring compliance, or tertiary prevention, to improve quality of life and reduce disease symptoms. Accordingly, the assistive or supportive function can be monitoring (e.g., behaviour or vital signs), detecting or recognising (e.g., disease exacerbations, risks or falls), preventing (e.g., disease worsening or accidents), assisting (e.g., in the daily-life activities), training or supporting rehabilitation (e.g., through exergames). The focus of such function (“Function focus”) can be any of the various assets of health and wellbeing, including psychophysical conditions, with vital signs (e.g., cardiovascular parameters, breathing patterns) and emotional state (e.g., stress, anxiety, fatigue, or depression), lifestyle (e.g., physical activity, and food intake) sleep hygiene, sensorimotor and cognitive conditions, social interactions or the Activities of Daily living (ADL). The “Output and feedback” that can be provided to the end-users include alarms and alerts (e.g., when a fall or an exacerbation event is detected), suggestions or advices (e.g., on lifestyle), engagement in social interactions or exergames, reminders for drug intake, support in performing any task of daily living, support in out-door activities. The “Ambient settings” can, indeed, be indoor or outdoor. We also include in the scheme “Other actors”, including in this category the other professionals that intervene in the caring or assistive process. Figure 11 details the various sub-categories listed above.

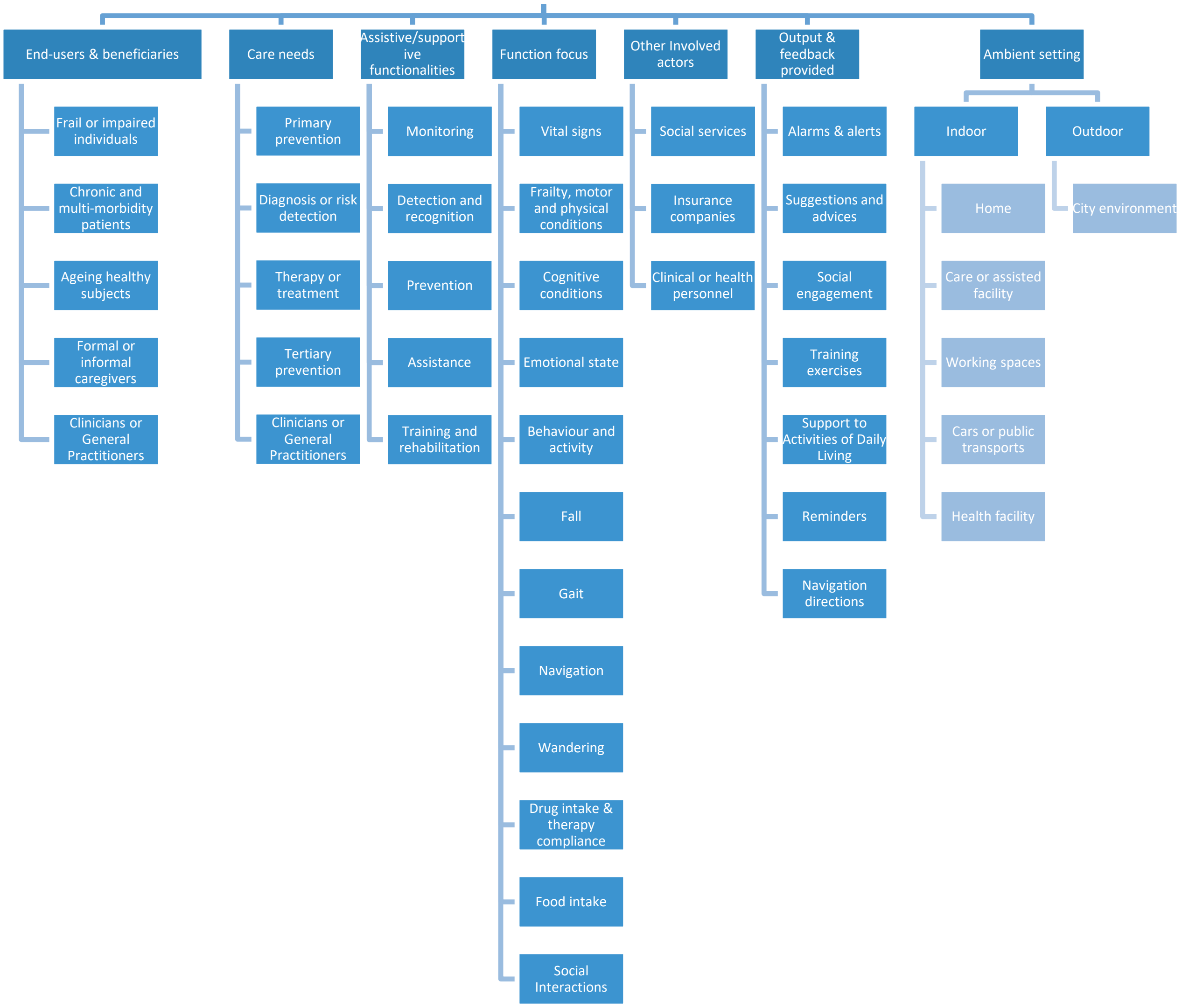

Figure 11. Detail structure of the subcategories under "Assistive & supportive scope"

An AAL system or application can encompass various options of those included in the taxonomy: for instance, it may address older adults suffering from Moderate Cognitive Impairment, by monitoring their vital signs and emotional state, engaging them in cognitive-training games, supporting them in ADLs and providing them with suggestions on physical activities and food intake. This might be done by integrating various sensing modalities and exploiting several data interpretation methods.

In this report, to support the activities of GoodBrother, we concentrate on AAL applications based on video and audio data. In this respect, we document the recent advances in the technological underpinnings related to video and audio data sensing and processing (Section 3). Afterwards, we provide the reader with the most successful assistive and supportive functions, based on video and audio data (Section 4), namely:

- lifelogging and self-monitoring
- remote monitoring of vital signs
- emotional state recognition
- food intake monitoring
- activity and behaviour recognition
- activity and personal assistance
- gesture recognition
- fall detection and prevention
- mobility assessment and frailty recognition
- motor rehabilitation.

# 3. Technological underpinnings of audio- and video- based AAL

This section summarises the state of the art of the technological underpinnings of video- and audio-based AAL. Firstly, an overview of the various video-based sensing technologies is provided in Section 3.1, by describing the main types of sensors used in AAL, especially for 3D acquisitions. A similar overview of the sensing technologies for audio data is presented afterwards in Section 0. Hence, the methods for analysing and understanding audio and video data for AAL purposes are summarised in Section 0. Finally, solutions adopted to combine multimodal sensing are overviewed in Section 0.

## 3.1 Video-based sensing technologies

Nowadays, monitoring and behaviour analysis tasks can be carried out by a variety of image-based technologies. These technologies can be classified by the place where they are installed, such as indoors or outdoors, wearables or embedded systems, but can also be divided by their mechanical capacities, such as bullet type, omnidirectional or pan-tilt-zoom cameras, or by their in-built features, such as motion detection or night vision.

Apart from the shape, angle and environment also the underlying camera technology varies. Not only RGB (Red, Green, Blue) cameras, which are highly controversial for their usage in AAL due to their lack of privacy protection, but also more privacy-aware solutions such as depth and infrared cameras have evolved in this area. While depth sensors measure distances from objects to the sensor, infrared (IR) sensors compute the temperature generated from objects. In the upcoming subsections, different image-based technologies are explained in more depth. Some of them are depicted in Figure 12.

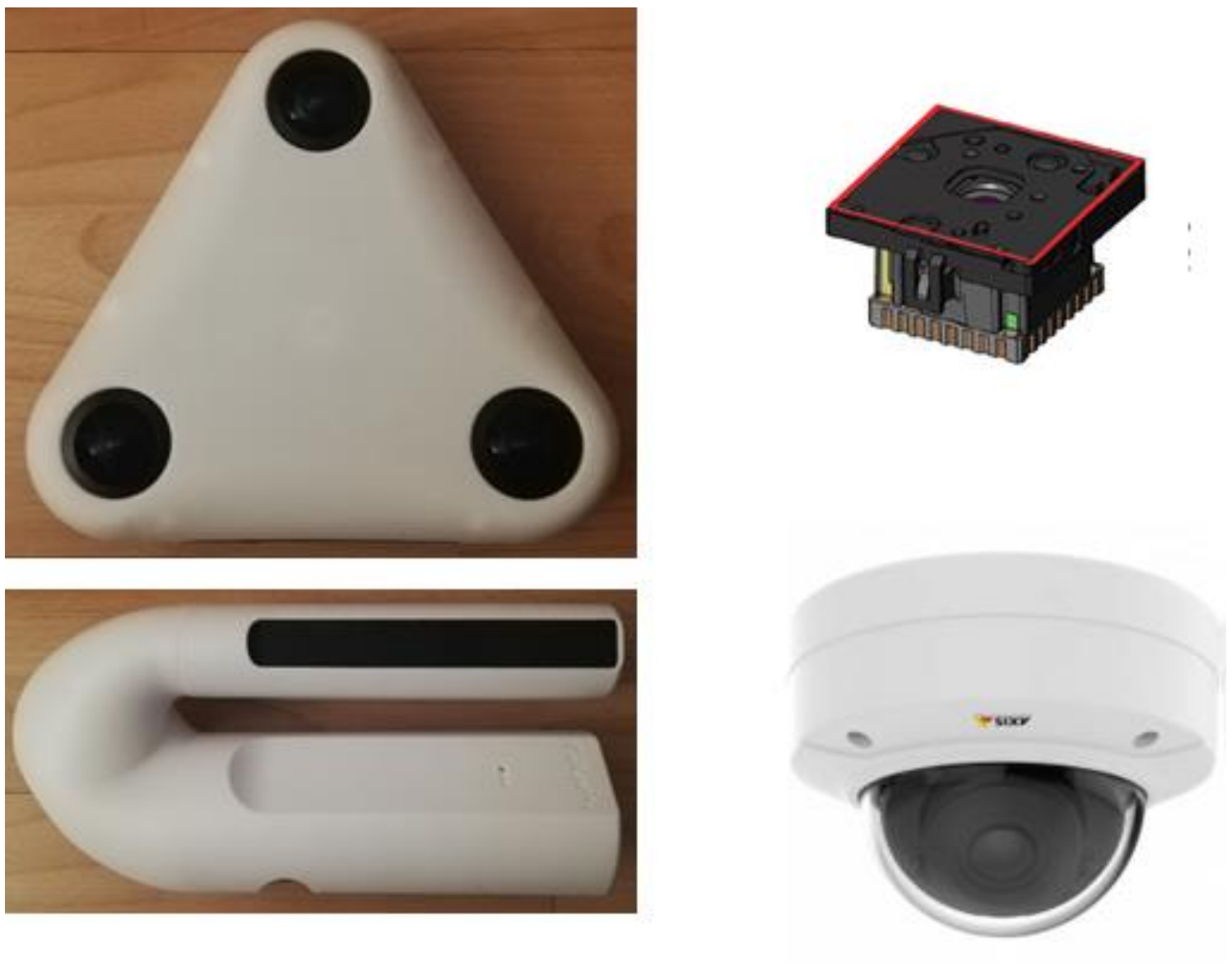

Figure 12. Different camera types used in AAL applications: top left - Omnidirectional camera, top right - Thermal camera, bottom left – depth sensor, (d) bottom right – RGB dome camera.

### 3.1.1 RGB cameras

For humans, images and video are straightforward to interpret, as they reproduce the way we see the world. Digital cameras replace the light-sensitive film of analogue cameras of the past, by a digital matrix of sensors, either as a charge-coupled device (CCD) or an active-pixel sensor (CMOS). These matrices contain sensors encoding the RGB signals received when the shutter is open for a fraction of a second. This process works similar to how the human eye forms images in the retina, in the back of the eye, using physiological structures called cones and rods, which are specialised in capturing different wavelengths of light.

The product of such a process in a digital camera thus produces images with a value representing the intensity of the light received in the picture element (pixel) at each coordinate of the sensor matrix. This is what the computer 'perceives' from a camera connected to it, just a bunch of numbers, representing pixel intensities. This type of representation is very high dimensional and has no semantics associated with it, therefore analysis is very complex. CV aims at making it possible for computers to interpret such rich information and infer useful knowledge from it. In the past, this was done via filtering, hand-crafted feature extraction and clever algorithms, such as Support Vector Machines (SVMs) or Random Forest (RF) among others. In the present, neural networks are used for this process, which have proven to be very good at this task, often improving previous attempts, and even outperforming humans in some specific scenarios.

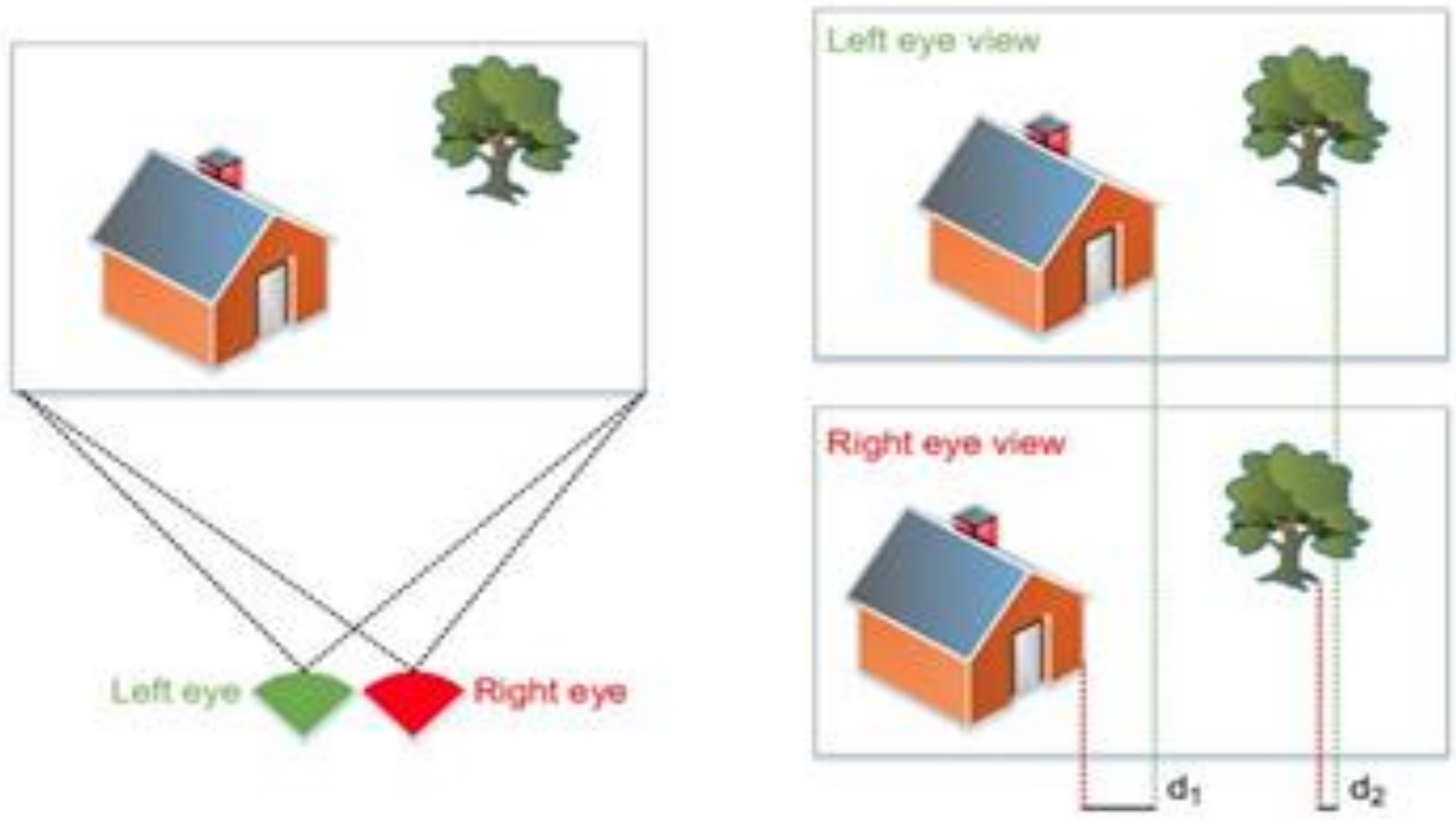

Figure 13. Human eyes perceive differences (d) in the relative horizontal position of objects, which helps understand which objects are closer or further away in the scene (adapted from [III.4])

Furthermore, humans have the ability to estimate depth thanks to the use of binocular vision, namely having two eyes separated by some distance. The different appearance of objects on the images, mainly in the relative horizontal position, received by the brain from each eye, helps it determine the distance of relevant objects, as shown in Figure 13. Likewise, in stereoscopic CV it is possible to use several calibrated cameras to obtain pairs (or sets) of images and estimate the so-called "disparity maps" of objects appearing in the scene via "triangulation" of matching points. Another possibility is to use structured light patterns and check the deviation of the light as it rebounds in observed objects [III.2]. These light patterns can be visible (see Figure 14), or be emitted by an infrared (IR) light source, in which case, an additional IR sensor is required in the setting. Using this setup (common in Microsoft Kinect, ASUS xtion, Orbbec Astraand similar devices), an additional "channel" is received, thus forming RGB+D (for depth) images.

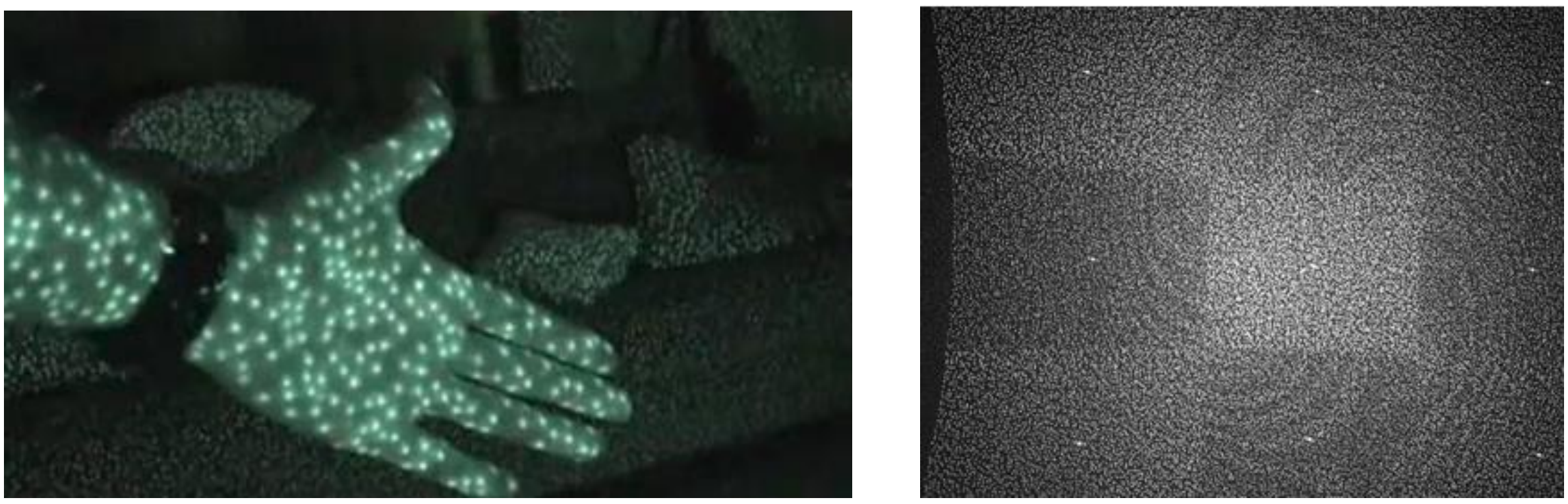

Figure 14. (left) Visible structure light pattern forming vertical strips, along with reconstruction; (right) the "statistical" infrared light pattern emitted by Microsoft Kinect v1

CV has historically held strong connections with Robotics, and therefore initial applications were related to helping robots to (*i*) navigate the world and map it (leading to "Simultaneous Localisation and Mapping" - SLAM); (*ii*) understand context by recognising objects; and, (*iii*) interact with humans, (*iv*) understanding their behaviours (e.g., actions, activities). This resulted in a plethora of different fields, focusing on the specifics of each of these tasks. However, current applications extend far beyond Robotics and include:

- industrial visual inspection of products
- semantic segmentation of images, for self-driving cars
- human pose estimation for interactive video games and natural Human-Computer Interaction (HCI)
- human action recognition for Ambient Intelligence (AmI) in smart homes and buildings

They support AAL technologies for people with functional diversity (older people, people with disabilities), among many others. Within the AAL field, the most relevant applications as collected in Planinc *et al*. [III.3] are:

- behaviour analysis or understanding: specific applications include detecting initiated activities to follow along and give cues for correct finalisation when the person is disoriented or leaves the activity due to forgetfulness, or detecting patterns of long-term behaviours in the performance of activities of daily living (ADLs).

- fall detection and prevention (gait analysis): detecting falling people, but also analyse gait as an important marker for cognitive impairment, and physical decline, leading to falls in the future.
- motor rehabilitation: programmes for older people and physical injury recovery therapies, aided by pose estimation, to check the correctness of exercise performance, as well as evaluate the range of mobility.
- vital sign and remote monitoring: including but not limited to heart rate monitoring via amplification of colour changes in the skin as part of normal blood circulation.

These applications are all revised in this report in the following Section 4.

Until recently, RGB+D and stereoscopic vision were seen as crucial to assist in certain tasks, such as 3D shape reconstruction, e.g., RGB+D made it possible to retrieve skeletal data about the body pose of a person as a set of 15-20 body joints, depending on the sensor model and brand. Nonetheless, with the use of deep neural networks it has been possible to use monocular images for these tasks. Examples of this are the Openpose [III.4] and LCRNet [III.5] networks, providing skeletal data as body joint coordinates (in 2D and 2D/3D respectively); or the Densepose [III.6] and Frankmocap [III.7] networks, developed at Facebook, Inc. The former provides a IUV image, consisting of three channels, each providing information at the pixel level about the body part index (I), and the (u, v) coordinates within the body part surface of each pixel (U,V channels). By contrast, the latter uses the SMPL (Skinned Multi-Person Linear Mode[4]) human body model [III.8], [III.9] developed at the Max Planck Institute and fits it on the detected body pose and shape (i.e., body variations, such as waist circumference, and so on). Example outputs of these networks are shown in Figure 15, which includes frames from the Toyota Smart Homes dataset by Das *et al*. [III.10].

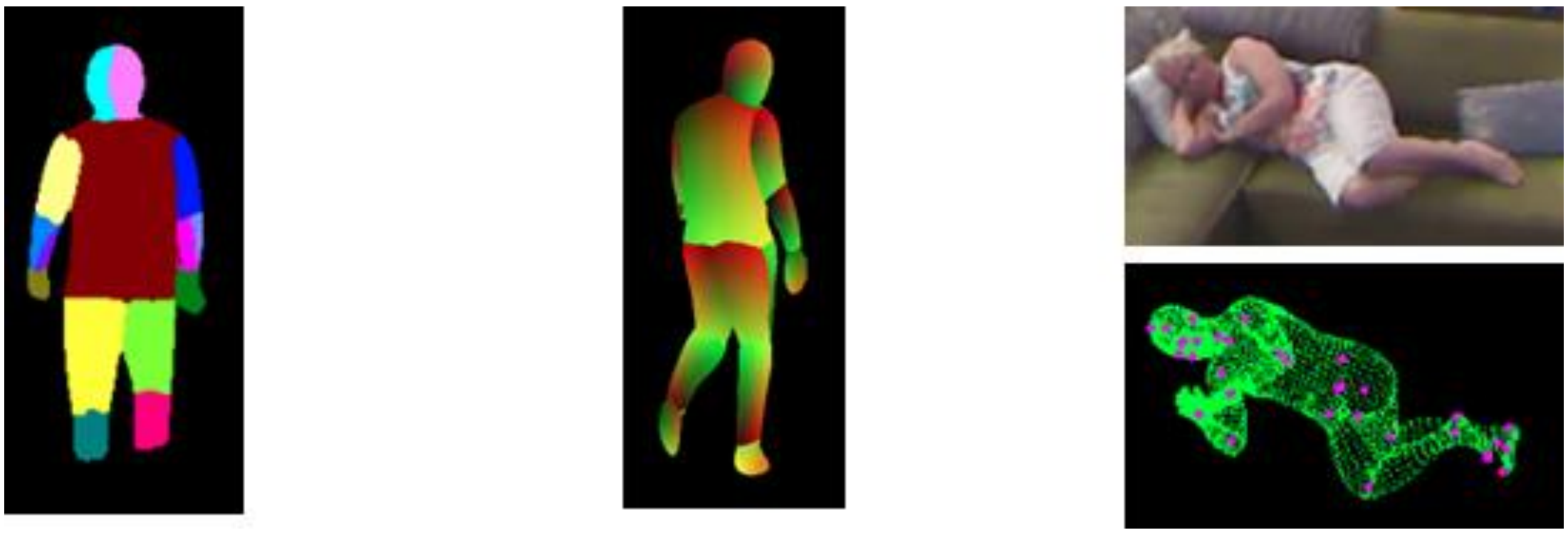

[4] https://smpl.is.tue.mpg.de/ (last accessed: 23/02/2022)

Figure 15. (left) Body part indices (colorised for improved visualisation) of a Densepose network output image; (centre) the original output of Densepose also contains U and V coordinates of the inferred body part surface (shown as red-to-green gradients); (right) a person lying on a sofa, and the same person as a fitted 3D avatar including thousands of surface points (green) and estimated body joint positions (purple)

### 3.1.2 Depth sensors

As pointed out by Planinc *et al*. [III.3], advances in the development of 3D sensors motivate their use instead of cameras for wearable sensors, since they provide advantages like privacy protection and improved robustness when it comes to behaviour modelling, gesture recognition or activity recognition. In contrast to RGB based analysis, depth-based approaches do not process RGB colour images, but depth of range images measuring distance from objects to the sensor. Figure 16 shows an RGB camera image together with its corresponding depth image: depth images do not visualise the scene with colours, but the grey level indicates the distance of the objects and its surroundings to the sensor. The darker the colour, the closer the object is to the sensor. On the other hand, brighter colours are used for objects at a higher distance. Black holes within the depth image indicate reflecting or absorbing areas where no valid depth measurement is available and are caused due to the functionality of the sensor. In contrast to RGB based approaches, only the silhouette is detected and thus no conclusions whether the person is wearing clothes or the emotional state can be obtained since neither the clothes nor the face are visible.

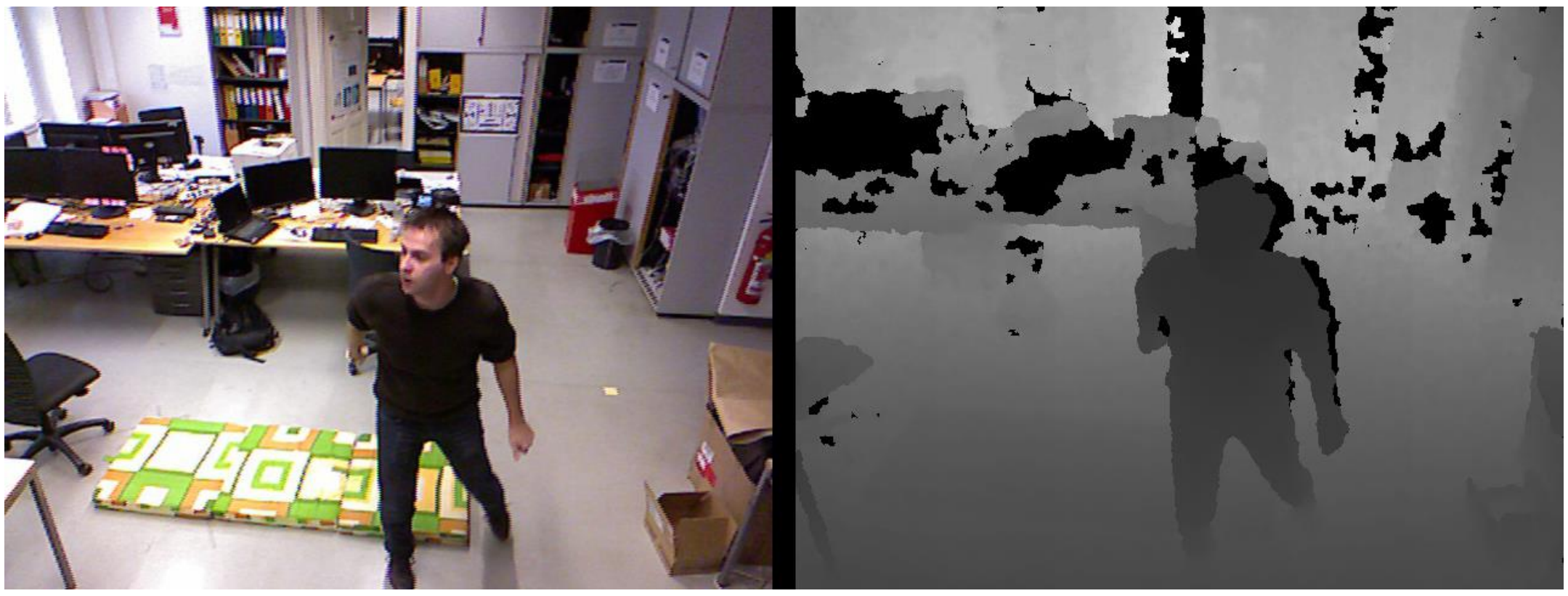

Figure 16. RGB Camera image and its corresponding depth image (Image from [III.3])

Hence, the appearance of the person is fully protected. However, this is only the case if processing is solely based on depth images. RGB-D based approaches combine depth information together with appearance information in order to obtain and combine more details. Although these approaches limit the privacy of elderly, colour information can be taken into consideration, thus permitting to perform a more in-depth analysis. With the introduction of the Microsoft Kinect in 2010, depth sensors have become more popular. The functionality of the Kinect is based on structured light imaging, where the projector emits a pre-defined IR light pattern to the scene [III.11], [III.12]. Due to the spatial arrangement of the pattern and its varying sizes as well as distortions depending on the distance to the camera, the depth camera captures the light pattern and an on-board chip calculates a depth map.

The main advantages of depth-based sensors, especially within the context of AAL, can be summarised as follows:

- No additional light source needed: due to the use of infrared light, sensors also work during the night (e.g., when falls of elderly people occur)
- Sensor is robust to changing lighting conditions: switching the lights on and off does not affect the results of the depth images. However, direct sunlight interferes with the projected infrared pattern and thus, no depth value can be calculated. This restricts the use of the sensors to indoor environments only
- No calibrated camera setup is needed: in contrast to the use of a calibrated multiple camera setup in order to calculate a 3D reconstruction, no calibration is needed
- Standard algorithms can be applied to depth information: standard algorithms for CV (e.g., foreground/background segmentation, tracking) can be applied to depth data directly
- Protection of privacy: if only depth information is processed, privacy is protected since the appearance of the person is not recognised in depth images. However, if a combined analysis of RGB and depth images is performed, privacy is not protected

Different types of information are extracted from depth data in order to carry out specific tasks. Skeletal data, for instance, is used to extract body parts and skeleton joints in order to estimate a person's pose [III.13]. Methods operating on depth maps frequently utilise histogram-based features and supervised learning for person detection [III.14]-[III.16]. Another approach is to first re-project depth map pixels to world coordinates and to operate on the resulting point cloud. One application for this method is the estimation of a person's proportion based on their height [III.17]. A reason why methods operating on point clouds are slow is the large number of points (up to 307200) and the fact that clustering is a complex task. Subsampling the point cloud can alleviate this problem [III.18]. An alternative are so-called plan-view maps, two-dimensional representations of the scene as viewed from the top and under orthographic projection [III.19].

### 3.1.3 Low-resolution thermal sensors

In the last decade, thermal vision systems based on low-cost IR array sensors provoked the interest of the researchers and became attractive in many AAL scenarios. These scenarios focus on

1. looking at people in their home for detecting possible emergency situations (see Figure 17)
2. monitoring indoor people's presence (occupation) in buildings
3. detecting stationary position or tracking moving people in a room or building

4. fall detection and detection of unusual activity for elderly persons in toilets, bathrooms, changing rooms, etc.

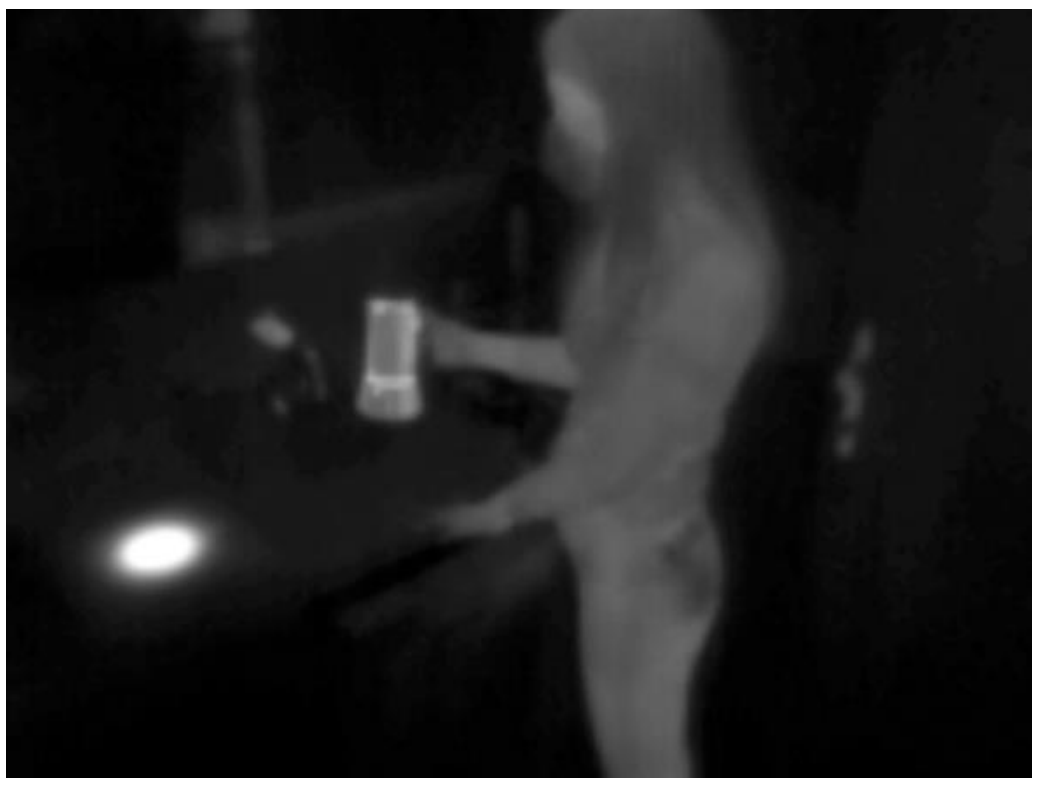

Figure 17. Raw thermal image showing a switched-on stove and a person in the kitchen

The use of low-resolution IR arrays comes with the following benefits: privacy preservation, low power consumption (passive device), low price, insensitivity to ambient lighting level and changes, operation in total darkness, fast response time, easy deployment and easy image processing. These sensors are less invasive and more convenient to use in indoor environments. The IR array sensors measure the heat generated from the human body or other objects and visualise it on a low-resolution IR matrix. Low-resolution IR thermal arrays typically have a resolution of 8×8, 4×16, 16×16 or 24×32 pixels. Several studies have proposed combinations of different sensing modalities, methods, processing techniques and machine learning approaches for detecting, localising and counting people inside rooms or buildings, utilising low-cost (8x8 pixels) IR sensor arrays set-up. In [III.20], Jeong *et al*. proposed a probabilistic method with image pre-processing and post-processing techniques for human detection using heat signatures from a low-resolution thermal IR sensor array system. Similar works have been presented by Basu and Rowe [III.21] and Trofimova *et al*. [III.22] reaching an accuracy up to 97%.

Apart from single human detection, low-resolution IR sensor arrays are used to estimate the flow of people by Mohammadmoradi *et al*. [III.23]. They utilise a threshold-based technique and temperature filtering technique and report an average accuracy of 93%. Maaspuro, in [III.24], studied an application of IR sensor array as a doorway occupancy counter using Kalman filter tracking algorithm. Detecting the number of people in an environment and tracking their position is also done by Singh and Aksanl [III.25]. They have demonstrated that with careful selection of algorithms up to 100% accuracy in detecting user presence could be obtained. In addition, fixing the sensor on the room ceiling yields the best results. They considered position tracking as static activity detection, where the user's position does not change while performing activities, such as sitting, standing, etc. By employing pre-processing on the data, including normalisation and resizing, and machine learning methods four static activities are detected successfully with accuracy up to 97.5%.

In the AAL context, several fall detection methods using thermal IR array sensors have been developed aiming at preserving privacy of people. In [III.26], Hayashida *et al*. captured falling using a thermal camera on the side and developed a heuristic algorithm to perform recognition. Spasova *et al*. in [III.27] present a novel fast, real-time and privacy protecting algorithm for fall detection based on geometric properties of the human silhouette and a linear support vector machine. For detection of humans, the proposed algorithm uses infrared and visible light imagery. A simple real-time human silhouette extraction algorithm has been used to extract features for training of the support vector machine. The authors reported achieved sensitivity and specificity of the proposed approach over 97%. As a continuation of [III.28], in [III.29] the researchers developed a Machine Learning (ML) algorithm to detect the fall of a person using a thermal sensor with resolution of 24x32 pixels. Figure 18 shows from different camera angles what images with this resolution look like. The algorithm is based on a pipeline composed of the following steps: pre-processing, feature extraction and classification. The collected dataset includes two sensor-mounting scenarios: overhead and sideways placement.

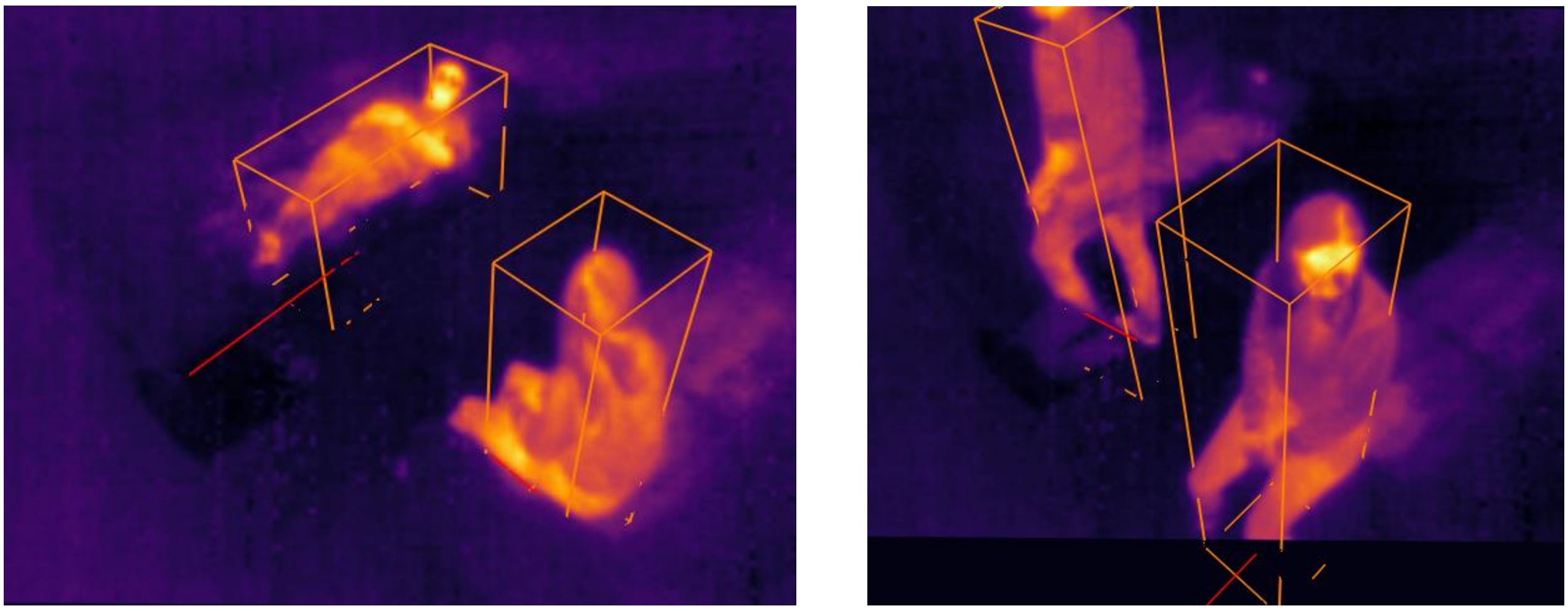

Figure 18. Example visualisations of raw IR data of a person lying on the floor (left) and standing (right)

### 3.1.4 Wearable cameras: first person or egocentric vision

As explored above, cameras can be placed in the environment, or, conversely, they can be placed on the user, i.e., as wearable cameras, mounted on a belt around the chest, or a lanyard; or as part of "smart glasses" or similar head-mounted devices. Methodologies dealing with footage from such cameras receive multiple names, such as "egovision", "egocentric vision" and also "first-person video" (FPV). One advantage of this video modality is that the torso of the person or other large objects present in the room do not occlude the action being performed [III.30]. In a setup with wall-mounted cameras, if the user opens a refrigerator door or another moving part of an appliance, furniture, this would cause those parts of the body of the user are occluded during the performance of the action. These types of occlusions can be avoided when cameras are worn by users. Furthermore, objects being manipulated as well as the hands are visible, which is an advantage when taking into account the fact that most ADLs involve the upper extremities.

Historically, methods for activity recognition from wearable cameras have been divided into “object-based” (context of visible objects being used to recognise the activity), and “motion-based” where physical features (magnitude, angle, frequency) of motions are used to determine the types of actions being performed. Both methods have disadvantages, as noted in the review by Nguyen *et al*. [III.30], namely, missing detections of relevant objects and types of actions involving small movements of the limbs implied in the action, respectively. It is worth noting, however, that both types are not mutually exclusive and can be combined.

The way these two proposed modalities of action recognition are combined can be via a hierarchical framework, as done by Betancourt *et al*. [III.31], in which “hand detection” or “motion pattern detection” are at a lower level of “basic scene understanding”, and higher levels correspond to subject-object interactions involving detections of relevant objects, and other contextual information. Nonetheless, this split between modalities was more relevant or clear before the advent of end-to-end differentiable (trainable) deep neural network methodologies (see next Section 0), in which all elements (subnets) of the model may contribute to the inference of the type of action being performed. Some branches of the model might be specialised in certain tasks, but the contribution to the final decision is not clear-cut, or at least, not as much as with previous methodologies. The reader is referred to a recent review by Bandini *et al*. [III.32], which shows examples of this.

Finally, for further reading of the latest advancements in action recognition for the purposes of lifelogging, from wall-mounted as well as wearable cameras, Climent-Pérez *et al*. [III.33] offers a literature review exploring existing video-based technologies in the AAL context, with a focus on methods whose outputs can be assembled into a lifelog for the user, that is then able to share it, according to their needs, with healthcare providers, caregivers or social workers of their choice.

## 3.2 Audio-based sensing technologies

Though CV has brought the largest advances in AAL, it often requires installation of video cameras at user’s homes, which may be considered intrusive and thus refused by the targeted population. Furthermore, video processing is extremely sensitive to light conditions that may vary substantially at different positions in a home [III.34]. In this context, audio-based technology can be seamlessly integrated into the user's environment and has a potential to become one of the most important modalities for interaction with AAL systems. It has a large range of sensing, does not require physical presence at a particular location and it is physically intangible [III.35]. Using voice is a more natural way of interaction than tactile interfaces. Despite these advantages, audio technologies still remain rarely deployed in real settings, partly because the audio analysis in the presence of ambient noise is a challenging task [III.36]. Nevertheless, with the recent advancement of audio technologies and their integration in mobile and Internet of Things (IoT) devices, the use of audio signals in AAL becomes feasible [III.37], with applications in human activity recognition (HAR) [III.34], [III.35], fall detection and fall prevention [III.38], [III.39], food intake monitoring [III.40], [III.41], emotional state recognition [III.42], [III.43], and so on.

Monitoring and behaviour analysis in the context of an acoustical environment can be realized these days by two different technologies, i.e., conventional microphones and their arrays or specific acoustical sensors. These technologies can be divided into two categories according to a frequency range, which they operate in, namely a human audible frequency range (20 Hz - 20 kHz) and a human inaudible frequency range (all frequencies outside of the human audible frequency range). When it comes to the specific acoustical sensors, they are represented by surface acoustic sensors, floor acoustic sensors, ultrasonic sensors and throat microphones or laryngophones. In the upcoming sections, the technologies deployed for monitoring and behaviour analysis in the context of an acoustical environment are described in more detail.

### 3.2.1 Microphone and microphone arrays

A standard microphone (Figure 19) is an electroacoustic device, which converts sound waves into an electrical signal. The sound waves produced by an audio source propagate through the air and cause slight changes in air pressure around the microphone, whose task is to detect them. The general principle has many similarities with a human ear, as they share the same objective. The sound waves produce motion of the outer part of a microphone (i.e., diaphragm). In the next part of a microphone, this motion is converted to an electrical signal proportional to the sound waves.

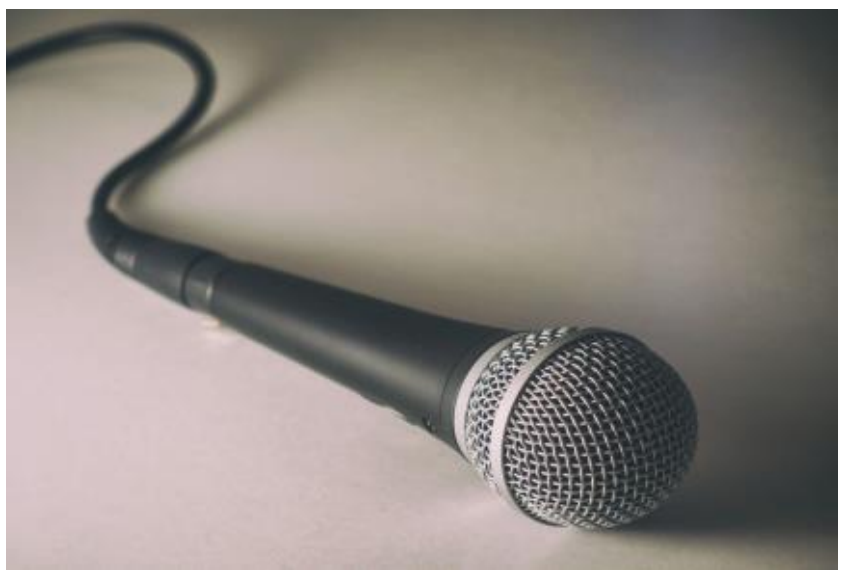

Figure 19. Dynamic microphone (photo: Martin Vorel, CC0)

The way a microphone converts motion into an electrical signal depends on its type. Two main types are dynamic and condenser microphones. Electromagnetic induction is used in dynamic microphones. The diaphragm is attached to an induction coil, which is placed in a magnetic field. The coil movement induces an electrical current representing the audio signal. The dynamic microphone design results in an inexpensive and robust device. The condenser microphone is based on the change of electric charge. The diaphragm is in the form of a plate of a capacitor. The diaphragm's movement alters the distance between plates and consequently changes the electric charge, which presents the audio signal. The condenser microphones have better sensitivity due to their design. They require a power source needed for the capacitor plates to operate.

The level of small air pressure changes also corresponds to the low electrical signal level produced on a direct output of a microphone. Thus, additional signal processing is necessary, where a separate amplification unit (i.e., preamplifier) is applied. The goal is to increase the level of the captured electrical signal without achieving the same effect on the coexistent audio noise signal.

Microphone polar pattern is an important parameter (Figure 20), which shows its directional sensitivity projected on a polar plane. It depends on the device design. Frequent polar patterns are omnidirectional, bi-directional, cardioid, subcardioid, hypercardioid, supercardioid, and shotgun. Which polar pattern is the most suitable one highly depends on the usage scenarios? Different AAL applications and systems prefer various microphone polar patterns.

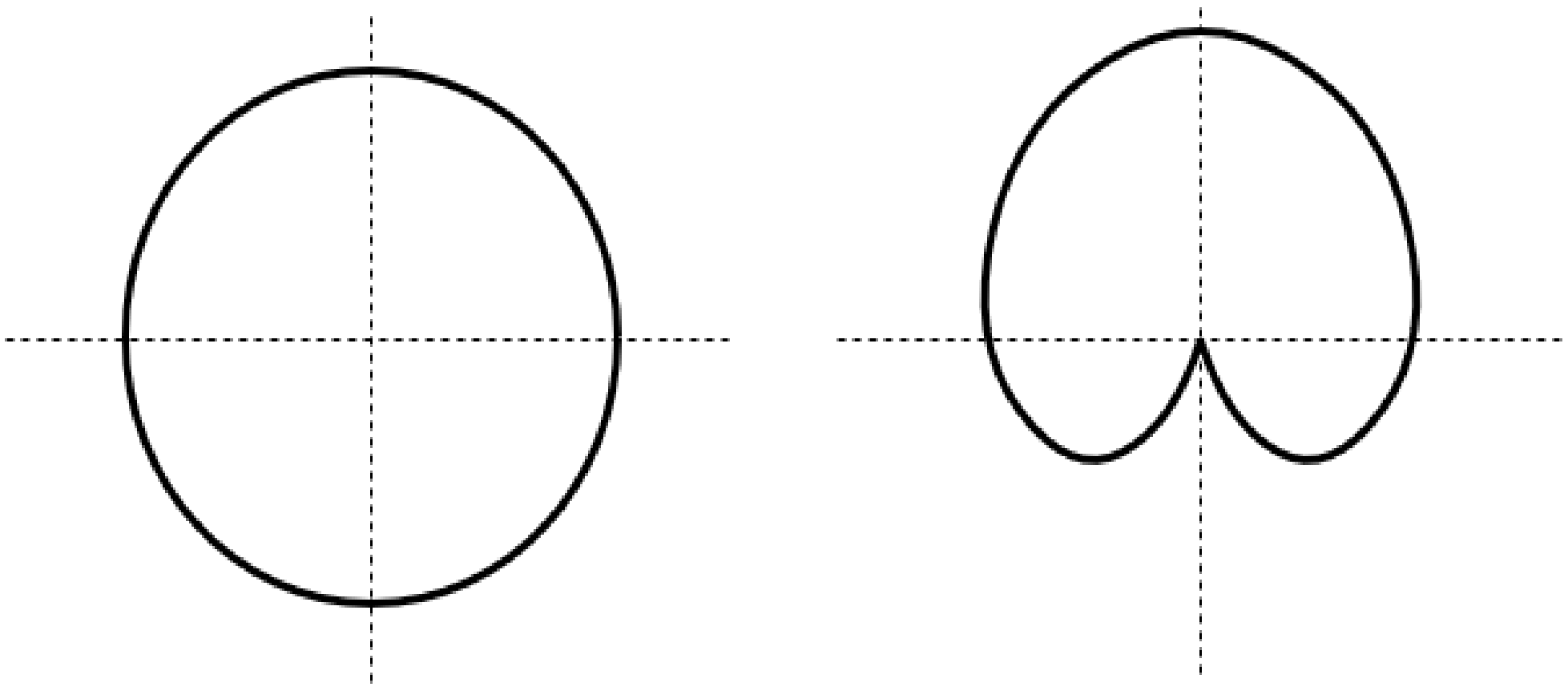

Figure 20. Omnidirectional (left) and cardioid (right) microphone polar patterns.

Another essential characteristic of a microphone is its quality. If advanced audio technologies are used in an AAL system, audio quality plays a significant role. Such examples are spoken language technology interfaces. Some parameters defining the microphone quality are a frequency range, sensitivity, level of total harmonic distortions, etc.

Microphones can be found in many devices, as speech presents a natural way of human interaction. One of the most omnipresent devices nowadays are mobile phones. They usually have one primary microphone and additional secondary microphones to improve the captured audio signal. In other scenarios, two microphones are also used for stereo recordings.

An audio device hosting more than one primary microphone is called a microphone array. Several microphones are operating in parallel, all capturing audio signals simultaneously, producing a multichannel signal on direct output. Microphones in an array are placed in different patterns, which also influences the task they are best used for. The raw audio signal needs to be processed to benefit from the array setup. Digital signal processing is carried out on a DSP board or a computer. Modules responsible for this provide the system with advanced functionalities such as virtual polar patterns, beamforming, source localisation, local echo cancelation, noise reduction, and so on so forth. Microphone arrays are essential for various AAL applications as they can better capture audio in a real-life environment.

### 3.2.2 Specific acoustic sensors

Surface acoustic sensors detect the mechanical waves that propagate through or on the surface of the solid materials. They typically use piezoelectric effect to convert the wave into an electric signal, such that changes made to the mechanical wave are reflected in the output electric signal. Surface acoustic sensors mostly work as passive sensors, which makes them power-efficient and reduces the production and operation costs. However, they are sensitive to coupling between surface material and the sensor and may require periodical sensor calibration. Surface acoustic sensors can be used for fall detection or for detecting the activities of daily living in a household that cause vibrations on the ground surface, such as object dropping, falling [III.44] or walking [III.45].

Floor acoustic sensors are special types of surface acoustic sensors used for detection of human falls. They are positioned on the floor surface, in the vicinity of the sound source and capture direct sound waves without reflection from the surface. They usually consist of an inner resonant container with a microphone located inside, that captures the vibration of the surface. A membrane at the bottom of the container touches the floor and enables acoustic coupling with the surface. Inner container is surrounded by an outer container (capsule) and acoustically isolated, to decrease the intensity of acoustic waves that propagate through air [III.46].

Ultrasonic sensors are piezoelectric transducers that actively send and receive ultrasonic pulses and measure the signal propagation time to and from the measured object. They operate by converting an electrical signal into mechanical vibrations and vice versa. Since the operation frequency is outside the human audible range (i.e., from 20 kHz to 200 MHz, typically at 40 kHz), they can be considered unobtrusive. Depending on the sensor and object properties, the effective sensor range is between several centimetres and several meters. The sensor that works at the frequency of 40 kHz can typically sense objects at a distance of up to 2m [III.47]. Ultrasonic sensors can be used for fall detection [III.48], detecting activities of daily living such as standing or sitting [III.49], or even in-air gestures [III.50] and respiratory rate [III.51] when operating in a close range mode.

Throat microphones, or laryngophones**,** are contact microphones that absorb audio vibrations generated by the larynx directly from the user's throat using the neck-mounted sensors. While standard microphones pick up sound vibrations that propagate through the air, capturing at the same time background noise, throat microphones capture only vibrations from the throat, ignoring background noise and wind turbulence. Therefore, they can be used in extremely noisy or windy environments (e.g., while riding a motorcycle). Moreover, they are able to capture whispers, making them suitable for use in situations where silent communication is required (e.g., military or law enforcement operations). Since they are positioned on the human’s neck, they can be used without obstructing helmets or respiratory protection. However, the nasal sounds and sounds generated by the tongue and lips are muffled. In the context of the AAL applications throat microphones are used for monitoring of chewing and swallowing [III.52] and in food intake monitoring [III.53].

## 3.3 Data processing and understanding for AAL

The data acquired by the video and audio-sensing layer are processed to perform specific tasks, such as detecting and recognising relevant objects, detecting movements and incidents, measuring vital parameters, or analysing gait. In this respect, AAL solutions relies on methods and technologies coming from several connected disciplines working on data processing and understanding as well as inference and problem solving, such as CV, Signal Processing, Internet of Everything and IoHT, Computational Physiology, Machine Learning and Artificial Intelligence.

As already sketched within the functional architecture of an AAL system shown in Figure 6, the typical pipeline for data processing can be summarised as shown in the Figure 21 [III.54], [III.55].

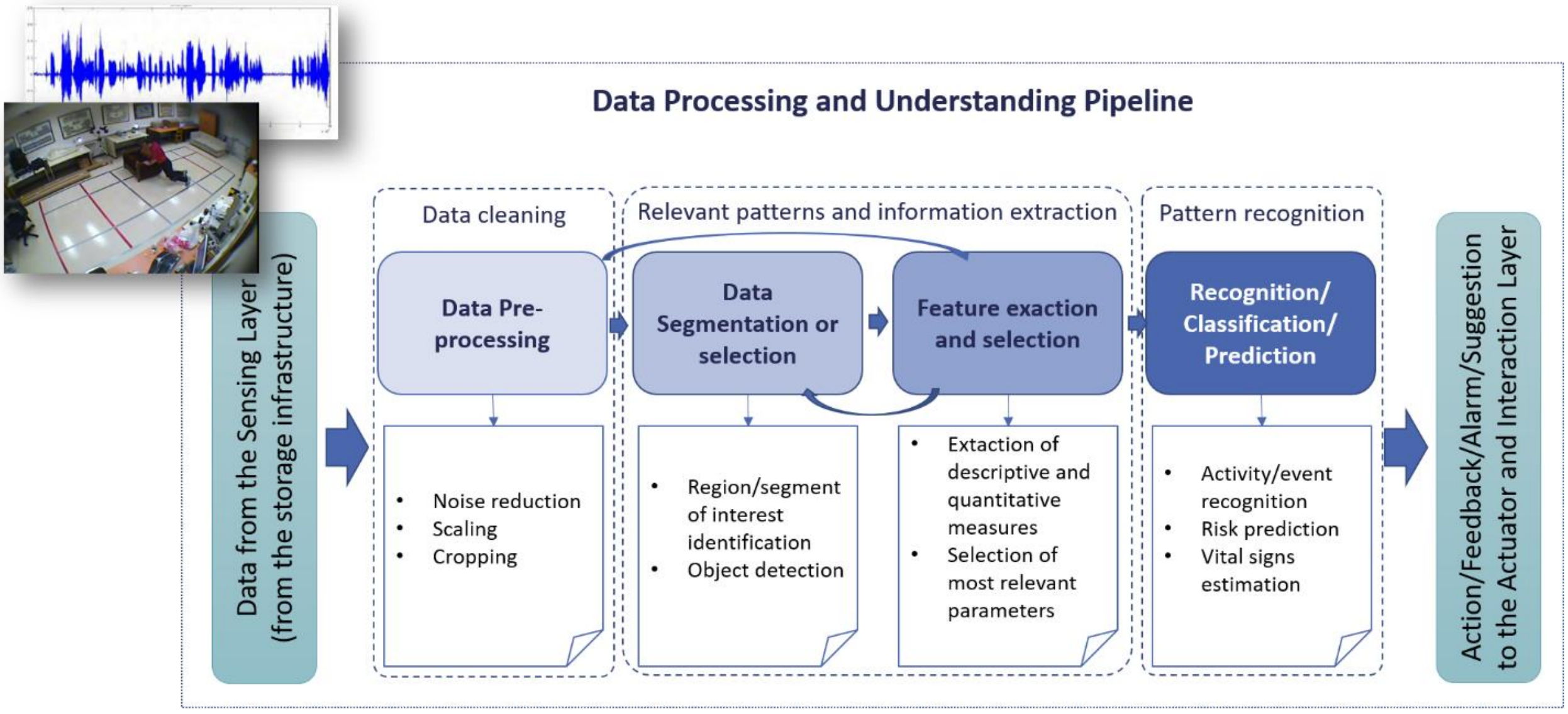

Figure 21. A simplified version of the video and audio-data processing pipeline

The data-cleaning step aims to improve the quality of the collected data by removing noise, harmonising the data or selecting specific parts of the data. This step is also usually referred to as a pre-processing step. To extract the relevant information or patterns from the data, a segmentation step is often performed. This consists in identifying and selecting only the regions of interest in an image (e.g., a person's face or silhouette) or a signal (i.e., a word or a sentence in an audio signal). Once the relevant patterns are selected, these are usually described by computing a number of parameters, called *features*, which express and measure significant characteristics of the pattern. Such features are also usually referred to as *hand-crafted* features, as they are selected according to the knowledge of task to be performed. In some cases, the features are extracted before segmentation to perform the segmentation itself. In other cases, the features are extracted without segmenting the data. The features are processed by "intelligent" algorithms that are able to recognise the detected patterns to perform a specific task, such as recognising a fall or a vocal command. In some easy tasks, this latter step can rely on simple rules, based on threshold applied to the features' values. In most of the cases, it requires more complex solutions based on advanced statistical or *data-inductive* models [III.56]. As said, this pipeline is a high-level and simplified vision of the data-processing chain. In practice, more iterations of the steps might be required (e.g., activity recognition may require person recognition and tracking) or each step should be further adapted to respect the data typology (e.g., when processing time-series or multimodal data). For instance, the last recognition step, based on data-inductive methods, might be required to recognise the relevant objects in an image or sentences in an audio signal.

Data-inductive methods are those methods able to *learn* directly from the data the significant pattern relationships that are relevant to perform a specific task. These methods gather under the discipline known as Machine Learning (ML). Among the several definitions proposed for ML so far, the most renowned one is by Mitchel and reads as follows "ML is the science that is concerned with the question of how to construct computer programs that automatically improve with experience" [III.57]. The peculiarity of ML methods is their ability to discover patterns and information from a set of data or examples. This ability enables them to learn autonomously how to recognise categories, perceive stimuli, predict trends, make decisions and apprehend behaviours. Simply speaking, ML methods are not specifically programmed to solve a task, but learn how to solve it based on the data available. ML is at the core of Knowledge Discovery and Data Mining as well as the most recent Big Data analytics.

In AAL, some of the most commonly used ML methods comprise Support Vector Machines (SVMs), $k^{th}$-Nearest Neighbour (k-NN), Artificial Neural Networks (ANN), Naïve-Bayes (NB), Decision Trees (DTs) and, among the latter, the Random Forest (RF) [III.58]. The listed methods are all supervised methods, which are trained by providing them with the expected outcomes (e.g., the expected class for a given example). SVMs and k-NN are instance-based methods, as they group data into sets of similar cases and any new instance of data is compared (using a similarity measure) against the representative case of each set to find the best match and make a classification or prediction. NB is a Bayesian method that requires the knowledge of a-priori and conditional probabilities of the data to apply the Bayes' Theorem. DT and RF are tree-based methods, which build a sort of decision-making diagram, in the form of a tree, based on conditions on the data values. The prediction for a new instance of the data is obtained following the tree structure until a leaf is reached. The RT is an ensemble version of DT, which combines a huge number of DTs to solve a task. ANNs are connected networks of nodes, whose inspiration comes from the biological networks of neurons. ANNs are structured in layers of nodes (neurons) and the output of each layer goes in input to the following layer. Each node applies an *activation function* to the weighted sum of its input. Moving from the first layer to the last one, an ANN maps the input data to the output label of a class [III.59].

In the last decade, boosted by the data deluge, ML has made significant steps forward. A particular type of ANNs, the so-called Convolution Neural Networks (CNNs), has demonstrated to perform unprecedentedly well, especially when solving perception tasks, such as vision, object recognition and natural language processing [III.60]. Considering the depth in terms of number of layers of this type of networks, the term Deep Learning (DL) has emerged when referring to them. DL has emerged in opposition to the shallow learning, which is the traditional learning approach of ML methods. DL methods are able to learn the main features that allow for the distinction of the relevant patterns, in this respect they are also referred to as Representation Learning. More precisely, DL methods and CNNs permit to simplify the data-processing pipeline seen before in Figure 21, as they do not necessarily require to segment the data and extract significant features, but they directly process the input (pre-processed) data to provide a result. In practise, they do not require the extraction of handcrafted features, as they identify on their own the relevant structures in the data and the characteristics of these structures. Being deep, CNNs and DL methods require seeing many data (i.e., a lot of examples) in order to be accurate and effective, as they need to set the values of a huge number of internal parameters (namely, the network architecture and the parameters of training process). Most of the more recent works on video- and audio-data processing for AAL rely on the use of DL methods as they ensure very high performances (see [III.61] for an extensive review).

The various methods strongly depend on the task that should performed (e.g., fall detection, vital signs monitoring, activity recognition, …), as partly illustrated in the previous Section 3.1. For this reason, revising the various solutions proposed in the literature is a too broad corpus of work for the scope of this section. The most recent advances will be extensively presented in the corresponding section of the following Section 4.

In the following, we discuss briefly some of the most common challenges that affect the data processing methods for video and audio understanding.

### 3.3.1 Main challenges for data processing

The previous sections on visual and aural data sensing have already highlighted some of the technical challenges related to each specific data modality. In this section, the most cogent challenges related to the task of interpreting the sensed data are discussed [III.56], [III.62]:

- **Lack of available datasets:** this has been affecting the AAL for years, until the recent few years when some datasets are being publicly appearing and shared within the research community [see GoodBrother report from WG4]. Nevertheless, this is true for video-based AAL, while for applications based on audio the challenge is still there. Moreover, the datasets available either are of limited size or are obtained by simulating real-life events (e.g., falls are simulated in the lab or using dolls in case of audio-based datasets). Therefore, it is necessary to develop systems that are able to work with insufficient training data, using techniques based on oversampling, semi-supervised learning, anomaly detection or one-class classification
- **Need for data annotation and labelling:** another problem that affects data is the availability of quality annotation that can be used as ground-truth to train the ML methods. In this respect, unsupervised or semi-supervised techniques might be useful to overcome the issue, as they are based on ML methods that autonomously discover how data gather into homogeneous classes [III.59]. Unsupervised methods are particularly beneficial when the problem is framed as an anomaly detection problem, such as, for instance, when detecting an abnormal usage of home appliances, detecting a fall, or detecting the worsening of a vital parameter tracked over time. Nevertheless, in this respect, such methods may require a huge amount of "normal" data representing all the possible normal situations, which might be difficult to obtain in experimental settings. This can cause the method to produce a high number of false alarms, when deployed in practice

- **Class imbalance:** risky or adverse events are usually and fortunately less common that normal conditions. This means that the most relevant events that are to be detected are usually less frequent and represented in the datasets. This may challenge the recognition methods, unless specific approaches are adopted. These may rely on data augmentation, class weighting techniques or anomaly detection methods
- **Privacy concerns:** audio and video data acquired for AAL purposes are negatively perceived by the assisted person with respect to privacy preservation. The feeling of intrusion also depends on the location of the audio and video sensor. Privacy-preserving methods based on specific processing techniques as those introduced in section and WG2 report are highly desirable.
- **Ethical concerns:** most of the newest methods to understand the content of audio and video data in AAL are based on DL. These methods ensure very high performances in terms of accuracy in the recognition, prediction and classification tasks. However, they can be very complex models, with a huge number of layers and complex computing modules, whose inner functioning is not easily understandable. For this reason, they are often referred to as "black boxes" or "opaque models". The European Commission and the High Level Expert Group appointed by the Commission have produced important guidelines to steer an ethical development of DL and AI models, especially when the impact of such models may affect directly citizens, as in the eHealth and AAL domains [III.63]. One of main principles promoted by these guidelines pertain to the transparency of decision based on AI and DL methods, which directly entails the possibility to explain why the methods have provided a certain output. This has led to a renewed interest the so-called "explainable AI". Several methods have been proposed in the literature and will be further revised in the Section 0. Explaining the results from an AI method is the first step to make this method trustworthy and, hence, promote its acceptability by end-users.

## 3.4 Multimodal data fusion for AAL

In this section, we introduce two technological approaches for fusing multimodal sensing data. Firstly, we survey how thermal and depth data can be combined. Secondly, we show how radar is used for data fusion.

### 3.4.1 Combining thermal and depth data

Depth and thermal sensors have advantages over traditional cameras in application fields such as AAL, which involves continuous monitoring of people that should be unobtrusive and privacy-preserving [III.64], [III.65]. This is because depth and thermal sensors require no external illumination, working even in darkness, and do not expose colours or textures, making it harder to identify people. Images of depth sensors encode the scene geometry whereas thermal images encode surface temperatures. Both technologies have been described separately in Sections 3.1 and 3.2, but they are also complementary for solving person-centric vision tasks, as people are clearly visible in thermal images (see Figure 22), facilitating detection and segmentation. Yet there is little research on this matter, presumably because thermal sensors that are inexpensive and produce images of sufficient quality have become available only recently.

Pramerdorfer *et al*. [III.66] address this research gap by first assessing the data quality of two common off-the-shelf depth and thermal sensors and deriving empirical noise models. Their main contribution is a method for synthesizing realistic depth and thermal images that utilizes 3D data modelling and the derived sensor noise models. This method is employed to create a dataset of 40k image pairs.

The value of synthetic data can be demonstrated by training CNNs [III.67] for scene state classification and evaluating their generalisation performance on real data. Pramerdorfer *et al*. [III.66] present their own dataset comprising 8k image pairs. Both datasets were made publicly available, with the aim to promote further research on (synthetic) depth- and thermal-data-based vision tasks[5].

The authors of [III.65] were among the first to highlight the potential of combining synthetic depth data and machine learning, by example of human pose estimation. [III.64] and [III.68] follow a similar approach for fall detection and hand pose estimation, respectively. Improving the realism of synthetic data and consequently generalisation to real data requires sensor noise modelling. Khoshelham [III.69] derived a basic noise model for the public Kinect depth sensor, which was extended by [III.70] . There are few works that utilize thermal data. [III.71] use a thermal sensor for CNN-based fall detection. Kniaz *et al*. [III.72] use a GAN to generate thermal images from colour images. The approach of Pramerdorfer *et al*. [III.66] requires no colour images and, based on visual inspection, appears to produce more realistic images. They state that the only works they are aware of that utilize both depth and thermal data are [III.73] and [III.74]. Both cover person tracking using particle filters in real data, while [III.66] focus on synthetic data and image classification using CNNs.

There is currently no hardware available on the market that combines depth and thermal sensors. Thus, [III.66] designed and assembled such a device on their own using off-the-shelf sensors and 3D printing. Figure 23 shows an illustration of a sensor combining RGBD and thermal data.

[5] https://cvl.tuwien.ac.at/research/cvl-databases/sdt-icip/ (last accessed: 23/02/2022)

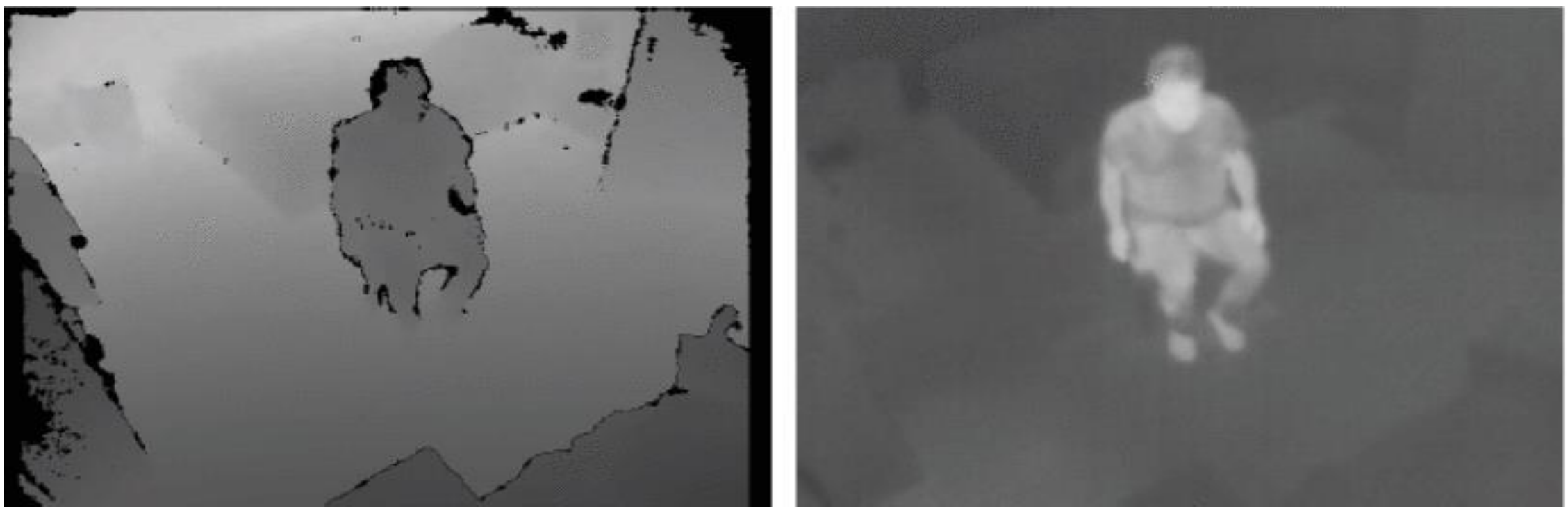

Figure 22.  Depth (left) and thermal (right) images of a scene (Image from [III.66])

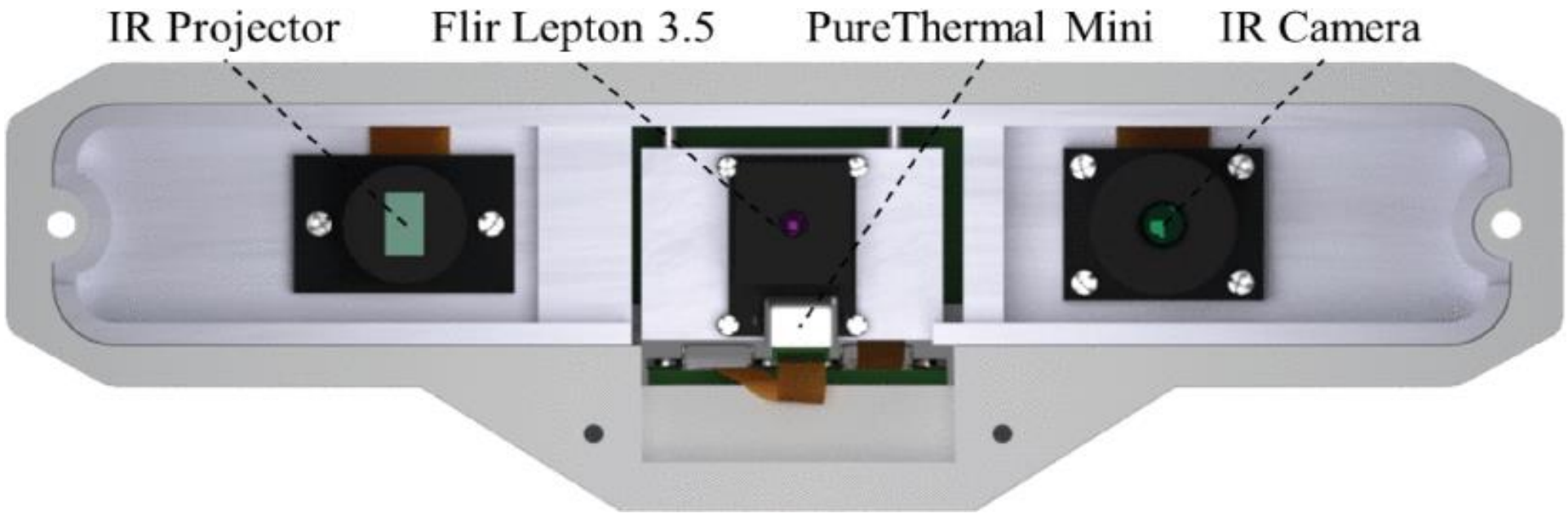

Figure 23. Illustration of a Depth-Thermal Sensor Unit and its arrangement of internal components (Image from [III.66])

## 3.4.2 Radar data visualization and fusion

In addition to the typical image-based technologies such as RGB, depth or thermal (see Section 3.1), radar data can also be visualized like an image, using point clouds to illustrate the rough shape of a person (see Figure 24). Cippitelli *et al*. [III.75]  present a review on radar and RGBD sensors for fall detection in which challenges of both individual technologies are investigated.

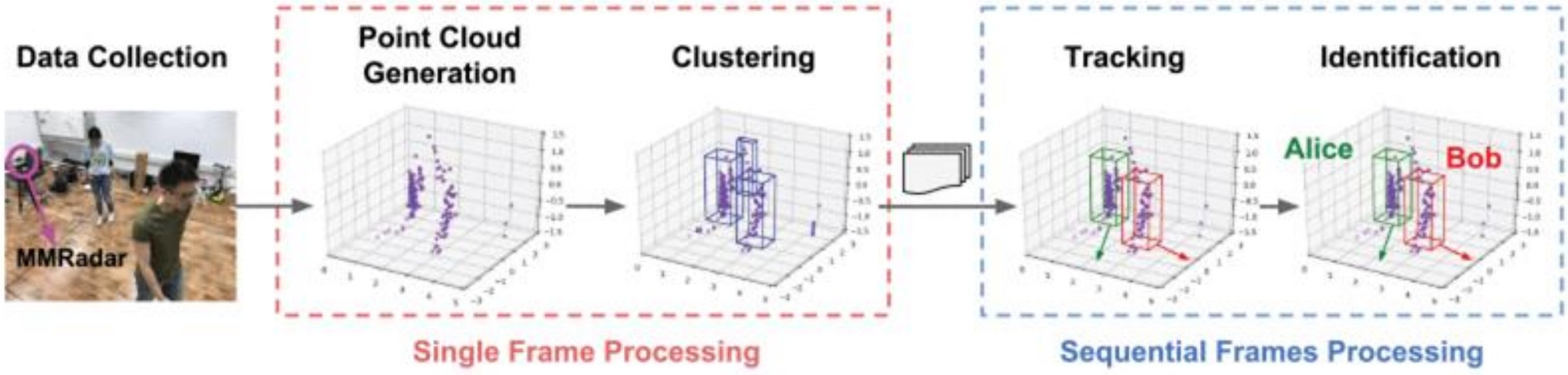

Figure 24. Point Cloud Example of two persons detected by an mmWave radar (image from [III.76])

Challenges related to radar technology include:

- strong scatterers and clutter indoors, generating multi-path and ghost targets or obscuring the person to be monitored; can also be a problem in RGBD
- pets can cause false alarms, can also be a problem in RGBD
- compliance of selected radar wave forms with telecommunication directives, leading to constraints in band width and transmitted power
- location of radar sensor can change attenuation of radar doppler signature; less problems on the ceiling than on the wall

Application areas of radar vary from multi-human detection [III.77] [III.78], gesture recognition [III.79] [III.80] [III.81] [III.82], localisation [III.83], to sleep and vital signs monitoring and much more. Unlike (standard RGB) cameras, radar is able to detect falls in a privacy-preserving way. This facilitates the use of radar technology in the end user's homes as well as in care facilities and hospitals. Moreover, radar allows to measure vital signs of new-borns and to detect burned people due to its non-invasive nature [III.84].

Diraco *et al*. [III.85] achieve sensitivity and specificity greater than 97% and 90% in fall detection by using a micro-motion signature and unsupervised learning. The highest accuracy is achieved in correspondence of ADLs/postures without too much movement (not further specified), such as, sleeping/resting, post fall, and watching TV. Thus, activities like cooking (standing posture) are more difficult to detect in comparison to the other ADLs. The same applies, although at a lesser extent, in the case of the eating activity, due to some occurrence of chest oscillations. Some differences are found also in dependence of the monitored subject's orientation. When more people are present in the sensor's FoV (in addition to the monitored subject), the movement compensation strategy is robust enough as long as the distance between the monitored subject (i.e., the person closer to the radar) and the other people is greater than 0.5 m.

Erol *et al*. [III.86] use range information integrated with a fall detection algorithm to distinguish an actual fall from a sitting motion. Both movements cause a high Doppler frequency, so distinguishing between them leads to a reduced false alarm rate. Although the Doppler time-frequency signatures of the two motions can be similar, the range extent of a fall is considerably higher than that of a sit. This varies according to the type and the depth of the base of the furniture the person is sitting on.

# 4. AAL applications: recent advances in successful assistive and supportive functions

Numerous AAL applications and systems have been developed in the last couple of decades addressing diverse of the scope seen in Section 0. The researchers and developers have attempted to address a large range of issues faced by those in need as well as their carers. This section discusses different application scopes or functions that are highly relevant to the activities of GoodBrother as based on audio and video data and covered by a big corpus of research.

We initially examine lifelogging solutions in conjunction with vital signs monitoring technologies, emotional state recognition and food intake monitoring. We then focus on activity and behaviour recognition along with activity assistance, gesture recognition and fall detection. Looking at more serious conditions, we then examine frailty recognition solutions along with mobility and gait analysis and motor rehabilitation technologies.

## 4.1 Lifelogging, quantified self and self-monitoring

Advances in wearable computing, with a myriad of products in the market (e.g., wearable cameras and smart watches, wristbands and glasses), increased functionalities of mobile devices and apps for health and wellbeing, and easier installation of cheaper home automation systems are supporting the adoption of remote healthcare and assisted living services by a larger population. For instance, *lifelogging* (also known as *quantified self* or *self-tracking*) technologies may enable and motivate individuals to pervasively capture data about them, their environment, and the people with whom they interact [IV.1], [IV.2]. Acquisition and processing of physiological signals (e.g., heart rate, respiratory rate, body temperature, and skin conductance), motion and location data, performed activities, images seen, and sounds heard, are the basis for the provision of a variety of cutting-edge services to enhance people's health, wellbeing, and independence. Examples of these services include personalised healthcare, wellness monitoring (physical activity, dietary habits), support for people with memory impairments, improvement of social participation, mobility, support to formal and informal caregivers, predictive systems (against decline in cognition, aggressive behaviours and for fall prevention) [IV.3]-[IV.8].

Lifelogging represents a phenomenon whereby individuals can digitally record their own daily activities in varying amounts of detail and for a variety of purposes. In a sense, it represents a comprehensive black-box of a person's activities and offers great potential to mine or infer valuable knowledge about their life. Typically, early adopters of lifelogging considered it to be an activity that was engaged in by the individual for their own benefit, and many would even counsel against sharing lifelog data. However, if lifelogging becomes more pervasive, one can imagine that many users would be willing to share aspects of their lifelog. Although there are many definitions in literature, lifelogging may be understood as *a form of pervasive computing which utilises software and sensors to generate a permanent, private* *and unified multimedia record of the totality of an individual's life experience and makes it available in a secure and pervasive manner*. This definition includes the idea that the process of lifelogging should consist of not only data gathering, but also storage, analysis and access. In the following three sections, more details are provided about the recent advances in specific lifelogging axes, namely, assessing, tracking and logging vital signs (Section 4.2), emotional state (Section 4.3), and food intake (Section 04.4).

Here it is worth to note that, in contrast to the advances in lifelogging technologies development, the knowledge about the human factor regarding the willingness to adopt such technologies and to be supported by digital services is still considerably underdeveloped [IV.9]. This lack of understanding has significantly reduced the transfer of these developments to innovations having a social and economic impact. Health technology, especially in the homecare and rehabilitation sector, in which a less technology-savvy and diverse population is the target group, can only fully deploy its huge potential for greying societies, if ergonomic, usability, and acceptance issues are adequately considered. As healthcare technology is increasingly integrated in private spheres and captures highly sensitive personal data, these developments may cause concerns about privacy and loss of control.

In May 2018, the General Data Protection Regulation (GDPR) established by the European Commission, entered into application. This legislation sets the obligation for technologies to meet the principles of data protection by design and data protection by default. Hence, lifelogging technologies must consider privacy by design methodologies in order to protect the fundamental rights not only of the users but also of other people interacting with them or by-standers, particularly if video or audio data are captured, which are easily understandable by anyone [IV.10].

## 4.2 Remote monitoring of vital signs

The remote monitoring of vital signs is a key step for assessing the health status of an individual or tracking over time the evolution of a disease. Vital signs, such as heart rate, heart rate variability, respiration rate, and blood oxygen saturation, carry relevant information to understand the patho-physiological conditions of non-communicable and chronic diseases, but also to derive clues on the overall condition of an individual, also in terms of wellness, fitness and attention.

Conventionally, the monitoring of cardiovascular and respiratory activity is achieved by using adhesive sensors, electrodes, leads, wires and chest straps, which may cause discomfort to the monitored individual. In some cases, when used for long periods, these sensors may cause skin damage, infection or adverse reactions on people with sensitive skin. An additional issue comes from an economic standpoint, as the monitoring leads and electrodes are supposed to be for single use only and require disposal.

In the last decade, there has been a growing interest in low-cost, non-contact and pervasive methods for measuring physiological information, as already introduced in the previous section on lifelogging (Section 0). In this respect, video- and image-based as well as audio-based methods have opened new perspectives. They enable the assessment of multiple parameters without any contact with the monitored individual, thus being beneficial for a long-term, contactless, continuous and comfortable heath monitoring. In this respect, several methods have been developed in the last years to ensure unobtrusive care experience in clinical or long-term care environments, to advance medical diagnosis and prognosis, and to manage well-being and chronic diseases thus guaranteeing a high quality of life. Overall, contactless vital signs monitoring has a broad range of applications and it is being also proposed in working and driving settings, to prevent accidents [IV.11].

Various review papers have overviewed the diverse contactless methods, approaches and applications presented so far to assess cardio-respiratory vital signs. One of the most recent and comprehensive ones introduces the driving principles of imaging-based measurements and surveys the approaches presented until 2016, by categorises them according to the imaging-sensor modality [IV.12]. The authors of [IV.13] carried out a systematic review of the current availability and performance of image-based monitoring methods. In the last couple of years, two review papers have been published on the topic. They both reviewed the works done by using cameras ranging from the visible to infrared (IR) [IV.14], [IV.15].

So far, no attempts have been made to comprehensively survey the wide range of contactless methods proposed in the literature to evaluate vital signs based on video and audio data.

In the following subsections, we firstly overview the most common vital signs that are measured within AAL applications and discuss the diseases that are most commonly addressed in this respect.

Afterwards, we survey the methodological approaches used to compute the various signs, by firstly considering the most common ones and, then, moving to non-conventional parameters.

### 4.2.1 The most common vital signs

The vital signs that are typically monitored using vision-based systems include respiratory parameters, cardiac parameters, blood parameters, body temperature and facial characteristics. Some works seek to monitor vital signs to monitor a specific disease or patient status, for example, post-operative monitoring, while other works do not relate the vital signs monitoring to any specific situation. In the following, an overview of the works done for each vital-sign category is provided.

#### Respiratory Parameters

The most common respiratory parameter that is monitored remotely is the breathing, or respiratory, rate (RR) [IV.15] [IV.16] [IV.17], typically in terms of breaths per minute. In some works, respiratory rate is estimated by specifically monitoring the nasal airflow or thoracoabdominal motion. Some works make use of the thoracoabdominal motion or other means, to estimate the tidal volume (TV) which is the amount of air that is inhaled or exhaled in or out of the lungs, respectively, with each respiratory cycle [IV.15].

### Cardiac Parameters

One of the most important vital signs is the heart rate (HR) which refers to the number of cardiac beats per minute, which is typically measured in beats per minute [IV.12]. Some works estimate the heart rate by modelling and denoising the blood volume pulse (BVP) signal [IV.18]. Another cardiac parameter that is often measured remotely is the heart rate variability (HRV), which is the beat-to-beat variation in the time interval between consecutive heartbeats, typically measured in milliseconds (ms) [IV.19]. Remote vision-based methods may also detect various cardiac arrhythmias [IV.20].

### Blood parameters

$SaO_2$ is the arterial oxygen saturation, typically measured by an invasive arterial blood gas test. Peripheral arterial oxygen saturation ($SpO_2$) is an approximation of $SaO_2$, obtained by means of a pulse oximeter, which is typically clipped to a peripheral body part, such as an ear-lobe or fingertip, to non-invasively measure the peripheral oxygen saturation by shining two light-emitting diodes having different light wavelengths into the body. These wavelengths are absorbed differently by blood containing oxygenated and deoxygenated haemoglobin thus resulting in different amounts of transmitted light for each wavelength. The ratio of these transmitted signals, corrected for the pulsatility of the signal, is used in Beer-Lambert's law to estimate the $SpO_2$ [IV.21]. The extracted pulsatile signal also provides a plethysmograph waveform. This principle is exploited in vision-based vital signs monitoring for image photoplethysmography (iPPG), which exploits the ambient visible light to estimate peripheral arterial oxygen saturation ($SpO_2$) [IV.21]. Image photoplethysmography is also referred to as remote photoplethysmography (rPPG). Tarassenko *et al*. [IV.21] explain that despite visible light has limited penetration in skin tissue; pulse oximetry based on light reflectance from the forehead has been shown to have a better correlation with the measurement of oxygen saturation by arterial blood gas than finger-probe transmittance pulse oximetry.

Image photoplethysmography has also been shown to have lower Signal-to-Noise Ratios (SNR) for the pulsation signal when the free flap is disconnected from the *blood supply* compared to when it has a good blood supply. This indicates that iPPG is a potential method for non-invasive monitoring of postoperative free flap in breast reconstruction surgery. Similarly, a reference pulsatile blood flow signal is computed over a large skin region and the time shift between the blood pulse wave extracted in a local region with respect to the reference signal, thus generating a spatial distribution of the time shift permitting the analysis of microcirculatory flow patterns [IV.22].

### 4.2.2 Most Common Diseases for Video-Related Assisted Living Technologies

This section details the diseases most commonly studied in publications that discuss remote monitoring, more specifically by video or audio methods.

The elderly population is a very popular study population for assisted living technologies [IV.22]- [IV.25]. In general, life expectancy is increasing in most countries, especially within Europe, and with birth rates not increasing at the same rate, populations are ageing, placing increased strain upon the care facilities available in many countries [IV.26], [IV.27].

Although advancing age is a natural life process and not per se pathological, it is a fact that a significant risk factor for many pathologies is age [IV.28]. Elderly patients are at higher risk of heart disease, cognitive impairment, mobility issues, neurological conditions and several other problems [IV.29]. Video monitoring of patients can enable them to be monitored by relatives and healthcare workers while they continue to lead normal lives inside their homes, with the added security and peace of mind that any issues will be caught early and aid can be sent in a timely manner if required [IV.30]. Video monitoring can be used to monitor both patents' vital signs, especially in those known to have comorbidities which may cause unpredictable and dangerous changes in parameters, as well as to monitor activities and general wellbeing of the elderly person, and provide assistance in case of a fall or in case a potentially hazardous situation is noted in the person's surroundings (such as a pot left forgotten on the stove in the case of someone with mild cognitive impairment) [IV.31], [IV.32].

Arrhythmias are abnormalities of the heart's beat with regards to rate or rhythm. The heart's beating pattern may be too fast, too slow or simply not regular in periodicity [IV.33]. They are another class of diseases that have been widely studied in relation to video monitoring [IV.34]-[IV.36]. Remote photoplethysmography has been used to obtain satisfactorily accurate results for heart rate and rhythm monitoring, both in laboratory settings and in hospital populations [IV.37]-[IV.39]. Arrhythmias are common disorders, with Atrial Fibrillation (AF) being the most widespread of them and resulting in a significant burden of morbidity and mortality [IV.40], [IV.41]. This arrhythmia is caused by disorganised contraction of the upper chambers of the heart, the atria [IV.42]-[IV.44]. Thromboembolism is a worrying result of AF, as blood pools in the atria as a result of their disordered movement and forms clots which then shoot out into the pulmonary or systemic circulation, causing ischaemic events such as strokes [IV.45]. More recently, evidence has also been found suggesting that the occurrence of strokes in itself perpetuates AF [IV.46]. It is estimated that the risk of AF during the average person's lifespan is between 22 and 26 percent [IV.47]. AF, like many other arrhythmias, often occurs in episodes, which may be short-lived and not produce any noticeable symptoms [IV.48]. Detection therefore requires prolonged monitoring, which at this time is achieved by using twenty-four hour or week-long Holter monitoring [IV.49],[IV.50]. Although this is a great improvement over having to admit patients to hospital for monitoring, it can be uncomfortable to wear the device for prolonged periods of time, and any detachment of leads or other technical issues will cause a loss of data, reducing the test's sensitivity [IV.51]. Data collected this way is also retrospective in nature [IV.51].

Blood pressure can also be monitored by remote wireless monitoring [IV.52]. To date, prolonged monitoring of blood pressure would require the patient to be admitted to a hospital or wear long term monitors which may be uncomfortable [IV.53]. Patients may also suffer from white coat hypertension, meaning that anxiety in medical environments is the cause of their elevated blood pressure [IV.54]. Allowing patients to perform their usual tasks while being monitored in a remote, non-contact manner enables a more realistic approach, with potential triggers for hypertension being discovered and white coat hypertension potentially eliminated, enabling a much better treatment plan suited to the patient's needs to be drawn up [IV.55].

Epilepsy is a disorder characterised by chaotic electronic discharges of the neurones of the brain, resulting in seizures which may be partial, manifesting themselves as uncontrollable movements in a part of the body, complex partial, which include alteration in conscious level, or generalised, usually with loss of consciousness and generalized convulsions [IV.56]-[IV.58]. The aetiology is not fully known, and the disorder affects up to one percent of the population worldwide, making it one of the commonest chronic neurological disorders [IV.59]. Patients may be aware that a seizure is impending or they may not [IV.60],[IV.61]. Video monitoring of seizures is important both in diagnosing the specific type of seizure that is occurring, measuring its duration, and instituting appropriate help if required [IV.62]. Specialised treatment with drugs such as benzodiazepines is required for seizures lasting longer than five minutes in order to prevent potential neurological injury – this condition is termed status epilepticus [IV.63]. Video monitoring of epileptic patients may therefore allow help to be readily available in a situation when the patient is rendered unable to call for it himself.

Mobility disorders, the most commonly studied being Parkinson's disease (PD), are ideal for video monitoring as this enables clinicians to recognise for signs and symptoms of the disease such as bradykinesia (slowness of movement), rest tremor (fine trembling of the hands while at rest) and hesitancy (difficulty commencing actions such as getting out of a chair) [IV.64]-[IV.67]. Patients with PD may suffer from side effects of medications such as dopamine agonists prescribed to lessen their symptoms, such as on-off effects and dyskinesia (uncontrollable muscle movements) [IV.68],[IV.69]. These problems can also be monitored via video, enabling the clinician to diagnose the problem and discuss remedies with the patient in the comfort of their own home, as many such patients find traveling to unfamiliar clinics particularly challenging [IV.70]. Labile parameters (unexpected changes in vital signs not explained by physiological processes) are also a symptom of PD and continuous monitoring of parameters can help in early recognition of dangerous variations [IV.71],[IV.72]. Other neurological conditions such as cerebrovascular accidents (CVAs), commonly known as strokes, are also extremely common and patients suffering from these conditions may also benefit from remote assisted living in similar ways [IV.73],[IV.74].

Renal haemodialysis patients are a population targeted in some studies, although these studies tend to focus on cardiorespiratory parameters rather than renal function [IV.21], [IV.75]. This population may be ideal for studies since most haemodialysis treatments take several hours, allowing time for ample data collection [IV.76]. Patients usually have underlying comorbidities such as hypertension, diabetes and related heart conditions, and dialysis in itself may cause variations in vital signs, but overall, these patients are mostly stable during treatment and are able to move in the bed and perform some activities such as carrying out conversations and using devices [IV.76]-[IV.79]. Larger movements such as walking, however, are not generally possible during a dialysis session [IV.76].

Obstructive sleep apnoea is a respiratory disease commonly studied in relation to non-contact video monitoring [IV.80]-[IV.82]. Obstructive sleep apnoea is a condition strongly associated with obesity in which obstruction of the airway occurs when the patient is unconscious [IV.83],[IV.84]. Multiple episodes of partial awakening are observed leading to disrupted sleep, excessive daytime somnolence with its associated reduced productivity and general quality of life, increased tendency to traffic accidents, and more long-term, increased risk of cardiovascular disease among others [IV.85]. The diagnosis for this condition at present relies on polysomnography, which measures blood oxygen saturation, heart rate, brain waves and movements while the patient is asleep; it is usually performed in equipped clinics, but a version with reduced monitoring may be performed at home [IV.86]. There are issues with patient discomfort, with some patients having difficulty falling asleep in an unfamiliar environment. Any equipment dislodgement during the test period will also affect the accuracy of the result [IV.87]. Video monitoring provides an alternative, allowing patients to remain in their usual environment which is conducive to more realistic results, and eliminating the issue of equipment becoming displaced when the patient moves [IV.88].

The conditions described above are the commonest studied in published papers related to video monitoring in a remote manner; many other diseases may be studied by these means in future enabling many more patients to benefit from such technology. Some of them rely on the recognition and monitoring of motion and activities and will be discussed in the forthcoming sections (e.g., Section 0 "4.5 Activity and behaviour recognition"). In the following, an overview of the computational approaches to estimate diverse typologies of vital signs are overviewed.

### 4.2.3 Vital signs estimation approaches

As introduced in the previous sections, various types of vital signs can be estimated remotely, by using video and imaging data. In the following, an overview of the most common approaches to estimate each type of parameters is reported, framing the approaches in a global picture and, then, summarising the most recent works in the field. More precisely, the types of vital signs considered are:

- cardio-related vital signs (Section “Cardiac parameters”)
- respiratory-related vital signs (Section “Breathing signs”)
- blood-related parameters (Sections “Blood oxygenation” and “Blood pressure”)
- energy expenditure (Section “Energy expenditure”).

### Cardiac parameters

Cardiovascular signs, most commonly HR and HRV, can be assessed and measured via several contactless modalities that take advantage of magnetic induction, the Doppler Effect, or imaging data acquired via a thermal or a video camera (Figure 25). These modalities are conceived to observe physical and physiological variations including skin colour, temperature, impedance changes, head motion, arterial pulse motion, and importantly, thoracic and abdominal motion due to the activity of the cardiovascular systems. In this report, we concentrate on methods based on imaging data, while encouraging the reader to refer to more extensive reviews, such as [IV.12], for an overview of the other computational approaches (e.g., the Doppler-based ones) as well as of the physiological phenomena that enable the contactless assessment.

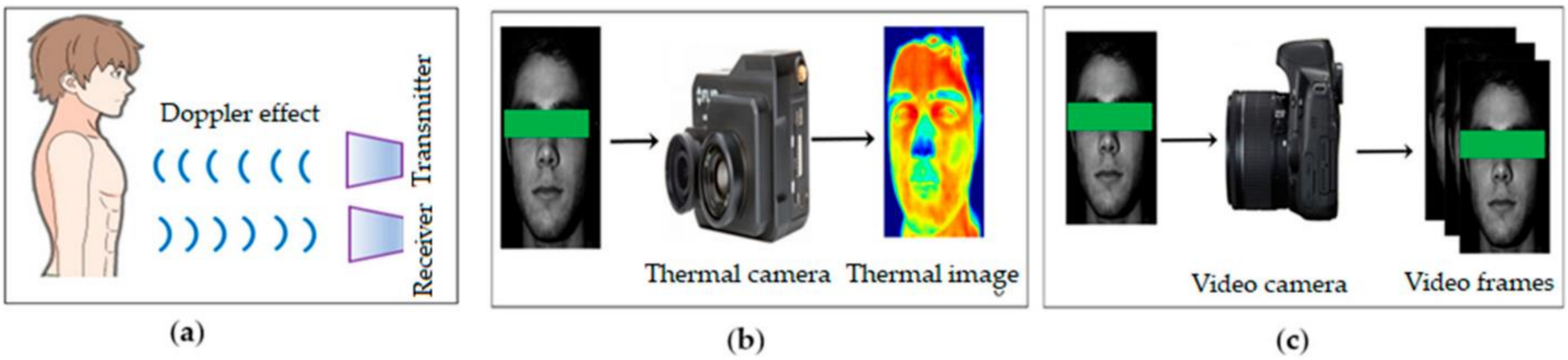

Figure 25. Contactless measuring methods: (a) The Doppler Effect; (b) thermal imaging; (c) video camera imaging. Image from [IV.14]

Imaging-based techniques are usually based on infrared/thermal imaging and video-camera imaging data, and are described in the following.

*Thermal imaging* is a passive noncontact method that can detect the radiation emitted from particular parts of the human body in the infrared (IR) range of the electromagnetic spectrum. Thermal imaging has served mostly the assessment of breathing-related parameters. Nevertheless, it has been also used to measure HR, by considering the heat differences due to pulsating blood flow in the main superficial arteries, such as the carotid artery in the neck and temporal artery in the forehead [IV.89]. However, thermal imaging-based approaches have several issues that make them less robust and viable. Indeed, they are susceptible to noise and motion artefacts, and constrain the movement of the subjects due to the high cost of the sensor, preventing saturation sampling of the environment. Their relatively low resolution limits the detection range and specificity to one subject.

*Video-camera imaging* is the most common approach to measure cardiovascular signs. Digital cameras offer high resolution along various axes, namely, spatially (i.e., the number of pixels per degree), temporally (i.e., the number of frames per second), with respect to intensity (number of bits per pixel) and in the wavelength spectrum (at least, three visible channels, with multispectral options increasingly common). Imaging data are processed to assess various physiological signals, mostly HR and HRV, from several body regions. The assessment is based on two physical principles. The first principle relies on skin colour variations caused by the cardiovascular activity and it has been already cited above as the underlying principle of PPG. This phenomenon can be assessed also via non-contact camera, and then, as said, it goes under the term of *imaging* photoplethysmography (iPPG), photoplethysmography *imaging* (PPGI) or *remote* phoplethysmography (rPPG). In this case, the cardiac signs are measured by detecting the variations in reflectance of human skin from videos, as these is detectable as variations in brightness values of the imaging data. The second principle relies on the cyclic motion of the body, due to the cardiorespiratory activity. This phenomenon is exploited by the so-called *motion-based* methods. In this frame, the most common regions whose motion is tracked comprise the head, the thoracic and abdominal regions, and the arterial pulse. As a general consideration, methods based on camera-imaging data appear more robust, safe, reliable, and cost-effective. Actually, a large body of research has concentrated on rPPG, starting from the first seminal work [IV.90] in the field, which showed that the PPG signals have different intensity levels in RGB channels and that HR could be measured from the facial video sequences. In [IV.91], the authors compare the accuracy of HR measurement from the skin surface of different body parts, and finally reported that measurements from the facial skin regions are the most reliable. In [IV.92], Poh *et al*. employ the Independent Component Analysis (ICA) method based on the joint approximate diagonalization of Eigen matrices (JADE) algorithm to decompose the raw signals of the three RGB channels into three independent source signals, and successfully extracted HR from the second component. By means of a multi-objective optimization method, the work in [IV.93] consisted in the application of a semi-blind source extraction method for HR estimation. For a comprehensive overview, we point the reader to some recent survey papers that summarise the various approaches in the field, such [IV.14] and [IV.94].

In the following, we briefly summarise some of very recent works presented in the literature to estimate HR and HRV.

In a recent paper [IV.96], Zhang et al. propose a method to overcome one of the challenges of rPPG, namely the distortion to the HR pulse caused by artificial signals caused by the ambient illumination variation and facial motion. They propose the conversion from RGB colour space to LAB colour space, to separate the luminance signal, and the smoothness prior approach is employed to remove the stationary artefacts in the raw signals. On this basis, the simple combined signal is introduced to extract the HR pulse signal.

In [IV.97], Perepelkina et al. propose HeartTrack, one of the few DL-based methods for the video-based remote tracking of HR heart rate. The authors trained a CNN on synthetic data to estimate accurately HR in different conditions. The synthetic data include videos featuring moving and talking subjects, different types of lighting, as well as various equipment. Training the CNN on the synthetic data and then testing it on a public available dataset (i.e., UBFCRPPG), the authors obtained state-of-the-art results.

Another recent work using a DL model is presented in [IV.98] by Gupta *et al*. They propose a method to estimate HR from face videos recorded using commercially-available webcams. The proposed method consists of a CNN trained as a regression model to estimate HR and HRV from face image detected in the video frames. The approach was also evaluated on a dataset of people with different ethnicity.

Yu *et al*. present in [IV.99] a method to remotely estimate HR in geriatric patients. The method is based on rPPG on video sequences recorded using a monochrome camera operating in the near-infrared spectrum and a colour camera operating in the visible spectrum. PPGI waveforms were extracted from both cameras using superpixel-based regions of interests (ROI). A classifier based on bagged trees was trained to automatically select artefact-free ROIs for HR estimation. HRV was calculated in the time-domain and frequency-domain.

Van Gastel *et al*. developed a system based on rPPG to estimate cardiac parameters (i.e., HR, RR and oxygen saturation) in obstructive sleep apnoea (OSA) patients [IV.82]. Cameras and an infrared lighting system were used to record data from patients in a specialized sleep centre. The processing pipeline was based on classic image and signal processing methods and its results were compared with the polysomnography sensory data, in order to assess their accuracy.

### Breathing signs

The methods used to estimate the breathing rate depend, to some extent, on the imaging modality used, such as thermal infra-red cameras, depth cameras or visible/IR cameras. One of the earlier works using thermography was that of Murthy et al. [IV.100], who observed that the expired air normally has a higher temperature than the image background, although the difference is low and continuously changing. In order to extract the breathing rate from this signal, Murthy et al. first acquire a number of subject-dependent breathing cycles by manual placement of regions of interest (ROIs) below the nose such that the inhaled/exhaled air is captured with the wall serving as a backdrop. Each pixel is then modelled as a mixture of Gaussian distributions which is then used to assign each pixel to the inspiration or expiration class, and to classify the period as inspiration or expiration according to the majority class. On the other hand, Goldman [IV.101] manually placed ROIs on the skin around the nasal, rib-cage and abdomen regions and showed that thermal variations provide breathing rates that are very closely correlated to nasal pressure swings measured at the nasal orifice. More recently, Kwon et al. [IV.102] used thermal imaging for respiratory rate estimation of spontaneously ventilating patients in a post-anaesthesia recovery unit. The region around the nostrils was manually selected as a ROI, nasal temperature changes from frame to frame (at 30 fps) were measured to estimate the respiratory rate. The Pearson correlation coefficient between direct measurement and thermography was 0.95 ($R2$ = 0.90). Limits of agreement analysis revealed a bias of −0.139 and limits of agreement of 2.65 (2.14 to 3.15) and − 2.92 (− 2.41 to 3.42) between direct measurements and thermography.

Bartula *et al*. [IV.103] make use of a visual camera to extract the breathing rate by projecting the intensities along each image row to obtain a one-dimensional profile from a selected ROI. The profile obtained from the current frame is cross-correlated to a profile from an earlier image to identify the location of maximum distortion, indicating the chest position. Massaroni *et al*. [IV.104] also make use of the intensity variation of the ROI and a similar vertical projection. Specifically, the user manually selects a point in the pit of the neck to extract a rectangular region, and the RGB intensities of each row in each frame are summed, effectively providing a vertical projection as mentioned in Bartula *et al.* [IV.103], and the top 5% of the rows having the highest standard deviations are used to obtain a mean pattern for every frame. This pattern is then band-pass filtered and normalised according to the standard deviation after mean removal, from which the respiratory rate is extracted using power spectral density estimation and using time-domain breath-to-breath respiratory rate. The highest resolution sensor (1280 x 720) yielded perfect agreement between the average breathing frequency values obtained using the proposed frequency domain method monitoring and the reference instrument were in perfect agreement; the time domain breath-by-breath monitoring obtained an average mean absolute error (MAE) of 0.55 breaths/minute with Bland-Altman showing a bias of −0.03 ± 1.78 breaths/min. For the lowest resolution sensor (640 x 480), an accuracy of 95.6% was obtained for average breathing rate, and the MAE was 1.53 breaths/min; the Bland-Altman of the breath-to-breath monitoring had a bias of −0.06 ± 2.08 breaths/minute.

Tarassenko *et al*. [IV.21] obtain respiratory rate estimates from patients undergoing haemodialysis in a clinical setting. To achieve this, they had to mitigate the effect of interfering signals, in particular flicker from artificial lights. Facial features are detected and tracked together with a background reference, allowing the image to be segmented into the face, the upper body and background. Tarassenko *et al*. [IV.21] carry out spectral analysis of the intensity signal using autoregressive (AR) modelling, however they note that the dominant frequency is due to the cardiac frequency, whereas the breathing synchronous motion modulates the intensity waveform resulting in a respiratory frequency peak but much smaller than the cardiac peak. In order to remove the flickering effect of artificial lights - which appear as aliased frequencies in the video - Tarasenko *et al*. [IV.21] cancel the poles corresponding to the aliased components by removing them from the AR model denominator, resulting in a new model which represents the transfer function for the light intensity reflected from the subject's ROI without the aliased components. To extract the respiratory rate, the reflected intensity signal is band-passed using a filter with a sharp transition band to separate the breathing frequency component from the much stronger cardiac component. The filtered signal is subsequently modelled using an AR model specifically designed for the respiratory signal and from which the respiratory rate is extracted.

Janssen *et al.* [IV.105] observed that the chest movement due to respiration is independent of other movements and therefore proposed to automatically extract a respiratory ROI and signal using motion factorisation. Jannsen et al showed that the proposed video-based respiration monitoring (VRM) method provided a suitable non-contact respiratory measurement method as long as the subjects have restricted motion. Regev and Wulich [IV.106] propose a method that detects points of interest on the subject's body using the Lucas-Kanade algorithm from which the optical flow is obtained. Those interest points that are determined to be moving in a harmonic fashion are used to track their harmonic frequency, and fused to estimate the breathing rate.

Godratigohar *et al*. [IV.107] proposed a method for respiratory rate estimation that permits subject movement during monitoring. ICA is applied on the RGB channels of facial videos from which rPPG source is selected. The Complete Ensemble Empirical Mode Decomposition with Adaptive Noise (CEEMDAN) scheme is used to decompose the selected ICA component into its Intrinsic Mode Functions (IMFs). Using a Machine Learning algorithm, the IMF that best reflects the respiratory rate is identified. The method results in improved estimation of respiratory rate for stationary subjects, while it yields an RMSE of 2.30 breaths/minute for moving subjects.

Fiedler *et al*. [IV.108] automatically detects faces and localises the forehead by making use of the face features. After some initial pre-processing of the intensity signal to remove specular reflections, the normalised green channel was used for rPPG extraction. As mentioned earlier by Tarassenko *et al.* [IV.21], the respiration signal modulates the rPPG signal; more specifically, the amplitude modulation is attributed to a reduced stroke volume during inhalation due to intrathoracic pressure changes, the baseline modulation is attributed to changes in tissue volume, and the frequency modulation is attributed to respiratory sinus arrhythmia, which is an increase in the heart rate during inhalation and a decrease during exhalation. Features are extracted from the rPPG waveform in accordance to the different modulations, normalised and the dominant frequency of the normalised signal determined. These measurements are then fused to yield the final respiratory rate measurement. This method has been shown to yield good results even on subjects in motion.

### Blood oxygenation

A literature review shows that numerous scientists have worked actively in the field of non-contact oxygen saturation measurement for the last fifteen years. Non-contact measurement has many similarities to the conventional (contact) pulse oximetry, but they also have some significant differences.

The most suitable sensor for remote $SpO_2$ measurement is a video camera as it is easy to use and is used contactless. $SpO_2$ can be estimated by processing consecutive video iPPG frames and extracting two signals corresponding to two wavelengths in the visible light region of the electromagnetic spectrum.

The measurement algorithms require the following sequence:

1. Identification of a ROI on the body surface
2. ROI stabilisation
3. Extraction of PPG signals from images (iPPG)
4. $SpO_2$ estimation.

**ROI identification**

The localisation of a ROI in non-contact $SpO_2$ measurement is essential. Some of the first publications [IV.109],[IV.110], which analysed the possibility for non-contact SpO2 measurement, quoted the inner arm as a relative stationary region of interest. Later studies [IV.14], [IV.111], [IV.112] are focused on the subject's face because of the limited motion ability of some parts of the human face. The process of face recognition is challenging because of the ability for free motion of the subject during the measurement. Qiang *et al*. [IV.112] suggest that the region between cheeks and nose is the most appropriate, followed by the forehead. In [IV.14], authors used face registration algorithms [IV.114] for faces in front and profile view. Casalino *et al.* [IV.115] used an OpenCV Python library for face extraction from video frames. In [IV.112], [IV.116], a Viola-Jones algorithm is employed for face detection. The facial region in the initial frame is detected using the cascade face detector, which returns the bounding box of the face.

**ROI stabilisation**

The PPG signal normally contains noise due to subject movement or camera movement. ROI tracking algorithms are developed to minimize the influence of these motion artefacts [IV.117]. In [IV.115], authors applied the Kanade-Lucas-Tomasi (KLT) tracking metho [IV.118], [IV.119] to a set of key points extracted from the ROI. The obtained transformation matrix enables finding the coordinates of the key points in consecutive frames. In [IV.112], KLT algorithm is employed again, in combination with Speeded Up Robust Features (SURF) [IV.120] to detect the tracking points.

**Extraction of iPPG signals**

In contact pulse oximetry red and infrared wavelengths are used to calculate the $SpO_2$. After analysing the spectral characteristics of oxygenated haemoglobin ($HBO_2$) and deoxygenated haemoglobin (HB) in the visible light region of the electromagnetic spectrum [IV.112], authors concluded that absorption coefficients of both components are same in the blue range of the spectrum (450 nm). Because of this, in approximately all video-based systems acquiring RGB images, the blue wavelength is used as an alternative to the infrared wavelength. One exception is presented in [IV.121]. The authors found that the combination of orange (λ = 611 nm) and near infrared (λ = 880 nm) light provides the best SNR for the non-contact video-based detection method.

Different approaches for the extraction and pre-processing of the red and blue iPPG signals from RGB images have been presented. Kong *et al*. [IV.122] used two CCD cameras, each with a narrowband filter (at 660 and 520 nm) mounted in front of the camera, but the complicated hardware limits the measurement convenience. Casalino *et al*. [IV.115] separate each ROI (from consecutive frames) into three channels – R, G, B. The applied post-processing consists of few steps:

(1) FIR filtering to suppress noise due to motion and fluctuations in ambient light
(2) chrominance-based method to further reduce the sensitivity to motion and illumination variations
(3) linear interpolation – to guarantee uniform sampling of the signal
(4) Power Spectrum Density Computation (Welch's method) – to select the most informative ROI.

Bal [IV.116] used an algorithm from Shadinrakar *et al*. [IV.117] to combine three colour channels in order to receive a stronger resultant signal. The post-process includes: Dual-Tree Complex Wavelet Transform and band-pass filtering. To extract the clean signal, Qiang *et al.* [IV.112] applied a multi-step post-process combining independent component analysis and wavelet de-noising.

**$SpO_2$ estimation**

Similarly to the conventional contact pulse oximetry, the "ratio of ratios" method [IV.123] is used for the calculation of oxygen saturation in non-contact video systems, according to the formula:

$$SpO_2 = A - B\frac{AC_{\lambda_1}/DC_{\lambda_2}}{AC_{\lambda_2}/DC_{\lambda_2}}$$

where: *A* and *B* are empirically determined coefficients [IV.122]; $AC_{\lambda i}$ and $DC_{\lambda i}$ are the corresponding amplitudes of the pulsatile and non-pulsatile components for two different wavelengths (red and blue). Subsequent processing [IV.14] [IV.116] is performed to obtain DC using a moving average with a specified window length (10 seconds), followed by averaging the peak-to-trough heights in each 10-second window to obtain the AC component.

*Accuracy*

Bland – Altman analysis is used to compare two non-contact video-based approaches [IV.112], [IV.115] with contact pulse oximetry, and the results are - an average difference of 0.07% within the limits of agreement of ± 2.58%. Three studies [IV.116], [IV.121], [IV.14], announced correlation range $r^2$ = (0.64–0.94) with conventional pulse oximeters. RMSE is reported in three studies (1.3% in [IV.121], 1.15% in [IV.124], 0.91 in [IV.112]), that is significantly lower than that required (<4%) by standard of the International Organisation for Standardisation.

### Blood pressure

The number of publications dedicated to video blood pressure (BP) measurement is significantly smaller than those presenting data on camera-derived measurements of cardiac and respiratory parameters [IV.125]. Non-contact determination of blood pressure values is performed by measuring the propagation time of the pulse wave between two parts of the body (pulse transit time – PTT) and correlating this time with the blood pressure values determined by a reference method. Because PPT is a parameter derived from PPG and by analogy with iPPG, the pulsation transition time is denoted as image PTT (iPTT).

In [IV.126], authors measured the iPTT between three different body parts - mouth, left and right palms. The experimental study was performed using various digital cameras: Logitech colour webcam, Pike black and white camera and Pike colour camera. The measured average values of iPTT between mouth and the palm areas, correspond to the values reported by other researchers, for PTT between ears and fingers measured by contact sensors.

In [IV.127], authors try to determine whether the contactless PPG signal received from digital camera recording can be used to measure blood pressure. The pulse transit time was obtained as the time delay between R peak of ECG signal and foot of the iPPG signal registered from the forehead. On the base of obtained results (mean and standard deviations of 9.48 ± 7.13 mmHg for systolic and 4.48 ± 3.29 mmHg, for mean arterial pressure), authors concluded that this method is appropriate for trends and fluctuations monitoring rather than for measurement of absolute values.

Time measurement of iPTT between two ROI (wrist and ankle) by a single camera is presented in [IV.128]. The identification of iPPG peaks in both regions was performed applying low-pass filtering and phase delay compensation. The authors declared a correlation coefficient of 0.88 between PTT and blood pressure.

The aim of the study presented in [IV.129] was to assess the degree of correlation between iPTT and BP "in vivo". The changes in blood pressure were provoked by short exercises (cycling). The authors used a high speed camera at 420 frames per second to register and calculate the time delay between iPPG peaks in two areas - head and palm. Blood pressure was measured by a verified monitor simultaneously with iPTT registration. The reported range of correlations between systolic BP and iPTT is from 0.632 to 0.960 with $p < 0.05$ for most of the participants.

One different approach for blood pressure estimation is proposed in [IV.130]. The iPPG is extracted from the one-minute face video record. The characteristic peaks, foots and dicrotic notches are used to calculate 21 time and frequency domain parameters based on which the blood pressure is predicted. Polynomial kernel regression is applied for data processing. For verification authors used clinically approved digital BP monitor. Reported results cover the standard allowable error limits mentioned by SAMI for Mean Absolute Error < 5 mmHg and Error Standard Deviation < 8 mmHg.

### Energy expenditure

Daily energy expenditure is usually estimated using sealed respiratory chambers, gas sensors in breathing masks, or inertial sensors. It can also be inferred from activity recognition systems using activity-to-energy-expenditure conversion charts (see Section 0 “4.5 Activity and behaviour recognition” for more details).

This Section reviews those methods that intend to estimate energy expenditure directly from the images.

Edgcomb and Vahid [IV.131] employ a single camera to estimate activity level, following three steps: foreground segmentation to detect the moving object (person) in the image, calculation of the minimum bounding rectangle (MBR) around the moving object, and extraction of features from this MBR. These features are MBR height, width, vertical velocity, horizontal velocity, combined velocities (speed), vertical acceleration, horizontal acceleration, combined accelerations (change in speed), horizontal work (mechanical physics-sense; assuming constant mass), vertical work, and combined work. Regression models are then built to convert feature values to energy expenditure (measured with a body-worn device).

Tsou and Wu [IV.132] proposed tracking 10 body joints obtained using the Microsoft Kinect. A multiple regression model was used to estimate calories consumption based on the velocity and acceleration of each joint. Ground truth was obtained with a heart rate monitor.

Tao et al. have proposed different frameworks for estimating energy expenditure from RGB-D data in a living room environment. In [IV.133], as in [IV.131], features are extracted from the bounding box, mapping the features directly to calorie estimates via a monolithic classifier, and adding a cascaded and recurrent classifier to capture temporal dependencies. Tao *et al*. [IV.134] propose an activity-specific pipeline to estimate energy expenditure utilising both depth and motion features as input. It uses pyramidal temporal pooling to obtain the final feature representation, capturing both short- and long-term temporal changes and summarise them to represent the motion in the video and recognise the activity taking place. Then, it uses a model for each activity built on a recurrent sliding window approach to predict energy expenditure. This work validates a new dataset, RGB-D SPHERE-calorie dataset, which covers a wide variety of home-based human activities comprising 20 sequences over 10 subjects. The ground truth a ground truth is obtained with gas-exchange measurements. Metabolic Equivalent Task (MET) lookup tables are also included for each activity. The proposed method demonstrates its ability to outperform the widely used METs based estimation approach.

Building on these previous works, [IV.135] combines the inputs from an RGB-Depth camera and two inertial sensors (worn on the wrist and waist). Different fusion methods at the feature level and decision level are proposed and compared. The proposed fusion method demonstrates its ability to outperform the METs estimation approach and the use of single modality systems.

## 4.3 Emotional state recognition

Understanding the emotional status of a subject is a widely explored topic in several research areas, from psychology and neuroscience to computer science, being the emotional status a key point in human social relations, communication and both human-human and human-machine interactions.

In 1989, Thoits [IV.136] defined the emotion as *a combination of four main factors: subjective experience in assessing the current stimulus or context, emotion expressions with gestures, culture and physiological responses*. From the physiological point of view, each internal or external stimulus is processed by the Autonomic Nervous System (ANS), known to act involuntarily and reflexively. In particular, ANS reacts through the Sympathetic Nervous System (SNS) by stimulating body functions, or inhibiting them, usually through the Parasympathetic Nervous System (PNS) [IV.137].

This means that every emotion is accompanied by both physical and/or physiological reactions, and consequently changes in physiological values with respect to a baseline measurable at rest conditions and in the absence of any stimulation. Examples of such changes are facial expressions, variations in heart rate and respiration, and sweat glands activity.

Based on these considerations, many approaches have been explored for emotion detection in the literature. Among the others, there are various imaging techniques, especially used in cognitive neuroscience (e.g., functional Magnetic Resonance Imaging (fMRI) [IV.138] and Positron Emission Tomography (PET) [IV.139]. Along with the imaging techniques, a non-invasive method involves the use of videos to detect verbal and facial expressions [IV.140], body postures and gestures [IV.141]. Also, the analysis of changes in the human physiological signals, such as Electroencephalography (EEG), Electrocardiography (ECG), Galvanic Skin Response (GSR), Electromyography (EMG), Heart Rate (HR), and Skin Temperature (SKT) is extensively used for recognizing the human emotional status [IV.142].

Among the modalities mentioned for emotion recognition, the fMRI and PET represent the most invasive techniques. Concerning the less-invasive techniques, the videos, along with the photos one, focus on the visible cues, expressed by an individual in response to a specific situation or stimulus. However, it is known that some people conceal their emotions better than others do, resulting in an emotion recognition modality highly affected by the subjectivity. For this reason, features extracted from the physiological signals try to bridge this gap of perceived emotion to the actual one felt by the individual [IV.143]. This means that we can assume physiological analysis as an objective measurement of emotion recognition.

Emotional status detection is useful for a large number of humans daily-life applications. An example can be the self-regulation of an individual's mental health for stress [IV.144] or the use of wearable sensors for driver distraction detection, which is an important issue for long-distance drivers [IV.145]

In the next sections, we review recent works on emotion recognition in AAL applications. We firstly survey the approaches based on video and image data and, then, move to briefly overview the most common strategies involving the wearable devices, as they supply very useful information on physiological signals and are largely used in AAL practices.

### 4.3.1 Contactless emotion recognition and monitoring

In AAL settings, unobtrusively placed cameras or audio sensors can provide a lot of information about the emotional state of the individuals living or working in an ambient space.

Computer vision techniques may support the estimation of emotion-related physiological parameters, such as HR and RR as seen in Section 0, as well as the analysis of facial expressions and body language evoked by emotions [IV.146]. Indeed, assessing individual psychological wellness requires the detection of complex mental states (e.g., fatigue, pain, depression, and mood), which require the integration of multiple cues: physiological, acoustic, and visual cues. The combined use of physiological parameters computed by sensors and expressive features is documented in [IV.147]. The visual clues that reveal affective states comprise facial expressions, micro-expressions, and other features, such as eye gaze and head orientation. Nevertheless, it should be noted that interpreting the facial codes to detect the underlying emotional states is still an open problem. Coding schemes such as the Facial Action Coding Scheme (FACS) [IV.148] are mainly related to primary affective states (e.g., anger, fear, happiness) rather than to complex conditions. Additional information about complex states may derive from *micro-expressions.* These are very brief low-intensity facial expressions (lasting no more than 1/25 second), which are believed to correspond to repressed feelings and emotions [IV.149]. Being usually very subtle, they are hard to detect. For this reason, local dynamic appearance representations are usually extracted from high-frequency video [IV.150]. The video-based analysis of multiple visual clues, including head, eyebrows, eye, and mouth movements, led to a technique for the identification of stress and anxiety presented in [IV.151], [IV.152].

Audio-based emotion recognition is most commonly based on speech analysis [IV.153], [IV.154]. Actually, speech signals are the most natural, intuitive and fastest means of interaction for humans and are known to carry much more information than text or spoken words [IV.155]. The affective arousal has a powerful effect on the voice, causing major changes in the acoustic parameters in correspondence of diverse affect states [IV.156]. Though, emotion recognition from speech signals has been studied for decades, delivering effective solutions able to generalize well to real-world applications is still a challenging task. Nevertheless, following the emergence of speech-based virtual assistants that provide readily available platforms for speech-signals acquisition and analysis [IV.157], a renewed interest in speech emotion recognition has lately fostered a growing body of research in the field.

In general, moving emotional analysis in AAL settings based on video and audio data is an exciting though extremely challenging task. Daily monitoring of the emotional state can be implemented to evaluate stress, fatigue, and anxiety in different conditions, including resting states or a wide variety of everyday tasks (e.g., watching TV, listening to music, using a PC, or doing homework). In addition, AAL favours the correlation of monitoring data with life-style habits, for both preventing and reducing psychophysical health issues and illness, while improving the quality of life [IV.158].

For a detailed survey of the most recent progress on contactless human emotion recognition, we point the reader to some recent and very comprehensive survey papers [IV.159], [IV.160].

In the following, we provide a brief overview of advances in contactless emotion recognition in AAL settings, by focusing on two different sceneries, the home/work and the car settings. These two have attracted a lot of attention in recent years.

### Monitoring at home and work

Some of the works done so far in AAL have estimated the emotional state of the monitored individuals in their living or working environments aiming at understanding their holistic wellbeing for disease prevention and lifestyle improvement [IV.161], [IV.162]. One's wellbeing actually relies on a combination of physical, mental and social health [IV.163]. In [IV.161], a wellness index is defined by including the evaluation of emotional state based on emotion-related physical parameters (i.e., HR, RR and HRV) as well as facial clues. The estimation of both types of parameters derived from image and video data acquired by a high-resolution RGB camera. The emotional states considered correspond to adverse mental conditions, such as stress and anxiety, as these represent risk factors for various diseases (e.g., cardiovascular diseases). The facial clues are obtained from a video-based analysis of micro-expressions related to head, eyebrows, eye, and mouth movements [IV.151]. Emotion-related physical parameters were extracted based on the so-called remote-PPG, already introduced in Section 0, based on video data [IV.164], [IV.165]. The work presented in [IV.162] addresses the estimation of emotional states based on speech-signals recorded from a mobile phone. The "emotion recognizer" adopted Mel Frequency Cepstral Coefficients (MFCC) features and an SVM as ML algorithm to classify emotions as happy, sad, angry and neutral. This information is incorporated in a wider estimation of one's wellbeing state.

Among the adverse emotional states, stress is undoubtedly the most commonly addressed condition, with several contactless methods proposed in the literature. In [IV.166], the authors propose a system that collects rPPG signals using the computer's webcam. An SVM classifier is used also in this case to detect stressful conditions. Facial Action Units (FAUs) extracted from videos are used in the study by Viegas *et al*. [IV.167]. In [IV.168], an analysis of DL techniques to stress assessment from video data is presented. Several works have also focused on estimating stress from thermal imaging [IV.169], [IV.170], [IV.171] demonstrating the viability of this sensing modality. For an extensive collection of the works on stress detection, we point the reader to a recent survey paper [IV.172].

The topic is particularly relevant within the working environments, where the emotional state affects the workers' performance, safety and well-being. In this respect, in [IV.173], the authors propose a CNN capable of classifying facial expressions, based on real-time images to detect early sign of stress in workers. Another recent work [IV.174] describes a system able to estimate individual's emotions based on multimodal indoor environment data, collected through the environmental sensors. The system does not make use of image or vital data, but relies on ambient parameters, such as temperature, humidity, illuminance, loudness, odour intensity, human detection, distance, CO2 concentration, dust concentration, to predict the emotions of the working individual. Vital data are used as ground truth to validate the "indirect" estimation (see Figure 26 for an illustration of the system proposed).

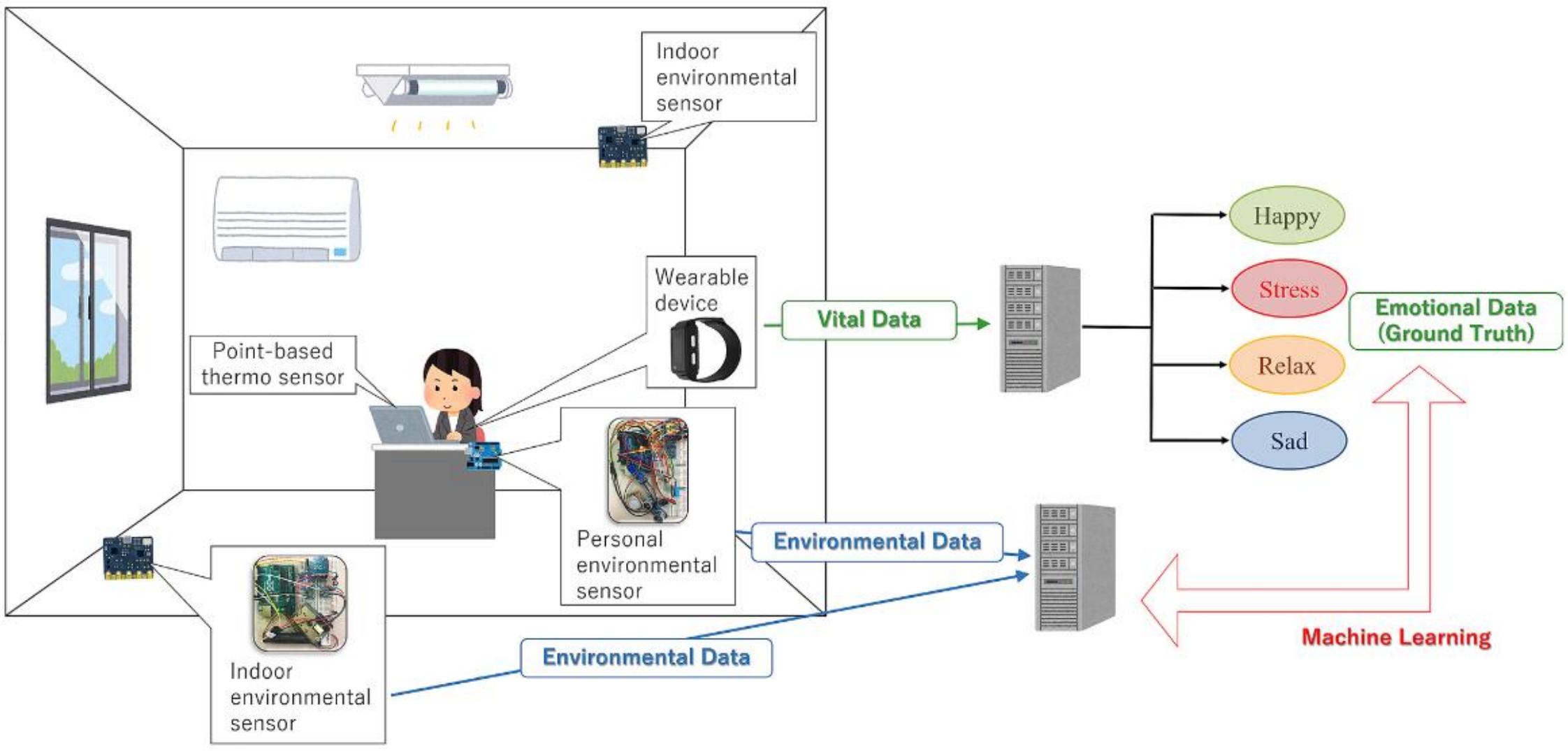

Figure 26. Individual emotion estimation from perception-based sensor data from [IV.174]

Monitoring the emotional state is also of paramount importance when assisting individuals with special needs, such as older adults suffering of Mild Cognitive Impairment (MCI) or mood disorders as well as children with Autism Spectrum Disorders (ASD). Special-need individuals may actually communicate their emotions and feeling via non-standard and non-verbal channels that challenge their caregivers [IV.175]. This is particularly true with ASD children in care and learning environments; for this reason in [IV.175], the authors propose a system able to recognise basic emotions of ASD children and provide feedback to caregivers. They employ a Siamese Neural Network [IV.176] as a meta-learning algorithm to recognize emotions from static image data. When considering older adults affected by MCI, depressive symptoms are widely common and can even cause a deterioration of the mental and cognitive abilities [IV.177], [IV.178]. In [IV.179], the authors propose a method for recognizing the emotional status based on facial expressions and its integration into a smart monitoring system for subjects suffering from neurodegenerative diseases. In [IV.180], the authors propose a first attempt to move from basic emotions detected in visual data to more complex emotions, such as depression. With respect to mood disorders, in [IV.181], the authors evaluate the performance of voice acoustic features extracted from read speech in predicting clinical depression scores in men and women. Results show the feasibility of their approach in diverse settings, as their method does not require a strict control on the recording environment. In [IV.182], the authors present a neural network model to predict stress, anxiety, and depression using FACS. In [IV.183], the authors propose a multimodal method based on ML techniques to assess depression severity using facial expressions, speech and clinically validated Patient Health Questionnaire.

Emotion recognition is also particularly relevant in human-computer interaction and can be instrumental to steer and regulate the interactions with a social and assistive robot. In this respect, the work presented in [IV.184] describes a framework able to recognise human emotions through facial expression for human-robot interaction. Features based on facial landmarks distances and angles are extracted to feed a dynamic probabilistic classification framework. In [IV.185], the authors investigated whether emotions observed from one modality could be transferred to another modality through a common representation. The idea is to observe some high-level features, such as speed, intensity, extent and regularity, in one modality (such as gait), and associate them with emotions. For example, sadness will have low intensity and speed in gait. Then, in a different modality (such as audio), the same high-level features will be used to detect emotions, hence, a sad voice will also be expected to have low intensity and speed.

Besides all the cases presented above for specific AAL applications, a big corpus of research has attempted to estimate emotions from multimodal clues derived from contactless modalities. In [IV.186], the authors survey the recognition of emotions from varying body gestures. In [IV.187], the authors survey and compare several approaches to emotion recognition based on gait and posture. The work in [IV.188] presents a method to recognise emotions from facial movements and gestures, which uses a complex deep neural network able to process spatio-temporal hierarchical features.

### Monitoring in cars

Recently, several works has addressed the problem of monitoring the emotional state of drivers for safety and assistance purposes. This is accomplished, as a part of an automated driver assistance system, by deploying several types of sensors within the driving environment to monitor the driver's fatigue and attentiveness that result from his or her emotional state. When considering contactless sensors, cameras are usually placed inside the vehicle and continuously look at the driver's face to periodically assess the driver's emotional state [IV.189].

One of the first works in the field used a thermal camera to solve the issue of face detection within the car environment [IV.190]. In [IV.191], a camera was mounted inside the dashboard of a car to collect images of the driver face for the detection of stress. An SVM trained on a public facial expression dataset was used to detect six facial expressions, among which the combination of anger and disgust was considered as a sign of stress.

More recently, a method based on DL is presented in [IV.192]. The authors extract appearance and geometric features with two CNNs and then combine them to accurately recognize the emotions. Based on the recognized emotional state of the driver, the driver can be made aware of his emotional state in case necessary.

### 4.3.2 Wearable devices for GSR data

Among the physiological signals collected for emotional status recognition, recently the GSR has gained huge interest, and nowadays it is one of the most involved signals in emotion research. GSR, named also as Electrodermal Activity (EDA) or Skin Conductance (SC), is a biometric index reflecting changes in the electrical properties of the skin [IV.193]. As above-mentioned, when individuals are exposed to stimuli such as images [IV.194], sounds [IV.195] and physical efforts [IV.196], the SNS with no conscious control induces a small sweat reaction.

By using two electrodes positioned on regions of the skin surface where there is an high density of eccrine sweat glands (e.g., fingers, hand, wrist and foot palm), the fluctuations of the skin’s electrical properties can be measured [IV.197].

The development of Internet of Things (IoT)-enabled wearable devices with wireless technology support has allowed and facilitated the shift from the wearable measurement of GSR in laboratory settings, usually with bulky wired instruments, to minimally-invasive, comfortable and real-time recordings, in free-living conditions [IV.198] with devices capable of streaming their data to a cloud-based repository.

Obviously, the positioning of the device influences the measured quantities. In fact, from the literature it is possible to observe that the positioning of the devices is not unique among the different studies and can vary for the same type of device [IV.199]. There are two main groups of GSR monitoring devices: wrist-worn and fingers-worn (placed around either the middle or the index fingers). The first type attracted the attention of many researchers, entrepreneurs, and tech giants in recent years, being comfortable and usable in people daily life in free-living conditions. Instead, the second one is generally involved in lab studies and used as the gold standard instrumentation to validate the first type, which can be referred as "consumer wearable devices" for sake of simplicity.

### Consumer wearable device

Wearable monitors are low-cost, easy to use, small and even stylish, making them widely widespread. Indeed, the market is abundant with wearable devices aimed at both physiological and activity monitoring, but with less extent at the emotion detection. Their correct positioning is fundamental to provide a good skin contact and, therefore, an accurate measurement. The different models, supporting a wireless connectivity and remote monitoring (especially for GSR), are listed as follows.

*Empatica E4*: This device is clinically validated (Class IIa Medical Device according to 93/42/EEC Directive), and equipped with four monitoring sensors, namely the 3-axes accelerometer, the photoplethysmography (PPG) sensor, the optical thermometer and the GSR sensor. In particular, this device captures the electrical conductance (in μS)) across the skin, by passing a minuscule amount of alternating current (at a frequency of 8 Hz, with a maximum peak-to-peak value of 100 μA) between the two Ag/AgCl electrodes located on the bracelet, in contact with the bottom wrist. The GSR signals were collected with a sampling frequency of 4 Hz, a dynamic range of 0.01-100 μS and a resolution of 900 *p*S. It is capable of measuring and logging data continuously in two different modalities: streaming and recording mode, with the battery life declared as >20 hours and >36 hours, respectively. The device is intended for research and comes with a desktop application (E4 Manager) to transfer data to a cloud repository, a web application (E4 Connect) to view and manage data, and mobile applications (E4 RealTime) to send (via Bluetooth Low Energy), stream and view data in real-time on mobile devices. It has been used in research, to measure the impact of auditory emotional stimuli [IV.200]; also, to recognize the user emotions while watching short-form videos [IV.201].

*Empatica Embrace*: This device, cleared by EU CE and by the US FDA for adults and children ages 6 and up, is equipped with four sensors, namely the 3-axial accelerometer, the thermometer, the 3-axial gyroscope and the GSR sensor. In detail, Embrace measures the GSR magnitude in μS, with a dynamic range of [0-80] μS and resolution of 900 femtoSiemens, at a sampling frequency of $f_s$=4 Hz. Specifically, three electrodes located on the E4 bracelet and in contact with either the ventral (inner) and dorsal (outer) wrist are passed by alternating current of maximum frequency of 4 Hz. Data are continuously sent via Bluetooth®, and analysed in real-time to identify unusual patterns in movement and skin conductance. When Embrace detects unusual patterns, through the Bluetooth®, a message is sent to the Alert app, which initiates calls and texts to designated caregivers, using the smartphone's data or Wi-Fi connection. Such smartband has been mainly involved in studies for seizure tracking and epilepsy management [IV.202].

*Gobe 2*: This smart band has four embedded sensors, which are bioimpedance sensor, accelerometer, piezo sensor and GSR sensor. For what concerns the individual's emotion, Gobe2 measures your emotional tension, strictly associated with feelings and mood, by detecting the changes in cutaneous sweat glands. After evaluating these changes, the smart band notifies the user about the emotional state, with a device vibration and a message on the display saying "EMOTION". The manufacturer declares that GoBe can detect the human stress, by using derived measurements, namely heart rate, previous night's sleep quality and personal details (i.e., weight, height, sex and age). A fully charged GoBe2 can transfer data, through BLE, to a dedicated app (named Healbe GoBe App) and work for up to 48 hours. Any further details about the GSR sensor are available, for the best of our knowledge.

However, recent validation studies conducted at Foods for Health Institute at the University of California Davis (USA) and the Red Cross Hospital (Guangzhou, China) confirmed the ability of GoBe2 to track dehydration and rehydration of the human body accurately, and proved that calorie intake is measured with 89% accuracy. More information on the study can be found at this link.

*Moodmetric*: It is a ring device specialised for measuring the EDA. The Moodmetric ring collects skin conductance levels and transforms the signals into a scale ranging from 1 to 100, where higher values indicate higher arousal that can be either positive (e.g., excitement) or negative valence (e.g., stress). Approximately, the scores in range from 1–20 reflect state of deep relaxation (e.g., meditational state); from 21–40, regular relaxation (e.g., walking); 41–60, states during mild activities (e.g., talking); 61–80, arousal during elevated activity (e.g., working under mild pressure); and 81–100, high arousal such as strong emotions. Such ring has a small data storage capacity, so the data are transferred by Bluetooth connection to computer for permanent storage. This device was involved in a study to examine the role of threat magnification as an underlying mechanism [IV.203].

The list of commercially available devices is presented in Table 1.

Table 1. Consumer wearable devices for GSR data collection

| Product and Producer | Purpose | Position | Sensors | Wireless Technology | Clinical Validity |
|---|---|---|---|---|---|
| E4 wristband, Empatica Inc., Cambridge, MA, USA | Collecting physiological and movement data | Wrist | PPG, GSR, Infrared Thermopile, 3-Axis Accelerometer | BLE | Yes (Class IIA Medical Device according to the 93/42/EEC Directive) |
| Embrace, Empatica Inc., Cambridge, MA, USA | Collecting physiological and movement data | Wrist | GSR, Thermometer, 3-Axis Accelerometer, 3-Axis Gyroscope | BLE | Yes (Clinically validated and FDA approved) |

| Product and Producer | Purpose | Position | Sensors | Wireless Technology | Clinical Validity |
|---|---|---|---|---|---|
| GoBe 2, Healbe, Corp., Moscow, Russia | Emotional monitoring, calorie intake and burned, body hydration | Wrist | GSR, Impedance, 9-Axis Accelerometer, Piezo sensor | BLE | No |
| Moodmetric Smart Rings, (Vigofere Ltd, Finland) | Emotional monitoring | Finger | GSR | BLE | No |

### 4.3.3 Open challenges for emotional state recognition

This section has reviewed and presented the recent advances in contactless and wearable technologies for emotion recognition. In the last years, research in the field has made considerable progress, moving from the exploitation of lab-acted databases to real-life scenarios and employing more complex models to better exploit the sensor data available [IV.160]. The annual competitions (e.g., AVEC and EmotiW) have promoted research and development in multimodal affect recognition systems, encouraging researchers to assess both the reliability and accuracy of their methods against standard benchmarks. Despite the current development and progress of electronics and technologies, there are still big challenges for effective solutions and their uptake in real-life settings. The first challenge comes from the physiological and social processes underlying human emotions, as these are highly variable (i.e., intra subject variability) and subjective (i.e., intra subject variability) as well as difficult to recognise also by humans. In addition, the emotional conditions of interest within the AAL settings are highly complex feelings that are difficult to discern. From the technological viewpoint, challenges arise from several issues, some of which are listed below [IV.160]:

- Real-life settings: data coming from realistic, also named wild, settings usually introduce many sources of variation that challenge emotion recognition methods. Variations include occlusions, complex illumination conditions in different settings, spontaneous behaviour and poses, as well as rigid movements. Audio may contain background noise, unclear speech, interruptions, and other artefacts
- Temporal dynamics: The temporal dimension is an integral element of emotion and feeling display that is not always considered in recognition methods. Temporal dynamics can be rich with contextual information that could be captured by models, for instance, to distinguish between displays that are similar in the short term, and to assess the relative importance of specific segments
- Multimodal sensor fusion: in social interactions, expressing and sensing emotional states encompass multiple modalities. Similarly, automatic recognition methods can strongly benefit from the used of multimodal sensors to collect simultaneously several physiological signals, as research in the last years has demonstrated [IV.204], [IV.205]. Audio-visual data fusion can increase model robustness and accuracy, but it is still unclear how and at which level of abstraction these modalities need to be fused.

- Reduced availability of ground-truth data: the number of factors that regulate emotion-recognition variability outpace the cardinality of labelled datasets. This makes training ML and, particularly, DL methods very difficult. Approaches based on transfer learning, unsupervised or semi-supervised learning are being adopted to overcome this problem, but further advances are necessary in the field.

## 4.4 Food intake monitoring

In recent times our society is experiencing a rise of health-related problems caused by inappropriate nutrition habits and dietary. Related diseases include diabetes, obesity, cancer and heart issues with several studies showing that correct nutrition decreases the risk of these illnesses. The process of support and observation of patients' eating habits can be facilitated by modern CV solutions and image processing.

The task of dietary support focuses on documenting what is being eaten and the amount of food intake. Such information is used to calculate daily calorie and macronutrient (carbohydrates, fats, protein) data which play a significant role in a person's diet. Considering a CV system for nutrition monitoring, we can extract a couple of tasks building such a framework with the major goal being to identify the content of the acquired image and what food can be detected from it. Additional challenges include volume estimation, visual attribute recognition (e.g., distinguishing between fried vs. baked), multiple instances of items, segmentation of items, open-world recognition (unbounded number of categories), fine-graded recognition (differentiating between subcategories of food items), hierarchical label spaces (handling related labels), occlusion (unlikely to convey complete compositional information from visual data), assumptions (such as background, calibration targets) reduce usability or lead to unrealistic arrangements. We herewith review a number of available technologies for automatic food intake observation.

One of the first prominent works in CV focused on dietary assessment is Mobile Phone Food Record (mpFR) by Zhu *et al*. 0. It was developed at the Department of Foods and Nutrition at Purdue University, US in 2010. It is an automatic system for food recognition, segmentation and volume quantification for images captured using mobile phones. The biggest constraint of this solution is that the image capturing method was designed for specific scenes with black background and high level of segmentation between products. Obviously, such an approach is far away from the real-life environment, but still it was used in many researches and studies during the years. In the original work, authors use hand crafted descriptor extraction and Bag-of-words (BoW)-encoding before classification. In further work, they focus on segmentation task using Normalised Cuts.

More recent research of Zhu *et al*. [IV.207] redefined method based handcrafted features and BoW-encoding. It is presenting an iterative multi-pass flow segmentation step and recognition step, using the recognition estimation for refinement of the segmentation step, and creating multiple hypotheses of the segmentations.

A system based on an android mobile application, was created at the University of Taiwan by Chen *et al*. [IV.208]. It performs quantity estimation using 3D sensors and food recognition. The second task is done by descriptor extraction, encoding and classification using SVM algorithm using a dataset containing 50 categories of Chinese meals represented by 100 images for each. The descriptors for texture are SIFT and Local Binary Pattern (LBP). Third texture descriptor is the concatenations of the means and the variances of the Gabor filter. For the colour, descriptor authors used histograms for each RGB channel computed for 4 x 4 grid on image. The overall recognition rate is 68.3%.

In 2011, the European Commission started the 7$^{th}$ Framework Programme with the founding project of Marie Curie Industry-Academia Partnerships and Pathways named GoCARB. Its main goal was to perform automatic estimation of carbohydrate intake with error less than twenty grams per meal. It is done by mobile application using the phone's camera in a controlled environment and the framework consists of plate detection, food segmentation, food recognition, 3D reconstruction and volume estimation for each segment. The application is assuming that the only one dish is visible, the meal is inside a round plate and there are no occlusions among food. Those make it inapplicable in a real-world environment where food can be mixed or chopped together and finally served in various ways.

The dataset and project activities led to several studies. Initial work focused on food recognition using hand crafted features. First prototype of Anthimopoulos *et al*. [IV.209], included dish detection and segmentation. Recognition was based on one feature for texture description LBP and a colour feature – histogram of 1024 most dominant food colours, SVM classifier with RBF kernel to distinguish between 6 possible classes of foods; the overall accuracy is 87%. In the follow-up work [IV.210], authors compared fourteen different image descriptors in an increased number of classes to eleven. The best performance of 77.6% accuracy is with BoW dictionary with 100.000 patches for each of the eleven food classes, using 10.000 visual words in the dictionary, the SIFT¬_HSV descriptor and SVM linear classifier.

The first evaluation of a complete system is presented in Anthimopoulos *et al*. [IV.211]. It includes a full framework with segmentation and volume estimation. Image classification is using the same architecture and descriptors like in previous work [IV.209]. Carbohydrate estimation is done by defining nine food categories: pasta, potatoes, meat, breaded food, rice, green, salad, mashed potatoes, carrots, and red beans. Image capturing conditions were fixed and the same for every dish. Estimated values are compared to weighted grand truth. Average mean error is 6 grams (10%) with standard deviation 8 grams. Result fits the starting criteria of 20 grams variance.

Christodoulidis *et al*. [IV.212] explores convolutional neural networks (CNN) as an alternative for hand crafted features. Segmented food is classified by CNN and different architectures are being compared. Best score was achieved with 4 layers CNN and the overall accuracy is 84.9% compared to the previous value of 82.2% [IV.209].

In the study of Rhyner *et al*. [IV.213], evaluation of the GoCARB system with self-reported carbohydrate intake is presented. There were 19 adult participants with diabetes that were observed for 10 days. Every day patients had six meals with three different food categories for each. Participants did one assessment of all six meals on one randomly chosen day, therefore on some days there was more than one individual performing the assessment. The results of the comparison is an average absolute error of about 12 grams (26%) of carbohydrates for the GoCARB system versus about 28 grams (55%) error of the participants' self-assessment. The framework performed well with 81% of the estimations obtaining the aimed goal of a ±20 grams variance. In case of self-reported estimations, only 59% of them reached this criterion. The automatic segmentation performed with 75% accuracy and the recognition was correct in 60% of the cases for all three food categories, in 36% two out of three were correct and in 4% only one was recognised. The algorithm is based on their previous work [IV.209] combining two image descriptors, LBP and a histogram of the 1024 most dominant colours and classifying the concatenation of the two vectors directly with an SVM.

FoodCam presented by Hoashi *et al*. [IV.214] is another food recognition work based on handcrafted image descriptors, trained mainly on Japanese meals with addition of some popular foreign fast food dishes. In the original work authors are comparing classification using different descriptors like SIFT, colour histogram in RGB colour space, averages of 24 Gabor filters and Histogram of Oriented Gradients (HOG). Their best performance is an accuracy of 60.87%.

Later in the work of Kawano and Yanai [IV.215], a mobile phone prototype for food recognition was implemented. It connects automatic recognition with semi-automatic segmentation where users manually use bounding boxes for this purpose and input size of dishes. Top 5 results of classification are displayed to the user for selection and confirmation. Detection is performed by SVM with several combinations of descriptors like HOG, SURF, colour histogram, Fisher Vector (FV) encoding with best result of 59.6%. However, major work here is focused on computational efficiency to make the system operate smoothly on smartphones.

In the work of Kawano and Yanai *et al*. [IV.216], a CNN was explored for feature extraction. After normalisation SVM algorithm is running the classification. Comparing the HOG and colour descriptors with FV encoding there is an improvement from 65.32 to 71,80% on the UEC-FOOD100 dataset, but the best result is performed by a combination of DCNN with FV where 77.35% is obtained. More recent work by Yanai *et al*. [IV.217] implements Network In Network [IV.218] [13] structure to boost computational time and accuracy with a result of 78.8% of classification in UEC-FOOD100 and satisfactory execution time of 66 milliseconds for image using iPad Pro.

Im2Calories [IV.219] from Google is a mobile system for food recognition that has a goal of macro nutritional composition estimation based on a single image. Authors define two problems in their research: the first is detection focused on food images from a public restaurant's menu and the second is work on available dataset (FOOD-101). In the first approach, GPS provides data to find the closest restaurant. It allows recognition of menu items from captured scenes. For classification, CNN based on GoogLeNet [IV.220] is used. However, this experiment failed and the final scores were not realised due to the poor performance on the test data. Comparing their model on the original FOOD-101 dataset brought a top accuracy at that time of 79%. Size estimation is done first by a segmentation algorithm named DeepLab [IV.221], which is based on estimated classification scores and low-level information of pixels and edges, using a Conditional Random Field graph. For depth prediction AlexNet [IV.222] is used and modified into multi-scale CNN in the way that was proposed by Eigen and Fergus [IV.223]. Data is converted into voxels for volume calculation. However, there was no evaluation of estimated calorie values due to the lack of existing databases and nutritional composition information.

Another modification of GoogLeNet [IV.220] was presented in [IV.224] where the proposed network incorporates semantic relationships between food classes. For that purpose the authors defined proper semantic groups with hierarchy. This hierarchy is used in the learning process by basing the loss function of the used CNN on a multitask learning and membership of parent structures are considered during classification. Their work brought improvement in accuracy comparing to the original GoogLeNet [IV.220] architecture from 69.64% to 72.11%,

Menu match presented by Beijbom *et al*. [IV.225] is a project from Microsoft Research focused on calorie estimation in nutrition oriented at specific restaurant meals. Location information like GPS signal is used to reduce the selection of images from nearby restaurants. The method does not consider varying sizes or ingredients, but it is designed for food items included in restaurants menu. The dataset consists of meals from three restaurants and as a grand truth, calories were estimated for each of the images by dietitians. Recognition is using hand crafted features and the best results were achieved with colour features. Classification is finalised with linear SVM. Final results of this system on the generic dataset used in work of Chen *et al*. [IV.208] was 77.4% bringing improvement of around 3% to the original work. Absolute error of calorie estimation was 32 ± 7.2 kcal (mean ± standard error).

Another deep learning approach is presented in study by Poply and Jothi [IV.226]. They use CNN network called Mask R-CNN [IV.227] to perform segmentation task. It is a network connecting two models where one is responsible for object detection and second for semantic segmentation. Masked area of visible items is used for further calculation of surface size. This data alongside the value of calorie per inch provides final estimation. Results on UNIMIB 2016 Food Database are promising with mean average precision of 93,7%. On 40 random images from dataset used as a test data overall accuracy is 95,45% for a single food and 93,4% for whole meals

Food intake recognition and calorie estimation is a complex topic that consists of multiple subtasks. The first stage of every process involves food segmentation. Images are often captured in conditions equal to the laboratory setting, for simplifying the task, but performances drop a lot when the algorithms are moved to process images from real environments. Of interest is the semi-automatic approach where the user is able to manually select the bounding box surrounding items. After segmentation, each visible product has to be classified for correlating with calorie and macro-nutrition information. At the end, volume estimation is performed. Nowadays, classification performance is rising. Initially, the most popular techniques used for classification were based on hand-crafted features extraction (e.g., BoW); but recently this task is overtaken by CNNs for both feature extraction and classification. However, even with perfect models running ideal classification and volume estimation, we are far from the grand truth. This is due to the complexity of some dishes, like wraps, tortillas or soups, which can include various ingredients not accessible for CV systems. Being aware of this, with adjustments to the way how food is served, we can still push the boundaries and get closer to perfection.

## 4.5 Activity and behaviour recognition

Human activity recognition (HAR) is a persistent problem in a number of applications such as health care, surveillance, human-human interaction (HHI) and human-computer interaction (HCI). The fast development of novel technologies and approaches for HAR in the past decades has led to a number of new components and systems that either have already found wide application or are still in the development phase. However, several challenges still exist, especially when applying current HAR methods and systems in real-word applications and realistic scenarios.

Audio and video-based HAR play a significant role in AAL applications. Especially vision-based activity recognition has gained particular attention given its potential high accuracy. However, a number of factors such as the interference with the environmental light, shadowing, different angles, and privacy protection narrow its application. Vision-based activity recognition can be combined with other methods (e. g. audio, WiFi, wearables and various sensors) in order to achieve better robustness and performance, as well as to extend the fields of application. In this section, we provide an overview and classification of various video- and audio-based approaches for activity/behaviour recognition, addressing the possibilities of combining these methods with other technologies and approaches, particularly in the scope of AAL systems.

### 4.5.1 Video-based Activity/Behaviour Recognition

Selected state-of-the-art techniques for video-based activity recognition are presented and analysed from a performance perspective in [IV.228]. The authors focus on the following features:

1. Hand-crafted motion features: interest points, optical flow, spatio-temporal pyramid, etc.
2. Depth sensor: RGBD information
3. DL: CNNs, Long Short-Term Memory (LSTM), two-stream architecture, etc.

The methods using *hand-crafted motion features* focused on visual and temporal information such as:

- Temporal visual features comprising static image features and time information that can be represented using approaches such as key-frame, bag-of-words, interest points and motion-based.
- Visual features (edges, corners, interest points, etc.) that can be used to form the so-called optical flow feature. The optical flow calculates the motion between two image frames at two instant points. Approaches for visual features are: HOG for human pose representations, time series of Histogram of Oriented Optical Flow (HOOF) to characterise human motion, Supervised Principal Component Analysis (SPCA) to speed up the recognition process without sacrificing accuracy.

The methods using *depth information* are based on the combination of colour image sequences together with depth maps in real time. Depth Motion Maps (DMM) are generated by projecting depth maps into three orthogonal Cartesian planes and accumulating global activities through entire video sequences. Then, HOG descriptors extracted from the depth motion map of each projection view (front, top, side) are combined as DMM-HOG, which is used to represent the entire action video sequences. A SVM classifier is employed for the recognition stage. The histogram of oriented 4D surface normals (HON4D) is a descriptor for video activity recognition that captures complex joint motion cues at the pixel level. This is done by using a two-layer BoVW model, motion-based and shape-based STIPs to distinguish the action as well as Multi-directional Projected Depth Motion Maps (MPDMM).

The methods using DL*,* which have demonstrated a great performance in image classification, are improving in video-based activity recognition, but they have not yet reached the performance of image classification. Diverse DL architectures and approaches have been proposed in the literature, comprising different connectivity patterns or diverse approaches to treat and model spatio-temporal information. For a more detailed list of the methods proposed so far, the reader is referred to [IV.228].

A comparison of the results obtained from different methods for several datasets is carried out in [IV.229]. The considered datasets include: Weizmann, MSRAction3D, HMDB51, UCF50, UCF101, Sports-1M, ActivityNet, Something Something and AVA. The depth information-based methods use the MSRAction3D as a benchmark dataset. The best result is achieved by the M3DLSK+STV method. For the hand-crafted feature methods, the Weizmann, UCF101 and HMDB51 datasets are employed as the benchmark datasets. In this case, the HOOF NLDS method outperforms the rest for the Weizmann dataset. The deep learning methods have been compared over the same benchmark datasets being the Two-StreamI3D, pre-trained the one outperforming the rest, when evaluated over the UCF101 dataset.

In [IV.230], the authors propose using a home environment consisting in a network of cheap low-resolution visual sensors (30x30 pixels) to analyse the long-term behaviour of an older person in an uncalibrated environment. This in-home monitoring system includes 3 processing layers. The low-level processing layer creates a feature vector by computing foreground pixels with the aim of tracking the motion level in each visual sensor. The mid-level processing layer consists of a Hidden Markov Model (HMM) that uses the feature vector as observation sequences to estimate state sequences, being the states of the different locations within the residence. Finally, the high-level processing layer uses the spatial (i.e., location) and temporal contexts to infer the ADL parameters using a rule-based approach. This rule-based method discovers potential activity classes from unlabelled sensor data, preventing from the costly process of data labelling imposed by supervised approaches, and considering a wider set of ADLs to recognize than when a predefined set is modelled. This technique was tested against data corresponding to 10 months of real-life recordings using a network of 10 low-resolution visual sensors working at 50 fps, where ground truth was collected by mixing the information contained in diaries with a visual inspection of the images. HMMs to detect location of the older inhabitant had a mean absolute error (MAE) of 29.82, 7.62, 13.83, 29.59 and 8.83 minutes detecting the time spent on sofa 1, sofa 2, kitchen, dining table and bathroom respectively, and relative absolute errors (RAE) of 8.32, 21.73, 18.54, 25.41 and 24.63%; Spearman's rank correlation coefficient obtained values of 0.947, 0.963, 0.942, 0.883 and 0.878 related to the ground truth.  Regarding activity discovery, the approach was able to discover 13 ADLs; the overall MAE was 9.39 minutes, while the overall RAE was 18.27%. This method enables tracking behavioural information over time analysing mobility patterns of the inhabitants, but no automatic behavioural modelling method to detect decline is proposed.

The performance of human action recognition systems is addressed and assessed in [IV.231] by putting a particular attention to fall recognition by using infrared cameras. The robustness of the system and the recognition accuracy is increased by means of a supervised CNN, specially designed for this use case und supported by a large dataset comprising 5278 random image samples from thermal videos, 30 seconds long each, which represent six activity classes such as walking, sitting on a chair with or without a desk in front, standing, fallen/lying on the desk and fallen/lying on the ground. This approach achieved a classification accuracy of 87.44%.

In [IV.229], the authors present an extensive survey of vision-based HAR methods and classify a plethora of different handcrafted features-based or learning-based HAR approaches. Handcrafted features-based approaches are further categorised into spatial-temporal or appearance-based approaches; they indicate that the spatial-temporal approaches either represent the human activities spatially if these are static (by means of body models, image models or spatial statistics), or temporally if the human activities are dynamic (by means of action grammars, action templates or temporal statistics). The appearance-based approaches, on the other hand, exploit 2D or 3D depth data for activity recognition, and represent the human activity by extracting shape (such as contours or silhouettes) or motion (such as optical flow) features. In comparison, the authors categorise the learning-based approaches as relying on either traditional methods (such as genetic programming, dictionary learning or Bayesian networks), or more recent deep-learning methods (such as Generative Adversarial Networks -GANs, RNNs or CNNs). This review also supports that a general HAR system is often composed of three main steps:

1. detection of body parts
2. tracking between successive image frames, and
3. recognition of the activity itself.

They proceed to review many different methods to achieve each of these steps, such as the use of the skin colour for body part detection, particle filters for tracking, and neural networks for classification. An overview of the different activities that are typically addressed by HAR methods in the literature is also provided. These range from hand gestures, to individual behaviours (e.g., smiling) and interactions with other persons or objects (e.g., cooking or writing), and actions that can be performed in groups (e.g., cuddling). Some of these activities such as sitting and eating can be recognised from static images taken from a single viewpoint, whereas the recognition of more complex activities might require the acquisition of video or the use of multiple viewpoints. Many open datasets may be used for evaluation purposes; these datasets may be categorised into four main types depending on the activity information:

1. actions (e.g., running, walking or skipping)
2. behaviours
3. interactions (e.g., preparing food and practicing sport) and
4. group activities (e.g., hugging).

Several limitations of current HAR methods are detected. Variations in lighting is one of the main difficulties, since it affects the quality of the captured images and, hence, their analysis. Similarly, the perspective at which the images are captured can also create issues related to object- and self-occlusion, which can hinder the visualisation of activities. Methods that rely on the appearance (such as the colour or form) for body part detection can also be easily confused by nearby objects having a similar appearance. The acquisition device can also introduce issues, such as noise. The availability of benchmark datasets is a further concern that the authors highlight, especially if the dataset in use has not been made public to other researchers. Daily life activities and fall detection are two such areas that the authors mention not being covered by sufficiently large datasets for training. The authors further mention that the use of vision-based systems for applications, such as surveillance, elderly assistance and patient monitoring, give rise to societal challenges, particularly those relating to acceptance and privacy. They mention the use of smartphones as one particular workaround, because the captured data would then be contained on the user's own device. However, this may give rise to even more challenges, such as the device's computational resources and battery life, and the capability of the user to operate such a system. The authors also allude to the challenges of detecting and understanding complex activities in long-term videos, in distinguishing between the starting and ending times of successive activities, and also in discriminating between involuntary and voluntary actions. Finally, the authors refer to open gaps when dealing with missed parts of video, recognition of multiple activities being performed by the same person at the same time, and the early prediction and recognition of actions, especially in crowds. The recent introduction of deep-learning methods has also given rise to further challenges relating to memory constraints, the collection and fusion of large multimodal data for training, the tuning of large numbers of parameters, and the deployment of different architectures on mobile devices, such as smartphones and wearables.

The problem of early recognition of human actions using multiple video cameras, as one of the emerging research areas relating to human activity recognition, has been addressed by the authors of the paper [IV.232]. The authors argue that, while the use of multiple cameras can mitigate the limitations of a single camera setup in fully capturing the progression of actions, multiple cameras require more bandwidth and processing power. A possible workaround can be the temporal down sampling of video frames (or in other words, frame dropping), but this uniform resource allocation across all cameras would not be optimal since some cameras might be capturing more salient information than others. Within a home environment, for instance, this can happen if the human activity of interest is happening in one particular room. The authors proposed a framework that is based on recurrent neural networks (RNNs) to model the dynamics of human actions, but also to account for the variable frame rate of each input video sequence, by estimating any missing video frames based on the previously observed video frame and the time elapsed from the last observation. Information from multiple cameras is integrated for view selection and action recognition by a weighted average of the RNN outputs (one RNN per camera), penalising those camera views for which the elapsed time from the last observation is large. The proposed method has been tested on three multi-view or multi-modality datasets: the NTU RGB-D, the IXMAS and the nvGesture datasets. Of particular interest are the results obtained in handling missing frames and performing view selection. The handling of missing frames is tested by classifying different actions in two scenarios, for three camera views: in the first scenario the frame dropping rate ranges from 20% to 70%, whereas in the second scenario the entire frame sequence is observed. The authors report action classification accuracies of (70.53, 65.38, 60.99) % for each camera view in the first scenario, which are rather close to the corresponding classification accuracies of (75.40, 71.04, 66.02) % in the second scenario. In testing the view selection policy, the authors consider yet another three metrics:

1. classification accuracy when only the first 40% of the action is observed
2. classification accuracy when the full action is observed, and
3. the average early recognition performance.

They report the respective classification accuracy values of (54.55, 79.62, 58.01) % for their learned policy. Their results are comparable to the classification accuracies, (59.90, 81.28, 61.77) %, obtained when all camera views are used, and exceed the results obtained when using the same view, randomly choosing a view, or cycling across all views successively.

The problem of robust activity recognition for the aging society is addressed in [IV.233]. The human pose is captured by means of a skeleton from sequences of RGB images. The authors propose an algorithm, which they refer to as the Interframe Matching Algorithm, which serves to filter out any erroneous skeleton key points that belong to non-target objects based on their spatial and temporal continuity. Subsequently, they propose a method to encode each frame pertaining to the specific activity under consideration into a feature vector based on the distances between key points, and then arrange all vectors that comprise the same activity into a matrix. The resulting matrix is converted into both a grayscale and a colour image, and subsequently used as input to a modified AlexNet for activity classification. Three datasets were considered to test different aspects of the method: the KARD dataset, which contains 18 types of activities (10 gestures and 8 activities), where each activity is performed three times by 10 persons; the Florence 3D Actions dataset, which contains RGB-D images and complete skeleton data; and the Noisy Activity dataset, which includes non-ideal conditions, such as obstruction and the presence of non-target objects. The Interframe Matching Algorithm was tested on the KARD and the Florence 3D datasets, reporting 100% accuracy on both. The full method was then tested on the KARD dataset, reportedly achieving 77.5% and 85.8% accuracy on the gestures and actions, respectively, when the greyscale images were used as input to the modified Alexnet, and 100% accuracy for both gestures and actions when the colour images were alternatively used as input. On the Florence 3D dataset, the full method reportedly achieved 89.4% and 100% accuracy, for the grayscale and colour images, respectively. On the Noisy Activity dataset, the proposed method achieved accuracy values in the range of 53%-92.5% for different actions using the grayscale images as input, while the colour images achieved 100% accuracy for all actions in the dataset.

In [IV.234], the authors propose a method that classifies activities based on skeletal data acquired by a Kinect sensor. This method processes a skeletal data sequence in order to extract eight statistical features, which include the mean, the standard deviation, and the skewness of the distribution of data, among others. This is followed by the application of Principal Component Analysis (PCA) to reduce the dimensionality of the extracted features. The classification of the activity under consideration is then carried out either by a K-nearest neighbour (K-NN) classifier or by a Support Vector Machine (SVM), both of which are trained on the dimensionality reduced data. The method has been tested on the KARD dataset, which consists of 18 different actions performed thrice by 10 subjects. The results reveal that the SVM classifier performs better than the K-NN in classifying the different actions. Specifically, the SVM classifier achieves an average overall accuracy of 98% across all three actions, which is 3.5% higher than the accuracy achieved by the K-NN classifier.

### 4.5.2 Audio-based Activity/Behaviour Recognition

Audio can be used as a standalone, or more likely as a supplementary source of information for monitoring and recognition of various daily activities performed in a domestic environment. Such a monitoring system should be able to detect the activities and make a decision whether the user's behaviour deviates from the regular behaviour. The problem is challenging given the huge number of observations produced over time, heterogeneous environmental sounds and the lack of a clear way to relate the audio data to observed activities.

Applications are various, ranging from detection of walking, cooking (boiling sounds, frying sounds), eating (sounds of cutlery, chewing), toilet flush, water usage, or a washing machine, in order to estimate how the user copes with daily house duties; to recognition of human nonverbal communication to estimate person's emotions, e.g., in people with dementia. Moreover, identification of a user's health status is possible by processing audible events, such as coughing, dyspnoea or snoring that can be used to indicate symptoms of a disease [IV.235],[IV.236].

Technologies for audio-based activity/behaviour recognition include acoustic sensors integrated in AAL systems. They can require a direct, explicit interaction interface with the user's environment or use the intangible, implicit interaction to perceive the environment. The sensors can measure either mechanical or acoustic waves that propagate through the materials. Active acoustic sensors typically detect sound waves that travel through the air and convert mechanical wave energy (sound wave) into electrical signals via electromagnetic induction (as in dynamic microphones) or using electrostatic principle (as in condenser microphones). The biggest advantage of this type of acoustic sensor is that they don't require an interface between the sensor and the source of sound. However, due to the nature of their operation, they may raise privacy issues by collecting speech information, making the user identifiable [IV.237]. Ultrasonic sensors are transducers that actively send and receive ultrasonic pulses and measure the signal propagation time to and from the measured object. Since the operation frequency is outside the human audible range (typically at 40 kHz) they can be considered unobtrusive. Finally, passive acoustic sensors measure mechanical waves transmitted through the solid materials, e.g., vibrations on the ground surface caused by walking (surface acoustic sensors).

### 4.5.3 Activity Recognition Chain

Human activity/behaviour recognition can be described by a chain of steps referred to as activity recognition chain (ARC), that include pre-processing, segmentation, feature extraction and feature selection, and classification.

**Pre-processing** assumes processing of raw data to obtain representations appropriate for use in the subsequent steps of ARC. It includes steps such as resampling, in case data is acquired at different sampling rates; silence removal and voice activity detection to isolate only relevant signals; and noise reduction to remove or reduce the ambient noise. Design of a system for activity recognition should furthermore consider the influence of the distance between the sensor and the source of sound on performance, especially with regard to ability of the system to generalize to conditions in a new environment, different from the one used for model training [IV.235].

**Segmentation** is used to divide signals into segments of meaningful size for the feature extraction and classification steps. The choice of the segment size may impact the capacity to extract relevant features from the signal. Short segment duration improves the system response rate, but might cause relevant features to be missed, leading to poor recognition. On the other hand, long segment duration may lead to co-occurrence of multiple events within the same sound segment [IV.235]. Typical segment sizes range between 1s and 10s. Short segments with 1s duration are optimal for discriminating between periods of activity and inactivity, duration of up to 4s are applied for recognition of simple activities (e.g., walking or falling), whereas longer segments are used for detecting complex activities (e.g. activities of daily living, such as eating, bathing, cooking etc.) [IV.238]. Analysing overlapping segments may improve the classification performance, but also increases the computational costs.

**Feature extraction** typically requires domain specific expertise, although many of the features are shared across different audio signal processing tasks, such as temporal features (zero-crossing rate, root-mean square energy, etc.), spectral features (band-energy ratio, spectral roll-off, spectral flux, spectral centroid, spectral slope, etc.), cepstral features (Mel-frequency cepstral coefficients (MFCC), gammatone frequency cepstral coefficients) and time-frequency representations (spectrograms, mel spectrograms) [IV.239]. Learning feature representations from data has largely replaced manual feature engineering lately, using approaches such as Non-negative Matrix Factorisation [IV.240], dictionary learning [IV.241] and deep learning, mostly based on Convolutional Neural Networks (CNN) [IV.242]. Nevertheless, feature extraction for audio-based activity recognition remains a difficult task since the audio signal is highly sensitive to acoustic conditions in the environment.

**Feature selection** is a process of identification of an optimal set of features. Discarding features with low importance reduces the computational complexity, but can in certain cases also improve the performance of the model. However, it is an iterative process that generates candidate feature sets, which may also have negative computational effects. The initial set of features before selection is still handcrafted. Deep learning bypasses feature extraction and selection by learning representations directly from raw data or low-level signal representations, requiring no manual feature extraction [IV.243].

**Classification** assumes training the model to recognize the activity using the features extracted from the audio signal, such that the model can be further used for classifying new data, previously unseen during the training process. Classification algorithms can be categorised into stateless and stateful algorithms. In stateless algorithms acoustic events are mutually independent (e.g. Support Vector Machines,

Random Forests, Feed-Forward Neural Networks, Gaussian Mixture Model), while in stateful algorithms the events are dependent on each other, meaning that the output is related to the previous inputs in the timeline, i.e. states (e.g. Bayesian Networks, Hidden Markov Models and Recurrent Neural Networks) [IV.244].

The use of deep learning approaches for learning feature representations, such as CNNs, has merged the feature extraction and classification into a single pipeline, requiring a training step to learn robust features, but on the other hand introducing the risk of the cold-start problem, when the system is not able to infer correct classes for the items without the sufficient information in the training dataset. This can be mitigated using transfer learning techniques, by pre-training the model in the domain where data is more available and easier to gather [IV.242]. Large-scale, open-source audio collections can be used for model pre-training [IV.245], such as AudioSet dataset recently released by Google, which contains over 2 million audio clips classified into 632 audio events, covering human and animal sounds, musical instruments and common environmental sounds [IV.246], including sounds of a large number of daily activities (e.g. bathing, washing hands, flushing the toilet, brushing teeth, shaving, walking, chopping food, frying food, using a microwave oven etc.) relevant for human activity recognition.

### 4.5.4 Challenges for activity recognition

The use of audio signals for activity/behaviour recognition, or integration with video and wearable sensors, is still an understudied area with many challenges to be resolved. Some of the challenges are specific for the audio modality, whereas the others are shared across different modalities. We provide here a non-exclusive list of challenges that the human activity/behaviour recognition community should respond to in the incoming years.

- **Lack of available datasets** is one of the biggest challenges, since data collection and annotation is a time-consuming and expensive task. Publicly available datasets either have limited size, limited set of activities or activities only in a specific domain (e.g., Kitchen20 dataset), or they are more general datasets containing various environmental sounds, not specially designed for the human activity recognition task (e.g., AudioSet). Therefore, collecting a large-scale dataset is a definite prerequisite to be able to use state-of-the-art deep learning models for robust activity recognition systems.
- **Class imbalance** is a common problem in activity recognition, since certain activities naturally occur less frequently than the others. For example, communicating will typically occur with a higher frequency than brushing teeth, while sleeping may be collected over much longer duration, leading to higher number of data instances. This causes problems from a machine learning point of view since most of the algorithms expect balanced class distributions [IV.247]. There are different ways how to address the class imbalance problem, such as cost-sensitive learning that weighs the training samples proportionally to misclassification costs; or using synthetic minority over-sampling technique (SMOTE) [IV.248] or borderline-SMOTE [IV.249] to create the synthetic samples of minority class data instances, thus performing oversampling.

- **Feature extraction** is challenging for activities with similar characteristics (e.g., walking and running), leading to difficulties to produce distinguishable features that uniquely represent similar activities [IV.250].
- **Concurrent activities** represent multiple activities taking place simultaneously (e.g., watching TV while communicating over a phone) [IV.250]. The performance of the human activity recognition algorithm will highly depend on the ability to separate the activities.
- **Composite activities** are sequences of simple atomic activities (e.g., washing hands can be composed of turning on the tap, soaping, rubbing hands, turning off the tap). The performance of the human activity recognition algorithm will highly depend on precise data segmentation techniques to decompose the composite activity.
- **Controlled experimental settings** used in the majority of studies, where the users are instructed to perform a predefined set of activities, may cause the trained model to perform poorly in real-world settings. The reason is that human activity patterns depend on behavioural context, which changes in different environments, especially if the user is aware of being under surveillance. Therefore, it is necessary to incorporate in-the-wild experiments for robust human activity recognition, where sounds are captured with microphones found in real-world devices (e.g., smartphones) and recorded in real-world environments [IV.251].
- **Privacy issues** need to be properly addressed in case of continuous audio recording for HAR. Although audio or speech signals can provide complex semantic information relevant for detecting activities, they also raise privacy concerns. An intuitive way to cope with this problem is degrading the signal quality to the point that speech and other acoustic markers become unintelligible, but this usually comes at the cost of decreased HAR performance. Speaker identity can be protected using speaker de-identification techniques, e.g., via voice transformation [IV.252]. However, the content of spoken language is still preserved. A solution to this problem might be to analyse the audio information on the fly, without saving the raw audio signals [IV.253].

In [IV.254], the authors propose a human activity recognition model based on audio features. A dataset of 7 activities (brewing coffee, cooking, using the microwave oven, taking a shower, washing dishes, washing hands and teeth brushing) complemented with noises that do not correspond to those activities is used; sounds were recorded in different dwellings to consider heterogeneous spatial environments, audio reflections and background noises. Data was trimmed to 10-seconds audio clips with a sample rate between 8 and 44 KHz (only mono information was used); from these clips (a total number of 64) a feature vector is computed. The first step of the method consists in selecting the most relevant features using a genetic algorithm, forward selection, and backwards elimination. This process resulted in a feature vector consisting of 9 features with potential classification power: trimmed mean, standard deviation, 95th percentile, and 6 Mel-frequency cepstral coefficients (MFCC). The resulting feature vector was used to train a Random Forest and a Neural Network. The Random Forest outperformed the Neural Network, reaching averaged accuracy values (8 activities) of 0.814, 0.856 and 0.842 when trained with the 9 selected features, with all initial features, and with only the MFCC features, respectively. Obtained results indicate that MFCCs accurately describe the behaviour of an audio signal to recognize human activities; furthermore, Random Forests are envisioned as potentially integrable with mobile devices for quasi real-time activity recognition. Interestingly, the Neural Network approach dramatically decreased its performance when using the complete set of features.

A comparison of sequential vs. non-sequential models for on-line HAR using audio and home automation sensors is presented in [IV.255]. The considered sensor types are binary presence detectors like Presence Infra-Red sensors, continuous microphone signals and temperature measurement. Streams of raw sensor data and inferred information are summarised as vectors of features, each of which corresponds to a temporal window of a duration T. The feature vector comes together with an activity label that is generated from the ground truth by taking the activity having the longest duration within Wn as the label. Two different datasets of eight Activities of Daily Living (cleaning, dressing/undressing, eating, hygiene, phone use, reading/computer/radio, sleeping, unknown) are selected. These activities are modelled under the six methods considered for comparison: Hidden Markov Model (HMM), Conditional Random Fields (CRF), Markov Logic Network (MLN), Support Vector Machine (SVM), Random Forest (RF) and non-sequential MLN. The datasets are:

- The multimodal SWEET-HOME dataset, part of the SWEET-HOME corpus (SH dataset), which includes the recording of 21 persons (including 7 women) from the data retrieved from all sensors in a daily living context, recording 26 hours of data. About 18 hours were kept for an activity recognition experiment while the remaining time was retained for specific audio analysis. Data were annotated with the 7 classes of activity using the Advene software.
- The HIS corpus monitors the activity of a person living alone at home, with the aim of helping geriatricians to evaluate the dependency level of various elderly people. Seven activities were selected to be classified automatically. 15 healthy and nonelderly subjects (six women and nine men) participate in the dataset. In total, about 13 hours of data have been acquired in the HIS flat. Data were annotated using the same conventions as for the SH corpus.

The method used to evaluate the classifier was based on Cross-Validation but used a specific type namely Leave-One-Subject-Out-Cross-Validation (LOSOCV). Results show that the CRF approach reaches the highest accuracy in most of the cases, namely in 3 out of 4 conditions (i.e., SH dataset without and with the *Unknown* action and HIS dataset without and with the *Unknown* action). The HMM approach shows the best accuracy for the HIS without including the *Unknown* class. MLN ranked always as the second or third method. SVMs were always the worst classifiers, under all the considered conditions, whilst the HMM on the SH dataset and the RF for the HIS dataset (even if this was amongst the best for the SH dataset). The reader is referred to [IV.255] for the precise accuracy values.

### 4.5.5 Multi-modal Activity and Behaviour Recognition

An extensive review and classification of various approaches and technologies used for activity recognition related to smart home and AAL applications is presented in the paper [IV.256]. The authors mainly used the data from the Web of Science (WoS) database and reviewed 92 relevant research works. Technologies used for activity recognition in smart homes and for AAL applications are grouped into four groups, namely those using smartphones, wearables, video, and electronic components. Additionally, two emerging technologies were identified: IEEE 802.11 (Wi-Fi) and assistive robots. These technologies have been related to six applications such as monitoring health conditions, health care, emotion recognition, fall detection, posture recognition, localisation, occupancy, mobility, and generic activity recognition applications. Assistive robots and Wi-Fi can be combined with others commonly used to expand the spectrum of applications for activity recognition in smart homes and AAL, Methods for activity recognition based on devices such as infrared cameras, ultrasonic sensors, other video and audio capture devices, RGB-D sensors, optical sensors, wearable sensor bands, smartwatches, smartphones, temperature and proximity sensors, magnetometers, EEG, and heart rate monitors can be combined with assistive robots and Wi-Fi to expand the spectrum of possible applications. In particular, the authors indicated that video-based technologies and those using audio and video processing in smartphones and smartwatches are very popular choices for implementing systems for activity and behaviour recognition. For example, video-based approaches are used by 28 % of proposed solutions, while smartphones count to 26% and wearables to 33%, of which smart bands and smartwatches are by far mostly used [IV.256]. Regarding the technology used for video-based systems, the authors stated that 60% of the reviewed works used RGB-D sensors. Sometimes, RGB-D cameras are combined with Vicon Systems cameras [IV.257] or thermal cameras [IV.258]. Sometimes, thermal cameras are used in combination with smartphones [IV.259], wearables [IV.260] and infrared cameras [IV.261].

Various activity recognition systems are reviewed from a taxonomy of applications in [IV.262]:

- **Active and assisted living systems for smart homes**: the analysis of active and assisted living systems for smart homes shows that domestic (including audio and video based) or wearable sensors are very common for HAR, combined with machine learning. RFID and NFC are also considered although more focused on eHealth than on HAR.
- **Healthcare monitoring applications:** in the healthcare monitoring application domain, the analysis reveals that systems are made of several HAR systems, such as fall detection, human tracking, or security alarms. Once any of these systems detects the need of help, an alarm is launched either via SMS or email. The most common sensors used in this domain are those accelerometers and gyroscopes, but also those based on video technology (including multi camera views). Floor vibration and sound sensing are also relevant, especially for fall detection.

- **Monitoring and surveillance systems for indoor and outdoor activities**: in the field of security and surveillance applications the use of video-based systems is the most widely extended approach.
- **Tele-immersion:** these applications allow users to share their presence in a virtual environment and interact with each other in real time as if they were present in the same physical place but in different geographical environments. Tele-immersion applications have to resort to compression methods so as to reduce the amount of data being transference.

Then, based on these categories, the proposed systems are classified according to the methodology used for HAR:

- **Visual**: this is the most popular application. In the field of active and assisted living, the most relevant approaches are those based on RGB-D cameras, the use of histograms of 3D joint locations (HOJ3D), the use of probabilistic reasoning (Bayesian and Markov processes) or the use of segmentation techniques such as the Gaussian mixture models combine with other techniques such as image moments. In the field of healthcare monitoring systems, visual systems are not that popular. They have been mainly applied to fall detection systems. Security and surveillance systems, on the other hand, have extensively used visual systems. Approaches such as Scenario Recognition based on knowledge (ScReK) or bag-of-words framework to detect fight events using Space–Time Interest Points (STIP) and Motion SIFT (MoSIFT) action descriptors are some of the most relevant works. Sports and outdoors activities can benefit from the use of visual systems. These can be aimed at improving performance or analysing game plans.
- **Non-visual**: in active and assisted living and smart home systems sensors such as light, sound, contact, or motion, are the most common ones. Nonetheless, the inconsistency and unreliability of the sensors is one of the major challenges faced for HAR. Annotated datasets are very relevant in this field. Other indoor environments use non-visual systems for purposes different from ADL recognition, such as detect theft in museums or track and localize passive targets.
- **Multimodal sensor technology**: this approach involves the combination of visual and non-visual sensors to recognize human activities. In the smart home environment, due to privacy concerns, non-visual sensors are preferred over visual ones. In the healthcare domain, on the other hand, multimodal systems are used as observation tools, to monitor health conditions of patients. For outdoor environments, such systems are employed for the recognition of human activities in outdoor environments. Finally, the improvements on the field of mobile devices have turned them into a very sophisticated device for activity recognition as a source of multimodal information.

Finally, this work summarises the most relevant datasets (publicly available) for HAR as follows:

- Video-based
  - KTH data set. KTH video database
  - Weizmann data set
  - INRIA Xmas Motion Acquisition Sequences multi-view data set
  - UCF sport data set
  - YouTube action data set.
  - i3DPost multi-view human action data sets.
  - MOBISERV-AIIA database
  - IMPART data sets

- Non-video-based datasets
  - CASAS.
  - Benchmark (Van Kasteren) data set
  - Sweet-Home data sets

Although Microsoft's Kinect technology can provide high accuracy and is able to solve the environment light problem, it does not function well in the presence of obstacles, in crowded rooms or behind walls. In [IV.263] the authors propose a hybrid system named HuAc, which combines Kinect and Wi-Fi activity recognition approaches to achieve an increased robustness of the system for applications in indoor environments with presence of weak light and different perspectives. The system uses the Wi-Fi Channel State Information (CSI) together with crowdsourced skeleton joints. In the Wi-Fi module, the moving variance of CSI is used to roughly recognize the start and the end of activities, while the distribution of CSI can be applied to obtain details about the activities. The $K$-means algorithm is used for clustering effective subcarriers in order to improve the stability of the system. To evaluate the performance of the proposed approach, the authors created a dataset that contains 16 activity classes, each performed 50 times by ten people. The KARD dataset contains RGB video, depth video, and 15 skeleton points. Then, they used the k-NN, RF, and DT algorithms to verify the effectiveness of the dataset. The authors have shown that the proposed approach achieves 93% of average recognition accuracy in the considered dataset.

The question on how to select both appropriate sensors and data fusion approaches to monitor a number of activities of daily living (ADL) has been addressed in [IV.264], the authors explore a set of 8 ADLs (i.e., transferring, dressing, toileting, feeding -breakfast, lunch and dinner-, bathing and continence) performed by older occupants in a Smart Home. In this work, authors' focus is based on a 3-dimensional view: localisation is achieved by presence sensors (i.e., Presence IR), posture (lying, sitting and standing) either by inertial sensors or image processing, and time by a central clock. Their approach is sensor-agnostic, with a first step of dividing the captured signals into time segments of equal length and computing the trend of each dimension. Then, this trend of information vector is compared with a reference matrix that stores the expected values for each activity that wants to be recognized, creating a new matrix that contains the distance to each ADL. Then a level of achievement vector is computed for each ADL to finally take a classification decision under the criterion maximum a posteriori. Proposed method was tested against data coming from 7 young and 4 older volunteers that executed the ADLs on demand (not real-life data was used for the validation). Overall performance was found better for the group of older persons, with a sensitivity of 86.9% and a specificity of 59.3% than for the younger, which obtained a sensitivity of 67.0% and a specificity of 52.6%. The main weakness of this work is the high dependence with the reference matrix.

To model user interactions and perform a multi-user activity recognition, a method based on the information collected by a multimodal wearable sensor platform and Coupled Hidden Markov Models (CHMMs) can be used, as proposed in [IV.265]. Activity information consists in user movement, location, interaction with objects (i.e., object touched and sound), interaction with other humans (i.e., voice), and environmental data (i.e., temperature, humidity, and light). To register audio information, a commercial audio recorder was used (no further details provided). Collected information is aggregated to compose a 47-dimensional observation vector every second per user wearing the sensor platform. Information contained in this feature vector is diverse given the multimodal nature of the data acquisition system; in terms of audio features, the vector contains the following: temporal variation of the audio signal (i.e., standard deviation over the window, dynamic range, and zero-crossing rate) and frequency domain characteristics (i.e., centroid of the spectral power distribution and bandwidth). For each user, an HMM is trained with a sequence of feature vectors that represent the observable variable, and the activity label representing the hidden one. HMMs corresponding to different users are then connected afterwards in a CHMM. To estimate the parameters associated to the CHMMs, first individual HMMs are trained using the maximum likelihood method, and then they are coupled with their inter-chain transition probabilities learnt during the training dataset, so CHMMs do not need to be re-trained. To validate this approach, authors applied a 10-fold cross validation method, measuring performance with the time-slice accuracy to represent the percentage of correctly labelled time slices; the length of the slice was set to 1 second. Single-user (i.e., brushing teeth, washing face, brushing hair, making pasta, making coffee, toileting, ironing, and making tea) and 2-user activities (i.e. making pasta, cleaning a dining table, making coffee, toileting -with conflict-, watching TV, and using computer) were tested. An average accuracy of 85.46% was obtained for the CHMMs, which performed better in multi-user activities. The reason behind that can be linked to the fact that in multi-user ADLs, observations from the users are useful in the model since they are all doing the same, so coupling single-chain HMMs is effective; however, when recognizing single-user activities, observations from other users will become background noise.

In [IV.266], authors focus on the recognition of daily activities within home environments where multiple sensors, actuators and home automation equipment coexist. To this aim, MLNs are proposed to overcome typical problems associated with statistical methods: lack of a formal base to represent uncertainty and interpretability. With this logical approach, expert knowledge can be added to the models. Information coming from 2 sources, namely raw audio signals and the home automation system providing symbolic information, is processed. Three types of abstraction are considered: localisation, speech/sound recognition, and activity level. For each temporal window, a 69-dimensional feature vector is calculated. The proposed method was applied on data collected in a smart home equipped with 7 radio microphones installed in the ceiling (2 per room except for the bathroom, that only had 1), switches and door contact detectors connected to a KNX network, 2 infra-red movement detectors also connected to the KNX network and electricity and water meters. Twenty-one persons participated in the study (38.5±13 years, 7 women). Subjects were asked to perform several daily activities, but no specific instructions on how to perform them were provided apart from uttering 40 predefined casual sentences on the phone (users were free to speak any sentence they wanted). The experiment registered 26 hours of information that were properly annotated to train the MLNs (a window size of 1 minute and a 25% overlap were set to recognize activities every 45 seconds). NB and SVMs were used as a baseline for comparison; a leave-one-out strategy was used for learning. MLNs obtained an overall accuracy of 85.3%, higher that those obtained by the NB (66.1%) and SVMs (59.6%). Moreover, a deeper analysis of the results indicate confusion when activities share a common location (e.g., tidying up, sleeping, or resting).

The problem of group activity recognition using a multimodal deep learning approach that fuses video and audio data is tackled by the authors of the paper [IV.267]. In this case, since models are dealing with multiple individuals, it is not enough to identify what an individual is doing but it is necessary to detect the combinations of activities and interaction amongst them; moreover, recognizing objects within the scene may help differentiate between group activities being carried out. The proposed method consists of two steps. First audio-visual features representing the activity are extracted; to this aim, 2 streams are used, a visual model to process visual cues, and an audio model to process the audio signal. The second step consists in combining the extracted visual and audio features; to this end, a Long Short-Term Memory (LSTM) network is used. The visual network used as input a video feed of size 144x144, down sampled from the original higher resolution source of 720x576; the encoding bit rate was 23.000 kbit/s with a maximum interval of 25 frames between 2 consecutive MPEG key frames. The basis of this visual model is represented by a deep residual network consisting of five main bottlenecks, each including a convolution (with 3 layers) and a residual block (ResNet-50). The audio signal was down sampled from 48 kHz to 16 kHz, and made available in WAV format; in addition, each audio file was smoothed out with a lowpass filter, and an automatic energy threshold was applied to distinguish between speech and silence. Mel spectrograms were extracted from the audio files, taking the logarithm for stabilisation. The resulting log Mel spectrogram further used for feature extraction had a size of 144x144. Authors also used residual learning for processing audio spectrograms. For the fusion network, 2 unidirectional LSTM layers with 128 cells each are proposed. The AMI Corpus dataset (100 hours of recordings) was used for the validation of the models, considering an 80/20 ratio between training and validation. Audio-visual sequences took place in an office environment, and were annotated according to 3 classes: presentation, meeting or empty. Obtained results indicate that when activity recognition was performed based on video streams, an accuracy of 99% was reached, being models more likely to misclassify a meeting as presentation than vice versa. On the other hand, when only audio information was used, models did not perform anywhere close. Deeper networks were used to evaluate the extent to which performance could benefit, but data were overfitted leading to a lower accuracy in the validation set. However, transfer learning (no random initialisation of the network weights) showed that it could boost performance for activity recognition, even on the audio dataset. Finally, the proposed multimodal architecture, which fuses the video and audio streams achieved an accuracy of 100% on the AMI dataset.

A multimodal, machine learning-based approach called WiVi, which is based on the fusion of information from WiFi-enabled IoT devices and an RGB camera has been proposed in [IV.268]. The proposed method exploits the Channel State Information (CSI) from the WiFi physical layer, which describes the propagation of WiFi signals between the transmitter and the receiver. It was experimentally found that different human activities perturbate the WiFi signals differently, and this information can be extracted from CSI measurements. A CNN-based WiFi sensing module that consists of a WiFi feature extractor followed by a WiFi classifier was designed, and fed with WiFi frames constructed from CSI phase difference data from 114 subcarriers. The vision sensing module, on the other hand, comprises a Convolutional 3D (C3D) CNN that is pre-trained on the Sports1M dataset. The unimodal activity estimations of both sensing modules are then fused via ensemble learning. The outputs of the two modules are concatenated and fed into a deep neural network (DNN), which provides the final activity inference. The proposed method was tested on data collected by the authors themselves, which includes three common activities, sitting, standing and walking, performed by two volunteers at arbitrary locations inside a conference room. A cross-validation accuracy of 97.5% was reportedly achieved by the WiVi method, in comparison to the 95.83% and 95% accuracy achieved by the individual WiFi and vision sensing modules. Under poor illumination conditions, the proposed method is set to rely on the WiFi module alone, which retains a 95.83% accuracy, in comparison to the vision module whose accuracy degrades to 46.67%.

### 4.5.6 Summary of Different Approaches

This section has covered the use of video, audio and multimodal approaches for recognizing human actions and behaviours. A number of original research articles and literature reviews have been analysed. Of those, 7 are video-based approaches, 2 audio-based and 4 tackle the problem multimodally. Table 2 summarises all the works studied in this section, while Table 3 depicts the datasets found in this state-of-the-art.

The list of methods in Table 3 indicates an increasing uptake of deep learning techniques in recent years. Such techniques have not only been applied to image data [IV.231], but also to multimodal data that combines video and audio [IV.267], and video and WiFi [IV.261]. This was to be expected since deep learning has been gaining increasing popularity over the past few years, and has found its application within many different domains.

The results of the reviewed methods tend to suggest that several of the audio and video-based methods in Table 2 perform comparably [IV.231], [IV.233], [IV.254], [IV.255]. In this case, the choice of which modality to work with could be made on the basis of which one offers the highest privacy and robustness for the environmental conditions under consideration. For instance, for users that would prefer not to have their images captured, especially within home environments, the use of audio alone could reduce their privacy concerns. Audio data is also not susceptible to variations in illumination conditions as opposed to image data, but outdoor environments with considerable background noise could pose a significant challenge to activity recognition methods that rely on audio data alone.

The use of multimodal data, on the other hand, could potentially improve the activity recognition accuracy [IV.267], [IV.268]. Of particular interest are the methods of Florea and Mihailescu [IV.267], and Zou *et al*. [IV.268], which fuse video and audio, and video and WiFi, respectively. The use of WiFi for activity recognition is an attractive proposition because it does not give rise to the privacy concerns that are typically associated with video and audio data. The use of multimodal data could also offer a more robust solution, since the method could rely on any one of the modalities under challenging conditions. An example is the method of Zou *et al*. [IV.268], which relies on WiFi alone under poor illumination conditions when the accuracy of the vision module degrades.

Most importantly, however, is to exercise caution when interpreting the results of different methods. While we have inferred several conclusions from the reviewed methods, it is also important to note that these methods utilise a variety of test datasets, which makes the comparison of their results a challenging task. The lack of available datasets is, indeed, one of the biggest challenges of video and audio-based human activity recognition since publicly available datasets either have a limited size or a limited set of activities, or the collected data is owned by the researchers and is not made publicly available. Hence, the collection of a suitably large dataset that enables appropriate benchmarking between different methods is certainly one of the prerequisites for advancing the state-of-the-art.

Table 2. Overview of the approaches for activity recognition and their main characteristics

| Audio, Video, or Multimodal | Description of the method or model | Type of algorithm | Interaction type and level | Performance (accuracy, robustness, ...) | Reference |
|---|---|---|---|---|---|
| Video | IR camera / thermal images / supervised Convolutional Neural Network (CNN) | supervised Convolutional Neural Network (CNN) | Monitoring/ action recognition/ HCI | Accuracy of the classification between different action classes: 87.44% | [IV.231] |
| Video | A hybrid system named HuAc that combines Kinect and Wi-Fi activity recognition | n.a. | Monitoring/ action recognition/ HCI | Average recognition accuracy in the considered dataset: 93%. | [IV.263] |
| Video | Handcrafted features and learning-based approaches | n. a. | Monitoring/ action recognition/ HCI | n.a. | [IV.229] |
| Video | Network of cheap low-resolution visual sensors (30x30 pixels) | n. a. | Monitoring/ action recognition/HCI | Overall MAE of 9.39 minutes | [IV.230] |
| Video | Following an identification of correct skeleton key points based on their spatial and temporal coherence, the skeleton key points in a sequence of image frames are encoded into a | Vision-based activity recognition:<br>Image classification using a modified AlexNet model | Action recognition | Grayscale input: KARD gestures - 77.5% KARD actions - 85.8% Florence 3D Actions - 89.4% NAD - 53%-92.5%<br>Colour input: KARD gestures - 100% | [IV.233] |

| Audio, Video, or Multimodal | Description of the method or model | Type of algorithm | Interaction type and level | Performance (accuracy, robustness, …) | Reference |
|---|---|---|---|---|---|
| | feature matrix based on a distance measure, and converted into grayscale and colour images to be input into a modified AlexNet model for activity classification | | | KARD actions - 100% Florence 3D Actions - 100% NAD - 100% | |
| Multimodal | Sensor-agnostic approach | Maximum a posteriori criterion in relation to the reference matrix | Monitoring/ action recognition/ HCI | Sensitivity: 86.9% older adults and 67% younger persons<br>Specificity: 59.3% older adults and 52.6% younger persons | [IV.264] |
| Audio | Embedded microphones in regular smartphones | Feature vector with relevant parameters (trimmed mean, standard deviation, 95th percentile, and 6 Mel-frequency cepstral coefficients) and a random Forest classifier | Monitoring/ action recognition/ HCI | Accuracy values of 0.814, 0.856 and 0.842 when trained with the 9 selected features, with all initial features, and with only the MFCC features | [IV.254] |
| Video | Multi-camera selection based on recurrent neural networks (RNNs) | Recurrent Neural Networks (RNNs) | Monitoring/ action recognition/ HCI | Accuracy for three camera views when 20% to 70% of the video frames are randomly dropped: (70.53, 65.38, 60.99) %<br>Average early recognition performance for their learned view selection policy: 58.01% | [IV.269] |

| Audio, Video, or Multimodal | Description of the method or model | Type of algorithm | Interaction type and level | Performance (accuracy, robustness, …) | Reference |
|---|---|---|---|---|---|
| Audio | Sequential and non-sequential models applied (and compared) to information retrieved from multiple sensors, actuators and automation equipment coexist, including audio sensors | Hidden Markov Model (HMM), Conditional Random Fields (CRF), Markov Logic Network (MLN), Support Vector Machine (SVM), Random Forest (RF) and non-sequential MLN | Monitoring | CRF achieves an accuracy of 85.43 (without the unknown action) 83.57 (with the unknown action) for the SweetHome dataset and 75.85% and 69.29% respectively for the HIS dataset | [IV.255] |
| Multimodal | Exploits information coming from in user movement, location, interaction with objects (i.e., object touched and sound), interaction with other humans (i.e., voice), and environmental data (i.e., temperature, humidity, and light) | Coupled Hidden Markov Models (CHMMs) to model user interactions and perform recognition | Action recognition | An average accuracy of 85.46% was obtained for the CHMMs, which performed better in multi-user activities. | [IV.265] |
| Multimodal | Problem of group activity recognition using a multimodal deep learning approach that fuses video and audio data | Deep learning, two steps: RestNet-50 + LSTM | Group activity recognition | 100% accuracy | [IV.267] |

| Audio, Video, or Multimodal | Description of the method or model | Type of algorithm | Interaction type and level | Performance (accuracy, robustness, …) | Reference |
|---|---|---|---|---|---|
| Multimodal | Multimodal fusion of WiFi data, obtained from Channel State Information (CSI) measurements that describe signal perturbation due to different human activities, and vision data from an RGB camera | Multimodal approach based on CNNs and DNN, Vision-based activity recognition and Wi-Fi | Sitting, standing and walking | Accuracy of 97.5% from the multimodal approach | [IV.268] |
| Video | Features extracted from Kinect data are reduced in dimensionality, and fed into a K-nearest neighbour (K-NN) classifier and a Support Vector Machine (SVM) to investigate the best classifier between the two for activity recognition | Vision-based/Kinect/K-nearest neighbour classifier/Support Vector Machine | Action recognition | K-NN: 94.5% average accuracy across all actions SVM: 98% average accuracy across all actions | [IV.234] |

Table 3. Datasets used for activity recognition

| Audio, Video, or Multimodal | Short description of datasets | Reference |
|---|---|---|
| Video | An extensive list of relevant datasets. | [IV.229] |
| Video | Data corresponding to 10 months of real-life recordings using a network of 10 low-resolution visual sensors working at 50 fps, where ground truth was collected by mixing the information contained in diaries and visual inspection of the images | [IV.230] |
| Video | The dataset consists of 5278 random image samples from thermal videos, 30 seconds long each, which represent six activity classes such as walking, sitting on a chair with or without a desk in front, standing, fallen/lying on the desk and fallen/lying on the ground. | [IV.231] |
| Video | NTU RGB-D dataset: collected by three Microsoft Kinect cameras, and comprising 60 different action classes: 40 from daily living, 9 health-related and 11 two-people actions.<br>IXMAS dataset: RGB images featuring 11 action classes from 10 actors, and captured by four side-view, and one top-view, cameras.<br>nvGesture dataset: a collection of depth images recorded by a SoftKinetic depth camera and RGB images recorded by two (frontal and top-view) DUO 3D cameras, featuring 25 gesture classes for HCI. | [IV.232] |
| Video | KARD dataset, containing 18 activities (10 gestures and 8 activities), with each activity performed three times by 10 persons; Florence 3D Actions dataset, containing RGB-D images and complete skeleton data; and the Noisy Activity dataset, which includes non-ideal conditions, such as obstruction and the presence of non-target objects. | [IV.233] |
| Video | KARD dataset, containing 18 activities (10 gestures and 8 activities), with each activity performed three times by 10 persons. | [IV.234] |
| Audio | AudioSet[6] is a large-scale collection of more than 2 million manually labelled 10-second audio excerpts from YouTube videos with the total duration of 4971 hours, classified into 632 audio events, including human and animal sounds, music sounds and environmental sounds. 485 audio event categories have at least 100 instances. Dataset is collected by the Sound Understanding group at Google and used primarily for the audio event recognition task. Therefore, AudioSet is not specifically designed for activity recognition, but being a large dataset that contains a number of sounds for activities of daily living, such as (chatting, listening to music, watching TV, flushing toilet, frying food, | [IV.246] |

[6] https://research.google.com/audioset/ (last accessed: 23/02/2022)

| Audio, Video, or Multimodal | Short description of datasets | Reference |
|---|---|---|
| | bathing/showering, brushing teeth, etc.), it is used for an activity recognition task in a number of recent studies [IV.251], [IV.255], or as a part of dataset for an annual DCASE[7] task 4 challenge for a closely related task of sound event detection in domestic environments. | |
| Multimodal | Sweet-Home corpus[8]: collected within the Sweet-Home project for a new smart home system design based on audio technology. It is composed of 3 subsets containing audio, video and/or home automation data:<br>• Multimodal subset for HAR contains over 26 hours of audio, video and data acquired using the home automation sensors. It is used for recognition of the following activities of daily living: sleeping, resting, dressing/undressing, preparing the meal, feeding, doing the laundry, hygiene activity and communicating.<br>• Home Automation Speech subset for automatic voice command recognition in French language contains 2.5 hours per channel of short-spoken commands and 36 minutes of spoken words for acoustic adaptation (23 speakers in total). Eight audio channels were recorded in different rooms and in different conditions (e.g., in clean conditions without noise, with the radio turned on and with the classical music played in the room).<br>• Interaction subset for evaluating user interaction with the Sweet-Home system. The subset contains multiple repetition of 15 simple voice commands (e.g., open/close blinds, ask about the temperature, ask to call a relative, etc.), categorised into initiate commands, stop commands and emergency calls and spoken by 16 participants in the study.<br>The dataset is not designed for any commercial use and can be used for research purposes only. | [IV.253], [IV.255] |
| Audio | Sounds recorded in different dwellings to consider heterogeneous spatial environments, audio reflections and background noises. Number of clips and homes not specified. | [IV.254] |
| Video | WiAR dataset containing Wi-Fi Channel State Information (CSI) and Received Signal Strength Indicator (RSSI) data related to 16 activities performed 50 times by 10 volunteers (five females and five males, the height of human body ranges from 150 cm to 185 cm) together with crowdsourced skeleton joints. | [IV.263] |
| Multimodal | Data coming from 7 young and 4 older volunteers that executed the ADLs on demand (not real-life data was used for the validation). | [IV.264] |

[7] http://dcase.community/challenge2019/task-sound-event-detection-in-domestic-environments (last accessed: 23/02/2022)
[8] http://sweet-home-data.imag.fr (last accessed: 23/02/2022)

| Audio, Video, or Multimodal | Short description of datasets | Reference |
|---|---|---|
| Multimodal | AMI Corpus[9] dataset is a multi-modal data set containing 100 hours of meeting recordings. | [IV.267] |
| Multimodal | Dataset consisting of three common activities, activities, sitting, standing and walking, performed by two volunteers at arbitrary locations inside a conference room. | [IV.268] |
| Multimodal | Health Smart Home (HIS) corpus[10]: collected in the smart home that is composed of a living room, bedroom, bathroom, toilet, kitchen and corridor, and equipped with microphones, cameras, accelerometer, magnetometer, infra-red, temperature and hygrometry sensors, for monitoring of activities of daily living such as sleeping, resting (watching TV, listening to the radio, etc.), feeding, dressing/undressing, communicating, using the toilet and the hygiene activity (washing hands, brushing teeth, etc.). Fifteen healthy participants (9 men and 6 women) participated in the study, 4 of whom were non-native French speakers. There was no time limitation on the time spent in the smart home, which varied from 23 min to more than 1.5 h, depending on the participant. In total.<br>The dataset is not designed for any commercial use and can be used for research purposes only. | [IV.270] |
| Multimodal | Home Tasks Activities Dataset (HTAD)[11] contains audio and wrist accelerometer recordings of 3 participants in the study (1 female and 2 males aged between 25 and 30 years) and performing 7 home activities: mopping floor, sweeping floor, keyboard typing, brushing teeth, washing hands, eating chips and watching TV. Each activity was performed for approximately 3 minutes.<br>The dataset can be freely used for research and educational purposes. | [IV.271] |
| Multimodal | Dem@Care dataset[12] is a collection of 12 datasets acquired using multiple in-situ and wearable sensors as a part of the lab and home experiments carried out by the Greek Alzheimer's Association for Dementia and Related Disorders. It is a multimodal dataset that contains audio recordings, static RGB (colour) video data, static RGB-D (colour-depth) video data, accelerometer data and data acquired using sleep and motion sensors. Datasets are used for monitoring and assessment of cognitive, behavioural, emotional, mental and functional status of people with dementia. One of the use cases is related to situational analysis of daily activities of the people with dementia in their home environment.<br>The datasets are available for research purposes under the specific terms of use. | [IV.272] |

[9] https://groups.inf.ed.ac.uk/ami/corpus/ (last accessed: 23/02/2022)
[10] https://lig-getalp.imag.fr/health-smart-home-his-datasets-2/ (last accessed: 23/02/2022)
[11] https://osf.io/4dnh8/ (last accessed: 23/02/2022)
[12] https://demcare.eu (last accessed: 23/02/2022)

| Audio, Video, or Multimodal | Short description of datasets | Reference |
|---|---|---|
| Audio | Kitchen20 dataset[13]: audio dataset for human activity recognition in a kitchen environment. It contains 800 audio instances uniformly distributed into 20 categories: 10 categories related to human activities (eating, cleaning dishes, chopping vegetables, opening/closing cupboards/drawers, manipulating cutlery/plates, sweeping floor, etc.), and 10 categories related to kitchen appliances (running dish-washer/microwave, blender mixing, opening/closing refrigerator, water flowing, etc.). Each category is represented by 40 audio recordings with the duration of 5 s, sampled at 44.1 kHz. The recordings were partly sampled from the FreeSound[14] dataset (662 instances) or recorded in a kitchen environment (138 instances) using a smartphone by 8 participants in the study. | [IV.273] |
| Audio | Ok Google, What Am I Doing[15]: collection of voice-based interactions between participants and their Google Home conversational assistant that are used for recognition of activities of daily living. The dataset is collected at 3 locations (kitchen, living room, bathroom) in a home environment with a total of 19 distinct activities (peeling/chopping vegetables, boiling water, washing dishes, disposing garbage, keyboard typing, watching TV, washing hands, toilet flushing, bathing/showering, etc.) performed by 14 participants in the study. The system is triggered by a voice command 'Ok Google' or 'Hey Google' to make a 30-second audio recording with the question-answer interaction and background sounds, which serve as indicators of a certain activity. | [IV.274] |
| Video | The TST Fall Detection dataset v1[16] contains depth frames collected using Microsoft Kinect v1 in top-view configuration and can be used for fall detection. | [IV.276] |
| Video | The TST Fall Detection dataset v2[17] contains depth frames and skeleton joints collected using Microsoft Kinect v2 and acceleration samples provided by an IMU during the simulation of ADLs and falls. | [IV.277] |

## 4.6 Activity assistance and personal assistants

Digital solutions in mental, physical and emotional assistance are rapidly appearing because of two main reasons: firstly, the technology is rapidly improving and it becomes possible to develop such solutions, and secondly, the it has been prognosticated for already two decades that that there will not be enough

[13] https://github.com/marc-moreaux/kitchen20 (last accessed: 23/02/2022)
[14] https://freesound.org (last accessed: 23/02/2022)
[15] https://dataverse.tdl.org/dataset.xhtml?persistentId=doi:10.18738/T8/OCWAZW (last accessed: 23/02/2022)
[16] http://ieee-dataport.org/documents/tst-fall-detection-dataset-v1 (last accessed: 23/02/2022)
[17] http://ieee-dataport.org/documents/tst-fall-detection-dataset-v2 (last accessed: 23/02/2022)

people to work in the care sector in the near future, and this will be a significant fraction of all the workforce shortage [IV.278]. Healthcare digital solutions had been previously employed in the hospitals [IV.279], but now the technology is rapidly moving into the home environments and everyday lives [IV.280]. New emerging technologies and current public policy debate mutually impact each other and the digitalisation trend [IV.281], [IV.282] highlighting the role of the technology in the much desired independent living [IV.280],[IV.283]. To a great extent, digital solutions are leading rhetoric and are described as the ultimate solution to positively affect patients' health and well-being [IV.284].

This section considers AAL solutions developed to support various activities. First, the section looks at end-users and activities supported by activity assistance systems and personal assistants. Further two of the most common types of assistants (i.e., software and robotic) are explored along with other more complex solutions that integrate several technologies for the assistance. We also seek the answer to two questions: (1) how do end users and the supported activities impact technical solutions and (2) what design strategies and technical solutions can be used to mitigate non-technical challenges. At the end of the section, conclusions regarding these questions are given.

### 4.6.1 End-users of activity assistance and personal assistants in AAL systems

AAL systems in private homes are a new arena for digital care solutions for the citizens. Even with all the perceived benefits, there are fears that digital solutions may contribute to social isolation and lead to other unforeseen issues. This is largely due to the fact that target audiences for assistive technologies include socially sensitive groups, such as stroke survivors, elderly people and individuals with dementia, and children with autism spectrum disorders, creating a range of ethical concerns [IV.285]. Aspects of ethics appear in a vast range of disciplines and are relevant to all issues that involve human behaviour and human interactions [IV.286]. There are multiple groups that can benefit from personal assistants and activity assistance in AAL systems. Based on research quantity and technological solutions, as well as care duration and psychological aspects, four groups can be defined: (1) older adults; (2) people with permanent disabilities; (3) people with temporal disabilities; (4) people who are caregivers, such as family and healthcare providers.

We live in a time where we witness the rapid change in age groups in society. The distribution of population age becomes skewed towards older adults [IV.287] as the fraction of the older population grows progressively, while the fraction of younger people is decreasing. Reduced physical ability and reduced mobility is one of the most significant independent risk factors for morbidity and disability during the ageing process. Loss of these abilities leads to the need for activity and personal assistance [IV.288], [IV.289] making this one of the main target groups for activity assistance systems. As a result, development of new products with innovative solutions are often aimed at the needs of older adults [IV.290].

Another group of people are ones with permanent disabilities or states that require assistance, such as autism - even though this group is often viewed in the same context as older adults [IV.291], there are several differences. One of the differences between the older adults and people with permanent disabilities lies in the fact that the attitude towards technology can be more positive.

The same can be said about the last group - the people who need the assistance temporarily. This group, however, differs in the duration of care. Here, tools such as Vigo (aimed for the recovery from the stroke) can be mentioned [IV.292]. One can argue that due to different timeframe, the approach to developing technological solution can change as well, and such factors as cost and maintenance can play a smaller role in activity-assistance tool.

Introducing the technology in the domestic context is particularly complex and can be difficult due to resistance not only from the older adults or patients themselves, but also their relatives, or even from homecare personnel who also can be end users. When it comes to homecare staff, it is about how the technology has been introduced into their work routines by management, their experience of using technology, and how the technology affects the flow of their workload.  When it comes to relatives, it is about how technology affects their physical environment or whether technology is easy to use [IV.283], [IV.293]-[IV.295].

Overall, when talking about any case and any patient, reduced abilities influence not only the attitudes but also the interaction between a person and the AAL system. Based on multiple sources, Bilyea and colleagues [IV.296] concluded that a particular method should be based on a particular patient. Research shows that people with more impairment prefer switch-based keyboards while patients with less impairment prefer joystick. Similar conclusion regarding patient interaction was made by another review done on intelligent personal assistants [IV.297] - even though the voice activation is preferred by those who are able to speak, it is unusable to those with speech impairment.

This leads to the conclusion that willingness to use the technology is impacted by how well the requirements of each group were considered. The technological approaches to meet this challenge are discussed in the last subsection.

### 4.6.2 Range of supported activities

Activities supported by AAL solutions range from quite simple, such as reminding a person to take their prescribed medicine, to quite complex ones, such as, recognising changes in average daily amount of movement of an older adult affecting his/her health level [IV.298]. Increased attention should be paid to the needs of people with physical and cognitive impairments (e.g., older adults) by exploring different interaction modalities and technologies depending on the specific impairment and other additional variables, such as characteristics of the person and context [IV.299]. In [IV.291], it is stated that 70% of elderly people or people with physical and cognitive impairments cannot live independently and need activity assistance and healthcare from their caregivers. AAL technologies cannot prevent all problems and risks resulting from cognitive disability or other impairments, but they can improve quality of life by assisting and supporting older adults in various activities, e.g., health-care and rehabilitation, provision of safe environments, fall prevention, management of daily activities and keeping control over their own life, social inclusion and communication, outdoors mobility and movement tracking, and others [IV.285], [IV.298], [IV.300], [IV.301]. Moreover, outdoor activities, e.g., walking, shopping, meeting other people, and performing physical activities, can prevent a reduction of functional abilities [IV.302]. Mobility is an important aspect to perform most of daily activities, for instance ensuring that elderly people can continue their interaction with others to avoid social isolation, that they are able to do their shopping by themselves, or that they can move independently at home [IV.298].

Daily activities or sometimes called Activities of Daily Living, or ADLs are regular tasks that people do regularly. In general, ADLs can be divided into two main categories: Basic Activities of Daily Living (BADLs) and Instrumental Activities of Daily Living (IADLs) [IV.303]. BADLs usually represent activities that are necessary for basic functional living, such as bathing, dressing, toileting, personal hygiene, sleeping, eating, etc. [IV.298]. Also, taking care for personal devices (e.g., hearing aids or glasses), that supports normal daily living, belongs to BADLs [IV.303]. IADLs include more complex activities that allow individuals to live independently in a society, e.g., shopping, housekeeping, laundry, cooking, taking care of others (e.g., pets), monitoring of personal health, transportation, and managing finances [IV.303], [IV.304]. Although these activities are not necessary for functional living, the ability to perform them can significantly improve the quality of life of elderly people [IV.304]. IADLs typically require more sophisticated skills since performance of these activities demand use of social skills, experience with electronic devices, financial management, etc. [IV.303].

A fundamental concept in performing ADLs is self-determination, especially, in the context of asking for some help from others when needed. Independence does not mean doing something entirely independently. On the contrary, self-determination refers to the opportunity to make a choice on how and when daily activities are carried out [IV.285]. Considering the recent pandemic situation, for the people who are put into isolation or need assistance, AAL systems have become crucial for ensuring communication, safety, and security to people [IV.302]. With the use of AAL solutions (i.e., robotic or computer-based) in daily activities, there is not a choice between complete replacement of some activities and assistance provided to elderly people using these technologies. Rather, there are varying degrees of device operation controlled by users [IV.285].

These facts imply that not only must the system meet the necessary functionality for all end users, it is also crucial for the system to be usable - i.e., have a user friendly interface. This is another challenge that is discussed further.

### 4.6.3 State of art on AAL technologies for ADLs assistance

In the literature, there are commonly two types of the assistants that are available in AAL environments and that usually are researched separately: physically embodied personal assistants, also known as robotic assistants [IV.296], and software assistants, that are sometimes referred to as intelligent personal assistants [IV.297]. These two types of assistants are usually used for different purposes.

Robotic assistants usually are used to help with movement, walking as well as other physical tasks [IV.296], such as cleaning, laundry collection, trash removal, delivery services, and transportation of food or medicine carts. Besides, they can be applied to various interactive tasks, like reading books aloud, playing music and games, educating and entertaining people, alerting caregivers, and providing security [IV.305], [IV.306]. For the use of robotic assistants, it is suggested to choose social assistive robots (SARs) as opposed to simple manipulators, since they are considered more suitable in the long run, able to interact and communicate with humans [IV.306] [IV.307]. Research shows that voice control is used in about two thirds of reviewed studies on [IV.306] personal assistants, including commercial platforms, such as Alexa, Siri, etc. [IV.297].

In turn, software assistants are used for mental tasks, for example, as reminders, or for emergency situation monitoring [IV.297]. It is important to stress that these two categories are not mutually exclusive, even though the studies are usually done separately. Besides software and robotic solutions, there are also other types of assistants that combine both software and hardware solutions, a network of connected devices (e.g., household appliances or banking systems/tools) as well as include mobile devices to support mobility.

#### Virtual software assistants

In [IV.308], the MavHome (Managing an Intelligent Versatile Home) smart home environment, that uses multi-agent technologies, is proposed. Agents in MavHome are used to provide health care assistance. This solution can identify patterns indicating or predicting a change in health status and can provide inhabitants with needed reminder and automation assistance. Automation in MavHome is implemented using a large number of controllers and provides sensing for light, temperature, humidity, leak detection, vent position, smoke detection, CO detection, motion, and door/window/seat status sensors.

PAIR, a Personal Ambient Intelligent Reminder has been proposed in [IV.309] to assist people with cognitive disabilities, their caregivers and health professionals. PAIR is an intelligent environment with a reminder agent. The agent can provide the caregiver a list of activities for each day (e.g., having breakfast or phoning someone) and remind him/her about the activity to be performed by the patient when the time is approaching. The agent can be combined with an activity recognition module to supervise over the patient to check whether an activity in a specific time is performed. If not, the agent reminds the patient about it or alerts the caregiver.

The COACH system (Cognitive Orthosis for Assisting with aCtivites in the Home) is an emotionally intelligent cognitive assistant that supports older adults with dementia during the handwashing activity [IV.310]. The system uses a partially observable Markov decision process (POMDP) to autonomously guide users through the ADL and offer audio-visual prompts similar to those of a caregiver if the user needs assistance [IV.311]. The affective component is based on a model of the dynamics of emotion and identity called Affect Control Theory which models the emotions and behaviours of the person and the assistant. The POMDP policy produces an approximately optimal action for the system to take, again on both functional and affective dimensions. This prompt is delivered as a video of an embodied caregiver acting in a style that is consistent with the recommended affective action.

DALIA (DAily LIfe Activities at Homeprovides an integrated home system that supports older adults as primary end-users, offers support to their caregivers as secondary end-users. DALIA platform includes a Personal Virtual Assistant - a human-looking avatar endowed with speech recognition and speech capabilities. DALIA can be integrated in existing devices (smartphone, tablet and TV) to provide a wide range of AAL-modules that support elderly people in their daily life. DALIA provides a wide range of functions: communication with family and friends, shared planning tool for appointments and automatic reminders, emergency call and fall detection, medication reminders, localisation of lost things, motivation for physical or mental activity tasks [IV.312].

In the EU AAL Personal IADL Assistant (PIA) project, the assistance is provided through the use of narrated video clips. The main user experience is based on an Android application that runs on tablets and smartphones. The video clips available were recorded by dedicated caregivers, who subsequently uploaded them to a cloud-based repository. Once uploaded to this repository, these clips could then be linked to NFC tags. These tags are affixed to objects related to the IADL which assistance would be provided for (e.g., dishwasher, washing machine, microwave, coffee machine, etc.). When an inhabitant of an environment required assistance, they simply needed to request it by waking an android device from sleep, aligning it to an NFC tag and placing it in sufficient proximity to the tag. When performed successfully, the application would read the tag and then download and play the associated video. Next assistant versions deal with issues related to correctly aligning tags and the use of an overly complex user interface. To address these limitations a voice-based assistant is implemented which queries the video matching mechanism [IV.313]. Besides, automatic generation of metadata for video content has been introduced in the newest version called ABSEIL to work with the video repository generated by the PIA project [IV.314].

In [IV.315], an interactive personal (mobile) assistant ALFRED has been introduced to support elderly people during their daily activities both at home and outdoors (e.g., shops, city). It provides a human interface and services for the interaction through natural language. Users can ask questions and ALFRED gives them concrete answers. For example, ALFRED may launch a phone call or may answer specific questions such as "What time is it?" or "Where is the nearest bus station?" All the coordination and data exchange with the IoT sensor network is managed to capture user's environmental and personal data. The data is collected by unobtrusive wearable sensors monitoring the vital signs of older people[18]. ALFRED provides interaction with the cloud platforms to assess the user's health and predict the user's mood (the home lighting system is adapted to the user's mood).

An assistive system with the semantic web called HBMS (Human Behaviour Monitoring and Support) [IV.316] aims to support people in their daily activities at their homes. The system helps both in BADLs like dressing and IADLs like meal preparation, homework, and laundry, using the phone and other technical devices, managing finances and using on-line (business) processes. HBMS has a multimodal user interface that works with different media types (audio, handheld, beamer, Laserpointer, light sources, etc.). In addition, evaluations to test the observation interface with video-based activity recognition systems have been planned.

The DIANA (Digital Intelligent Assistant for Nursing Applications) project improves the life and safety of cognitive impaired persons 65+ while assisting nurses and caregivers. DIANA provides support in such tasks as monitoring patient's safety 24/7, controlling their walks during the night, reacting to alarms from existing sensors and preventing false alarms, supporting patients in ADLs and observing their health trends (e.g. water intake, changed behavioural patterns) and recording medical data. All these tasks are managed by a digital AI-based assistant. Currently, the virtual assistant guides people in very specific ADL - particularly in going to the toilet by employing a computer vision solution which detects the current pose of a person on the toilet based on privacy-preserving 3D depth data. Instructions of the next step the user needs to perform are shown on a screen [IV.317].

### 4.6.4 Social assistive robots

Pearl the Nursebot [IV.318] has been designed for the elderly to help them live a more independent life. This personal mobile robotic assistant can help the elderly with daily tasks (it reminds about eating, drinking, taking medicine) as well as some walking guidance, provide companionship and even help them to remotely communicate with physicians and caregivers [IV.319]. Decisions for the interaction with the user are made using a POMDP. The interaction is realized through speech synthesis, visual display onto a touch screen, and motions of an actuated head unit [IV.318].

The Baxter robot has been used to support the dressing of older people [IV.320]. The Baxter is used to dress one arm of a jacket onto a wooden mannequin by tracking the joint locations of the arm and calculating the trajectory. The Vicon Tracker System constantly monitors the joint positions of the mannequin with the help of markers and compares the position of the joints to the initial state. The interaction between the user and the robot is implemented using a simple fixed vocabulary for voice commands.

[18] https://www.alfred.eu/wp-content/uploads/Newsletter_ALFRED_1.pdf (last accessed: 23/02/2022)

In [IV.321], SARA ('Socially Assistive Robotic Solution for Ambient assisted living) is introduced. It is a complex distributed home automation and robotics system (based on Softbank's humanoid ''Pepper'' robot) providing health monitoring and (socially interactive) assistance in daily living tasks to the elderly (and their caregivers) at home to prolong their autonomy and delay their institutionalisation. It comprises multiple mobile and static robotic elements together with wearable health monitors, personal smart devices (phones and tablets) and various computational and information services. SARA is able to keep track of the location of significant objects; provide mobility support for the elderly as they move through their environment; monitor a user's health conditions; notify assigned caregivers of possible problems and incidents; enable doctors/caregivers and family members to remotely check in on a patient (e.g. on a request, SARA initiates a remote audio/video link); perform cognitive and physical rehabilitation exercises, as well as remind patients to take prescribed medication at allotted times.

A novel approach seamlessly integrating humans, devices, heterogeneous sensors, home appliances and assistive robots called ECHONET-based smart home environment has been proposed in [IV.322] for an elder-care. It relies on the Pepper robot which is endowed with human-like social skills (i.e., natural language processing, user emotion estimation, etc.). The robot is able to autonomously reconfigure its way of acting and speaking, when offering a service, to match the culture, customs, and preferences of the person it is assisting. In addition, it can gain access to the iHouse network and provide user-requested data and services through verbal and nonverbal interaction.

In [IV.305], an assistive robotic architecture which integrates a smart wheelchair and a robot manipulator. The user on a semi-autonomous Quickie Salsa R2 smart wheelchair can move in an indoor environment and can select with Baxter robot manipulator a desired object (e.g., cup of water) among those available in a dataset. The object is recognized by the robotic manipulator and, once the user is driven in front of the manipulator, the pick and place algorithm is realized using both an image classification and a specific kinematic controller to pose the object in a point that the user arm can reach.

A multimodal robotic system for a specific dressing task (i.e., putting on a shoe) has been proposed in [IV.323]. The central part of the system is a Barrett's 7-DOF WAM robotic arm equipped with an in-house developed gripper. Visual input is provided by two Microsoft Kinect cameras. In addition, crocs-type shoes with attached ribbon that can be grasped by the robot are used. The use of the system with the WAM robot includes: 1) user tracking, gesture recognition, and posture recognition algorithms relying on images provided by a depth camera; 2) a shoe recognition algorithm from RGB and depth images; and 3) speech recognition and text-to-speech algorithms implemented to allow verbal interaction between the robot and user.

### Other AAL solutions for assistance

A system to support the elderly and impaired people during the food preparation process is proposed in [IV.324]. The system named FoodManager is intended for the planning of weekly menus meeting nutritional needs, health conditions and personal preferences of an individual (with easy cooking guidelines); for managing the oven parameters wirelessly (hands-free operation); for food and shopping list handling. FoodManager's core (application logic) is hosted on the Home Gateway to provide users with eating and cooking-related services, the home automation network allows the control and communication with household appliances, sensors and actuators. The software with the user interface provides interaction with the system.

The ASIES (Adapting Social and Intelligent Environments to Support People with Special Needs) project [IV.325] deals with the problem of independent living for cognitive impaired people through the support of daily activities at home or work by using various solutions developed within the project. For the home environment, a multimodal dialogue system Mayordomo in combination with the Octopus system has been proposed to enable speech-based control of home appliances [IV.326]. In addition, AngryEmail application has been created by using Octopus to integrate emotional processing with mail applications and to control the real-world environment using voice [IV.327]. For the work environments, aQRdate and QRUMBS systems are proposed in this project. Both of them are based on the use of mobile devices and QR Codes to provide assistance to people with cognitive disabilities (adaptive manuals to guide the user while doing an activity or guidance to physically find resources at the working place). In addition, two complementary tools DEDOS-Editor and DEDOS-Player running on a multitouch surface are developed for the provision of learning activities to people suffering from Down's syndrome [IV.325].

The Bank4Elder project aims to develop innovative web and Automatic Teller Machine (ATM) interfaces involving older persons along all the design and development processes. Currently, most of older population usually do not feel safe when using online banking or ATM. The project therefore contributes to tackle this challenge by looking at ways to better personalise the way users interact with banking interfaces. The project has developed and validated with senior citizens 4 interfaces (TV, web, mobile and ATM) to access various banking services taking into account that there are two groups of users: high tech and low tech (people who have never used a computer nor a smartphone) [IV.328].

The Entrance platform (ENabling elderly people TRAvel and iNternet acCEss) aimed at older adults has been proposed for booking e-tickets and vacation packages, game systems, and navigation assistance [IV.329]. The home terminal consists of hardware (a silent computer to be used in living rooms) and software adapting to users with different levels of technology proficiency. The mobile platform comprises navigation software, which is also used in the serious game on the home platform.

In [IV.330], HoTAAL (Home of Things for Ambient Assisted Living) platform is described, where home appliances (refrigerator, microwave and bin) provide seamless social interactions with each other and older people to provide assistance in meal preparation tasks. RFID technologies are used for the refrigerator together with two cameras, bar-code readers and scales to identify and quantify food. The microwave is equipped with a camera and a non-contact IR thermometer located inside the cooking chamber. The bin uses two cameras, switches and weighing sensors. Twitter social computing platform is used as the messaging system.

OMNIACARE [IV.331] is a hardware/software system designed for healthcare and can be used by both caregivers and patients. The system is aimed at users who have discrete mobility capabilities and are self-sufficient in daily activities, but who may find themselves in difficult situations and may need support. By using the system, the caregivers are able to check the patient's situation, receive alerts in case of critical situations, and to provide remote assistance by communicating directly with patients. OMNIACARE is composed of the central server, home server, systems for vital signs capturing, webcam and Android smartphone. The smartphone displays reminders and alerts, provides navigation and GPS tracking of the assisted, if the localisation is needed (e.g., outdoors).

In [IV.332], WELI (Wearable Life) wearable application designed to assist young adults with intellectual and developmental disabilities in classrooms has been described. WELI is available for Android smartphones and Android smartwatches, serving as a complementary tool for assistants who send wearable notifications to the students using a mobile application. The features provided include reminders for focus, silence, participation, rewards, taking a break, survey and mood self-assessment. WELI provides more unobtrusive assistance without verbal interventions to neurodiverse students, minimizing disturbance to classmates and instructors and reducing the stigma of having an assistant visibly coaching students aloud in class [IV.333]. The notifications are delivered directly on the student's watch, starting with a short vibration to grab the student's attention followed by the display of a graphic and textual information. For interaction needs the smartwatch provides touch screen capabilities facilitating micro interactions (e.g., swiping, scrolling and tapping) for the students to quickly input their feedback [IV.332].

### 4.6.5 Summary of AAL solutions

All the reviewed AAL solutions developed to assist ADLs of various user groups are summarised in Table 1. In the table, for all solutions end-users and supported types of ADLs (either BADLs, IADLs or both) are identified, and summary of assisted activities as well as technologies used to provide assistance is presented.

Based on the information summarised in Table 4, it can be concluded that the most common target audience (i.e., end-users) of AAL assistants is older adults, who require support in daily activities. However, there can be found solutions that either require input from caregivers or assist them in the monitoring and interaction with their patients (in particular elderly people) since taking care of several people simultaneously can be quite complicated. A relatively small number of solutions are intended for other user groups, e.g., young people with intellectual and developmental problems (for example, with Down's syndrome) or people with temporal disabilities requiring assistance until recovery.

Most of the analysed AAL solutions aim to provide autonomy and improve the quality of life for the people requiring assistance in daily activities so that they are able to live independently without a need to rely on other people. Supported daily activities include both BADLs and IADLs. Assistance is provided to such BADLs as eating, drinking, dressing, toileting, washing, healthcare, and other activities. The supported IADLs are also very diverse, for example, cooking, laundry, taking care of plants, banking services, navigation, booking services, gaming, help with various devices, etc. In addition, the social inclusion of target users is facilitated by providing companionship and offering assistance for communication with other people, e.g., family members, friends, caregivers, and others.

The assistance in the AAL solutions reviewed is mainly provided based on different technological platforms and devices, as well as various AI-based approaches (e.g., agent technologies) and algorithms. Computer vision solutions are commonly used to monitor users, recognize their gestures, posture and activities performed in order to select the best actions for the provision of assistance in general. For the younger generation, the use of everyday devices (smartwatches or smartphones) is very common to assist them in social life and learning activities.

Table 4. Summary of AAL solutions developed for ADLs assistance

| AAL solution for the assistance | End-users | Support for BADLs | Support for IADLs | Activities supported | Technological basis |
|---|---|---|---|---|---|
| Pearl the Nursebot | older adults | X | X | BADLs reminder, walking guidance, companionship, communication | robot with speech synthesis and visual display |
| MavHome | home inhabitants | X | X | health care, needs reminder, automation assistance | multi-agent technologies, sensors, controllers |
| FoodManager | older and impaired people | | X | nutrition and shopping list planning, control of household appliances | home automation network with household appliances, sensors and touchscreen interaction |
| ASIES project | people with cognitive impairment | X | X | control of home appliances, guidance at work, provision of learning activities | speech recognition, emotion modelling, mobile technologies and touchscreen interaction |
| PAIR | people with cognitive disabilities, caregivers, health professionals | X | X | activity reminder, monitoring of the activity performance | agent-based technologies |
| Bank4Elder | older adults | | X | banking services | personalized web and ATM interfaces for various devices |
| COACH | older adults with dementia | X | | hand washing | emotion modelling, video tracking, audio-video prompts |

| AAL solution for the assistance | End-users | Support for BADLs | Support for IADLs | Activities supported | Technological basis |
|---|---|---|---|---|---|
| DALIA | older adults, caregivers | X | X | communication, appointments, reminders, emergency calls, motivational support, etc. | virtual assistant in devices with speech recognition and speech capabilities |
| Entrance | older adults | | X | e-booking services, gaming and navigation | silent computer and adaptive software, multi-sensory mobile interface |
| Personal IADL Assistant | older adults, formal and informal caregivers | | X | cooking, laundry, ironing, dishwashing, using TV and other activities | Android application, NFC tags |
| Baxter robot for the dressing | older adults | X | | dressing | robot with speech recognition, video tracking and markers |
| HoTAAL | older adults | | X | meal preparation, control of home appliances | RFID technology, cameras, bar-code readers, IR thermometer, scales |
| ALFRED | older adults | X | X | answers to questions, social inclusion, health care, games for brain training or physical activities | IoT Wireless sensor network, natural language interaction |
| ECHONET-based smart home environment | older adults | X | X | interaction with a user, control of home devices and home appliances | robotic solution and smart home, natural language processing, emotion recognition |
| OMNIACARE | patients, caregivers | X | X | patients' monitoring, remote assistance, alerting, communication, navigation | central and home servers, vital sign capturing, webcam, Android smartphone |

| AAL solution for the assistance | End-users | Support for BADLs | Support for IADLs | Activities supported | Technological basis |
|---|---|---|---|---|---|
| SARA | older people, caregivers | X | X | object tracking and localisation, mobility, health monitoring, notifications, remote patient's monitoring, reminding and rehabilitation | robotic solution with home automation, wearable health monitors, smart devices and ICT services |
| Barrett's 7-DOF WAM robotic arm | anyone requiring assistance | X | | putting on shoes | user tracking, speech, gesture and posture recognition |
| Quickie Salsa R2 smart wheelchair with Baxter robot | anyone requiring assistance | X | X | mobility, object manipulation | robotic solution together with sensor-based wheelchair |
| WELI | young adults with disabilities | | X | reminders for various class activities | Android smartphones and smartwatches, touchscreen interaction |
| HBMS | older adults | X | X | dressing, cooking, laundry, using technical devices, managing finances, etc. | multimodal user interface that works with different media types |
| DIANA | older adults with cognitive impairments | X | X | monitoring patients' activities and health, toileting | non-wearable 3D sensor, computer vision technologies |

For older adults, the interaction with the assistants is carried out in a simplified manner, i.e., with the use of natural language and speech recognition or various displays with touchscreen (e.g., TV, tablet, or smartphone) showing required information. However, only a small part of the reviewed AAL solutions is capable of adapting to the users (e.g., users with different levels of technology proficiency) - which has already been identified as one of the challenges.

Moreover, user emotion recognition and emotion modelling in the analysed solutions can be found to improve the interaction and believability of digital assistants (both virtual software and robotic). Some of the assistants use the emotions of the user and respond to user actions with their own emotions and actions.

Overall, the reviewed AAL assistants are ranging from simple stand-alone solutions, which can be used in mobile devices, to complex robotic solutions integrated into smart home environments, where robots are able to control household appliances and various other devices and objects (e.g., air conditioners or doors).

### 4.6.6 Challenges with AAL technologies for activity assistance

The development of AAL technologies supporting people in their ADLs faces not only technological challenges (e.g., data heterogeneity, data size, context acquisition, multi-user data) but also social, ethical, security, acceptability, usability issues [IV.291], [IV.302], [IV.303],[IV.334], [IV.335] that can be met by using technology solutions.

Human assistance is slowly but surely replaced with various solutions, such as domestic robots. Older people have a twofold view on the usefulness of these robots: as long as the robots meet the needs related to information management, it is accepted but when it comes to personal care or relaxation activities, older people prefer human assistance [IV.336]. Same study shows that older people with higher education and those living with partners are more optimistic about using digital assistance.

To mitigate the resistance, many factors, such as trust in the technology, ease of use for the software and hardware, as well as ease of maintenance must be considered [IV.295]. These factors are impacted by more practical AAL system designs: distribution of the tasks between users and technology, as well as considering the temporality and spatial dimensions of the home space. In this subsection, we discuss these issues and the technological solutions for meeting the challenges and subsequently mitigating concerns regarding AAL solutions. In this discussion, however, we set aside practical issues such as affordability and space dimension, as these challenges can be solved from sociological or economical perspective, however, there is not a lot to do from technological perspective. We see the willingness to use the system as an umbrella challenge, that is impacted by multiple components (Figure 27), and based on literature provide solutions to these components:

- privacy and autonomy of the user must be considered;

- multiple users must use the system;
- the system must be easy to use;
- the system must balance social interaction.

In the Figure 27, with lighter colour are marked those solutions that must be implemented in the design phase.

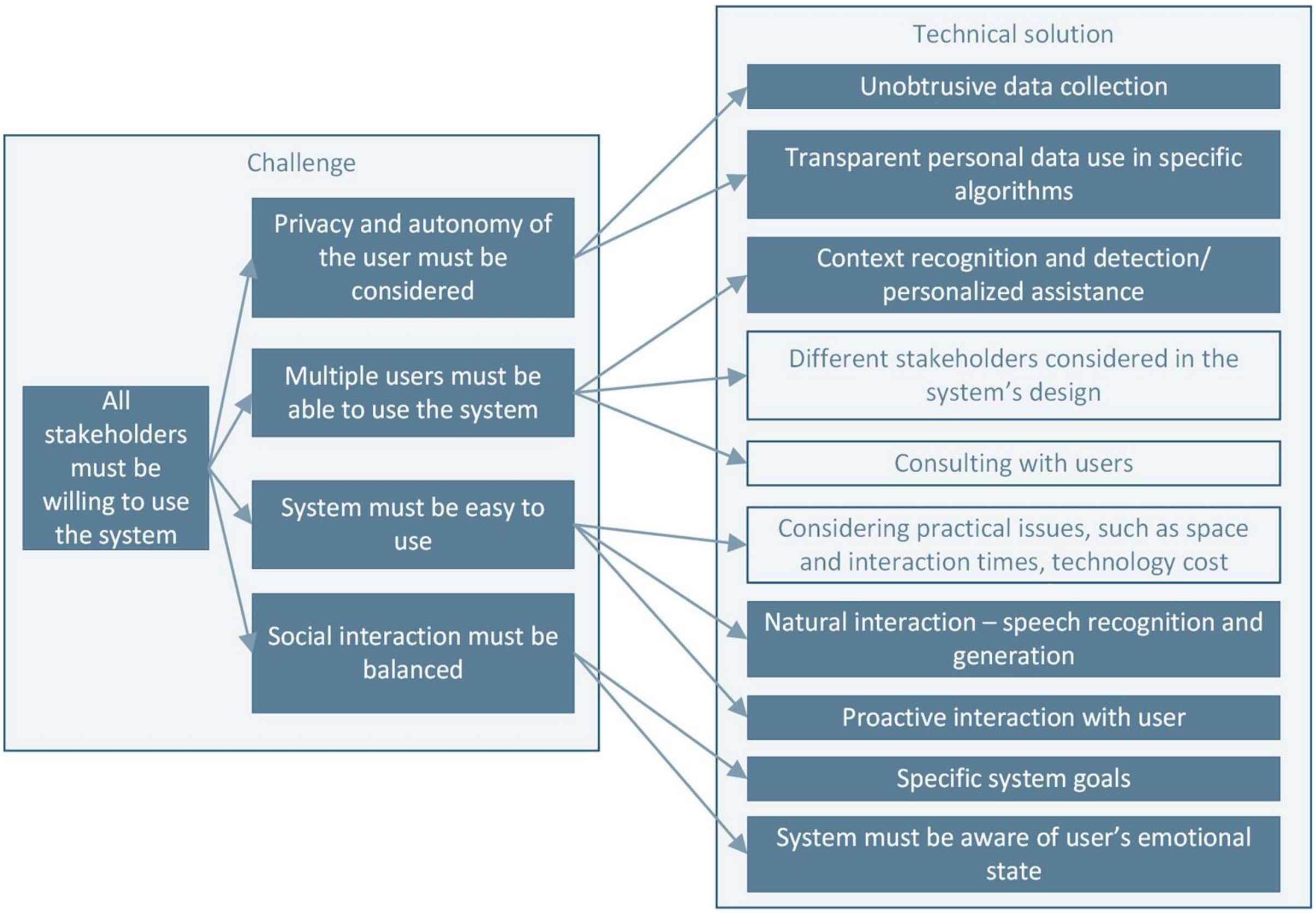

Figure 27. Challenges of AAL systems and corresponding technical solutions.

### The importance of autonomy and privacy

Ethical questions like protection of autonomy and privacy arise in such situations [IV.285], [IV.335]. One of the most ethically challenging areas in the development of AAL solutions is the use of monitoring and tracking technologies with elderly people, particularly those with cognitive disabilities [IV.285] - that is, data acquisition and use.

Monitoring of daily activities and physiological data, as well as recognition of environment and activity context, requires the use of personal data [IV.298]. For AAL technologies, privacy concerns can often reduce elderly people's willingness to use these systems [IV.337], especially if the assistance in ADLs requires the collection and mining of large amounts of daily activity data [IV.299]. Collection of data from the users and their environment, as well as intelligent processing of that data, are often seen as intrusive and violating rights to privacy [IV.338], [IV.286]. Achieving the goal of making the elderly accept and use the provided technology is a first, but a key step to performing more complex and interesting activities [IV.338]. Technologies must be robust, reliable, continuous, and unobtrusive, while algorithms must be sensitive and specific for the detection of various activities and their context. Furthermore, these systems must be secure, provide privacy, and help the delivery of better health care at a lower cost [IV.298].

Most people do not have problems with the use of new technologies, particularly those who use and actively recognize the benefits of mobile and ubiquitous technology [IV.302]. However, most elderly people usually do not feel comfortable wearing devices or seeing their homes equipped with intrusive sensors [IV.338]. There is also a negative view on these technologies, especially in the situations when physically assistive robots are used with dependent individuals. The main concerns are related to the involvement of robots in particularly intimate BADLs such as bathing or visiting a toilet [IV.337].

### Multiple users must be able to use the system

While it is essential to study the experiences of various groups of people in their everyday lives to manage their autonomy and independence, it is also equally important to consider their suggestions on how digital design could satisfy their needs. The involvement of older adults who are current technology users and future older adults who will become users eventually, is vital in the design process of digital solutions. Through their participation in the early design stages, the probability of using these solutions will also increase, as the studies show [IV.336].

Multiple users within the same environment only increases this range of challenges. Most of the current AAL developments are intended for the use with a single person in the particular environment. If there are several people requiring assistance with ADLs then it can increase the complexity of the assistance task. It can be difficult to provide correct support to the right person especially if each of them requires a different level of assistance or two or more people need to interact with the same objects at the same time [IV.303].

### Ease of use

While usability is important for all the groups that may use AAL technology, the discussion on the usability topic often shifts in particular towards older adults. From the psychological perspective, it is crucial to understand that older people are undergoing double transition [IV.339], [IV.340]: aging itself and rapid digitalisation. Aging is a psychological, biological and physical process, which requires adjustment, and, while digitalisation presents many benefits, rapid introduction of the possibilities of digital assistance in their daily lives on top of other adjustments, can be overwhelming [IV.340]. It is important to keep these processes in mind and give time to accept technology as a valuable, meaningful and manageable complement to their lives. In particular for older adults, the adverse reactions to digital technology are often linked to the feelings of being anxious about using digital tools and being intimidated by technology. The most practical strategy is to connect different digital assistance features with individual characteristics for different older adults [IV.341]. Technology for mental, physical and emotional activity assistance should be developed and designed based on the target user's specific personal, social and physical context [IV.342].

To improve interaction and acceptance of AAL technologies, from the technological side, several factors should be taken into consideration: the technical ease of use (i.e., interaction) and proactiveness of the technology - knowing user's goals and initiating interaction in accordance to them [IV.297], [IV.298], [IV.343]. Interaction with AAL technologies is often implemented through a range of modalities such as speech, gesture, touch, vision, affective and physiological responses. These interaction modalities can lack adequate visibility and affordances and require new forms of interaction that might be unfamiliar to users, thereby reducing the usability of AAL systems [IV.299]. The unfamiliarity with current technology, anxiety, and missing perception of the potential technology benefits leads to their avoidance [IV.338], [IV.344]. Besides, lack of skills in using technologies can also explain why elderly prefer human assistants [IV.336] - arguably, the ease of interaction with humans via well-known methods, such as talking, is one of the main factors. Gone are the days when computers were only used by technical specialists - now, as computing is becoming ubiquitous, it is necessary to develop natural interfaces that enable familiar interaction with technology. Besides, it is also necessary to consider the fact that with the progression of aging various sensory impairments appear, e.g., vision or hearing impairments, problems with speech, haptic perception, decreased muscle strength, and balance [IV.299].

Preferred interface also depends on interaction frequency. In general, voice control and dialogue-style nature is beneficial for interaction [IV.343]. In addition, one of the possible solutions could be to incorporate an ability for AAL systems themselves to track these abilities and thus respond to changes by adapting their interfaces [IV.299], [IV.345]. The use of adaptation methods and techniques allows the development of personal assistants able to guide each user according to their needs. However, personalized assistance is possible only when information about each person, such as his/her characteristics, needs, preferences, behaviour, or emotional state, among others, is available [IV.301], [IV.325]. Furthermore, the models of normality must be updated in an adaptive way since the needs and habits of people change over time as well as the parameters to be observed [IV.302].

Another part of natural interaction is what content a personal assistant communicates. This includes understanding the context of interaction [IV.346], as well as building dialogue which is a natural form of interaction [IV.343]. De Barcelos Silva and colleagues [IV.297] demonstrated that only approximately 10% of intelligent personal assistants are proactive in reaching user's goals.

**Balancing of social interaction**

In some cases, the use of assistive technologies instead of reducing negative impacts, even strengthens the adverse effects because the use of digital technologies can increase their isolation from society (digital divide) by reducing elderly people contact with other humans (e.g., caregivers, relatives) [IV.335], [IV.347], [IV.348].

There is also a high probability of an emotional attachment to robots in environments that otherwise may be lacking human companionship [IV.285], [IV.349]. This means that the interaction with the user must be mediated through the user's emotional state.

Believability and personification of robotic and virtual assistants have an important role in reaching their goals, however, there are studies that show that not all users prefer assistants with personality [IV.350]. This only supports the point made in Costa’s and colleagues' study [IV.346] of the necessity to adapt to user’s needs - as this adaptation approach has a potential to support better system development and use.

### 4.6.7 Conclusions

Currently, technologies for personal assistance and activity assistance have become available due to technological breakthroughs and general necessity for such technology. These solutions, however, are often custom-made and have been made in a particular project. To solve this and make virtual and robotic assistants mainstream, multiple non-technical challenges must be solved. Technologically, we see the the development of various assistance technologies in two main ways:

- Adaptability - which means adapting to a particular patient, to his/her emotional state, personality and preferred interaction style, as well as other factors, such as the timeframe when the assistance is needed, and the assistance activities themselves - this includes the “proactiveness” of the assistant;
- Safe data processing - a large reason for deciding not to use assistance is often based in mistrust, so safe data processing and security ensuring mechanisms is a key part of such systems.

There are also multiple purely technological issues that has been covered in this whitepaper, such as activity detection, algorithms for distributed systems etc., which must be explored for these systems to be usable, however, we have found that in the case of activity assistance, a lot of attention must be paid to non-technical issues as well.

## 4.7 Gesture recognition

Since AAL started, it has funded a lot of collaborative projects working on products and services designed to improve the quality of life of the elderly. These projects have achieved remarkable results and many of the resulting solutions have found their way into the market and are now being used across Europe to make a real difference in the lives of older people. Of these, projects and systems that include gesture recognition technologies to improve the quality of life of older people are noteworthy.

In this section, we will briefly present the main AAL research results and projects on gesture recognition; then we compare them. In this section, we also revise a relevant sub-topic of gesture recognition, namely sign- language recognition (Section 0 and subsection).

### 4.7.1 The most recent advances in gesture recognition

The project *JAME* started in March 2020 and aims to keep people who are chronically ill with disabling symptoms as active as possible, socially and at work, people affected by a tremor that has a significant impact on their lives. Combining advanced technologies (ML algorithms, high-quality sensors) with design principles the partners have developed a portable device that can monitor the movement and gestures of the upper limbs and, through an ad-hoc algorithm based on machine learning, can recognize the tremor and suppress it. The project includes an in-depth analysis to understand user attitudes and preferences, a clinical study aimed at optimizing technology and technological development activities. The aim is to create a non-invasive and non-stigmatizing device that can be worn proudly or discreetly, according to the user's attitude. The main expected result is the development of a device designed to control disabling symptoms in patients with chronic conditions, able to overcome the classic barriers of use and poor acceptability due to the extremely stigmatizing nature of currently available devices. JAME is expected to allow the user to become autonomous and interact with people without feeling embarrassed by their own illness.

The goal of the EU project *VIRTUAAL* started in May 2019 was to create immersive serious games and increase motivation in elderly patients living in nursing homes or daily centres and test the usability of the modern technologies in combating cognitive impairment [IV.351]. It focuses on several mental capacities that may be trained and measured (attention, executive functions). An analysis will be made, of both impact on cognitive status and on the feedback of the elderly users, to conclude if there's a sustainable business model behind VR & AR related therapies against cognitive impairment. A Spanish SME specializing in applications for the elderly care market has already started to develop games for cognitive stimulation. They aimed to test these therapies, especially those related to virtual reality, and assessed their usefulness for elderly users.

*BoMI* is a multimodal body-machine interface designed to help people with upper limb disabilities, using advanced assistive technologies, such as robotic arms [IV.352]. The proposed system uses a network of portable and wireless sensors that support up to six sensor nodes to measure the natural gesture of the upper body of users and translate it into control commands. The natural gesture of the head and upper body, as well as muscle activity, are measured using inertial measurement units (IMU) and surface electromyography (sEMG) using custom multimodal wireless sensor nodes. An IMU detection node is attached to a user-worn headset. It has a small size (2.9 cm x 2.9 cm) and low power consumption (max.

31 mW) and offers an angular accuracy of 1 multimodal patch sensor nodes, including both IMU and sEMG detection modes are placed over the working parts of the user's body to measure muscle movement and activity. BoMI runs on a Raspberry Pi. It can adapt itself to several types of users through different control scenarios using the movement of the head and shoulder, as well as muscle activity and offers a power autonomy of up to 24 hours.

The *gesture-controlled home automation using CNN* detects user gestures and controls the household appliance [IV.353]. The main goal is to provide portability and allow the blind, deaf and dumb to control easily various devices. Control methods are also needed due to the increasing number of industrial and household appliances that need to be controlled. The user gesture input is captured using an Android application and sent to the Raspberry pi (which acts as a microcontroller) and the Raspberry pi then operates on the respective functionality of the devices. The system uses a CNN algorithm to classify images. CNN is used in issues such as pattern and image recognition. They have more advantages compared to other techniques. Using the standard neural network that is equivalent to a CNN will have a much larger number of parameters, so the training time would also increase proportionally. Advantages of the proposed system:

- Low power
- Simple circuits because they do not require special hardware
- Devices can be controlled more comfortably
- Helps overcome situations where normal wiring is difficult as well as financially impossible
- Can be used in the home theatre system where short distance communication is required
- Suitable for people with physical disabilities to use the devices in the room

*SenseCap* is a technology probe that consists of an iPad Touch 4G mounted on a baseball cap and is designed to capture head rotations to improve the physiotherapy procedures used to rehabilitate patients' balance [IV.354]. The technological probe, SenseCap, with the help of mobile phone sensors captures in real time the physical movements of the head and records the exercises performed by patients but also other values that reflect how to perform the exercises, then they are sent to the physiotherapist. Based on this data, the physiotherapist can make informed decisions to adjust the treatment. Patients wear the baseball cap when they perform the exercises and take it off when they finish. SenseCap collects the following data: the number of sessions and exercises performed each day, duration of each session, average head rotation speed, rotations per second, average head rotation, dizziness before and after each session. Subsequently, these data are communicated to the physiotherapist and are presented through a dashboard on the iPad. The non-compliance rate of patients could reach 70%. However, SenseCap has collected data that may not be found in survey studies. For example, five of the patients exercised more than prescribed. One patient practiced even more, reaching 25 sets of horizontal exercises in one day, compared to the six sets prescribed.

The *gesture-based method for natural interaction in smart spaces* is based on two types of gestures ("individualisation" and "action" gestures) that allow the construction of a grammar to identify the object and the action to be performed on it [IV.355]. Gestures are recognized through two different devices: a Kinect device and a smartphone. Although the signals taken from them differ (the first device provides the position of the user's hand, while the smartphone provides acceleration of the user's hand movement), the paper demonstrates that the same recognition algorithm, based on Dynamic Time Warping, can work successfully for both entries. The grammar of the method states that the user must identify the object with which he interacts by performing an individualisation management primitive (the initial letter of the object), then using an action gesture primitive to indicate the action to be performed. The action may involve a second object, which will be identified as the first. The user will be able to customize the action gestures in the vocabulary and configure the commands assigned to the gestures, and the system will provide clues to the interaction so that the user does not feel lost. The main component of the system is a gesture recognition module based on an adapted Dynamic Time Warping algorithm. This module works abruptly at acceleration or position data inputs, and is suitable for implementing device-based or infrastructure-based gesture recognition solutions (i.e., smartphone or Kinect based). The average recognition rate is 93.63% for smartphone-based recognition and 98.64% for Kinect-based recognition, respectively. The paper also details the architecture and software tools that allow the interaction method to operate in a real environment.

The study *Elderly Care Based-on Hand Gestures Using Kinect Sensor* explored the feasibility of extracting hand gestures in real time using the Microsoft Kinect V2 sensor in three scenarios: finger counting, the built-in system provided by Kinect itself, and CNN-based deep learning [IV.356]. The proposed methods used the same practical circuit for each scenario, which reports that the correct SMS message sent to the smartphone of the care service provider was directly correlated with the results and accuracy of the recognition system. The experimental evaluation of the proposed methods was performed in real time for all participants in three different scenarios. The experimental results were recorded and analysed using a confusing matrix that gave acceptable results, making this study a promising method for future home care applications.

The *Brando* system consists of an integrated robotic device with a control scheme for the active / passive rehabilitation of the upper limb, through VR games with different exercises / scenarios [IV.357]. The robot consists of the integration of a robotic arm based with 6 degrees of freedom on the final effector and a passive arm, mounted together in a common platform. Three of the six degrees of freedom are operated, allowing elbow and shoulder mobility. The three degrees of passive freedom correspond to a 3-DoF cardan shaft for passive movements of the wrist joint. The passive arm provides complete gravitational compensation of the upper limb to help the patient perform movements without carrying their weight, which is often considered a useful configuration in manual physiotherapy. VR game applications have been designed to assign isolated movement tasks of pronation / supination (PS) or flexion / extension (FE) movements within constrained range-of-motion constraints and moderate but incremental resistance in motivating game scenarios at different difficulty levels.

### Comparison of approaches

A first criterion for comparison, particularly important, concerns the ability of the application/system to improve the user's health. Given that users are largely elderly, the aim is for the applications or devices they use to bring about visible improvements in their health. A second criterion is the use of virtual reality or augmented reality in the development of the proposed solution. These two can have a positive effect both in the treatment stage and in the user recovery stage, suggesting solutions that can be both fun and attractive. For the elderly, convenience is also a fairly important factor in choosing a product. When it can cause difficulties or complexities, the user will stop using that product. In this category can be integrated products that are easy to transport, which go practically unnoticed. Monitoring the user can be of real use to his/her family because they can find out, at any time, what he/she is doing and where he/she is. Also, the ability to monitor vital parameters is an important feature, which should be present as soon as possible in many applications of this type. Also important lately is the use of the convolutional neural network, a beneficial solution to achieve the correct classification of images according to class. The use of microcontroller-based solutions can also be an important criterion in making the comparison, as it entails both advantages and disadvantages. When the system has a large volume and cannot be easily integrated into the user's daily life, it has a major drawback and, even if the proposed solution is effective, it will not be of interest to users seeking convenience. The leap motion solution can also be a real help to both people with certain disabilities and the elderly, as it is easy to use and does not require the use of many objects and devices to perform a task. Finally, the improvement of living conditions is the most important criterion, as it reflects the main purpose of AAL solutions.

## 4.7.2 Sign language recognition

A large stream of research investigated sign language recognition. Sign languages carry meaning by visual-manual modality and are primary communication by the deaf and hard of hearing or their family members. Research into sign languages has been growing with the advances in computer graphics, computer vision, neural networks, and hardware devices. The new technology can help people learn, communicate, interpret, translate, visualize, document, and develop the various sign languages.

Concerning sign languages, we have identified three prominent areas of research:

- Interpretation and linguistics of sign languages,
- Sign language synthesis and visualisation, and
- Sign Language Recognition (SLR).

Interpretation and linguistics of sign languages is mostly concerned with the meaning that is conveyed using the sign language. With the recognition of sign languages as natural languages in the late 1970 and early 1980, linguist research took an in depth look into this field [IV.358]. Neural-aspects are considered for fully grasping the connection between sign and phonetic languages. Natural language processing is also concerned with interpretation, a task similar to the interpretation and comprehension problems that are present in the field of natural language processing.

Sign language synthesis and visualisation is an area that tackles the issues of visualisation of sign languages and creation of signed speech. This field researches and develops means of realistically “spoken” sign languages, using video and image sequences or digital characters [IV.359] [Goyal]. Sign language recognition is the scientific area responsible for capturing and translating sign speech using mostly computer vision and/or artificial intelligence techniques [IV.360] [Cooper]. The research benefited in the last decade by introducing general purpose sensors such as Intel RealSense, Microsoft Kinect, and LeapMotion [IV.361], [IV.362].

Each of the areas has a vast amount of valuable research that needs to be filtered and analysed. Although the field is relatively young, there are few literature reviews that have taken into consideration gesture as well as sign language analysis.

Cooper *et al*. [IV.360] conducted an extensive review on sign language recognition approaches and challenges. The publication focuses on all aspects related to SLR: sign language linguistics, data acquisition related to sign languages as well as approaches for sign language recognition. The linguistics portion describes some of the complexity that is present in sign languages, like body posture, non-manual features, among others. The data acquisition section enumerates the approaches for acquisition of sign languages available at that time (data gloves, images and video input, depth-based cameras). It also listed the prominent sign language datasets that have been obtained using the methods. Also, the review discusses the main approaches of hand pose acquisition, as well as individual finger spelling and other non-manual features. As for recognition, the authors describe the state-of-the-art approaches, focusing on individual or continuous sign recognition. Mainly, various approaches using various types of Neural Networks, Hidden Markov Models and its variations (like Parallel HMM), decision trees and self-organizing maps are utilized for various parts of SLR.

Toiba and Elons [IV.363] discussed the developments in sign language recognition, taking into consideration much of the results discussed by Cooper et al., and focused on neural network approaches for SLR. They report recent results (as of the time) near 90\% recognition for specific and carefully chosen recognition cases. The paper however does not present a methodology of the surveyed results and provides a more general overview of the field.

Er-Rady *et al*. [IV.364] ran a survey about automatic sign language recognition. In it they describe the approaches as well as the best methodologies for sign language recognition. In the paper, the authors explain the complexity of data acquisition, the features of sign languages, as well as other properties that the modality allows (e.g., facial expressions). They also describe the most common components that should be present in a system that uses Automatic Speech Recognition system (ASLR).

Sagayam and Hemanth [IV.365] compiled a survey about the use of hand posture and gestures recognition [Sagayam]. They describe the most significant research approaches as well as results from the recognition. Specifically, all the approaches mentioned (decomposition, decision trees, support vector machines, hidden Markov models, among others) are able to produce 90+\% of recognition rate. In the paper, the authors have a special section dedicated to studying hang motion analysis (HMA), mentioning HMM and its variations as the most common way to analyse hand motion.

Gesture recognition competition, called ChaLearn [IV.366], created the data set with 50000 gestures described with RGBD data. The gestures recorded varied a lot, including signalling for drivers, aircrafts, representing numerals, sport referees, among others. Here, the accompanying depth channel contributes significantly to the approaches taken by the competitors, although HMM based approaches took the podium for the recognition of gestures.

### The solutions for sign language recognition

In the following, we present a summary of studies that examined the application and results of ML and DL algorithms in improving sign language recognition and interpretation.

Barros *et al*. [IV.367] presented a fully automatic arm and hand tracker that detects joint positions over continuous sign language video sequences of more than an hour in length. Their framework did not require the manual annotation of that work, and, after automatic initialisation, performed tracking in real-time. The method was applied to 20 signing footage videos with changing background, challenging imaging conditions, and for different signers.

Xue *et al*. [IV.368] developed a Chinese Sign Language (CSL) recognition system using the portable and cost-affordable Leap Motion sensor and applying k-NN. The experiment result showed that such a CSL recognition system achieved static sign language interpretation with high accuracy. To verify its robustness, the total sample size was 5000 and the sample size for each gesture was 250.

Pfister *et al*. [IV.369] developed a tool that recognizes gestures in videos, including localizing the gesture and classifying it into one of multiple classes. The domain adaptation and learning methods were evaluated on two large scale challenging gesture datasets: one for sign language, and the other for Italian hand gestures. In both cases performance exceeded previously published results.

Ansari and Harit [IV.370] proposed a method for a novel, low-cost and easy-to-use application, for Indian Sign Language recognition, using the Microsoft Kinect camera. In the fingerspelling category of their dataset, they achieved above 90% recognition rates for 13 signs and 100% recognition for 3 signs with overall 16 distinct alphabets (A, B, D, E, F, G, H, K, P, R, T, U, W, X, Y, Z) recognized with an average accuracy rate of 90.6%. The authors used a vocabulary of 140 symbols collected from 18 subjects (5041 images in total).

Prateek *et al*. [IV.371] developed a system called Dynamic tool for American Sign Language (ASL). The system is a finger spelling interpreter which can consistently classify the letters a-z. It achieves a cross-validation accuracy of 98.66%. The system first converts the videos into frames and then pre-processes the frames to convert them into grayscale images. The Convolutional Neural Network (CNN) classifier is used for building the classification model which classifies the frames into 26 different classes representing 26 English alphabets.

Paulraj *et al*. [IV.372] designed a simple sign language recognition system employing skin colour segmentation and Neural Network. Experimental results showed that the system has a recognition rate of 92.58%. The feature dataset contained 3080 feature vectors. The network was trained with 60% of samples (N=1848) and tested with the remaining 40% (N=1232). This network model was used to classify the right- and left-hand signs.

Jasim and Hasanuzzaman [IV.373] proposed a computer vision-based hand sign gesture recognition system for sign language interpretation. The mean accuracy of SLR based on Linear Discriminant Analysis (LDA) on the Chinese numeral gesture dataset was 92.417%, while for the Bangladeshi numeral gesture dataset it was 88.55%. The mean accuracy of LBP based sign language interpretation on the Chinese numeral gesture dataset was 87.13%, and on the Bangladeshi numeral gesture dataset was 85.1%.

Chen and Zhang [IV.374] devised a method using the HOG and SVM algorithms with the Kinect software libraries to recognize sign language by recognizing the hand position, hand shape and hand actions. The average recognition rate was up to 89.8%. To verify the method, a special 3D sign language dataset containing 72 words was collected with the Kinect platform.

Hafiz *et al*. [IV.375] created the system that consists of three components: real time hand tracking, hand-tree construction, and hand gesture recognition. Its features include: (1) a simple way to represent the hand gesture after applying a thinning algorithm to the image, and (2) using a model of complex-valued neural network (CVNN) for real-valued classification. The results showed that the classification ability of single-layered CVNN on unseen data is comparable to the conventional real-valued neural network (RVNN) having one hidden layer. Moreover, convergence of the CVNN was much faster than that of the RVNN in most cases. The system was tested with 26 different gestures.

Quesada *et al*. [IV.376] created a system based on hand tracking devices (Leap Motion and Intel RealSense), used for signs recognition. The system uses an SVM algorithm for sign classification. Different evaluations of the system were performed with more than 50 participants; and high recognition accuracy was achieved with selected signs (100% accuracy for recognizing certain signs).

Dhruva *et al*. [IV.377] designed an algorithm for hand recognition using image processing and explored its effectiveness in security-based systems. The algorithm was tested for different gestures on over 50 samples and the accuracy of 95.2% was achieved.

Misra *et al*. [IV.378] introduced 18 new ASCII printable characters along with some of the previously introduced characters (i.e., A–Z alphabets, 0–9 numbers and four arithmetic operators - add, minus, multiply, divide). It has been observed that maximum accuracy achieved using the combination of existing and proposed features is 96.9%, as compared to 94.60\% accuracy achieved using existing features for classification of 58 gestures.

Fahn and Chu [IV.379] presented a human-robot interaction system that recognizes meaningful gestures composed of continuous hand motions in real time based on hidden Markov models. Experimental results revealed that their system achieved an average gesture recognition rate of 96% at least. They considered eight types of compound gestures, each assigned a motion or functional control command, including moving forward, moving backward, turning left, turning right, stop, robot following, robot waiting, and ready, so that users can easily operate an autonomous robot. To compose the gestures, they defined four basic types of directive gestures made by a single hand - moving upward, downward, leftward, and rightward.

## 4.8 Fall detection and prevention

As the population is getting older [IV.380] and the risk of falls increases with age [IV.381], a research division of AAL focusing on fall recognition and prevention has evolved [IV.382]. The European Next Generation Ambient Assisted Living Innovation Alliance (AALIANCE2) determines the application area of fall prevention technology to range from moving safely at home and outdoors to preventive motor training and system recognition of dangerous situations [IV.383]. Fall recognition technologies provide support by contacting care givers and informing them about a fall [IV.383].

When speaking about falls, the definition of the term is not uniform [IV.384]. The Kellogg International Work Group on the Prevention of Falls by the Elderly [IV.384] highlights the problem of unlike definitions to be the reason for inadequate comparisons between fall studies. Zecevic *et al*. [IV.385] show that seniors' associate falls with losing balance, while healthcare providers generally describe a fall in the context of its consequences. Researchers tend to think of it as the event itself [IV.385]. Although a unique definition may not be necessary, the communication of what is regarded as a fall is essential [IV.384]. A commonly used definition is the one of the Kellogg Group, which describes a fall as "an event in which a person unintentionally comes to rest on the ground or other lower level and other than as a consequence of the following: sustaining a violent blow, loss of consciousness, sudden begin of paralysis, as in a stroke or an epileptic fit" [IV.384].

### 4.8.1 Frequency of falls

About 28-35% of people aged 65 and over fall each year [IV.386], [IV.387]. This percentage rises to about 32-42% in people over 70 [IV.388], [IV.389]. Generally, the frequency of falls increases with age [IV.381] and frailty level [IV.390]. It is estimated that 20-30% of falls lead to mild to severe injuries, and form 10-15% of all emergency department visits [IV.391]. Over 50% of injury-related hospitalizations are among people over 65 years [IV.392]. Especially fatal falls (falls leading to death) are more likely to happen to elderly people. This trend is also reflected in other studies, such as of the Department of Health and Human Services of the United States [IV.393] and of the World Health Organization (WHO) [IV.394].

Kannus *et al*. [IV.395] investigate fall-induced deaths among the elderly population in Finland [IV.395]. They have found out that the number of fall-induced deaths for Finns above age 50 has more than doubled between 1971 and 2002. As the population becomes older, a rising number of similar deaths is predicted in Finland and other Western populations [IV.395]. Apart from health concerns, another major issue arising from this trend are costs [IV.394]. In 2015, approximately 50 billion US dollars were spent on medical costs resulting from fatal and nonfatal falls in the United States [IV.396]. Due to the ageing population, the costs are assumed to be rising [IV.396]. Kannus *et al*. [IV.397] predict that by 2030, fall-induced cervical spine injuries in Finns aged 50 years or older will be about 100% higher than they were during 2000-2004.

### 4.8.2 Importance of early help in falls

Gurley *et al*. [IV.398] perform a study about people found helpless or dead in their homes by the San Francisco emergency department. This paragraph refers to the work of Gurley and colleagues [IV.398]. A group of 367 persons with an average age of 73 years were found, of which 23% were found dead and 5% died in the hospital. The frequency of such incidents was found to increase sharply with age. Men above 85 years who were living alone had the largest share (123 per 1000 per year). The research team demonstrates that the mortality rate correlates with the time a person has been helpless for. For patients who have been helpless for more than 72 hours, the mortality rate was 67%, while it was 12% for patients who waited for help less than one hour. After being found helpless, the majority of patients was unable to return to independent living. Wild *et al*. [IV.399] focus especially on falls of elderlies at home. Over a period of one year, data of 165 people aged 65 and older who fell at home was collected and compared with a control group. Twenty people were unable to get up by themselves for more than one hour, and four of them stayed on the ground for more than six hours. The subjects were visited within seven days after the fall, and again after three and twelve months. After one year, 32 fallers had died compared with eight participants from the control group. Half of those who were helpless for more than one hour died within six months.

### 4.8.3 Fall Detection

Focusing on fall prevention and fall detection, different classes, based on the technology, can be distinguished [IV.400]. Fall detection methods can be classified in wearable devices, robots, audio-based approaches, 2D and 3D sensors [IV.400].

Wearable devices are widely employed in fields such as human activity recognition since this provides a low-cost and easy-to-deploy technology while at the same time it manages to preserve users' privacy. Inertial sensors (also known as IMU or inertial measurement units) are among the most common ones embedded in different wearables. This type of sensor provides quantitative information about acceleration, which is of great help for the purpose of fall detection and prevention. Gyroscopes and magnetometers are also commonly used in combination with accelerometers.

Different approaches found in the literature about where to place such sensors are listed in Table 5.

Table 5. Types of wearable sensors regarding their location

| Wearable type | Location | References |
|---|---|---|
| Wrist band | wrist | [IV.401] |
| Shoes | foot | [IV.402] |
| Using stickers | Over different parts of the body | [IV.403] |

Other methods for fall detection are based on the use of smartphone-included sensors in order to avoid an additional device that has to be carried around [IV.404]. Kau and Chen [IV.405] propose using an electronical compass and a triaxial accelerometer on the smartphone for detecting falls. With the tilt angle and waveform sequence as inputs, a feature sequence is generated and analysed using a classifier method. Hakim *et al*. [IV.406] use the inertial measurement unit with a machine learning algorithm to classify activities of daily living in order to reduce the number of fall alerts. A problem that arises through the use of smartphones as fall sensors is the limitations in battery, memory and real-time processing [IV.407]. Apart from accelerometer-based methods, smartphones can also be used to detect falls by using the integrated microphone [IV.404]. Khan *et al*. [IV.408] split the audio signals into frames for extracting the frequency spectral features. With a collection of footstep sound signals and the use of a one class support vector machine method, falls are distinguished from non-falls [IV.408]. The difficulties in this method lie in the generation of training data, since realistic fall sound signatures are difficult to design [IV.409]. Moreover, the proper use of acoustic and vibration sensors is restricted to a certain floor type [IV.404].

Pressure sensors are the most common type of ambient sensors [IV.404]. Ambient sensors are not wearable devices but attached to the surrounding area of a person, as for example at home [IV.407]. The low costs and the non-obtrusiveness of pressure sensors are in opposition to the low detection precision, which is below 90% [IV.410]. Passive Infrared (PIR) sensors use infrared signatures to detect falls [IV.404]. As the signal changes with motion of a hot object in front of the sensor, a person can be recognized [IV.411]. Falling and ADL like walking can produce similar signals [IV.411].

Hence, Yazar *et al*. [IV.411] use a combination of PIR and floor vibration for fall detection. For the analysis of signals from the vibration sensors, a feature extraction method based on a single-tree complex wavelet transform is used. The PIR is integrated for reducing the number of false alarms that can occur due to falling objects or slamming doors. A major disadvantage when using PIR sensors is the restricted area where falls can be detected [IV.404].

Doppler sensors can distinguish moving objects from background noise and have the advantage of being cheap as well as small [IV.412]. The drawback of this method is that Doppler sensors are less sensitive to movements orthogonal to the irradiation direction than to movements in the irradiation direction [IV.413]. Tomii *et al*. [IV.413] propose a method using multiple Doppler sensors to reduce the dependency on the movement direction. A support vector machine is used to classify the extracted features, which results in 95.5% of accuracy using three sensors [IV.413]. When using Doppler sensors for fall detection, the fact that the electrometric wave signals can penetrate apartment walls has to be considered because it limits the usage in a multi-party house [IV.404].

### Image-based fall detection

While stationary image-based devices can be installed almost everywhere, users might not take wearables with them while sleeping or taking a shower [IV.414]. Getting up at night poses a risk for falling just as well as wet bathroom floors [IV.414]. Additionally, when using stationary devices, the user does not have to remember wearing and recharging them [IV.404], [IV.414], [IV.415]. Image-based methods combining RGB with skeleton points or using depth images allow the usage in day and night condition [IV.414]. Furthermore, image-based methods can monitor more than one person at a time [IV.415]. Computer vision approaches can also more precisely distinguish between falls and ADLs [IV.414]. On the other side, the use of one single camera in a stationary environment can lead to a restriction in perspective [IV.416]. A further issue arising from the use of image-based methods is the aspect of privacy [IV.417]. RGB cameras are distinguished from depth cameras and infrared sensors [IV.404]. Processing RGB images within the system, sending warning signals instead of pictures [IV.414], using depth images [IV.418] or algorithms for hiding people’s identities [IV.415] are approaches to assure privacy.

A common image-based approach for fall detection is to train algorithms with large datasets, so that certain features are recognized and classified [IV.416], [IV.418]. Algorithms are based on the analysis of shape, inactivity or head motion [IV.416]. Fan *et al*. [IV.416] extract the human body using a background subtraction method. Six different shape features are measured to classify the human posture. A classification vector based on the squared first order temporal derivatives of the created slow features is generated. The research team uses a support vector machine to distinguish falls from other activities. Another method presented by ShanShan *et al*. [IV.419] is the Gaussian Mixed Model method to extract the human silhouette. Using semi-contour distances, the posture is quantified and classified by a support vector machine technique. Both Fan *et al*. [IV.416] and ShanShan *et al*. [IV.419] extract a person's silhouette from RGB video sequences. Once extracted, only the shape of the human and not his identity is visible. Another approach is the utilization of a depth camera with the purpose of applying privacy protection at the earliest stage. Planinc and Kampel [IV.420] use the Kinect as a 3D depth sensor in order to calculate the orientation of different body parts. With the use of the least square algorithm a straight line is fitted to the data points. Together with the floor orientation and the distance between the spine and the floor, a fall is differed from non-falls.

Zhao *et al*. [IV.418] focus on falls from bed and use a human upper body extraction method to make the algorithm work even when human-bed interactions happen. The Large Margin Nearest Neighbour classification method is used to detect a fall from the bed. The algorithm is implemented on depth image data, using the Microsoft Kinect and the Orbbec Astra camera. The research team highlights the advantage of depth cameras to be insensitive to illumination variation [IV.418]. A system based on the Kinect sensor and fast Fourier transformation is presented by Kong and Meng [IV.421]. At first, the received RGB and depth image from the sensor is transformed into a skeleton image. By using fast Fourier transformation, the image is then encrypted and sent via a carrier image, which has to be decoded. In order to detect a fall, machine learning is applied. When a person is detected by the sensor, the height and width of the created skeleton image are calculated. These parameters are the basis for the classification which is done with a k-Nearest-Neighbour/Support Vector Machine approach.

Ma *et al*. [IV.415] present a method using a thermal camera to determine and to extract facial regions. Visible light rays enter, controlled by a Spatial Light Modulator (SLM), an RGB camera, where only images with hidden facial regions are processed. In order to detect falls, this method combines a 3D convolutional neural network with an autoencoder. While the neural network is used for feature extraction, the autoencoder models' normal behaviour. A method presented by Kido *et al*. [IV.438] uses solely a thermal camera to detect falls in bathroom environments. Normal activities in the toilet room are distinguished from falls by performing discriminant analysis. Therefore, the thermal image is split into 81 areas with known average temperature. One disadvantage of this method is that it requires visual confirming by caregivers [IV.438]. Ozcan and Velipasalar [IV.417] propose a computer vision approach that is not restricted to a specific location. Instead, a wearable camera is used to ensure the detection of falls independent of the environment [IV.417]. HOGs are combined with Gradient Local

Binary Patterns to generate features. The classification of fall events is implemented by using a relative-entropy-based method.

### Audio-Based Fall Detection

Falls are one of the most frequent causes of injuries in elderly people, with 20% up to 40% of people over 65 years reporting falls at home.  Effects can be fatal due to the loss of consciousness or inability to call for help in a distress situation [IV.422]. Therefore, developing robust, nonintrusive, and socially acceptable fall detection system that is able to alert fast in case a fall happens is very important for the elderly population.

Fall detection systems can be split into two categories based on the sensor technologies they are using: environmental (microphones, cameras, floor vibration sensors, pressure sensors, infrared sensors) and wearable (accelerometer and gyroscope sensors). We focus in this section on the former category, in particular audio-based systems for fall detection.

#### Feature Extraction

The input audio signal used in fall detection systems typically originates from various sources; therefore, extracting relevant features for distinguishing between falls and other types of audio events is of utmost importance. The feature extraction follows two main directions: handcrafting the features that requires substantial domain knowledge, and learning more robust feature representations directly from raw data or low level signal representations.

Previous studies mostly rely on standard features in speech/audio processing tasks, such as spectrograms, Mel Frequency Cepstral Coefficents (MFCCs) and Linear Predictive Coding (LPC) features [IV.423] [IV.424], Gaussian Mixture Model (GMM) super-vectors, known to be successful for speaker recognition, speaker verification and emotion recognition tasks, have been used for detecting falls in [IV.425]. Universal background model (UBM) is first built on an entire audio dataset, and further used to determine the set of GMM super-vectors by adapting the mean vectors of the UBM with the maximum-a-posteriori criterion for each audio event (fall or non-fall). Acoustic local ternary patterns that represent an audio signal as a 3-value ternary pattern, calculated from the magnitude difference between the central and surrounding samples in the frame, and further quantized with 3 discrete levels (+1,0,-1), are shown to outperform MFCC for a fall detection task and furthermore, they are robust against signal rotation [IV.426].

An end-to-end unsupervised fall detection system based on a deep convolutional autoencoder is proposed for joint feature learning and fall detection task directly from the low-level audio signal representations, i.e. the normalized log-power spectrograms [IV.427]. A recent study reports the use of Siamese convolutional autoencoder, a network composed of twin convolutional autoencoders with the same topology and weights, and trained on the log-Mel coefficients extracted from the raw audio signal to learn more robust feature embeddings. During the training two convolutional autoencoders are exposed to two different samples from the dataset [IV.428].

### Classification

Classification assumes training the model to distinguish the human falls from the non-fall acoustic events using the features extracted from the audio signal. Based on availability of the data for model training, either supervised or unsupervised machine learning approaches are used.

If the dataset contains sufficient number of human falls, supervised techniques such as SVMs [IV.426], [IV.429], [IV.430], k-NN [IV.431], [IV.424] or ANNs [IV.423] typically achieve satisfactory performance. However, collecting large labelled datasets of human falls is a difficult task; therefore, most of the studies use simulated falls with the human mimicking dolls without any guarantee that the model will generalize well in real-world conditions [IV.428].

Other approaches rely on unsupervised techniques for anomaly detection, which require training the models on a large amount of non-fall activity data that is easier to acquire, while during inference the human falls are recognized as outliers. Examples of this are one-class classifiers, such as one-class SVM [IV.432], one-class k-NN and one-class GMM [IV.433]; or more recently convolutional autoencoders [IV.427].

A hybrid system with a large amount of data that represents the non-fall activity and only few examples of real human falls most closely resemble the reality.  Such approaches rely on one-shot learning techniques, such as one-shot SVM or one-shot Siamese autoencoders [IV.428]. Semi-supervised approach that combines one-class SVM with template-matching classifier is proposed in [IV.434]. In the first step, one-class SVM is used to distinguish anomalies (human falls) from the non-fall events, while in the second step the template-matching classifier makes a final decision whether the input is a human fall or not using the set of templates GMM super-vectors representing the audio events labelled as false positives by the user.

### Challenges

Some of the challenges related to the audio-based fall detection are specific for the audio modality, whereas the others are shared across different modalities. We provide here a non-exclusive list of challenges that the audio-based fall detection community should respond in the incoming years.

- **Lack of available datasets** is the biggest challenge, since data collection of human falls, especially for elderly people, is extremely difficult. Publicly available datasets either have limited size, the falls are simulated using human mimicking dolls, or by young people with appropriate protection, without any guarantee that the falls are correctly emulated and correspond to real-world situation. Therefore, it is necessary to develop systems that are able to work with insufficient training data, using techniques based on oversampling, semi-supervised learning, anomaly detection or one-class classification [IV.435]

- **Unsupervised or semi-supervised techniques** typically require only a large amount of "normal" non-fall data, and no or only few examples of falls to train the models. However, it is unclear how to correctly define the "normal activity". A big dataset composed of all possible "normal activities" is required, which is very difficult to acquire in a real scenario, and might cause an increased number of false positive events [IV.435]
- **Class imbalance** is a common problem in datasets with a large number of non-fall events and only a small number of real human falls, typically acquired for human fall detection. In these situations, over/under sampling or weighting techniques need to be applied in order to alleviate the class imbalance [IV.428]
- **Sensitivity to background noise** is another typical problem of audio-based technologies for fall detection, since the recognition as well as the user localisation ability is highly dependent on the acoustic conditions. Detecting the slow falls with minimally produced sound may be further challenging, but also the type of floor surface can limit the detection range of the fall detection system [IV.436]
- **Privacy concerns** related to the use of audio-based fall detection systems depend heavily on the type of sensor used in such systems. Since the ultrasonic sensors are operating outside of human audible range, they can be considered unobtrusive and do not raise privacy issues as much as other acoustic sensors. The feeling of intrusion also depends on the location of the sensor. Intuitively, most of the users would prefer sensors not to be placed in private places, such as bathrooms, but on the other hand the risk of falling might be biggest there. The collected data is very sensitive, since beside the falls, it can expose other personal data, such as information about users' daily activities and routines [IV.437]. Privacy issues should not stand in the way of the potential benefits of fall detection systems, but on the other hand, privacy should not be sacrificed for these benefits. Therefore, the privacy protection mechanisms, such as data encryption, should be implemented to make such systems socially acceptable [IV.436].

### 4.8.4 Fall Prevention

A part of fall prevention deals with recognizing balance and motion abnormalities to assess the fall risk of a person [IV.439]. Another approach is to detect a person leaving the bed without assistance and sending alarms to the caregivers for helping the person [IV.440]. In the following work, the event of a person getting up from bed is referred to as "getup event" or "getup".

A method to prevent patients falling from bed are restraints such as bed rails [IV.441]. But when applying restraints, the risk of injury has been found to increase [IV.442]. Beds mounted close to the ground as well as floor mats in front of the bed are one of the solutions developed to prevent falls and to minimize the extent of injury. A drawback of those devices is that they cannot alarm nursing staff [IV.443].

Although pressure-sensitive mats with the functionality of sending alarms when the pressure exceeds a certain threshold exist [IV.444], [IV.445], they can pose a risk for tripping [IV.446].

In general, one can group fall prevention systems into those that detect a person in danger of falling and those that help to improve a person's balance and strength [IV.447]. One example for the latter is a robot developed by Maneeprom *et al*. [IV.448]. The robot shows videos about how to prevent falls, giving advice on choosing appropriate shoes and walking assistive devices. Additionally, daily voice messages and exercise reminders are provided.

Lee *et al*. [IV.449] use ten piezo-resistive pressure sensor pads positioned in a bed mattress to measure the pressure applied on the sensors. By evaluating the pressure values, a person's position and activity status is determined. When a person is about to exit the bed, an alarm is sent via a mobile application.

A method based on a body-worn accelerometer is presented by Wolf *et al*. [IV.450]. An accelerometer positioned with a tape on the user's leg calculates the orientation of the sensor and the amount of movement to distinguish between lying, sitting and standing. Ribeiro *et al*. [IV.451] develop a mat consisting of a pressure sensor and a 3-axial accelerometer to detect movements on the bed or armchair. A control system analyses arm as well as torso movements to detect a person getting up from bed or from a chair. A warning level loop with different states is used to offer the opportunity to manually deactivate a visual warning light before sending an acoustic alarm.

#### Image-based fall prevention

Not only in fall detection, but also in fall prevention, image-based sensors like RGB or depth cameras do play a role [IV.452]. They are used to measure balance, stability, reaction time, diversions from typical activity patterns and other physical parameters [IV.446], [IV.447]. Moreover, they are used for games and activities operated via a camera with the intention to improve a person's balance and strength [IV.452]. A product commonly used as a tool to assess the fall risk of elderly is the Kinect sensor from Microsoft [IV.439], [IV.453], [IV.454]. Kampel *et al*. use the Kinect to automatically analyse the Timed Up and Go (TUG) test. Two different approaches to calculate the start and end time of six TUG phases are presented. When working with skeleton data, the trajectory of the spine-shoulder joint is used for the computation, while it is the centre of mass of a person when using depth data.

Dubois *et al*. [IV.439] present a method to automatically evaluate a person's risk of falling by carrying out eight different balance tasks in front of the Kinect. In order to assess the stability during a task, the body centroid is calculated. The horizontal dispersion of the pixel cloud is used to determine whether a person uses the arms to maintain the balance. For the final assessment of the individual fall risk, apart from the balance tasks, the age, average physical activity and results from the TUG are considered too.

Rantz *et al*. [IV.454] use the Kinect to design a system for the home-environment of elderly people in order to continuously analyse gate parameters like velocity, step and stride length. The use of automatic gait assessment methods has the advantage of a regular tracking of fall risk parameters [IV.454]. In case of an increased risk, nursing staff can be alerted. By including a Doppler radar, the problem of occlusion is avoided. Similar to the previously described bed exit alarm system Bucinator, cameras are also used to inform health workers when a patient with an increased risk of falling is about to leave the bed [IV.443], [IV.455]. Ni *et al*. [IV.455] combine the Kinect's depth sensor with its RGB camera to monitor patients and to alert the nursing staff in case of need. The region around the bed is defined as region of interest and split into eight rectangular blocks. Multiple motion and shape features are extracted and combined using a multiple kernel learning framework. The system Ocuvera, developed by Bauer *et al*. [IV.443] uses the Kinect's depth sensor only. After detecting the floor and the bed, machine-learned shape models are used to find human shapes. For every person found in the scene the likelihood of bed exit is predicted in order to trigger alarms before a person exits the bed. Problems that still need to be addressed are the integration in existing alarm technology of hospitals and the improvement of the algorithm in order to minimize the time between the event and the alarm [IV.443].

## 4.9 Frailty recognition

Modern society faces economic and social challenges caused by population aging: higher age increases the possibility to be frail, including impaired mobility and cognitive decline, which causes ADL to become a challenge and increases the need for assistance [IV.456]-[IV.458]. The probability to become hospitalized is four times higher for frail people and twice as high for pre-frail people than for healthy people [IV.458]. The recognition of frailty and pre-frailty is of particular importance since the progress of frailty can be damped or even reversed by application of appropriate treatment [IV.459]. This Section describes frailty itself and tools to assess it, as well as its correlation with mobility. Ways to assess a person's mobility status are explained and how these measurements are supported by technology using sensors and automated analysis methods.

### 4.9.1 Definition of Frailty

Until 2013 there existed no consensus definition of frailty [IV.460]-[IV.462], but various individual descriptions have emerged: Lang *et al*. [IV.463] describe frailty as a lengthy process with increased vulnerability and predisposition to functional decline that may lead to death. According to Sales [IV.464], frailty represents multiple comorbidities over time and Fried *et al*. [IV.465] describe frailty as the reduced resistance to stressors. In 2013 a consensus group of major international societies defined physical frailty as medical syndrome caused by multiple contributors and characterised by reduced endurance, strength, and physiologic function resulting in increased vulnerability to develop dependency or death [IV.459]. They define frailty further as an indicator for a higher vulnerability to stressors than non-frail persons, causing adverse health issues and functional deterioration, however, it might be reversible due to interventions [IV.459].

Frail people are restricted in executing ADL or have limited mobility; they show higher fall risks and hospitalisation probabilities [IV.466]. Adults older than 65 years have a higher probability to become frail as Bandeen-Roche *et al*. [IV.458] show evaluating 7,439 non-nursing home persons. The 15.3% of the persons analysed were frail and 45.5% pre-frail [IV.458]. The vulnerability of a frail person in comparison to a healthy person is illustrated in Figure 28, showing the impact of a minor health problem on healthy and frail elderly persons. The event has more effect on the health state of frail persons, which may cause those persons to become dependent, immobile, or delirious [IV.457]. Since frailty can be prevented or delayed [IV.459], [IV.467], it is necessary to identify the risk of frailty for early application of preventive interventions, which leads to a demand for methods to detect frailty in a valid and reliable manner [IV.457], [IV.467]. There are two major assessment tools, which have emerged in the past decade [IV.460], [IV.461], [IV.468]: (1) the frailty phenotype of Fried *et al*. [IV.466], and (2) the frailty index of Searle *et al*. [IV.469].

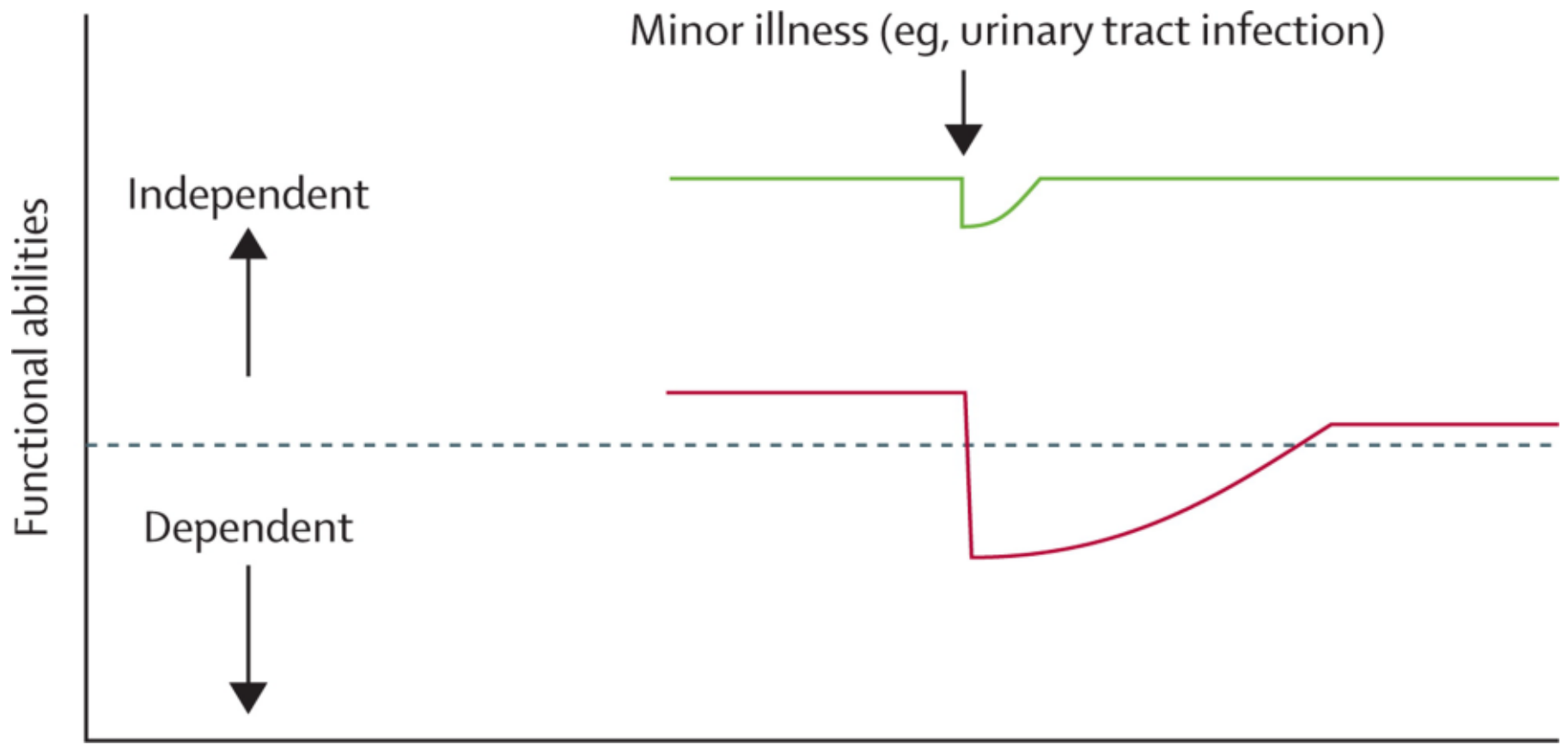

Figure 28. Vulnerability of elderly people to minor health problems: red line shows a frail elderly person becoming dependent due to this event, whereas the same event has less effect on a healthy person. (Image readapted from [IV.457])

### Frailty Index

Searle *et al*. [IV.469] describe an index to measure frailty, which is based on social factors, physical, and psychological parameters. The Frailty Index combines multiple clinical state variables into a single number, which quantifies the possibility of adverse events [IV.470]. It is defined by Searle and Rockwood [IV.471] as

$$\text{Frailty Index} = \frac{number\ of\ deficits}{number\ of\ considered\ variables}.$$

The chosen variables are not predefined but must follow five criteria: 1) chosen variables have to be health-related, 2) their prevalence has to increase with age, 3) they should not saturate too early (e.g. age-related eye lens change), 4) the variables must cover a range to ensure a general frailty index (otherwise it is only e.g. a cognitive index), 5) if several people are tested with the same index, the variables are required to remain the same [IV.471].

The higher the number of deficits, the higher is a person's frailty index. Following list shows suggestions for deficit variables from Searle *et al*. [IV.469]: need help getting in/out of a chair, need help with housework, need help walking around a house, walk outside, lost more than 4.5kg in last year, feel everything is an effort, had heart attack, has arthritis. The number of considered variables should range from 13 to 100, although the optimal number lies between 40 and 50 variables as Davis *et al*. [IV.470] investigated. Jones *et al*. [IV.472] describe the construction of a Frailty Index based on the Comprehensive Geriatric Assessment [IV.473], which allows to use variables obtained routinely by geriatricians and are reliable for risk stratification of possible adverse events. Due to the inclusion of deficits with relation to adverse events, the frailty Index is considered to be more sensitive as a predictor for adverse health issues [IV.474].

### 4.9.2 Correlation with Mobility Parameters

As shown by these assessment instruments, multiple variables and criteria, such as diseases, mobility decline, or impairments in ADL, have to be observed in order to identify a person as frail. However, covered variables show high correlation with frailty in general: Davis *et al*. [IV.470] evaluate 1,295 people of 70 years and above having a Frailty Index between 0 and 51% to investigate the relation between frailty and mobility. They created 5 classes of 259 persons each, with ascending frailty indices: Q1) 0-8%, Q2), 8-12%, Q3) 12-17%, Q4) 17-23%, and Q5) 23-51%. In Q5 37% showed difficulties with mobility, while none of Q1 had mobility impairments. Davis *et al*. [IV.470] states that although mobility impairment only is not sufficient to classify a person as frail, it is a predictor for frailty due to its correlation. Delbaere *et al*. [IV.475] also show a correlation between mobility and frailty: the avoidance of activities due to fear of falling increases physical frailty. By reducing the ADL and outdoor activities, the muscle strength is reduced, which increases the risk of falling and slows the speed of walking [IV.476]. The correlation between falls and frailty is also shown by Davis *et al*. [IV.471]: persons of Q5 reported 3 times more falls than those in Q1. This corresponds with the results of Bandeen-Roche *et al*. [IV.458], who describe that the fall risk is three times higher for frail people than for non-frail people. Montero-Odasso *et al*. [IV.477] define gait speed below 0.8m/s as pathological and show a correlation between this slow walking speed and the occurrence of falls: persons (70+ years) with pathological gait speed fall twice as much as persons with normal gait speed. Studenski *et al*. [IV.478] describe a correlation between gait speed and mortality at a certain age, based on nine cohort studies with data of 34,485 individuals. They showed the relation between age, gait speed and life expectancy: the lower the walking velocity, the higher the risk of early mortality (see [IV.478] for an illustration of this effect). Reasons for that are the need for energy and movement control during walking [IV.480]. Further, walking requires interaction of multiple organ systems in the body, e.g. lungs, heart, circulatory, nervous system, and muscles [IV.478]. Damages in those systems can result in slow gait speed, hence gait velocity is a summarizing indicator for vitality [IV.478].

A significant association between gait velocity and persons' balance is reported by Montero-Odasso *et al*. [IV.477]: 79% of 92 observed persons with pathological gait speed had problems maintaining balance on one leg, while only 40% of persons with normal gait speed had these problems. Balance deficits indicate decreased power of the lower body muscles, which correlates with higher fall risk and results in problems when standing up [IV.481], [IV.482]. Cheng *et al*. [IV.481] examine the stand-up duration of 105 persons (young, non-fallers, fallers) and show that the time required to stabilize oneself after a Sit-To-Stand (STS) movement is 0.83s longer on average for fallers than for non-fallers, and hence, also the average of the total STS duration is 2.13s longer [IV.481]. Zhang *et al*. [IV.483] examine STS times of 948 adults (60+ years) using the 5 times STS test (5tSTS), which consists of 5 successive stand-ups without use of upper extremities. They conducted further a 3-year follow-up showing following results: For persons incapable of performing the 5tSTS the probability of falling was 4.22 times higher at the follow-up than those having the fastest stand-up times (<11.2s). The probability to fall was 1.09 times higher for the slowest group able to perform the test (>16.6s) than the fastest group. Further, persons not able to perform the test showed 24.70 times higher probability to develop impairments performing ADL, than the fastest group. Hence, problems performing STS movements are a significant predictor for ADL-related disability and increased fall risk [IV.480].

## 4.10 Gait analysis and mobility assessment

Mobility parameters and fall risk correlate with a person's frailty state [IV.484], [IV.485] and are thus used as predictors for frailty [IV.486], [IV.487]. In medicine, mobility is measured using mobility assessment tests [IV.488]. There exists technology assisted versions of these tests as shown in the next Section. Six examples of the medical assessment tests are described in Table 6. The tests are selected since they require no special equipment, and their conduction requires only short time (<30min) [IV.488]. In the Timed-Up-and-Go-Test (TUGT) [IV.480] and the Six-metre-walk test [IV.489] the persons are told to conduct the test at "habitual speed", instead of "as fast as possible", which means performing the movements like in everyday life. Performing geriatric tests at habitual speed partially mediates the inverse function between disability of elders and cognitive function [IV.490]. Further, there is a significant correlation between changes in habitual walking velocity and cognitive decline [IV.491].

Persons still change their behaviour and do not execute the movements at habitual speed as in daily life, due to the created test situation [IV.492].

Table 6. Examples of mobility assessment tests

| Test | Description | Reference Values |
|---|---|---|
| TUGT [IV.480] | The subject has to stand-up without assistance, walk 3m, turn around, walk back and sit down. The time needed is measured. Examines: gait velocity, balance, lower extremity. | • 60 - 99 years: 9.4s [IV.493] |
| Half turn test [IV.494] | The subject walks a few steps, then has to turn 180 around. The number of steps needed to turn around is counted. Examines: mobility, balance. | • 74 - 98 years: 4.5s [IV.488] |
| Alternate-step test [IV.494] | The subject has to place both feet on a step, beginning with left and right alternating. The time needed to complete 5 steps is measured. Examines: lateral stability. | • 74 - 98 years: 10.8s [IV.488] |
| 5tSTS test [IV.495] | The subject has to perform five stand-ups and sit-downs consecutively from a chair without arm rest. A variation of this test is the STS 1, where the subject has to stand-up only once [IV.488]. The total time needed is measured. Examines: lower limb strength. | • > 60 years: 11.4s<br>• > 70 years: 12.6s<br>• > 80 years: 14.8s [IV.493] |
| Six-metre-walk test [IV.489] | The subject has to walk at normal walking speed for 10m in total, 2m before and after the 6m walk to ensure a constant speed. The time needed for the distance is measured to calculate the velocity. Examines: gait velocity, fall risk | • > 60 years: 1.36±0.21m/s (men)<br>• > 70 years: 1.33±0.2m/s (men)<br>• > 60 years: 1.30±0.21m/s (women)<br>• > 70 years: 1.27±0.21m/s (women) [IV.493] |
| Short Physical Performance Battery (SPPB) [IV.496] | The subject has to perform certain standing positions for 10s each. Further, either a 2.44, 3 or 4m walk has to be done twice, as well as five STS transfers. The durations to hold the positions, to walk the predefined path, and to perform the stand-up are measured and individually scored from 0 to 4 (12 points in total). Examines: balance, gait velocity, lower body strength, and overall physical fitness. | • > 70 years: 8.4±2.7 points [IV.497] |

### 4.10.1 Human Gait Analysis

As an indicator of a person's health condition, human gait is analysed for diagnosis and monitoring, as well as rehabilitation [IV.498]. One way to analyse gait is visual observation by a human of a person walking or running, and for repeated viewing without exhausting the patient, video recordings of the walk are recorded. This method depends highly on the subjective rating and personal experience of the observing clinician. Spatio-temporal measures (e.g., gait velocity) allow us to obtain quantitative parameters [IV.499]. In order to perform objective gait analysis, spatio-temporal parameters of gait cycles have to be obtained and analysed [IV.500]. Figure 29 illustrates a gait cycle of a healthy adult, which is defined as the interval between two consecutive floor contacts of one foot partitioned into two main phases: stance and swing. During stance phase the foot is in constant contact with the floor beginning with the heel, followed by the flat foot and ending with the push off from the ground, which itself starts from the heel to the toes. During swing phase the foot is brought to the front, while the other foot is in the stance phase. When the toes are pushed off, the other foot regains floor contact which causes both feet to touch the floor at the same time, called double support. The event when a foot touches the floor is defined as Heel Strike (HS) and the event when the foot leaves the floor as Toe Off (TO). Other gait parameters are derived from these gait events: from HS to the second following TO (the TO event of the same foot) represents a stance phase. The duration between two consecutive HS events is defined as the step time. The stride time is the duration between an HS event and the next HS event of the same foot, which corresponds to the total gait cycle duration [IV.501]. The phase of both feet touching the ground is defined as double support, while single support defines the phase of only one foot being in contact with the floor.

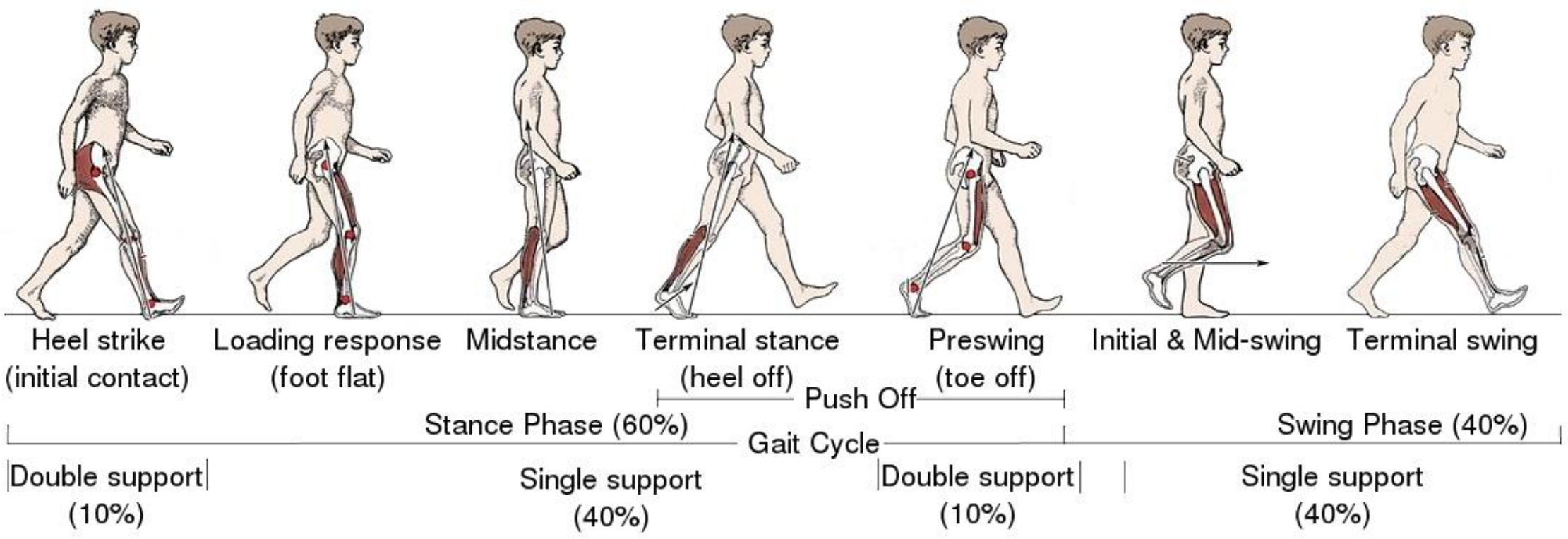

Figure 29. The gait cycle consists of two main phases: stance phase where the foot has permanent contact to the ground, and the swing phase where the foot is brought forward to begin the next step[19]

### 4.10.2 Sit-To-Stand Analysis

STS transitions consist of three phases, which are depicted in Figure 29: I) weight shift and begin of trunk flexion, II) knee extension and end of trunk flexion, III) lifting the weight, extension of the trunk flexion and full extension to standing position [IV.502]. The performance of STS movements is determined by an interaction of multiple physiological and psychological factors: knee extension and flexion, dorsiflexion strength of the ankle, foot reaction time, body weight, sway of posture, reported pain, sensitivity to visual contrast, proprioception of lower limb, anxiety, and general vitality [IV.503]. STS performance is further influenced by the chair height, position of the feet, as well as the use of armrests [IV.504]. Examination of the stand-up performance by recording the required duration is used in mobility tests, as described in Table 2.2: TUGT [IV.480], STS test [IV.495], and SPPB [IV.496]. These tests have in common, that the measurement begins, when the subject is advised to start, which includes the person's reaction time. Depending on the evaluation method, there exist different reference values for stand-up durations in literature ranging from 1.51s to 2.42s on average for healthy young adults [IV.505]-[IV.508] and 1.56s to 2.54s for healthy older adults [IV.507], [IV.509]-[IV.511]. Pathologic durations for STS transfers are specified from 2.73±1.19s to 4.32±2.22s [IV.505], [IV.508], [IV.512]. In contrast to the STS movement, the Sit-To-Walk (STW) transition is partitioned into four phases: 1) initiation: forward movement of the centre of mass, 2) seat-off and peak of vertical velocity: body rises, lower limb joints and trunk are extended, 3) initialisation of the gait and swing phase: unloading, weight shift, 4) stance phase: end of swing phase, walk [IV.513]. The centre of gravity moves higher but less forward in STS than in STW movements. To perform an STW motion, a forward impulse has to be created using the upper body (head, trunk, and arms) resulting in a higher horizontal velocity. The STW movement requires more advanced motoric control than the transition from sitting to standing [IV.514].

[19] Image adapted from https://www.physio-pedia.com/File:Gait-Cycle.jpg (last accessed: 23/02/2022)

Especially the momentum created in the third phase, when the transition to walking takes place, requires balance control [IV.515].

### 4.10.3 Fall Risk Assessment

Falls represent a serious obstruction for elderly to live independently and are the major reason for injury related deaths for adults of the age 79 and above. The inability to stay upright while standing or walking is the result of a complex system failure. To maintain balance, interaction of multiple muscles is required and controlled by commands from the motor cortex reacting to sensory signals from the spinal neurons. This activity is coordinated from the central nervous system, which reacts to the respective terrain conditions. Maintaining balance is a cognitively challenging task and confrontation with additional cognitive load results in higher fall probabilities also for healthy, young persons as shown by Templer and Conell [IV.516]. Hence, frail persons suffering from multiple physiological impairments and vulnerable to stressors are prone to falling when exposed to exogenous stress and having no physical assistance available in their vicinity [IV.517]-[IV.519].

Table 7 shows an overview of assessment tests, which are used to determine a person's fall risk [IV.520].

Table 7. Examples of fall risk assessment tests

| Test | Description | Reference Values |
|---|---|---|
| Get-up and go [IV.521] | The subject has to stand up from a chair and walk a short distance (e.g., 3m), turn around, walk back to the chair and sit down again. Observers rate the subject's performance on a 5-point scale. Examines: gait and balance | • Gait abnormalities: ataxia, apraxia, shuffle<br>• require arms for stand-up<br>• require assistive devices [IV.520] |
| TUGT [IV.480] | The subject has to stand-up without assistance, walk 3m, turn around, walk back and sit down. The time needed is measured. Examines: gait velocity, balance, lower extremity | • 13.5s: increased fall risk [IV.522] |
| extended TUGT [IV.523] | The subject has to perform the<br>TUGT with a 10m walking path.<br>The durations of multiple components are measured individually: stand-up, gait initiation, walk to turning point, turn, walk to starting point, sit-down. Examines: gait velocity, balance, lower extremity as in TUGT but more sensitive | • first walk is most sensitive<br>• 70 - 79 years: 1.18±0.15m/s (men)<br>• 70 - 79 years: 1.11±0.13m/s (women) [IV.523][IV.524] |
| Functional reach test [IV.525] | The subject has to stretch the arms out frontward and lean as far forward as possible without losing balance. | • < 0.15m: 4 times higher fall risk in succeeding half year |

| Test | Description | Reference Values |
| --- | --- | --- |
| | The maximum possible distance is measured. Examines: balance. | • 0.15 - 0.25m: 2 times higher fall risk in succeeding half year [IV.526] |
| Morse fall scale [IV.527] | The scale consists of six variables, which are rated using scores. The variables are assessed by observer(s): history of falls, number of medical diagnoses, ambulatory aids (e.g., wheelchair), application of intravenous therapy, gait, and mental status. Examines: fall risk on three level scale (no, low, and high risk). | • 45 - 55 points (and above): increased fall risk [IV.528] |

## 4.11 Motor rehabilitation

Technology is developing day by day and this development affects the lives of individuals. Although the connection of these developments with inactivity is worrisome [IV.529], the possibilities offered by technology bring with it an increase in the applications of active and active assisted living. Thanks to these visual and auditory-based applications, many gains can be achieved in terms of healthy or unhealthy individuals. In fact, it is stated that one of the factors affecting the success of tele-medicine applications is technology-oriented [IV.530]. Such e-health applications reduce economic and social costs and provide an opportunity to collaborate remotely [IV.531].

Technological applications have many benefits for motor rehabilitation in terms of e-health. It has been reported that audio and visual applications can be motivating and effective in increasing the physical activity of inactive mothers [IV.532], eliminating the movement deficits of individuals with upper and lower extremity movement disorders [IV.533], reducing upper extremity dysfunction in patients with breast cancer [IV.534], improving the range of motion of individuals with distal radius fractures and trapeziometacarpal arthrosis [IV.535], performing oropharyngeal exercises in patients with snoring problems [IV.536], and in the treatment of stress urinary incontinence in primiparas [IV.537]. The digital assistant named Vigo, developed by Epalte *et al*. [IV.538], supports therapeutic activities for paralyzed patients and their relatives, provides educational information and increases participation in therapeutic activities. Similarly, the Rehab@home system, which was developed specifically to assist neurological patients who perform rehabilitation exercises at home without a physiotherapist, provides support for active movement [IV.539]. In addition, the effectiveness of the Multimodal Guidance System, which provides robotic assistance in the execution of 2D tasks through tactile and vocal interactions, has been demonstrated for people with certain disorders that affect coordination, such as Down syndrome and developmental disabilities [IV.540].

One of the greatest achievements of technology in the field of motor rehabilitation is the combination of game-based therapies with virtual reality. Applications in this field, which are supported visually and audibly, are new and potentially useful technologies. Interactive video exercises on active and active assisted living are used with increasing frequency, thus enabling individuals to practice physical activity or exercise while playing active games. "Exergames", which is a combination of the words exercise and game, is defined as the use of interesting activities with virtual reality that cannot be safely performed in the real world [IV.541]. Physical activity is encouraged by pairing various visual and auditory stimuli with different activities to increase exercise behaviour [IV.542]. Low cost, wide availability, and the ability to combine multiple educational approaches have recently made it attractive to use video game systems such as Nintendo Wii, Sony PlayStation Move and Microsoft Xbox Kinect [IV.543], [IV.545]. In these systems, body movements are defined in games with the help of devices and applications can be used for therapeutic exercise in clinics [IV.546]. Individuals, alone or as a group, engage in active physical activity with the contribution of technology [IV.547]. These practices create changes in the cortical activation of the brain depending on motor learning principles such as dual-task, activation of mirror neurons, stimulation of limbic pathways, challenge, and visual and auditory feedback [IV.548], [IV.551]. Compared to traditional exercises, it provides advantages such as motivating individuals to practice, improving motor and cognitive skills by performing dual tasks, and focusing attention on the output of movement [IV.552]. Studies have shown such applications are used in multiple sclerosis [IV.553], cerebral palsy [IV.554], Parkinson's [IV.555], chronic stroke [IV.531], [IV.556], [IV.557], traumatic bone and soft tissue injuries [IV.558], juvenile idiopathic arthritis [IV.559], ASD [IV.560], and obesity [IV.546], [IV.561]. These contribute in terms of participation in physical activity [IV.553], [IV.562]-[IV.565], improving somatosensory system [IV.566], gross motor skills [IV.567], cardiorespiratory fitness [IV.568], psychological well-being [IV.546] social interaction and cognitive [IV.569]. Exergaming systems may offer a potential solution to improve compliance with exercise programs. Game-based applications are time efficient, easy to implement, and can assist in clinically effective programming [IV.531]. Given the rapidly growing interest in virtual reality-based therapy in rehabilitation, this new method could be widely applied in future research [IV.559]. It is a relatively new and promising option to increase physical activity [IV.531], [IV.553].

## References

### Section 4.1

### Section 4.2

Section 4.3

## Section 4.4

## Section 4.5

## Section 4.6

## Section 4.7

## Section 4.8

## Section 4.9

## Section 4.10

Section 4.11

# 5. Challenges and potentials for AAL real-world uptake

This Section provides the reader with an overview of the main challenges, hindrances and opportunities posed by the uptake in real world settings of AAL technologies. More precisely, procedural approaches based on co-design are revised in Section 5.1. User acceptance and ethical considerations are also debated in Section 5.2. Privacy preservation in video data is tackled in Section 5.3. The approaches to ensure transparency and decision explainability in data processing are presented in Section 5.4. The main technological details of data transmission and communication are outlines in Section 5.5. Finally, the potentials coming from the silver economy are overviewed in Section 5.6.

## 5.1 Co-designed approaches

In this section, we will describe user centred design and design thinking as enabling approaches in design of usable systems.

When designing and developing digital technologies with the aim of supporting some aspects of work or everyday life, the process can differ substantially. On the one hand, the process can be conducted with detachment from the end-users (i.e., those that are designated users of the digital technology which is the design object), or on the other hand, through engagement with end-users. The latter is the focus of this section and that particular focus builds on the vast literature on the importance of end-user participation and engagement in the design process [V.1]. In such a process, the focus is on empowerment of the end-users and making them co-producers in the design process. Involving end-users in the design process is not new, within participatory design that has been the guiding philosophy for five decades [V.2]. An important building block in the movement towards what became participatory design was written already in 1972, by Cross [V.3] in the preface to Design Participation: "Professional designers in every field have failed in their assumed responsibility to predict and design out the adverse effect of their projects. These harmful side effects can no longer be tolerated and regarded as inevitable if we want to survive the future. The increasing amount of protest against a wide range of dubious developments is an indication that many people are now not prepared to go on accepting the missing 'price of progress.' [...] There is certainly a need for new approaches to design if we want to arrest the escalating problems of the man-made world and citizen participation in decision making could possibly provide a necessary reorientation" (p.11).

Since participatory design and later collaborative design (which ultimately was termed co-design, which is the primary focus of this section) emerged in the literature, this particular design philosophy has often been targeted towards designing a specific service or one specific digital technology. However, as the diversity of digital technology increases and the world of digital technology now encompasses everything from small apps to large-scale infrastructures, the design situation also changes. We are no longer merely designing digital technology for or with end-users; instead, we are co-designing complete future experiences and digital technology that constructs cultures and new practices, which have large implications on both everyday life and work tasks [V.4]. Based on that, co-design outlines a collaborative creative activity, where the end-users, who are not trained in design work on beforehand, alongside the designers together engage with each other in order to further the design process [V.5]. The fundamentals of co-design as an approach to the design process therefore includes the notion that the end-users have a voice in the design process that ultimately will have an effect on their everyday life or work tasks [V.6], [V.7]. As pointed out earlier, the digital technology that is being designed can be of complex nature, and so can the group of end-users. The end-users can both be heterogeneous within their particular group (e.g., patients that have diverse needs but are a part of the same patient group) and moreover, the groups of end-users can also be several (i.e., patients, loved-ones and healthcare professionals) [V.4]. That indicates that the co-design process and end-user engagement can be of high complexity. In this section we present what the approaches user-centred design and design thinking contain to facilitate user participation in design processes [V.8].

### 5.1.1 User-centred design

Design with users and design for users are very important aspects in designing systems if user acceptance is central. The success depends not only on the methodology applied, but also on the consideration of the use context of the target group. Designers need the most appropriate approach for a user-centred development. To avoid the gap between the use and design of systems, socio technology has been utilised as a guiding approach [V.8], [V.9]-[V.11]. User-centred methods have proved themselves as very useful means of facilitating open and cooperative settings (see Figure 30).

*User-centred thinking* creates a direct link between the current and future users [V.12], [V.13]. Gould and Lewis [V.14] defined three principles for a user-centred design process: early focus on users and tasks, to gather knowledge about the social, cultural, personal, and all other types of characteristics of users; empirical measurement, gained by capturing and analysing user feedback; iterative design, based on iterations after each user feedback. The iterative process of user-centred design enables approaching a final product step by step, by reducing development risks and avoiding dismissing big parts of the achieved components or results.

*User-centred design (UCD)* helps to achieve not only understanding users but also involving them throughout the whole design and development process. The consideration of human characteristics and capabilities as central factors in the design process [V.15] facilitates the creation of better accepted, really used and sustainable systems.

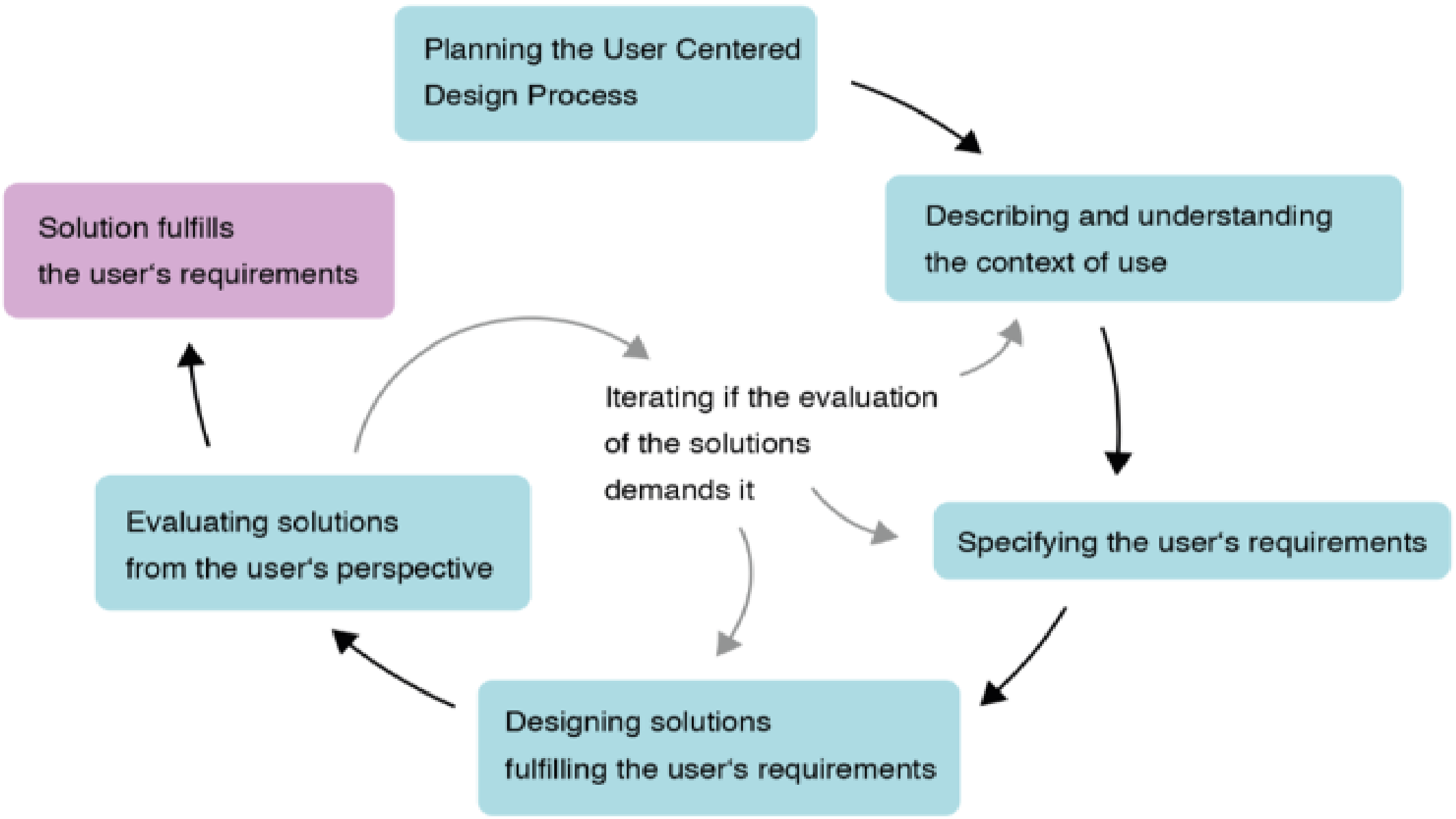

Figure 30. The iterative process of user-centred design [V.16])

### 5.1.2 Design thinking

Two decades ago *design thinking* was introduced as a cognitive process of designers [V.17], [V.18]. It was about understanding design creativity and improving design-thinking abilities. Today, design thinking is defined as "a complex thinking process of conceiving new realities, expressing the introduction of design culture and its methods into fields such as business innovation" [V.19], [V.20]. The most known design thinking models are: the 3 I Model (Inspiration, Ideation, Implementation) by IDEO [V.21], [V.22]; the HCD Model (Hearing, Creating and Delivering) again by IDEO; the model of Understand, Observe, Point of View, Ideate, Prototype and Test by Hasso-Plattner Institute [V.23]; the 4 D or Double Diamond design process model (Discover, Define, Develop, Deliver) by British Design Council (2005); the Service Design Thinking Model (Exploration, Creation, Reflection, Implementation) by Stickdorn and Schneider [V.24].

"Design thinking is a human-centred approach to innovation that draws from the designer's toolkit to integrate the needs of people, the possibilities of technology, and the requirements for business success" (Tim Brown, IDEO). The exploration of the role and potential of design thinking within organisations has changed the original objective of this research [V.25], [V.26]. So, "design thinking is not only a cognitive process or a mind-set, but has become an effective toolkit for any innovation process, connecting the creative design approach to traditional business thinking, based on planning and rational problem solving" [V.19], [V.20]. This shifted design thinking from design disciplines more and more to the fields of management and marketing.

Design thinking can be seen as a framework providing a set of methods that are used in user-centred design processes. If we take design thinking as an approach seriously and apply (all) its methods thoroughly throughout the whole design process, we can easily follow the goal of understanding the everyday practice and its actors. This would lead us furthermore to design systems that consider the context of use, user experiences and the needed technology support. Our objective in designing systems is being innovative and improving user experience. We think this can be done only by understanding the actors, their actions, their use context and of course by including them as experts into the design process.

In the user-centred design process, contextual inquiry, capturing ludic experiences and use context definition are needed to describe and understand the context of use and, furthermore, to specify the users' requirements. Later on, designing and evaluating solutions for users require the steps of interactive user-centred system and product design.

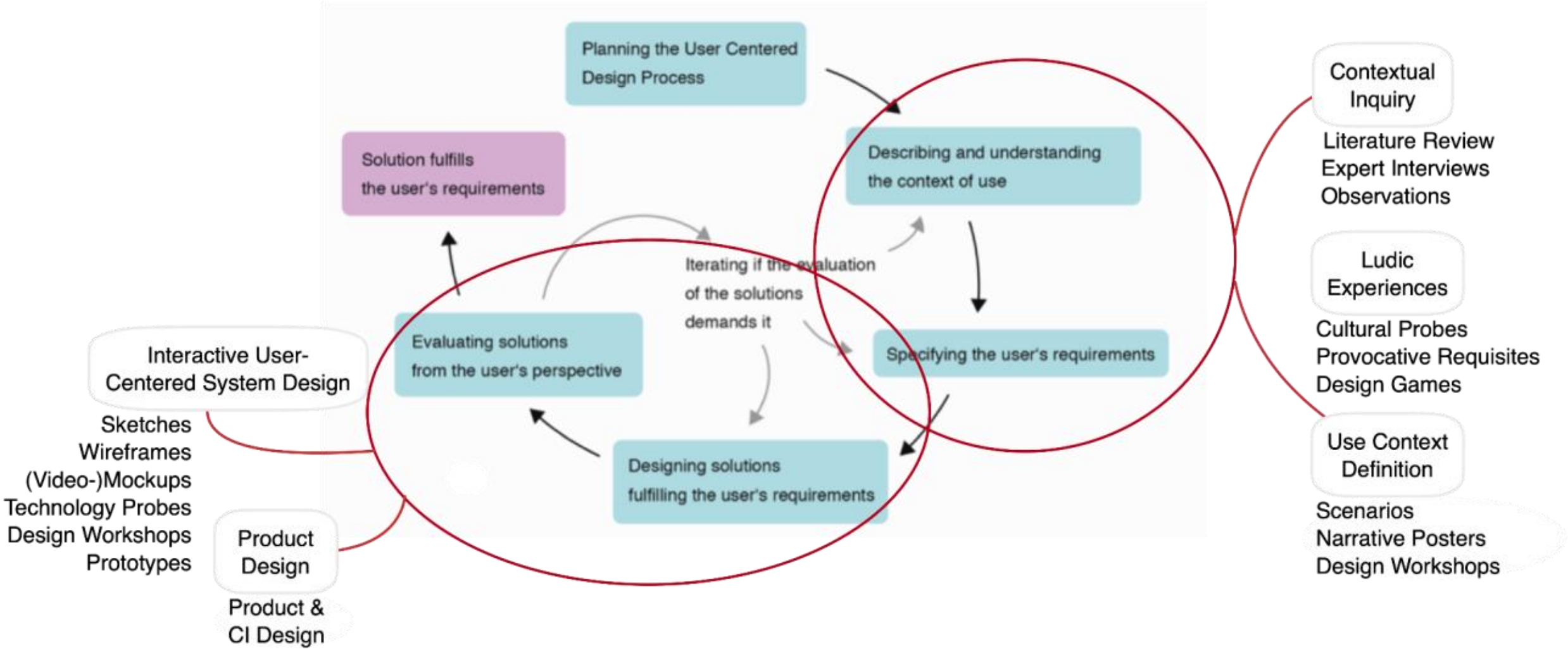

Figure 31. User-centred design with signed methods (based on [V.8]).

In this section, only such methods are presented in detail that are relevant for user involvement in the design process. These are the very first idea, observations with video analysis, cultural probes, provocative requisites, design games, scenarios, design workshops, sketches, wireframes, (video) mock-ups, prototypes, technology probes, and product and corporate identity. Others, which are mainly carried out by the design team without interacting with users directly, like literature review, interviews with experts, and narrative posters, are left out for space reasons (see Figure 31). For more details, the reader is referred to [V.8] and [V.27].

- *The very first idea* is a crucial step in a design process. There is always a driver of an idea. In most cases, the idea is vague and needs a lot of elaboration, which happens at different design stages in a project. The main method applied in that case is brainstorming, which needs to be well documented and archived. All associations even if they are not actually used in the current idea development or furthermore in the project need to be kept in a retrievable way for a later access and for a design discussion, e.g., in a design workshop, in case some options or ideas need to be opened again during the project for further consideration.

- It is not enough to imagine how a system will be used; it is necessary to really observe how the targeted users will use it. And, everyone is not the same. Tests make sure that designers understand their users and their use behaviours in a specific context with a specific piece of system or technology. *Observations with video analysis* help to understand real use scenarios. Before starting to develop solutions, the current situation needs to be captured in detail. How are people acting now? What tools are they using? Are there patterns of acting, individually or cooperatively? What types of help do people need and get? Where are the problems or gaps in handling? What can be the reasons for these problems in general and in specific situations? These questions and several more can be answered by just observing people in the environment of attention, like at a workplace, for which a design idea should be generated and further developed. This method creates a lot of video and audio material that must be analysed afterwards and probably used to create short pieces to illustrate certain situations to address in design teams. The effort needed for the afterwards work should not be underestimated, but it is worthwhile to carry out. Designers should be aware of the point of repetition of actions and occurrences, and stop observing the setting to avoid the additional unnecessary screening work.
- *Cultural probes* are very playful and multifaceted. This is the beginning of creating first ideas on colours or other aesthetic elements in the design process. These elements are mainly used in the design of the probes, but they also include several parts, which are normally kept for the further design of the system. On the other hand, designers consider important questions that might be thought of after analysing the videos and identifying the problems they want to tickle with the design. These questions and other additional data about the users and use scenarios can be captured by cultural probes in an asynchronous way [V.28], normally within at least two weeks. Users get the probes and fill in with their data, which are then handed over to the designers for further analysis. The handing over step is normally accompanied with a short interview in which designers clarify open questions or misunderstandings directly by the originator of the data. Cultural probes have to provoke inspirational responses from the target group in diverse cultural settings [V.29]. It is a way of experimental design in a responsive way. Cultural probes host personal dialogue between designers and target user groups by provoking with the probes as interventions, in both directions: the design ideas to the target user group; the way of doing or living of user groups in form of rather visual than textual material to the designers.
- *Provocative requisites* are characterised by their capacity for provocation and their experimental nature. Through ambiguity of the designed requisite, users' attention is led to certain playful and provocative statements or questions illustrated usually in a public space to make the requisite accessible for people. Designers capture the user reaction with the help of video and audio recordings, which then need to be analysed. Everything integrated into the provocative requisite has to have a relation to an aspect of the design idea, including optional aspects or questioned interaction elements connected to the design artefact. Parts of the requisite might help to evaluate some of the features of the design, probably yet at a very early stage, like layout, interaction, sound, or visual elements of the design.

- A *design game* enables exploration of imaginaries, interaction elements, and emphatic aspects of co-creation or co-design [V.30]. It is based on the assumption that designing is a social process consisting of communication, negotiation, and compromising interrelatedly. The process is as important as the final product [V.31]. User participation is prepared with certain pre-structured settings and tasks. The result of interactions is not a system or a design artefact, rather a co-created understanding of a context, user experiences, possible design ideas, and ideal situations or "dreams". Through gaming an exchange between users and designers can take place that has also a learning effect. The game elements and interaction mechanisms enable taking roles in the game and, through this, to experience other perspectives. These elements again stipulate the creativity of the game players. New innovative ideas may occur and be developed cooperatively in the game [V.32]. Design games help to create decision making situations which are very close to reality. Discussions and negotiations might occur in such situations, which are ensembled within rules, regulations, or conventions to ensure keeping the focus on the goals of the game. At the same time, all regulatory measures are loose enough to offer a free space for improvisations or creation of innovative combinations of ideas. A design game consists normally of a game board, several cards, dice(s), meeples, paper-based material, or additional digital elements. The game has to be played several times to get fruitful feedback from the players, who must be selected carefully. The games must be recorded for further detailed analysis. Design games are bridges between past experiences, current subjects, and future visions [V.33]. "To create scenarios that describe intended use situations" [V.31], design games are used by designers as tools, by game players as ways of (design) thinking, and by designers of the design game as structures with concrete design materials [V.30].
- *Scenarios* are used in UCD to support the tensions between reflection and action, as well as between typical and critical situations [V.34]. Since the last two decades, scenarios containing several use cases have become more abstract, pre-direct users' actions, and support prototyping in which users are very much involved. Later on, the role of scenarios in design processes evolved [V.35]-[V.38]: they became the "the basis for overall design, technical implementation, co-operation within design teams, and across professional boundaries, e.g., between users and designers or between usability people and technical designers and implementers" [V.20], [V.34] . The purpose of a scenario depends on two factors: on the type of situation the scenario is dealing with, and on the type of design situation that the constructors want to support. There are several reasons why scenarios should be used in design processes: use and exploration scenarios [V.38] help to present, situate, and illustrate solutions; explanation scenarios, on the other hand, help to identify potential problems. If designers want to have broad and conceptual answers, then they need open-ended scenarios. In case of gaining (more) detailed, specific answers, closed scenarios are very useful. Usually, scenarios are designed based on knowledge about typical ways of doing things. Nevertheless, it is also important to address specific, critical instances of the typical. Scenarios can provoke new ideas and are, therefore, very good in design processes. With small scenarios prototypes can be evaluated in a structured way. To evaluate prototypes which evolve vertically and horizontally during the design process, scenarios must be moved from typical ones to critical ones. Beside use cases, scenarios contain personas. Personas [V.39] represent potential target users of a system. They are created after studying the potential users intensively. Non personas show the limits of a system by identifying who would not be addressed by the design. The number of personas should be limited to primary and secondary personas. Personas, if used actively and consistently throughout the whole design process, ensure a clear communication between the design team members and focus on the intended design. Personas must include characteristics of persons they represent, their experiences, expectations, and limitations, their life or work situation which is relevant for the design, and some unique quotes showing the main design aspect formulated by the persona him/herself [V.40], [V.41]. To make the best use of scenarios in user-centred design processes, they must be attuned to the particular purposes of situations that they are to be used in, and to be very selective based on these purposes [V.12], [V.34]. To conclude, scenarios mediate the thinking and communication that takes place in design processes.

- *Design workshops* are creative meetings of design teams. They can be used at different stages of a design project. At an early stage, design ideas can be generated in a heterogeneous small group cooperatively, usually in an inspiring room populated by several artefacts and materials to encourage the participants to interact with each other or improvise and use material around to illustrate their ideas. At a later stage in a design process, such workshops can be very useful to detail the interaction with the mockup or prototype, by considering users’ requirements and abilities to use the system under development. Design workshops focus on a small number of features or properties of a system. Workshops need to be prepared and documented properly to benefit from all creative ideas that come out during the workshop, even if not all of them are really used in the current project. Design workshops are also places, in which design artefacts are generated, with different fidelity levels depending on the design progress. Design artefacts evolve through iterations while sketching, wireframing, creating (video) mockups, and prototyping.
- *Sketching* is a quick, easy, and cheap way of generating and discussing design ideas and concepts. Sketches are simple, ambiguous, inspiring, and by their nature never “finished” [V.42]. Even if they are seen as "throw away" artefacts, they can be used later on in a design project, e.g., to try out alternatives or combine different ideas. They are as such identifiable and show the form and function of the artefact or idea they represent. They are somewhat vague and a low-fidelity representation – only for the current stage in argument building. The fidelity must have the right size: too little, then the argument might be unclear; too much, then the argument might be over-done or the idea is already decided or completely worked out instead of suggesting.

- *Wireframes* are documents showing the design structures, hierarchies of information, elements of control and contents. They contain specifications, notes, meta-data, navigation, and interplay of interface elements. They play an important role in product design because they help to convey features of the product and the technological and business logics. Wireframes are blueprints of the product's functionality and can be of different types depending on their use and by whom they are used: almost a prototype of input and output interfaces and interactions [V.43], or a basic setup of interface elements without deeper functionalities. Wireframes can be reference zones, low- or high-fidelity wireframes, storyboards, standalone, or specifications, to name a few.
- *(Video) Mockups* are the first artefacts showing the look and feel of the design idea. They extend wireframes by experiential components, e.g., colours, graphics, materials, etc. without straying too far from the wireframes' definitions. Mockups are dummies which are scaled models or representations to present them to the users and other stakeholders. The input and output interfaces or interaction elements of mockups are prototypical without deeper functionalities. The main goal of mockups is to check requirements with users during the development process and to ensure decisions made so far on interaction and interface elements of the product.
- *Prototypes* are a simple and partly executable version of the product in development. "In fact, a prototype can be anything from a paper-based storyboard through a complex piece of software, from a cardboard mockup to a modelled piece of metal" [V.44]. In a prototype, the look and feel of the final product must be well-represented to make the interaction with it possible and the evaluation realistic enough for the final product development. If all interactions with the product are implemented in a prototype, then we talk about prototyping horizontally, whereas if only one specific interaction is implemented completely then these are called vertical prototypes [V.22]. Characteristics of a prototype can also be described with two terms: fidelity and resolution. Low fidelity prototypes are created quick and dirty, used mainly for early validation, enable open discussions, and require prompting. High fidelity prototypes, on the other hand, represent sharp opinions and concrete ideas, are self-explanatory, as well as well refined. Low resolution prototypes contain less details, focus on the core interaction, are created quick and dirty, and used for early validation. On the contrary, high-resolution prototypes show more details, focus on the whole, are well refined, and show concrete ideas.

- *Technology probes* help to find out the feasibility of the technology chosen for the final product. They might be a complex or simple implementation of a future technology that is planned to be used in the final product. Technology probes "... are a particular type of probe that combine the social science goal of collecting information about the use and the users of technology in a real-world setting, the engineering goal of field-testing the technology, and the design goal of inspiring users and designers to think of new kinds of technologies to support their needs and desires" [V.45]. So, they offer means of gathering and testing of interactions, technologies, and user feedback. They can be deployed in real use environments. They deliver real data (e.g., through logging) for further developments in the project [V.46]. This way they play a very relevant role in the decision making about the technology and interactions of the prototype and later on of the product.
- At the end of a user-centred design process, the *product and corporate identity* need to be finalized. The product definition is concerned with illustrating the product vision, including all of its interactions, physical properties, surrounding systems, marketing strategy, target groups, and target markets. It shows the product as a whole (e.g., purpose, cost, usage, handling, maintenance, etc.) and the environment it is embedded in. It is defined in a clear product language following a defined product vision, corporate design, communication, behaviour, and identity, containing aesthetics, target group features, logo, materials, colour scheme, etc. [V.20], [V.47].

### 5.1.3 Conclusions

In this section, we described user-centred design as a dynamic multidimensional process utilised by several design thinking methods and artefacts. The whole design process is an iterative circle of intertwined factors, namely of people (users, designers, other stakeholders), particular design phases, and artefacts as intermediaries or final results to represent certain design aspects and parameters. The iteration of a UCD process is accompanied with user studies for design and for evaluation, which methodologically needs different approaches in each design phase. In UCD projects, usability studies have to be seen as integral parts of design processes. This makes usability studies to activities that are responsible for the product and its future use [V.34].

Methods we presented in this section show how a UCD process can be established and how a design process can evolve from the very beginning until the definition and presentation of the product design. We showed the relevance and use of single methods by stressing out their strengths and weaknesses. There is no strict rule saying that all the methods presented here should be used in any type of design projects. Each project is unique and designers have to select the most suitable methods for their particular projects. This section only helps to show the possible and useful ways of doing user-centred design.

## 5.2 User acceptance and ethical considerations

The use of cameras and microphones is integral to AAL. AAL technologies are increasingly presented and sold as essential smart additions to daily life and home environments that will radically transform the healthcare and wellness markets of the future. Yet, some end users, as well as some professionals and caregivers, continue to justifiably perceive these devices as intrusive; they pose a risk to privacy, psychological well-being, and social behaviour of those exposed to and associated with AAL applications, potentially impacting current and future generations. An ethical approach and a thorough understanding of all ethics in surveillance/monitoring architectures are therefore pressing. AAL poses many ethical challenges raising questions which will affect immediate acceptance and long-term usage. Data monitoring, storage, and sharing have implications on privacy, consent, and respect for personal autonomy. Furthermore, ethical issues emerge from social inequalities and their potential exacerbation by AAL, accentuating the existing access gap between high-income countries and low and middle-income countries. Ethics should be incorporated at the AAL design stage taking all of these aspects into account and evaluating (i) beneficence, (ii) non-maleficence i.e., a risk/benefit analysis (iii) respect for autonomy, and (iv) protection of confidential information and data that may reveal personal and sensitive attributes.

Legal aspects mainly refer to the adherence to existing legal frameworks and cover issues related to product safety, data protection, cybersecurity, intellectual property, and access to data by public, private, and government bodies. The scope of General Data Protection Regulation (GDPR), its application and requirements are broad and are not directly concerned with ethics. The principle of data protection by design is recognized as a binding legal requirement, though it remains challenging to implement due to its context-dependent and evolving nature. Successful privacy-friendly AAL applications are needed, as the pressure to bring IoT and AI solutions quickly to market cannot overlook the fact that the environments in which AAL will operate are mostly private (e.g., the home). Another area pertinent to AAL is the Medical Device Regulation concerning health-related solutions. Other legal requirements exist and future regulation will emerge as technology advances.

The social issues focus on the impact of AAL technologies before and after their adoption. Some are rooted in the collective understandings of the technology at hand, whereby users can relate audio-video based AAL to activities such as surveillance practices. One of the prominent social and design challenges will be facilitation of the workflow and avoiding the sense of additional technological burden. Taking care of that will directly impact institutional and individual adoption of AAL. Future AAL technologies need to consider all aspects of equality such as gender, race, age and social disadvantages and avoid increasing loneliness and isolation among, e.g., older and frail people. Finally, the current power asymmetries between the target and general populations should not be underestimated nor should the discrepant needs and motivations of the target group and those developing and deploying AAL systems.

These differences could lead to governance challenges, serious ethical questions, and potential misuse of the technology.

Whilst AAL technologies provide promising solutions for the health and social care challenges, they are not exempt from ethical, legal and social issues (ELSI). A set of ELSI guidelines is needed to integrate these factors at the research and development stage.

Many studies that focus on developing video technology in the healthcare context show its positive effects on its usability and effectiveness [V.48]-[V.54]. The population demographics point out a rapid increase in people older than 65 years in the global context. Ageing is a process that leads to gradual loss of function, fragility and development of age-related diseases [V.55]. In contrast to the rapid growth of older people, the number of young people who can work in health and care activities is declining [V.56]-[V.58]. The researchers try to solve the current and future situation through interdisciplinary and interprofessional collaboration across the countries' borders in various intelligent technological solutions. These solutions focus on offering the elderly the opportunity to live at home as long as possible independently with the help of other people. Other payrolls focus on conducting medical observations of a person's state of health through various audiovisual technological aids [V.56]-[V.62]. The ageing population and the emergence of digital opportunities in health care also require systematic risk analysis in medical ethics. Therefore, we need to take a step back and refresh our knowledge of what medical ethics stands for.

Let us go through the four fundamental ethical principles. The four main principles of medical ethics are autonomy, justice, beneficence, and non-maleficence. Autonomy requires that care recipients receive complete care information, including risks, to take an active party to participate on their terms and make their own decisions without pressure. The principle of justice focuses on the equitable distribution of resources. The principle of charity is about doing well, and the principle of non-maleficence means not harming [V.63]-[V.66]. The four medical ethical principles have been developed to guide researchers involved in research in various ways. These principles were primarily intended for the medical profession. However, today, the research that involves people has become an arena for various scientific disciplines and professions. The involvement of various scientific disciplines in human research is necessary to find solutions to global ageing challenges [V.55],[V.67]. For instance, these principles are supplemented with an ethics framework related to digital solutions in the health and care arena.

The additional principle that has been added is the principle of explicability [V.68]-[V.71]. The principle of explicability is an umbrella concept that includes transparency and accountability. Transparency means that if the digital system fails, it must be possible to find out the underlying causes. This means that information on the development of algorithms may be available to ensure its generalizability in various health and care settings [V.69]. Accountability is about tracking who is responsible if any digital system makes a mistake [V.68]. The researchers' responsibility is to problematize possible risks when the ethical principles can be violated on a visible or invisible level. If we start from the principle of autonomy, interdisciplinary, cross-sectoral, national and international research actors have to find a common linguistic nomenclature to understand each other in terms of ethical principles. This applies to the eyesight of co-researchers if there are non-biomedical scientific disciplines. Another aspect is to work proactively that other essential stakeholders such as decision-makers also incorporate ethical principles in the policy document. The good intention with facial expression recognition is to identify how the patient is feeling and have it as a basis for various medical or nursing measures to minimize suffering. At the same time, it is crucial to problematize several risks when using face expression recognition and its purpose. The following questions must be answered: what is the indication to use face recognition, has the person whose facial accident is registered given his consent or his guardian, face recognition the only way to collect information about the person's health condition, which principal is behind the decision to initiate face recognition, how data related to facial expressions is transferred from the patent to the healthcare staff, there is a risk that data has become available to unauthorised persons, where data is used, who has access to the data and data can be misused or sold or in any way offend a person whose facial expressions are registered. Furthermore, another critical thing to problematize is how the available product for identifying and interpreting approach expressions is user friendly, and this maintenance primarily because of the updates. There is a risk that software will not be used due to these circumstances.

What happens if this service becomes available to patients in the form of apps that they can download themselves? First, there is a risk that when closing the app, many tracking services and involuntary monitoring are connected, which the patient has not approved. Almeida's research [V.72] has discovered significant monitoring problems when examining popular apps that users usually report about their health. Those apps were activated automatically, tracking services that are secretly installed without informing users. In that way, many apps are violating the General Data Protection Regulation (GDPR).

## 5.3 Privacy preservation

Recently, there has been far-reaching advancements in the development of video-based devices with improved processing capabilities, heightened quality, wireless data transfer, and increased interoperability with Internet of Things (IoT) devices. Computer vision gives the possibility to monitor an environment and report on visual information, which is commonly the most straightforward and human-like way of describing an event, a person, an object, interactions and actions. Therefore, cameras can offer more intelligent solutions for AAL but they may be considered intrusive by some end users.

Next, this Section provides some insights on privacy preservation in video data. A more detailed state of the art has also been published by GoodBrother Working Group 2 on privacy-by-design in audio and video data[20].

### 5.3.1 Privacy by design

In Europe, all the technologies that process personal data are governed by the General Data Protection Regulation (GDPR). According to the GDPR, the processing of personal data must adhere to fundamental processing principles such as lawfulness, transparency, fairness, data minimization, and purpose limitation (Article 5). Data protection by design (DPbD), as enshrined as a legal norm in Article 25 of the GDPR, requires that audio-and video-based technologies for AAL are conceived with the fundamental principles set out in the GDPR in mind. Broadly speaking, DPbD aims to implement all of the principles of data protection law through technical and organizational measures that are embedded ex-ante and throughout the lifecycle of the system [V.73] [V.74]. Article 25 of the GDPR states that data controllers must implement "appropriate technical and organisational measures, such as pseudonymisation, which are designed to implement data-protection principles, such as data minimisation, in an effective manner and to integrate the necessary safeguards into the processing in order to meet the requirements of this Regulation and protect the rights of data subjects." Moreover, the principle of data protection by default (DPbDf) (para. 2) requires that these measures ensure that "only personal data which are necessary for each specific purpose of the processing are processed."

[V.75] presented a methodology that separates DPbD into levels. In this methodology, it is key to consider the specific design elements which include: sensors, the models, the system, the user interface, and the user, discussed in more detail below. DPbD techniques should be implemented in order to meet legal requirements.

At the **sensor level**, privacy preservation techniques may prevent the capture of sensitive data in visual feeds using various software and hardware implements. These mechanisms can prevent the capture of sensitive content in the first place by the camera. This can also be implemented at the software level, as a filter to clear the captured images of protected content before the images are stored to disk.

At the **model level**, methods are created that preserve privacy for users while at the same time enabling models to infer information from data. Privacy-preserving data mining (PPDM) is emerging as an important set of techniques that can be used to maximize the benefits of large datasets while, at the same time, minimize the privacy impacts.

[20] https://goodbrother.eu/deliverables (last accessed: 07/03/2022)

For **system level** privacy preservation, it is required that the data used in the pipeline becomes secure, and that user consent for the use of the data in the pipeline is traceable. Traceability requires two components. The first is that personal data can be traced to when user consent for its use was recorded. Secondly, the flow of the data to various sources should also be traceable. This is essential because withdrawal of consent is an important facet of privacy laws like the GDPR; upon withdrawal of consent, actions have to be taken by the authorized administrator to comply with the request.

At the **user-interface level**, the functionality of an AAL application should be adjusted to the competence of the user in order to increase the transparency of data processing and to strengthen data privacy. ENISA suggests that privacy-friendly default configurations and settings, intelligible to this population should be used [V.76].

**User level** privacy measures empower users by helping them manage their data. These also help users understand the privacy risks involved with the sharing of their data, and also give them mechanisms with which they can control the disclosure of their data.

### 5.3.2 Visual privacy protection methods

Visual privacy protection methods can be classified under 5 major categories [V.77]. These categories, built upon the taxonomy developed by Padilla Lopez et al. [V.78] are presented in Figure 32.

**Intervention methods** interfere during data collection, limiting the amount of private visual information that can be collected from the environment. These methods physically intervene camera devices to prevent the acquisition of an image by means of a specialised device that interferes with the camera sensor. Intervention methods can be classified into sensor saturation, broadcasting commands, and context-based approaches [V.79]. Sensor saturation provides privacy by feeding the acquisition sensor with a saturating signal greater than the maximum input the sensor supports. This can be obtained by physical intervention, for instance stickers or webcam covers usually employed in laptops. Other methods locate retro-reflective CCD or CMOS in the vicinity and directs pulsing lights to distort the recorded images. Some devices are able to broadcast commands to disable input devices in the surroundings. For instance, a camera with inbuilt facial recognition might bur sensitive areas after receiving specific commands [V.80]. In context-based approaches, the input sensor is intervened based on context recognition, for instance, a specific location [V.81].

Algorithms under the **blind vision** category rely on secure multiparty computation (SMC) techniques from cryptography and which allow computer vision tasks to be executed without compromising on the privacy of the data used for computation, nor of the algorithm itself.

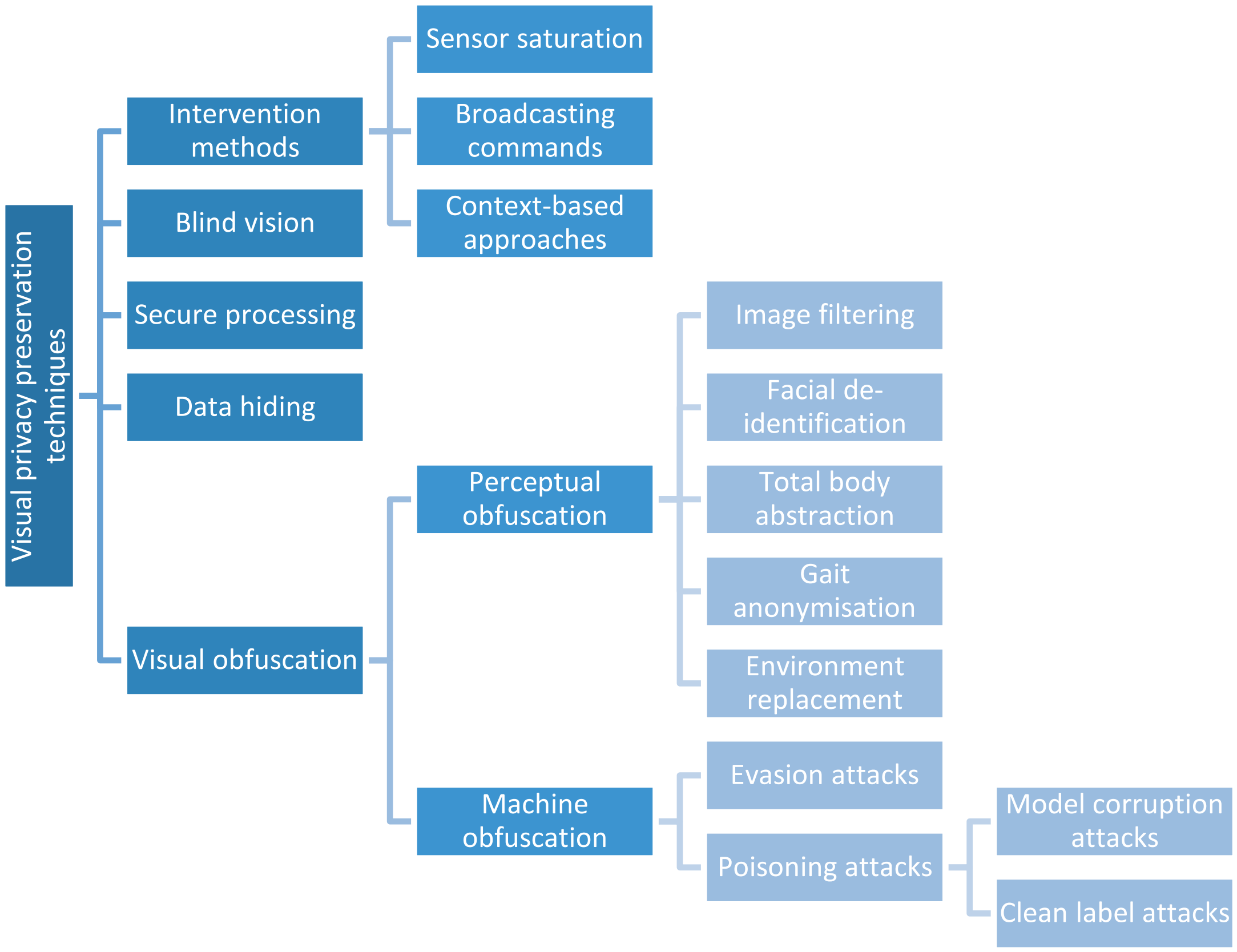

Figure 32.  A taxonomy of visual privacy preservation techniques for AAL (adapted from [V.77])

**Secure processing** methods allow processing private data in a unidirectional manner. This implies that the database queried would be a public one, but the query and its results are kept private. Secure processing algorithms do not rely on SMC techniques.

**Data hiding** methods refer to reversible privacy preserving algorithms where the information required for reversal to the original is stored in the original image itself, using techniques such as steganography, digital watermarking, and fingerprinting.

In **visual obfuscation**, different techniques are employed to hide sensitive visual information. These techniques can be divided into perceptual obfuscation and machine obfuscation.

In **perceptual obfuscation**, the aim is to provide visual privacy in order so that only human observers with appropriate privileges could access the visual feeds. The most common technique is the use of **image filters**, such as blurring or pixelating entire images or only sensitive areas. Other filters are morphing and warping that are usually employed in face anonymisation. Some more recent techniques involve changing the color of each pixel in an image using a palette of false colours. An interesting technique is cartooning, in which privacy-sensitive objects in images are replaced by abstract cartoon clip art.

**Facial de-identification** methods alter faces in an image by generating new ones and blending them into the original image. Traditional methods relied on the use of the *k*-same family of algorithms. Recent methods employ Generative Adversarial Networks (GANs) to generate realistic looking faces with similar expressions to those appearing in the original image.

**Total body abstraction** methods aim to provide privacy by replacing the entire body of a person appearing in an image. Most methods use semantic segmentation to extract humans in images, and then replace them with abstractions such as avatars. Other techniques use GANs to generate full-body replacements. Some methods also modify images so that sensitive objects and persons are removed. Inpainting methods are employed to substitute the removed person with a generated realistic background.

Gait is an important cue to identify individuals appearing in a video. There have been recently some promising techniques to generate **anonymise gaits** by using deep learning techniques, mainly GANs.

In **machine obfuscation**, the aim is to protect privacy from machine learning algorithms that may extract relevant knowledge by processing large datasets. There are two groups of techniques. In **poisoning attacks**, the objective is to disrupt the training of machine learning models through the introduction of "poisoned" images in the training set so that the models missclasify specific images, for instance those containing a person. In evasion attacks, the images that are going to be processed are transformed, by adding noise or removing specific areas, so that they are misclassified or people are misidentified.

### 5.3.3 Secure transmission and storage of video data

In order to ensure a high level of confidentiality, anonymity and privacy, a particular attention has to be paid to providing efficient or reliable methods for secure storage and transmission of multimedia data obtained and used by AAL applications. In this context, local storage systems and cloud infrastructures must provide an adequate information security at formal, informal and technical levels. Additionally, data security and integrity within the entire network infrastructure including wireless and wired communication technologies as well as local, metropolitan and wide area networks must be considered.

The rapid development of video technologies, increase in data rates and processing speeds, wide use of the Internet and cloud computing as well as highly efficient video compression methods have made video encryption even more challenging. The traditional ciphers such as the Data Encryption Standard (DES) and the Advanced Encryption Standard (AES) are mostly optimized and applied for encrypting text and binary data. Due to the fact that video streams contain a huge amount of data that need to be processed and transmitted almost in real-time, the traditional approaches are not very well suited for multimedia data. The use of these methods causes a large overhead and requires significant processing resources, which are usually not provided by energy-efficient, low-performance end devices or are very expensive to realize.

AAL applications rely on the IoT network architecture, which defines a simple skeleton for the network. Transferring a large amount of data between different nodes can cause security issues. The examples of attacks that may take place in the network layer are: the man-in-the-middle attack, the sniffing attack, the DoS/DDos attack, the hello flood attack, the wormhole attack, the Sybil attack, the sinkhole attack and the traffic analysis attack. Other attacks can occur at the application layer, for instance, deep fake video and facial manipulation in multimedia data, or buffer overflow, code injection, phishing attack and eavesdroppingin non-multimedia data.

An efficient and robust encryption of multimedia data containing video together with using efficient compression methods are important prerequisites in achieving secure and efficient video transmission and storage. These methods include protocols and algorithms for authentication and access control, cryptography, privacy preservation, as well as increasing the security at the device-to-device, i.e., end-to-end communication level. The main targets of these protocols and algorithms are on enhancing security and integrity of network communication and data storage, peer discovery, proximity services, and location privacy. However, a number of potential issues linked to diversity of interconnected devices, limitation of available resources and solution deployability must be taken into account when selecting the most appropriate method.

## 5.4 The need for transparency in data processing: Explainable AI

AI performs outstandingly in complex tasks. AI systems are being used in a wide array of applications, healthcare being one of them. But in some cases, increased performances come with increased complexity, creating “black-box” models whose inner workings are not easily understood. This has led to a renewed interest in explainable AI, or XAI.

The field of explainable AI focuses on the understanding and explanation of AI systems. In particularly sensitive domains, such as healthcare, understanding why a decision or prediction is made is paramount in ensuring the safety and fair treatments of users. As such, XAI has been the subject of several surveys in recent years. The work in [V.82] offers an extensive survey of XAI methods, as well as different ways of classifying those methods. Another work [V.83] also catalogues popular XAI methods, classifying them between Intrinsically Interpretable Methods, Model Agnostics Methods and Example-Based Explanations, and offers an illustration with a common case study. The work in [V.84] takes a slightly different approach for its classification (Perceived interpretability, Interpretability via Mathematical Structure and Others) and applies those categories to the medical field.

We first give a brief overview of different XAI methods that are used in AAL, then present AAL applications in which XAI is used.

### 5.4.1 Explainable AI Methods

#### Inherently Explainable models.

Inherently explainable models, or “white-box” models produce results that do not require additional explanations to be understood. This is an advantage in the domain of interpretability, but often comes with a cost in performance due to the complexity of modern applications

Linear regression outputs a predicted target as a weighted sum of the input features. The learned weights can then be used as a way of estimating the importance of each feature. Logistic regression extends this concept to classification problems by outputting a probability between 0 and 1. These simpler models are rarely used on their own in medical or AAL application.

Decision Tree models are another intrinsically transparent type of model, which can be used even when the relation between input and output is not linear. Simply following the path from the root to the leaf node (decision path) explains how the decision has been made. However, they can become harder to interpret when they have more depth. The fact that several trees can exist for the same problem and that a slight change in input can make a big change in output can also come in the way of interpretability.

Rule-based learners generate rules to characterise the data they are learning from. Simple Boolean rules of the IF-THEN-ELSE type are considered to be inherently interpretable. These rules can be generated via Deep Learning, but once the model has been created the decisions are rather straightforward to interpret. However, interpretability can degrade when the number of features increases.

#### Model Agnostic Explanation

Model Agnostic Explanations are separated from the machine learning model, allowing them to be used with any black-box model. They can either provide explanations for a single prediction (local methods) or for the behaviour of the entire model (global methods). Most of them do so by evaluating the relative importance or influence of the input features.

Partial Dependence Plots (PDP) [V.85] is a global explanation method which shows the average effect of one or two features on the predicted outcome of a model. However, it assumes that the features are not correlated to each other, which is often untrue and leads to an inaccurate approximation of the relationship between the feature and the output. Individual Conditional Expectation (ICE) [V.86] builds on PDP by plotting for each instance the variation in output against the variation in the feature of interest. The average of all lines in an ICE curve gives the PDP.

Feature Importance is another global method which represents the increase in the prediction error when we change the value of a feature. Analysis of variance (ANOVA) is a univariate feature test which works by comparing each feature to the target variable, to see whether there is any statistically significant relationship between them.

*Shapley Additive Explanation* (SHAP) is a method based on game theory which aims at evaluating the influence of each feature on one final prediction. Each feature can be seen as a "player" in a game whose Shapley value expresses their contribution to the total gain (i.e., the prediction). By computing how much the total outcome changes when the feature is excluded in the prediction, and repeating the process for all possible coalition of features before averaging the result, we can obtain the Shapley value of the feature of interest for this prediction. This value comes with the desirable properties of total accuracy, lack and consistency, but can be time and resource consuming.

*Local Interpretable Model-agnostic Explanation* (LIME) [V.87] is an interpretability technique which relies on creating an approximation of the model around a particular prediction. In order to do so, new samples are randomly generated around the input of interest and the original model is used to make new predictions. Those are then used to train a new, inherently interpretable model (for instance linear regression or decision tree). This does not offer a global explanation of the model but helps interpret one particular prediction

#### Explanation techniques for CNN

As one of the most effective Deep learning techniques for computer vision, Convolutional Neural Networks (CNN) have been used more and more frequently in Deep Learning. This, combined with their black-box nature, has led to the need of explainability methods specific to them. Class Activation Maps (CAM) [V.88] and their generalisation Grad-CAM [V.89] are visualisation techniques which can be used on CNN. By using the final convolution layer, it can produce a heat-map (i.e., highlight part of the input image) indicating which parts of the image participate the most in the classification.

### 5.4.2 Explainable AI in AAL

#### Activity Recognition

The work in [V.90] aims at recognizing ADLs in post-stroke patients. In order to differentiate between 19 ADLs (brushing teeth, vacuuming, washing dishes, etc.), 11 post-stroke patients are equipped with 5 wearable IMUs: one on each upper-arm, one on each wrist and one the hip. After testing several machine learning models (DT, RF, SVM, and eXtreme Gradient Boosting), eXtreme Gradient Boosting (XGBoost) is selected because of its superior performance based on AUC. SHAP is then used to determine which sensors are overall most instrumental in the classification task. This highlights that hip and wrists sensors participate more than upper-arm sensors in the final prediction.

The use of activity recognition for diagnosis is also the basis of [V.91], where a framework is proposed for the detection of cognitive decline in smart-home based on abnormal behaviour and locomotion patterns. Using the smart-home sensors (passive infra-red sensors, motion sensors, power sensors, etc.), activities and locomotion patterns are detected. Once this is done, a decision tree regression model (C4.5) is used in order to compute an anomaly level and an anomaly score. Additionally, a natural language explanation is produced for each anomaly level and anomaly score. This work also evaluates the quality of the explainability with clinicians, finding that 87% of participants either agree or strongly agree with the statements "The system with explanations would help me providing a more accurate diagnosis than the one I would provide without the use of the tool." and "The explanations would help me deciding whether the prediction of the system is correct."

Activity recognition is also used for the assessment of gross-motor skills in [V.92], where the video input is first pre-processed with OpenPose in order to extract skeletal key points before going through a CNN for activity classification. Grad-CAM is then used to colour areas of the skeletal image depending on how much they contribute to the classification. This visual explanation is assessed by experts in gross-motor skills measurement and evaluation who rate whether they agree with the reason for classification. This allows the authors to choose between several different CNNs.

The work in [V.93] proposes an approach for explainable activity recognition in the context of smart homes using environmental and wearable sensors. In order to identify ADLs, two classifiers (Random Forest and Support Vector Machine) are trained and the feature importance is computed for each of them. Additionally, each individual prediction comes with feature relevance (local explanation): a feature is relevant in a prediction if it's both important (in the classifier) and has a high value in the input. The feature relevance is presented to the user through a generated semantic explanation.

### Gait Analysis and Fall Detection

The work in [V.94] tackles the issue of explainable gait analysis through the identification of patients with sarcopenia using gait parameters. Sensors are placed on the insole of both feet of 10 patients with sarcopenia and 10 healthy subjects. Together with prior medical knowledge, these sensors are used to extract gait parameters (Cadence, stride length, step time, etc.), gait phases and descriptive statistical parameters. Once this is done, SHAP is used with XGBoost to determine which features are the most important. Once this is done, different machine learning models are tested (RF, SVM, Multi-Layer Perceptron, CNN and BiLSTM) and their performances are measured via accuracy. The performance is improved when using features identified as important by SHAP. Here, the importance of the feature is not used to generate natural language explanation but to have feature selection which improves accuracy.

In [V.95], the authors focus on explainable fall detection through video analysis. The OpenPoselibrary is used to compute body parts coordinates from which 1120 features are extracted. Feature importance is then computed via univariate analysis on an ensemble of trees to determine which body parts play the most important role on fall detection. Through their tests on the Le2i dataset and the URFD dataset, they are able to identify the mean acceleration of the ankle, along with neck mean speed and nose displacement and acceleration as being among the top 5 most important features.

### Remote monitoring

The two works [V.96] and [V.97] define and implement a framework for remote monitoring of patients with Chronic obstructive pulmonary disease (COPD) using both wearable devices( including ECG patch with movement monitoring, pulse meter, weight scale, sphygmomanometer for blood pressure monitoring ) and environment sensors (temperature, atmospheric pressure, CO2 measurement, weather station, etc.). This system classifies breathing performance as either High or Low using understandable if-then-else rules. The model is generated by a Switching Neural Network (SNN).

The work in [V.98] proposes an architecture to monitor patients' sleep using piezo-electric sensors and passive infra-red sensors in order to detect sleep disturbance related to COVID-19. While the complete framework isn't implemented, it does include an explanation phase using SHAP and LIME, as they are two of the most popular methods for explanations of black-box models.

**Risk Assessment**

Risk assessment in an Intensive Care Unit (ICU) can make the difference in whether a patient lives or dies. With this in mind, [V.99] aims at auditioning a black-box clinical risk calculator using several model-agnostic interpretability techniques. The trained model is a combination of the two predicted probabilities from the XGBoost and LightGBM models, but the interpretability techniques (Permutation Variable Importance, Partial Dependence Plot, Individual Conditional Expectation Plot and Shapley Values) can be used on any model. These different plots complement each other: PVI identifies the Glasgow coma score and the age of the patient as the two most important feature, PDP and ICE then helps visualize the evolution of mortality risk against these features while Shapley Values can exemplifies how the model operates differently depending on each patient profile.

The work in [V.100] is a retrospective study on early recognition of clinical deterioration in Brazilian hospitals. The vital signs (temperature, oxygen saturation, respiratory rate, blood glucose level and blood pressure) and other pertinent features (biological sex, age, ward, department, and length of hospital stay) are used to train several machine learning models (Naive Bayes Theorem, Logistic regression, Decision Tree, Ensemble methods, Random Forest, Gradient Boosting) to predict the patient's mortality during hospitalisation. This result is compared to the more widely-used indicators Modified Early Warning Score (MEWS) and National Early Warning Score (NEWS). Permutation Importance is then applied in order to generate explanation about why an alert was raised.

**ECG interpretation**

The interpretation of ECG is a difficult task with important repercussions on the treatment and prevention of cardiovascular diseases. ECG can also be used as an input in overall patient monitoring (as in [V.96] and [V.97]), hence the attention this area has been getting. The work in [V.101] studies the technical feasibility and practical usefulness of visual explanations for ECG classifiers. After training a CNN and a KNN model on the MIT-BIH arrhythmia dataset, the authors then apply LIME, SHAP and PSI in order to identify the most relevant parts of the ECG. The pertinence of the explanations is measured through numerical scores (Jaccard Index, performance decrease) and a user study involving three ECG readers.

The work in [V.102] uses the detection of Atrial Fibrillation from ECG as a case study for a post-hoc explainability framework of classification of biomedical time series. A lightweight CNN is trained on this problem with the 2017 PhysioNet Challenge dataset, after which the behaviour of the network is analysed through both global (Feature Importance as determined through ablation and permutation studies) and local (LIME and GradCAM) methods.

The approach taken by [V.103] is slightly different, as it aims at an automated interpretation of ECG to allow a lay person to detect QT-prolongation. A rule-based algorithm was developed to classify the QT-interval as 'normal' or 'abnormal'. This human-defined white-box model is then compared with an C4.5 decision tree classifier automatically generated through statistical machine learning.

## 5.5 Communication and network technologies for AAL

As the modern society is gradually getting older, the use of efficient and secure ambient and assisted living (AAL) systems become more and more important. Thus, assistance systems and medical care for the elderly have gained a particular interest of researchers, industry and the government. The overall framework of an AAL system comprises a functional loop consisting of four main phases: (i) observe: collecting information, signals and commands associated with individuals and their environment, (ii) contextualize: process, analyse and understand the obtained data to gather the contextual information, (iii) decide: according to the obtained contextual information the system can decide what action to perform, and (iv) act: the system act by either informing the user about the decision or performing a particular action. All these phases require different technologies and components and the information gathered must be exchanged between the system's components in an efficient and secure way. For this purpose, various communication technologies can be used. In this section, we will review different communication and network technologies by means of requirements set by different applications and their suitability for supporting assistance systems and medical care applications as well as other audio- and video-based AAL applications, especially those matching the needs of fragile and elderly people.

### 5.5.1 Communication Technologies for AAL Applications

AAL applies the information and communication technology to develop and provide components, systems, and services for improving the quality of life and providing a secure environment to older and fragile people. In this context, communication technologies play a significant role in achieving these goals, since all components of an AAL system have to be interconnected by a reliable, secure, efficient and high-performance communication network.

A variety of communication technologies have already been used or proposed for use in IoT and Smart Home applications [V.104]-[V.107]. Not all of them are well suited for application in AAL system. Especially video- and audio-based AAL systems demand high performance and security levels for data transmission and storage [V.104], [V.108], [V.109]. Figure 33 shows an overview of different communication technologies, protocols and transmission media as well as the main characteristics and methods for achieving a high level of security, privacy and dependability.

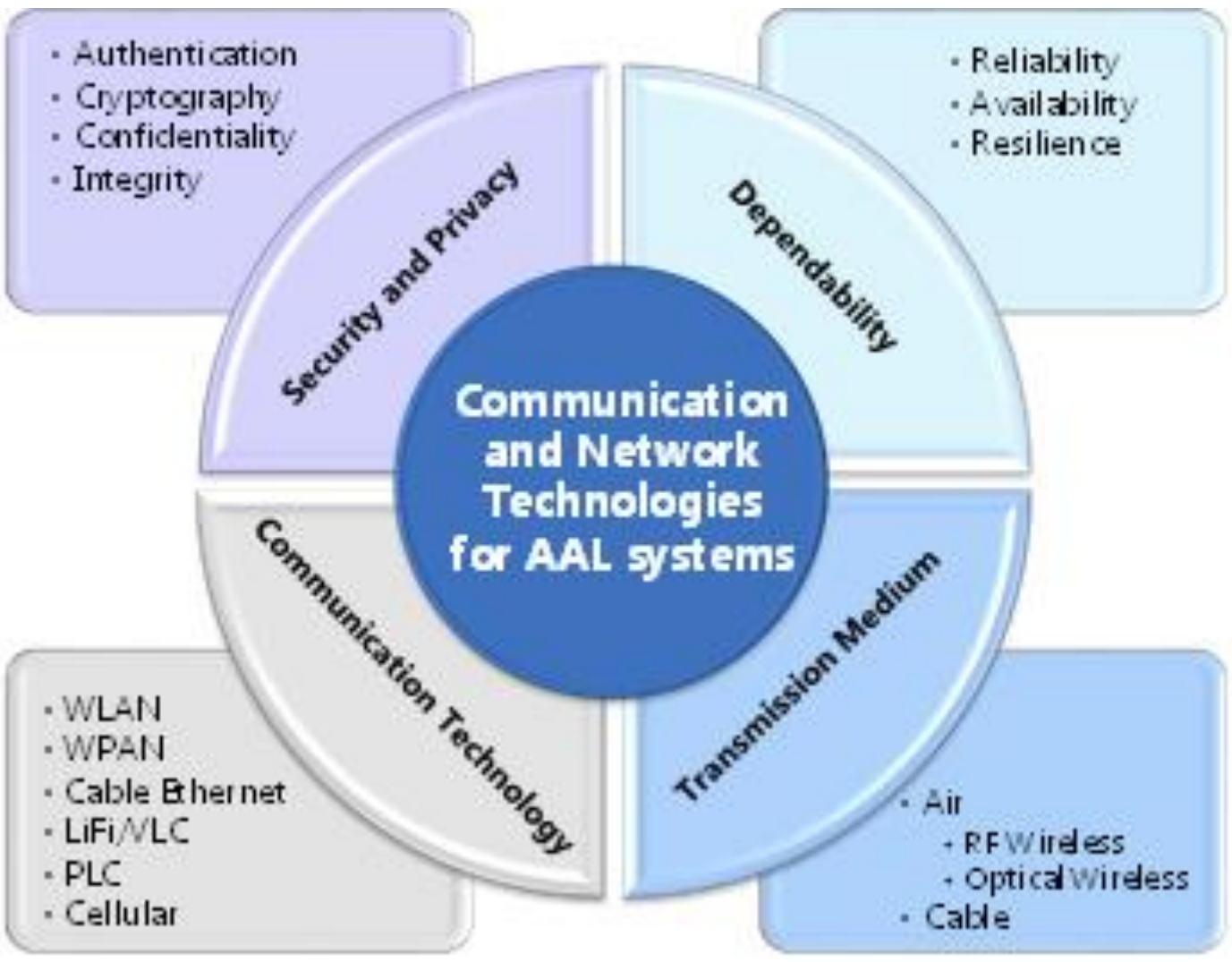

Figure 33. The main characteristics and requirements of communication and network technologies for AAL systems.

As can be seen from Table 8, different communication technologies provide different data rates and transmission ranges. While wireless personal area network (WPAN) technologies such as those specified in IEEE 802.15.4 support only relatively low data rates up to maximum 250 kbit/s and short ranges up to 100 m, the point-to-point Ethernet connections over a fibre-based optical cable can easily provide up to 10 Gbit/s data rate over a distance of several km or even up to 40 km.

WPAN technologies have primarily been developed with the goal of connecting a certain number of sensors or other end devices that have lower data rate and range requirements to each other or to a central aggregation point. Therefore, they are usually not very well suited for transmitting a large amount of data within short periods of time, which is mostly the case in video-based applications. There was an effort to develop and standardize the ultra-wideband (UWB) WPAN for the use in applications requiring high data rates over short distances. UWB WPAN should support a data rate of more than 100 Mbit/s over a few meters. However, even though the specification of this technology was initiated within the IEEE 802.15.3a group, it has never been finished. New specification of the Bluetooth technology provides a data rate of up to 2 Mbit/s over short distances. The reach can be increased to even 240 m, but in this case the guaranteed data rate is reduced to about 125 kbit/s, which is sufficient for interconnecting tiny sensors, but not end devices generating video streams.

WLAN or WiFi has a very wide penetration and is well suited for interconnecting various end devices within smart home environments. New generations of the WiFi standards such as IEEE 802.11ac and IEEE 802.11ax are able to provide high data rates of up to 1 Gbit/s. However, all radio frequency (RF) based wireless transmission technologies suffer from several limitations effects such as fading, limited and dynamically varying bandwidth and interference, which significantly degrade their performance in a practical system. Additionally, due to the fact that wireless transmission systems use a shared medium and the signal is usually emitted in all directions and can penetrate through walls, wireless systems are inherently insecure and must be secured by using sophisticated security and cryptographic methods. Both WPAN and WLAN use current and standard encryption methods mainly based on the AES-128 block cipher. Instead of or in combination with RF based wireless one can use an optical wireless technology such as visible light communication (VLC) or LiFi, which use the light generated by light emitting diodes (LEDs) as a carrier for the transmission of information. Since LEDs are currently used for lighting and widely deployed, it could be a convenient and efficient method to transmit video and audio data via the optical channel. However, even though very high data rates in the extent of 1 Gbit/s and above have been demonstrated using VLC and LiFi systems, the major impediments of a wide practical use of this technology are the relatively rapid drop-in data rate with the increase in distance and the sensitivity to sensitivity to noise from ambient light.

Another option to transmit video and audio data is to use a cable-based communication technology such as power line communication or cable based Ethernet. Both technologies provide high data rates over sufficiently long distances, but with limited mobility.

Finally, one can decide to transmit data of end devices directly via the cellular network. In this case, the end devices must be equipped with a 2G, 3G, 4G or 5G modem and a SIM card or an embedded SIM (eSIM).

Table 8. Characteristics of various communication technologies and protocols for the use in AAL applications

| Communication technology | Data rate | Transmission medium | Range | Encryption |
|---|---|---|---|---|
| WPAN (IEEE 802.15.4) | < 250 kbit/s | Air (RF spectrum) | Typ. < 100 m | AES-128 |
| WLAN (WiFi) | < 1 Gbit/s | Air (RF spectrum) | Typ. < 100 m | CCMP-128, AES-256 |
| Bluetooth (BT) | < 2 Mbit/s | Air (RF spectrum) | Typ.< 10 m | AES-128 |
| Cable Ethernet | < 10 Gbit/s | Cable (TP or fiber) | Typ. < 10 km | Higher layer protocols |
| Optical Wireless (LiFi/VLC) | < 1 Gbit/s | Air (visible and IR spectrum) | Typ. < 10 m | Higher layer protocols |
| Broadband Power Line Communication (PLC) | < 1 Gbit/s | Cable (power line) | Typ. < 100 m | AES-128 |
| Cellular 2 G (EDGE) | < 1 Mbit/s | Air (RF spectrum) | Typ. < few km | GEA2, GEA3, GEA4 |
| Cellular 3G (UMTS HSPA+) | < 42 Mbit/s | Air (RF spectrum) | Typ. < 1 km | UEA1, UEA2 |
| Cellular 4G (SAE/LTE) | < 300 Mbit/s | Air (RF spectrum) | Typ. < 500 m | 128-EEA1, 128-EEA2, 128-EEA3 |
| Cellular 5G | < few Gbit/s | Air (RF spectrum) | Typ. < 300 m | 128-NEA1, 128-NEA2, 128-NEA3 |

### Requirements on Communication Technologies set by AAL Applications

Various AAL applications set different requirements on communication and network technologies. Table 9 summarises the requirements on data rate (DR) and data volume (DV), end-to-end delay (E2E-D) as well as packet loss (PL) and bit error rate (BER) that are set by eHealth applications. As can be seen from this table, applications requiring high data rates are video and audio observation, multimedia conference and the transfer of medical data via the use of video, audio and images (x-ray and mammography). Some real-time observation and monitoring applications such as emergency multimedia conference, tele-robotics and vital data monitoring (class 0) require low end-to-end delays below 300ms.

Table 9. Requirements on data rate/data volume, end-to-end delay and data losses set on the communication and network infrastructure by some examples of eHealth applications [V.110]-[V.112]

| Intended application | Data Rate (DR) and Data Volume (DV) | End-to-End Delay (E2E-D) | Packet Loss (PL) and Bit Error Rate (BER) |
|---|---|---|---|
| Medical data transfer | Video DR: 32 – 384 kbit/s Audio DR: 4 – 25 kbit/s | Video: 150 - 400 ms<br>Audio: 150 – 400 ms | Video: PL < 1%<br>Audio: PL < 3% |

| Intended application | Data Rate (DR) and Data Volume (DV) | End-to-End Delay (E2E-D) | Packet Loss (PL) and Bit Error Rate (BER) |
|---|---|---|---|
| Video and audio observation / Multimedia conference | DR: 384 – 1544 kbit/s | 100 – 400 ms | Video: PL < 1%<br>Audio: PL < 3% |
| Emergency multimedia conference | DR: 384 – 1554 kbit/s | 0 – 150 ms | Video: PL < 1%<br>Audio: PL < 3% |
| Tele-Robotics | DR: < 20 kbit/s (sensor)<br>DR: 0.3 -2 Mbit/s (video) | < 330 ms | Video and Audio<br>BER < 10-3 |
| Blood pressure monitoring | DR: < 10 kbit/s | n/a | n/a |
| Digital audio stethoscope | DR: ~ 120 kbit/s | n/a | n/a |
| Vital data monitoring (ECG, EEG, etc.) | DR: 1 – 20 kbit/s | Class 0: 12 – 300 ms<br>Class 1-3: ~ 1000 ms | PL = 0%<br>BER < 10-3 |
| Ultrasound sonography | DV: 256 kbyte | < 5s | PL = 0% |
| Transmission of x-ray images | DV: 18 Mbyte | < 1 min. | PL = 0% |
| Mammography | DV: 24 Mbyte | < 1 min. | PL = 0% |

Figure 34 shows that both required data rates and complexity of communication are much higher for image, audio and video data transmission than for other eHealth applications. The relation between the communication technologies presented in Table I and the applications can be seen on the right-hand side of the figure. Accordingly, WPAN and 2G cellular networks can be used for eHealth applications which do not require high data rates such as vital data monitoring (e.g., EEG, ECG, …) or blood pressure monitoring, but not for multimedia conferences and transmission of high-volume data. On the other hand, WLAN, Ethernet over cable, VLC/LiFi, PLC and LTE can support both video- and audio-based ALL and other eHealth applications. Bluetooth is able to support high data rate transmission over short distances or low data rate applications over 100 m.

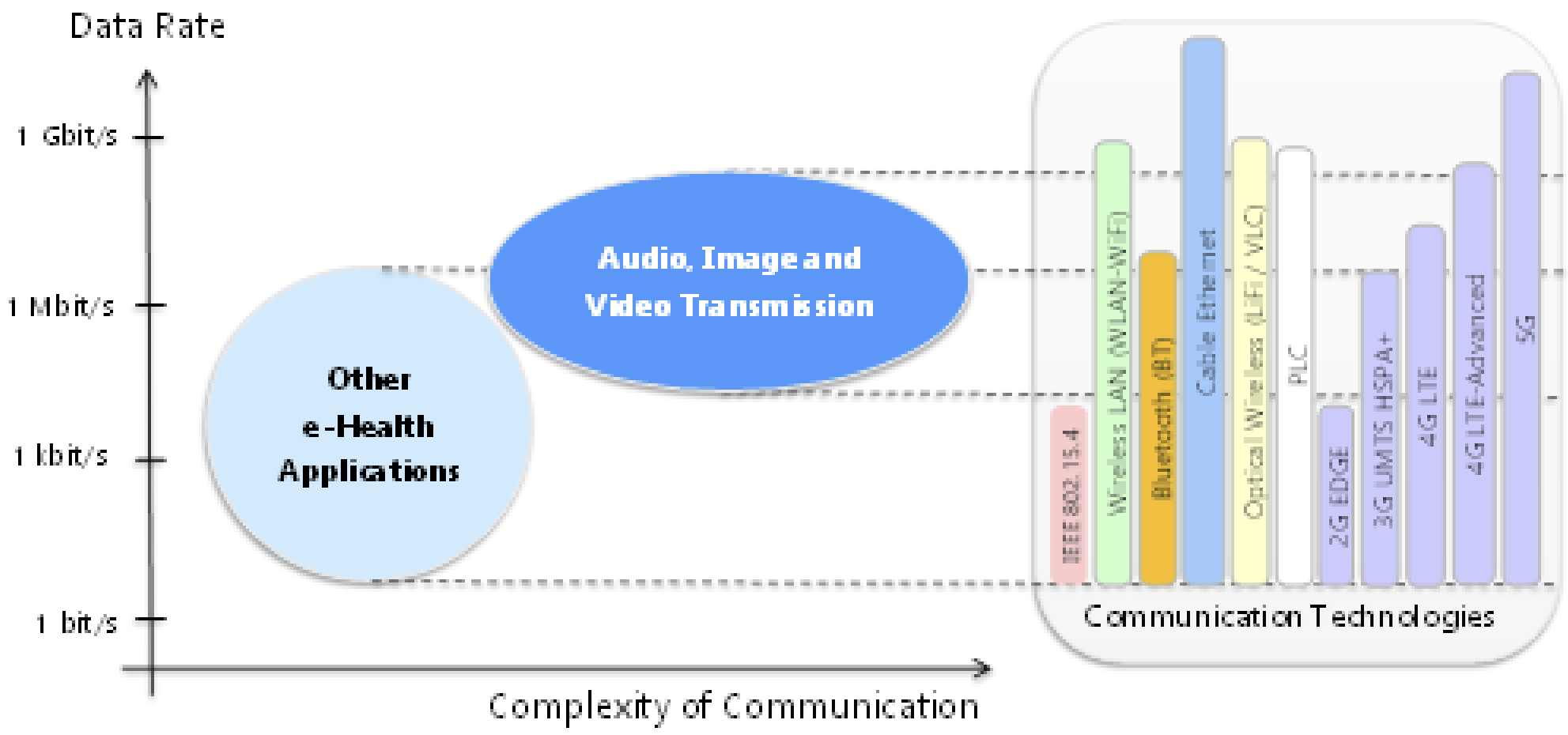

Figure 34. Communication technologies for e-Health and audio- and video-based AAL applications.

### 5.5.2 Camera Network Architecture

Many video-based AAL applications use a kind of camera network [V.106]. Depending on how and where visual processing is implemented as well as what type of camera or video sensors is used, there can be various camera network implementation options. These camera network implementation options can set different requirements on local and aggregation/core networks because different data rates and various latency security and confidentiality levels may be required. Figure 35, Figure 36, and Figure 37 show three possible architectures of camera networks. The first architecture (see Figure 35) envisages the use of simple cameras or any type of visual sensors. The data is neither processed in the device nor at the side, but simply multiplexed and transmitted to a remote care and service centre where both the visual processing and alerting/decision take place. Additionally, the video and audio data may be stored on a server, which can be either located at the premises of a third party (i.e., a cloud service provider), or located at the care and service provider side. Figure 36 shows the second architectural option, which also uses simple cameras or other visual sensors, but visual processing and alerting/decision functions are implemented within the local area network at the user side. The third option makes use of distributed processing by utilizing smart cameras or smart visual sensors comprising both visual processing and alerting/decision functions. The three presented options provide different levels of security and privacy and set different requirements on communication technology and network infrastructure. For example, the approach depicted in Figure 37 that provides local visual processing in combination with distributed smart camera networks has the potential of providing the highest confidentiality level while reducing the data traffic in both local and aggregation/core networks.

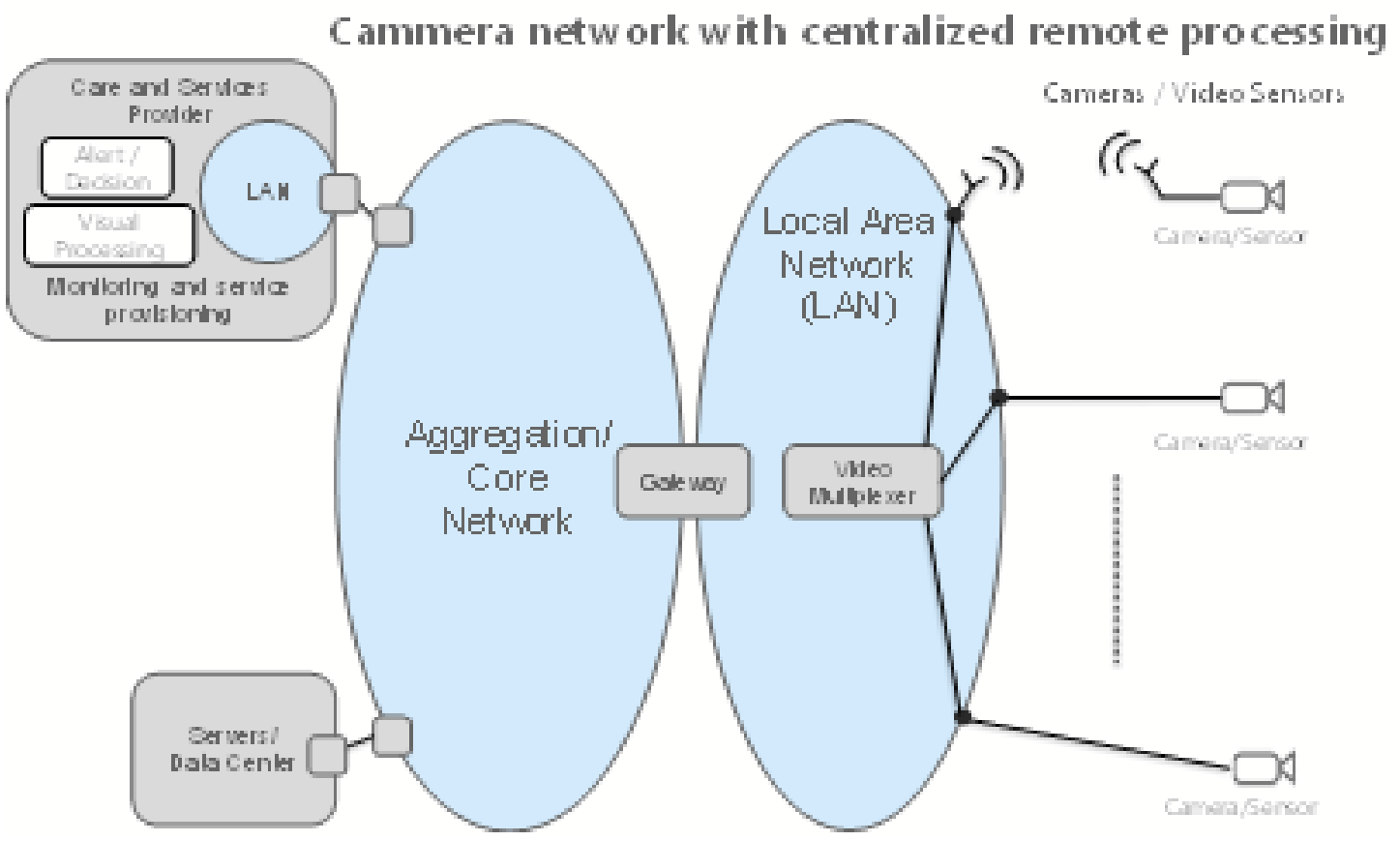

Figure 35. Camera network with centralized remote processing

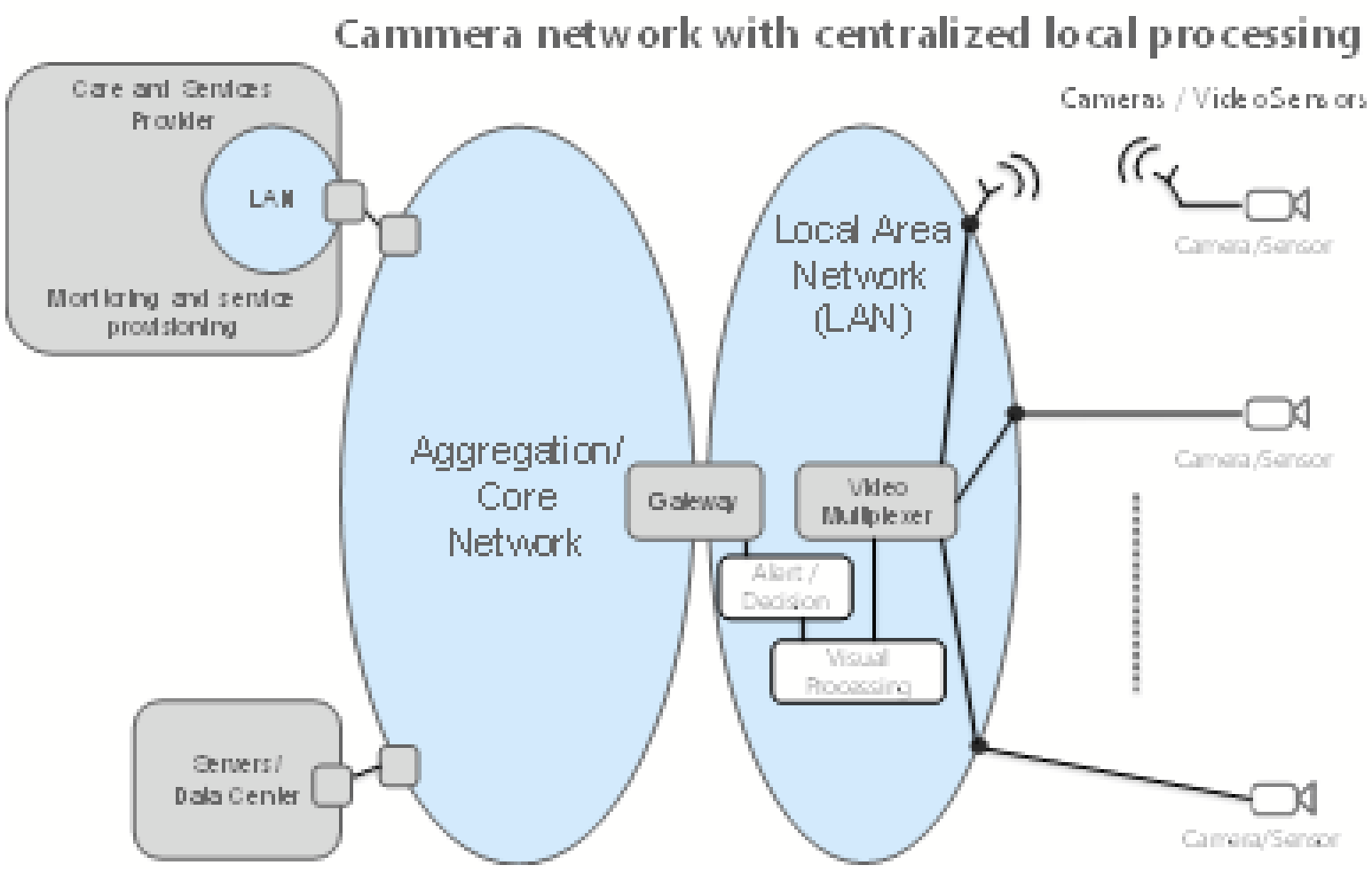

Figure 36. Camera network with centralized local processing

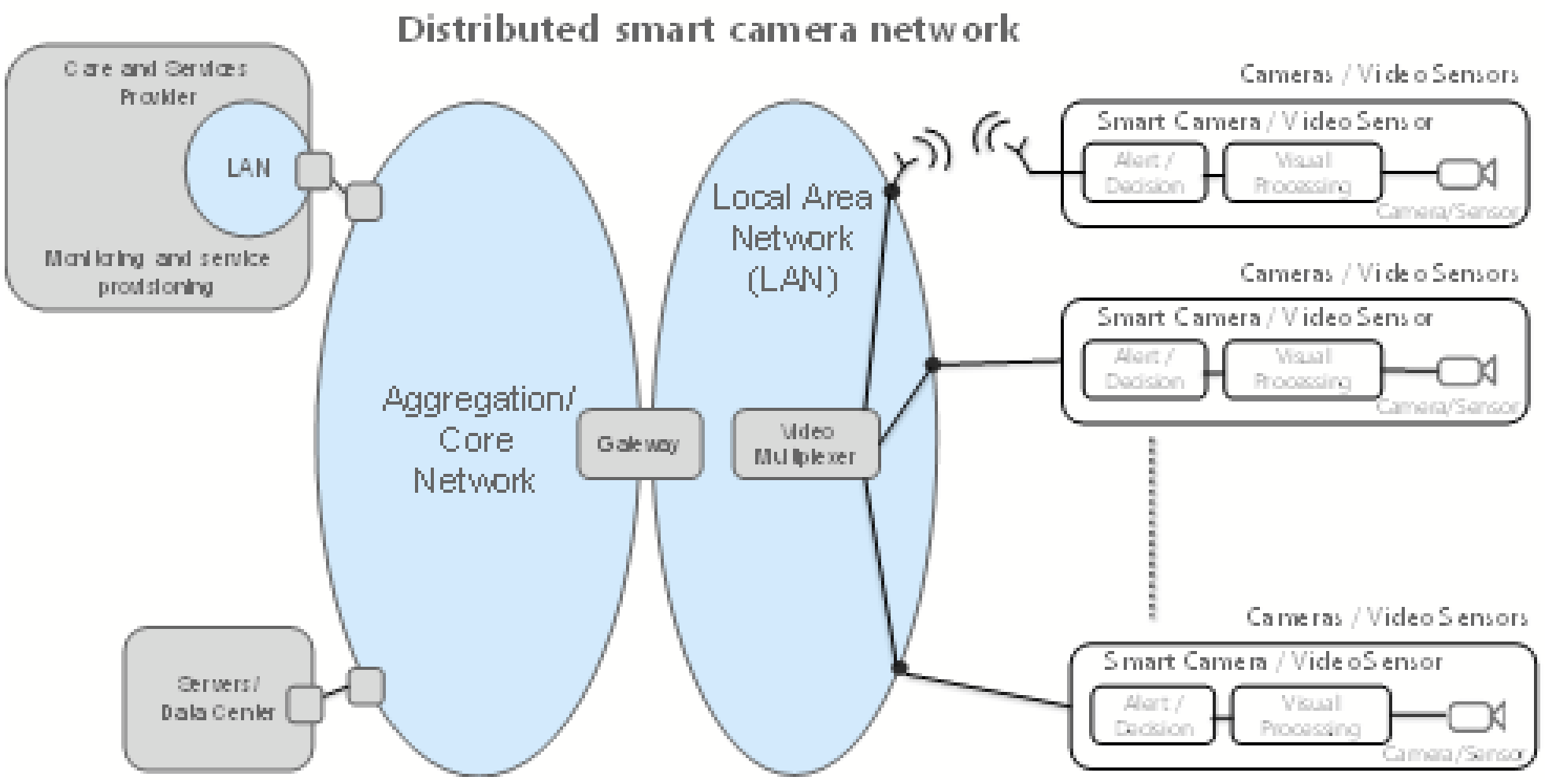

Figure 37. Distributed smart camera network

## 5.6 Health economics

The European Commission (EC) proclaimed at the turn of the twenty-first century that it is necessary to develop "constructive responses" to the challenge of an ageing population [V.113], [V.114]. In 2007, the term "Silver Economy" was used to refer to one of these reactions. The European Union (EU) Council urged the EC and EU Member States to expand opportunities for older adults to participate actively in society, with a special focus on volunteering, and to develop new markets within the silver economy, as described, responding to the growing demand for certain products and services among older people [V.115]. Additionally, the EU Council Resolution proposed the adoption of horizontal policies, which would include initiatives that cross-sectoral policy boundaries (such as health, labour, education, economic, and innovation policies), which aligns with the United Nations-recommended principle of "mainstreaming ageing," that also encourages the inclusion of topics related to old age and ageing into all public policies [V.116].

The uptake in real life of AAL solutions strongly relies on the opportunities offered by the Silver Economy. This motivates the following sections, which explain in more detail the prospects open.

### 5.6.1 Basic Definitions and Fields of the Silver Economy

#### Understanding of the Silver Economy

There is currently no universally accepted description of the silver economy [V.114], [V.117]. One may argue that the initiatives of the EC and the Council of the EU facilitate the societal creation of several interpretations of these phenomena, as is the case with the concept of "active ageing." The Organisation for Economic Co-operation and Development (OECD) [V.118], the Group of Twenty (G20) [V.119], the World Economic Forum (WEF) [V.120], and the United Nations Economic Commission for Europe (UNECE) [V.121] all presented positions and policy recommendations on the silver economy in the subsequent years. Thus, it may be argued that the term "silver economy" is intended to be a slogan, catchy phrase, calling word, or buzzword—a term that "creates excitement" by promoting a favourable picture of old age and an interest in the economic advantages associated with population ageing. Both older adults and older adults and public policy institutions who perceive the prospect of integrating measures to address the problem of an ageing population are urged to develop the silver economy's many definitions.

Nonetheless, the silver economy may be roughly characterised as the provision of products and services to older individuals and all people who are ageing and the extension of working lives and the promotion of volunteerism and active citizenship among older persons [V.115]. It may also be interpreted narrowly, as advocated in the EC's study "Europe's Demographic Future" [V.122], as a combination of favourable circumstances for providing products and services and the increased spending power of older customers. The EC [V.117] recently proposed a similar interpretation in its first comprehensive report on the state of the EU's silver economy, defining it as the sum of all economic activities that benefit people aged 50 and over (50+), including products and services purchased directly by these individuals and additional activities that generate expenses. Thus, the importance of considering direct effects (e.g., job creation and income tax), as well as indirect effects (e.g., region promotion, localisation factor, local activation, and revitalisation) as well as induced effects (e.g., further increase in expenditure due to an increase in regional revenues), has been underlined. According to the EC's strategy, the silver economy should be built on education development, commercialisation of scientific research, and robust and flexible markets. The EC emphasises in both definitions that the term "silver economy" does not refer to a single "silver market," but to a diverse range of products and services from a variety of already established sectors, including ICTs, finance, housing, transportation, energy, tourism, culture, infrastructure, local services, and long-term care.

The silver economy is increasingly being referred to alternately as the "longevity economy." This strategy has gained popularity in recent years in North and South America [V.123]-[V.126]. Nevertheless, this concept is not synonymous with the silver economy because it places a greater emphasis on the economic consequences of increased life expectancy and the process of "double ageing," which involves a faster increase in the percentage of the population aged 80 and over (80+; that is, the share of the "old-old" (75–89) and "oldest-old" (90+) age group) than in the overall ageing population.

Additionally, the term "silver economy" is sometimes incorrectly used synonymously with the more specific term "silver market" ("ageing market," "ageing marketplace," or "mature market") [V.127]. The phrase "silver market" first originated in Japan in the early 1970s in response to the progressive expansion in the provision of amenities for senior citizens [V.128]. The term "silver" appeared to allude to grey hair and emphasise respect for older persons prevalent in the Asia-Pacific region's nations. In a nutshell, the silver market is described as a market sector that includes goods and services aimed at rich individuals aged 50+ and unique solutions for trade between economic organisations (business-to-business; B2B) that allow them to adjust to an ageing workforce [V.129]. However, experts who study this issue emphasise that the silver market is not primarily dependent on marketing items aimed at older adults; hence, emphasising the users' and customers' ages should be avoided. Otherwise, the silver economy and the silver market may serve to exacerbate age discrimination (ageism) [V.129], [V.130]. At the moment, the silver market includes developments in the categories of "universal design," "transgenerational design," and "intergenerational design," as well as "barrier-free goods and services," that is, goods and services that are accessible to everyone, including individuals with impairments. Additionally, these terms represent economic entities' attempts to tailor the qualities of their products and services to the requirements of individuals of all ages (age-friendly) and with varying physical and cognitive capacities. Implementing these concepts seeks to engage and empower users and facilitate their social engagement and integration with other individuals.

In a broad sense, the silver economy may also be characterised as an economic system that prioritises balancing the production, distribution, consumption, and exchange of products and services essential for older individuals, as well as younger but already ageing generations [V.131]. The system's critical components are gerontechnologies, which are embedded in age-friendly goods and services. Additionally, older employees, senior entrepreneurs (dubbed "olderpreneurs" or "silver producers"), senior consumers, and senior investors are significant participants in this system [V.121]. It is necessary to mention at this point those older workers are often described as those aged 50+ owing to the reduction in physical capacities that starts around the age of 30 and peaks around the age of 50, as seen by employees' deteriorating capacity to do physically demanding duties [V.132]. Senior entrepreneurs are those who establish their own firm beyond the age of 50. It has been discovered that their businesses are less likely to fail during their early years on the market than those founded by younger persons [V.133]. Senior consumers are people whose primary expenditures are on health care, housing adaptations for senior living, recreation, and tourism. Pension money, real estate (e.g., retirement homes and communities), and other financial assets are often the focus of older investors.

### Sector Boundaries and Fields of the Silver Economy

Several typologies of silver market aspects have arisen in recent years, highlighting characteristics that should be fostered at the regional and local levels. Segmentation by P. Enste, G. Naegele, and V. Leve [V.134] is often discussed in the literature on the subject. This segmentation includes the following 14 areas: (1) ICTs in inpatient and outpatient care; (2) smart living, home modifications, and assisted living services, all of which are increasingly reliant on ICTs; (3) promotion of independent living, which is also more reliant on ICTs; (4) the health economy as it relates to older individuals, which includes medical technology and e-health, hearing and vision aids, dental prosthetics, and orthopaedics; (5) products and services related to education and culture, which might be consumed owing to increased free time in retirement and greater levels of education among older persons; (6) ICTs and media in conjunction with health, independence promotion, and security; (7) service and social robotics (or socially assistive robots; SAR), in conjunction with the advancement of independent living; (8) mobility and its promotion, for example, automobile traffic safety; (9) leisure, travel, culture, communication, and entertainment; (10) fitness and wellbeing, with increased health awareness, notably among the "young-old" or "third age" categories; (11) apparel and fashion; (12) daily-life services and other household services; (13) insurance coverage for age-specific "risks;" and (14) financial services relating to capital preservation, wealth management, and saving counselling. However, it should be highlighted that this typology is not universal since it is based on the experiences of industrialised nations and is focused on promoting high-tech firms.

Additionally, there are two more contemporary typologies. The first was provided by the team behind the "Mobilising the Potential of Active Ageing in Europe" (MoPAct) initiative in reference to the EU Member States [V.135]. Based on the following ICTs, scholars identified eight types of innovative goods and services as possible growth-stimulating areas: (1) telemedicine; (2) collaborative networks or software; (3) broadband access; (4) Internet; (5) intelligent homes; (6) assistive communication technology; (7) universal design; and (8) social services and media. These technologies, for example, may be utilised to develop telecare and e-health solutions and promote independent living and AAL. The EC's [V.117] second typology of silver economy fields summarises the silver economy's implementation in the EU and emphasises the importance of supporting technologies in ten areas: (1) "connected" (digital) health; (2) robotics and games; (3) silver tourism; (4) integrating care services and improving communication; (5) age-friendly construction; (6) an active and healthy lifestyle; (7) age-friendly universities; (8) self-driving automobiles; (9) older adults' entrepreneurialism; and (10) interactive platforms for quick product and service creation.

Segments of the silver market may be developed and operate differently in different nations. Thus, it is necessary to conduct research that will enable us to compare and contrast alternative institutionalising models of the silver economy. Models of this kind may be characterised in terms of continents, nations, and regions [V.131]. This grouping may include, but is not limited to, a higher degree of social and cultural variety, differentiation of the social category of older people, and various interactions and networks among public and private institutions, non-governmental organisations (NGOs), and informal communities [V.134]. For comparison, until 2014, the China International Silver Industry Exhibition [24] distinguished 11 segments of the silver market: (1) rehabilitation and nursing products; (2) auxiliaries; (3) food, pharmaceuticals, and dietary supplements; (4) household products; (5) real estate support; (6) cultural and educational services; (7) tourism and leisure services; (8) insurance services; (9) financial services; (10) consultation; and (11) vocational training in the silver industry. The majority of the segments in this typology align with previously described categorisations but place little focus on innovation. However, there are strong ties to low-tech businesses, such as food and nutritional supplement manufacturing, domestic items, and broadly defined counselling and vocational training in silver industries.

Another way to view the notion of the silver economy is via the categorisation of its sectors suggested by American experts. Moody and Sasser [V.127] define the term "ageing marketplace" and divide it into four "silver sectors/industries": (1) financial services; (2) healthcare; (3) tourism and lodging; and (4) retirement housing. Additionally, they say that this market involves the operation of so-called "age brands," which refer to goods and services marketed at older customers that are separated from other market offers by marketing and design professionals rather than by genuine and unique qualities. Scholars suggest that silver sectors are characterised by "new ageing enterprises," i.e., businesses and NGOs that target older persons to boost their feeling of subjectivity and activity [V.136]. The word "new" is used to separate the good characteristics of the stated entities from the "old," unfavourable management styles associated with ageing. These "old ageing businesses" include nursing homes, senior clubs, and senior centres, all of which face criticism from the political economics of ageing and critical gerontology viewpoints. According to these theoretical perspectives, older people are treated medically and objectified as commodities by society. Thus, Public policy can intentionally (or unintentionally) reinforce this approach and the actors who implement it through "age segregation" policies, such as supporting the delivery of services separated for older and younger people. According to this view, "old ageing businesses" are the outcome of overt or tacit differences, discrimination, stigmatisation, and isolation of older adults from other age groups.

### 5.6.2 Supporting the Development of the Silver Economy

According to Enste, Naegele, and Leve [V.134], governments and their areas should adopt at least nine sorts of activities to advance the silver economy, most notably through promoting older people's social inclusion and independence, as well as their safety and quality of life. These efforts should include the following: (1) expanding the range of products and services available to older adults; (2) sensitising and coordinating the actions of various public agencies; (3) expanding marketing to older adults; (4) meeting the needs of the poorest older adults; (5) empowering and strengthening the representation of older adults' interests; (6) producing products and services in response to customer feedback; (7) enhancing and developing current goods and services; (8) advancing user- and age-friendly design, as well as universal design; and (9) supporting consumer rights among older individuals. Additionally, we can add additional intervention directions: (10) focusing on mainstreaming ageing, i.e., integrating ageing into various development strategies at the central and regional levels; (11) enhancing governance by increasing access to information on various aspects of the silver economy and widely disseminating such knowledge; (12) assisting in the coordination of activities of public, private, and non-profit entities; (13) fostering technology transfer and conversation in the sphere of product and service design; and (14) examining the capacity of the aforementioned entities to achieve mutual advantages via the development of new gerontechnologies and social innovations.

Additionally, based on the assessments of E. Cox, G. Henderson, and R. Baker [V.133], some viable actions for implementing the silver economy concept at the local level, particularly in cities, may be identified. These include the following: (1) recognising age diversity in local policies and regulations; (2) viewing demographic change as an economic opportunity that can benefit the entire society; (3) fostering collaboration among various entities and joint financing of projects; (4) involving older people in strategic planning and public life; (5) emulating best practises; (6) enhancing analysis of the local demographic situation; (7) promoting continuous training and lifelong learning; (8) encouraging collaboration between local governments and entrepreneurs in programming development to capitalise on opportunities for economic growth that are geared toward the needs of an ageing population; (9) assisting older entrepreneurs; (10) partnering with local institutions to discover areas for ageing research innovation and commercialisation; (11) developing programmes for older employees and active labour market strategies for those facing unemployment; and (12) adopting age-friendly infrastructure in public spaces. This selection is by no means exhaustive of the options that local governments might collaborate with their partners.

## References

### Section 5.1

## Section 5.2

## Section 5.3

## Section 5.4

## Conclusions

The global population is ageing. Given that age is the primary risk factor for many diseases and coupled with the concomitant advances in healthcare technologies and modern medicine permitting individuals to live longer with chronic conditions, the introduction of AAL is inescapable. Whilst AAL technologies offer hope in addressing economic challenges to healthcare and the prospect of prolonging independent living thereby improving life quality, the underlying ethical concerns and legal issues cannot be overlooked.

It could be argued that as all technologies need to pass certain regulatory processes, the legal aspects pertinent to AAL have been addressed. The relevant regulations such as the GDPR cover collection, use, processing and sharing of personal data. Consumer protection laws deal with the safety of components used in these technologies. Yet as technologies advance, the laws must stay abreast of such progress. A framework for which further research can be performed is presented. It concludes that AAL technologies require a legal system that both promotes their development while at the same time safeguards against risks posed by the technology. Here, it is clear that the law is failing to provide a speed of adoption commensurate with the development of the technology. Not only are there serious uncertainties in the application of existing legal frameworks to AAL technologies but there is also a lack of appropriate legal restrictions and precautions to control some of the risks posed by lifelogging technologies. It is recommended that a more holistic approach to the regulation of AAL technologies is taken, one that integrates deeper technological and international perspectives than the current legal framework represents.

The more nebulous aspects of AAL application lie with the ethics of their implementation. AAL technologies are by their nature intrusive. They can also be deemed to be overprotective which can subtly erode respect for autonomy. The benefits brought by AAL need to be carefully weighed against the risks and the risk/benefit ratio assessed on a regular basis as individuals age and applications of technology change. An ethical approach and a thorough understanding of all ethical principles relevant to surveillance/monitoring architectures are essential. AAL poses many ethical challenges, raising questions which will affect immediate acceptance and long-term usage. Furthermore, ethical issues emerge from social inequalities and their potential exacerbation by AAL, accentuating the existing access gap between high-income countries (HIC) and low and middle-income countries (LMIC). Ethics should be incorporated at the AAL design stage taking all of these aspects into account and evaluating (i) beneficence, (ii) non-maleficence i.e., a risk/benefit analysis (iii) respect for autonomy, and (iv) protection of confidential information and data that may reveal personal and sensitive attributes

The social issues focus on the impact of AAL technologies before and after their adoption. Some are rooted in the collective understandings of the technology at hand, whereby users can relate audio-video based AAL to activities such as surveillance practices. One of the prominent social and design challenges will be facilitation of the workflow and avoiding the sense of additional technological burden. Taking care of that will directly impact institutional and individual adoption of AAL. Future AAL technologies need to consider all aspects of equality such as gender, race, age and social disadvantages and avoid increasing loneliness and isolation among, e.g. older and frail people. Finally, the current power asymmetries between the target and general populations should not be underestimated nor should the discrepant needs and motivations of the target group and those developing and deploying AAL systems. These differences could lead to governance challenges, serious ethical questions, and potential misuse of the technology.

Whilst AAL technologies provide promising solutions for the health and social care challenges, they are not exempt from ethical, legal and social issues (ELSI). A set of ELSI guidelines is needed to integrate these factors at the research and development stage.